\documentclass{aa}  

\usepackage{lscape}
\usepackage{rotating}
\usepackage{graphicx}
\usepackage[final]{hyperref}
\usepackage{rotating}
\usepackage{float}
\usepackage{txfonts}
\usepackage{float}
\usepackage{siunitx}
\usepackage{tabularx}
\usepackage{color,hyperref}
\usepackage{geometry,fancyhdr}
\usepackage{multicol,multirow}
\usepackage{arydshln}
\usepackage[flushleft]{threeparttable}
\hypersetup{colorlinks,breaklinks,
            linkcolor=blue,urlcolor=blue,
            anchorcolor=blue,citecolor=blue}
\usepackage{caption}
\usepackage{longtable}
\usepackage{comment}
\bibpunct{(}{)}{;}{a}{}{,} 
\usepackage{morefloats}
\usepackage{graphicx}
\usepackage{subfigure}
\usepackage{txfonts}
\begin{document}

   \title{Molecular inventory of a young eruptive star's environment}
   \subtitle{Case study of the classical FU~Orionis star V1057~Cyg}
   \author{Zs. M. Szabó
          \inst{1,2,3,\thanks{Member of the International Max Planck Research School (IMPRS) for Astronomy and Astrophysics at the Universities of Bonn and Cologne.}},
          A. Belloche\inst{1},
          K. M. Menten\inst{1,\thanks{This work is dedicated to the memory of Karl M. Menten, who proposed the idea of the line survey. His profound knowledge, thoughtful mentorship, unwavering guidance, and fruitful discussions 
          inspired us throughout this journey.
          This work is a tribute to his vision and lasting impact.
          }},
          Y. Gong\inst{4,1},
          \'{A}. K\'{o}sp\'{a}l\inst{3,5,7},
          P. \'{A}brah\'{a}m\inst{3,5},
          W. Yang\inst{6,1},
          C. J. Cyganowski\inst{2},
          and
          F. Wyrowski\inst{1}
}
    \institute{Max-Planck-Institut für Radioastronomie, Auf dem Hügel 69, 53121 Bonn, Germany\\
              \email{zszabo@mpifr-bonn.mpg.de}
    \and
    Scottish Universities Physics Alliance (SUPA), School of Physics and Astronomy, University of St Andrews, North Haugh, St Andrews, KY16 9SS, UK
    \and
    Konkoly Observatory, HUN-REN Research Centre for Astronomy and Earth Sciences, MTA Centre of Excellence, Konkoly-Thege Mikl\'os \'ut 15-17, 1121 Budapest, Hungary
    \and
    Purple Mountain Observatory, and Key Laboratory of Radio Astronomy, Chinese Academy of Sciences, 10 Yuanhua Road, Nanjing 210023, China
    \and
    Institute of Physics and Astronomy, ELTE E\"otv\"os Lor\'and University, P\'azm\'any P\'eter s\'et\'any 1/A, 1117 Budapest, Hungary
    \and
    School of Astronomy \& Space Science, Nanjing University, 163 Xianlin Avenue, Nanjing 210023, People's Republic of China
    \and
    Max-Planck-Institut für Astronomie, K\"onigstuhl 17, D-69117 Heidelberg, Germany}

   \date{Received ; accepted}

 
  \abstract
   {Studying accretion-driven episodic outbursts in young stellar objects (YSOs) is crucial for understanding the later stages of star and planet formation. FU~Orionis type objects (briefly FUors) represent a small but rather pivotal class of YSOs, whose outbursts are characterised by a rapid, multi-magnitude increase in brightness at optical and near-infrared wavelengths. These outbursts may have a long-lasting influence on the chemistry and molecular inventory around eruptive young stars. However, no complete line survey in the millimeter wavelength range exists in the literature for more evolved (i.e.,~Class\,\begin{tiny}II\end{tiny}) sources, in contrast to wideband coverages at optical and  near-infrared wavelengths}
   {We carried out the first dedicated wide-band millimeter line survey towards the low-mass young eruptive star and classical FUor V1057~Cyg, which has the highest observed peak accretion rate among FUors. This source is known to have a molecular outflow, and is associated with dense material, making it a good candidate to search for molecular species. 
   }
   {We performed a wideband spectral line survey of V1057~Cyg with the IRAM~30-m telescope from $\sim$72 to $\sim$263\,GHz (with spatial resolution between $\sim$36$\arcsec$ and $\sim$10$\arcsec$), complemented by on-the-fly maps of selected molecules. We also recorded additional spectra around 219, 227, 291, and 344\,GHz (with spatial resolution between $\sim$30$\arcsec$ and $\sim$19$\arcsec$) with the APEX~12-m telescope. We conducted simple radiative transfer and population diagram analyses to derive column densities and excitation temperatures. We constructed integrated intensity maps of the emission from several molecular species, including those that reveal outflows. These maps a $^{12}$CO (3--2) position-velocity diagram provide insight into the past outburst activity of the source.
   }
   {We identified mainly simple $C$-, $N$-, $O$-, $S$-bearing molecules, deuterated species, molecular ions, and complex organic molecules. Several molecular species (HCN, HC$_3$N, and HNC) trace large-scale ($\sim$2$\arcmin$) structures in the environment of V1057~Cyg with hints of small-scale fragmentation that remains unresolved by the single-dish data. The position-velocity diagram of $^{12}$CO shows concentrated knots, which may indicate past episodic outburst activity. We calculated the dynamical timescale of the outflow and found it to be on the order of a few tens of thousands of years (between 15\,000 and 22\,000\,years), similar to other eruptive stars, suggesting that the outflow cannot result from the ongoing outburst alone, since the source has been in the current outburst for less than a century.
   The population diagrams for species such as CH$_3$OH, H$_2$CO, HC$_3$N, indicate rotational temperatures ranging from 8\,K to 15\,K, and column densities ranging from 1.4$\times$10$^{12}$\,cm$^{-2}$ to 2.8$\times$10$^{13}$\,cm$^{-2}$.}
   {With over 30 molecular species (including isotopologues) detected, V1057~Cyg and its environment display a rich chemistry, considering the more evolved state of this source compared to well studied, but younger (i.e.,~Class\,\begin{tiny}0/I\end{tiny}) FUors like V883~Ori. 
   The results of our line survey show that V1057~Cyg is a good candidate for future interferometric observations aimed at resolving emission extents to constrain molecular freeze-out and searching for emission lines of water and additional complex organic molecules.  
   Our observations highlight the potential of mm line surveys to characterise the chemistry of eruptive stars and their environments, including more evolved sources, and so to complement optical and near-infrared studies in improving current statistics regarding the molecular inventories of these objects.
   }

   \titlerunning{Case study of the classical FU~Orionis star V1057~Cyg}
   \authorrunning{Szab\'o et al.}
   \maketitle
%
\section{Introduction} 
\label{sec:intro}

Young stellar objects (YSOs) undergo accretion-driven episodic outbursts. Studying the impact of these outbursts, in particular on the molecular composition of their environments, is fundamental for our understanding of star and planet formation. 
FU~Orionis objects (briefly FUors) represent a small but rather pivotal class of YSOs, whose outbursts are primarily characterised by a rapid, multi-magnitude increase in brightness at optical and near-infrared wavelengths, attributed to a sudden increase in the accretion flow from the disk onto the protostar, a process lasting for several decades, likely centuries \citep[for a recent review, see][and references therein]{fischer2023}.
The prototype of the class, FU~Orionis, went into outburst around 1936/1937 \citep{wachmann1954}. It was later considered as a new phenomenon different from a nova outburst \citep[][]{herbig1966}, and eventually the FUor class was defined by \citet{herbig1977}.
During an outburst, the accretion rate of a FUor rises from the average rate of a typical T~Tauri star of $10^{-9}$ -- $10^{-7}$ \(M_\odot\) yr$^{-1}$ up to $10^{-5}$ -- $10^{-4}$ \(M_\odot\) yr$^{-1}$ \citep[e.g.,][]{kenyon&hartmann1996,audard2014}.

The powerful and energetic FUor outbursts potentially have a long-lasting influence on the chemistry and molecular inventory of the disks around eruptive stars. The elevated temperatures can, temporarily, increase the abundances of complex organic molecules (COMs), which are carbon-bearing molecules with six or more atoms \citep[e.g.,][]{herbst-vanD2009,ceccarelli2017,van-t-hoff2018,lee2019-v883ori,jorgensen2020,ceccarelli2022-review}.
Improvements in observational techniques and the sensitivity of instruments allow the chemistry of FUors and FUor-like objects to be studied in a new way, covering different frequency regimes \citep[e.g., L1551~IRS5,][]{fridlund2002,bianchi2020,mercimek2022,andreu2023,marchand2024} and providing small scale views of the inner regions around these objects.
Among the most noteworthy results from interferometric observations is the direct evidence of the outburst affecting the location of the water snow-line, shifting it from a typical few au to $\sim$40--120\,au observed in the FUor V883~Ori \citep[see][and references therein]{cieza2016,van-t-hoff2018,leemker2021}. 
Observational studies of young eruptive stars in the (sub)millimeter regime targeting both molecular line emission and dust continuum radiation provide more in-depth views of the small-scale structures and kinematics of the disk, and the envelope surrounding these objects: transitions of $^{12}$CO and its rarer isotopic species $^{13}$CO and C$^{18}$O can be used to detect low- and high-density circumstellar material \citep[e.g.,][]{kospal2011b,kospal2016,kospal2017a,feher2017,cruz-saenz-de-miera2023}. HCO$^+$ and HCN emission can trace cavity walls, while SiO and SO trace the shocked material in outflows \citep{Hogerheijde1999}, and finally, the line wings of $^{12}$CO emission trace the outflow itself \citep{evans1994}. 

Despite the growing number of observational studies in the millimeter regime including FUors and FUor-like objects \citep[e.g.,][]{McMuldroch1995,fridlund2002,white2006-ch3oh-l1551,feher2017,lee2019-v883ori,bianchi2020,wendeborn2020,mercimek2022,ruiz-rodriguez2022,lee2024} and model predictions on molecules as outburst tracers in eruptive systems \citep[][]{visser2015,rab2017,molyarova2018,zwicky2024}, wide-band spectral line surveys at (sub)mm wavelengths of more evolved FUors do not yet exist.
The lack of studies could be (primarily) due to the distances of many of these sources, highlighted by \citet{audard2014} and revised recently by \citet{szabo2023a,szabo2023b}. Broad frequency coverage is needed to cover multiple transitions of a given species, which is crucial to determine the physical properties of the gas using population diagrams \citep[e.g.,][]{goldsmith1999}.

Historically FUors, especially classical prominent examples of these stars, have been well-studied at optical and near-infrared wavelengths. This resulted in comprehensive analyses that are utilised to identify and characterise new members of the class \citep[e.g.,][]{herbig1977,herbig2003,clarke2005,connelley2018,hillenbrand2018,szegedi-elek2020,szabo2021,nagy2022,szabo2022,nagy2023}. Millimeter wavelength 
observations of spectral lines and modeling have addressed their molecular content \citep[e.g.,][]{rab2017,molyarova2018,zwicky2024}.
Eruptive stars appear to be still close to their natal cores, whose chemistry can be investigated well through (sub)mm and radio wavelength observations \citep[e.g.,][and references therein]{szabo2023a,szabo2023b}, which indicate the presence of high concentrations of dense material in the vicinity of the surveyed protostars. 
A line survey covering wide frequency regimes, aiming to provide an in-depth view of the molecular inventory of the environment of an FUor can motivate interferometric studies, which may allow further study of the outburst effects on the chemical composition of the stars' close environment.

This paper focuses on V1057~Cyg, a classical FUor in the Class\,{\footnotesize II} evolutionary phase \citep[e.g.,][]{herbig1977,herbig2003}, the second FUor discovered. At a distance of
$897_{-20}^{+19}$\,pc \citep{bailerjones1,bailerjones2}, V1057~Cyg has the highest peak accretion rate among FUors \citep{szabo2021} measured to date. The source went into outburst in 1969--1970, brightening by $\sim$6\,magnitudes in the $V$-band \citep{welin1971a,welin1971b}. The only pre-outburst spectrum shows properties of a T~Tauri star \citep[TTS,][]{herbig1977,herbig2009}. The source displays variable and high-velocity wind features of P~Cygni profiles blueshifted by up to 100 -- 300\,km\,s$^{-1}$ and recently jet tracers were detected, usually rare in FUors, but more common in classical T~Tauri stars \citep[][]{magakian2013,takagi18,kospal2020,park2020,szabo2021}.
V1057~Cyg has exhibited multiple extraordinary flares in the 1720\,MHz OH maser transition, remaining the only eruptive star known to show this phenomenon \citep[][]{lo1973,lo1974a,winnberg1981}. 
We note that although the optical and near-infrared properties of V1057~Cyg suggest a more evolved evolutionary state \citep[e.g.,][]{herbig2003,szabo2021}, observations conducted in the millimeter regime have suggested a younger, Class~{\footnotesize I} state \citep[e.g.,~significant contribution of the envelope,][]{feher2017,calahan2024a}.

In this paper, we present the results of a wideband spectral line survey of V1057~Cyg complemented by on-the-fly maps of emission from selected molecules. The line survey 
was conducted with the IRAM~30-m telescope and has an almost continuous coverage from $\sim$72\,GHz ($\sim$4.1\,mm) to $\sim$263\,GHz ($\sim$1.1\,mm). Other selected frequencies, around 219\,GHz -- 227\,GHz ($\sim$1.3\,mm), 291\,GHz ($\sim$1.0\,mm), and 344\,GHz ($\sim$0.8\,mm), were observed with the APEX~12-m telescope. Together, this dataset provides an in-depth view of the molecular composition of this young eruptive star’s environment.
The paper is laid out as follows. In Sect.~\ref{sec:observations} we describe the observations as well as the data reduction. In Sect.~\ref{sec:results} we present the results and analysis  of the data, while in Sect.~\ref{sec:discussion} we compare our findings with results reported in the literature. Finally, in Sect.~\ref{sec:conc} we summarise the most important findings of this study.

\section{Observations and data reduction}
\label{sec:observations}
To acquire our main spectral line survey data of V1057~Cyg  we observed the star's position, 
($\alpha,\delta)_{{\rm J}2000}$ = 
$20^{\rm h}58^{\rm m}53\rlap{.}^{\rm s}73$, 
$+44^\circ15'28\rlap{.}''4$,
with the IRAM~30-m telescope, operated by the Instituto de Radioastronomía Milimétrica (IRAM) on Pico Veleta in the Spanish Sierra Nevada. The observations (project id: 060-23, PI: Menten) took place from 2021 August 4 to 6, with a total observing time of 35.5\,hours. We used the Eight Mixer Receiver\footnote{https://publicwiki.iram.es/EmirforAstronomers\#Overview} \citep[EMIR,][]{carter2012} in the E0, E1 and E2 bands, with the fast Fourier transform spectrometers (FFTSs), providing an instantaneous bandwidth of 16\,GHz and a spectral resolution of $\sim$200\,kHz. 
The survey was conducted in the position-switching mode with a reference position at the relative equatorial offset ($-$600$\arcsec$, 600$\arcsec$). 
In addition, mapping was conducted using on-the-fly (otf) mode, centred on the position of V1057~Cyg with a reference position at (300$\arcsec$, 300$\arcsec$).
NGC~7027 and K3-50A were used to obtain pointing and focus corrections regularly during each observing session.
The main parameters of the 30-m observations are summarised in Table~\ref{tab:obs-log-iram}. 
The main beam efficiencies were computed using the Ruze formula, with parameters taken from the IRAM webpage\footnote{https://publicwiki.iram.es/Iram30mEfficiencies}.
The data were reduced using the GILDAS/CLASS software developed by IRAM \citep[][]{pety2005-gildas}.
Platforming effects, materialising as intensity offsets in the spectra, were identified and corrected by subtracting first order baselines in each subband. In addition, we excluded scans, where careful inspection revealed remaining platforming, which generally stemmed from problems in the outer basebands, or any other issues (see also Fig.~\ref{fig:composit-iram}, \ref{fig:lines-example}, and \ref{fig:survey-73ghz-small} to \ref{fig:survey-242ghz-small}). 
\begin{table*}[htbp!]
\tiny
\caption{Information on the IRAM~30-m observations conducted with EMIR.}              
\label{tab:obs-log-iram}      
\centering                                      
\begin{tabular}{ccccccrrrrc}          
\hline\hline                 
\multirow{1}{*}{Date}         & \multirow{1}{*}{Mode} & \multirow{1}{*}{Band} & \multirow{1}{*}{Tuning} & \multirow{1}{*}{Frequency range} & \multirow{1}{*}{Beam}  & \multicolumn{4}{c}{Measured rms noise level$^{(*)}$} & Size of otf map \\
\cdashline{7-10}
\multirow{2}{*}{yyyy-mm-dd}  &   & & \multirow{2}{*}{(GHz)} & (LSB, USB)  & \multirow{2}{*}{($\arcsec$)} & LO & LI & UI & UO & \multirow{2}{*}{($\arcsec$)} \\ 
& & & & (GHz) & & (mK) & (mK) & (mK) & (mK) & \\
\hline \hline                                
2021-08-04  & onOff & E0       & $73$  &  $72-78.5$, $86.4-94.1$   & $34.4-26.7$    & $12.5$ & $8$ & $5.5$ & $11.2$ & $-$ \\      
2021-08-04  & onOff & E0       & $80$  &  $77.7-85.5$, $93.2-101.1$  & $32.2-22.8$    & $5.1$ & $3.9$ & $4.5$ & $4.7$ & $-$ \\
\multirow{2}{*}{2021-08-06}  & \multirow{2}{*}{otf} & \multirow{2}{*}{E0} & \multirow{2}{*}{$90$}  &  \multirow{2}{*}{$87.7-95.5$, $103.3-111.1$}  & \multirow{2}{*}{$28.2-22.6$}    & \multirow{2}{*}{$11.2$} & \multirow{2}{*}{$10.7$} & \multirow{2}{*}{$12.3$} & \multirow{2}{*}{$14.7$} & $300\arcsec$$\times$$300\arcsec$ \\
& & & & & & & & & & $315\arcsec$$\times$$315\arcsec$ \\
2021-08-04  & onOff & E0       & $94$  &  $92.7-99.8$, $107.7-115.5$  & $27.3-21.6$    & $4.5$ & $3.5$ & $4.4$ & $6.4$ & $-$ \\
\hline
2021-08-05  & onOff & E1       & $133$ &  $130.7-138.5$, $146.3-154.1$ & $19.1-16.31$    & $7.6$ & $5.8$ & $9.3$ & $12$ & $-$ \\
2021-08-05  & onOff & E1       & $141$ &  $138.7-146.5$, $154.4-162.1$ & $18.1-15.51$    & $7.5$ & $5.9$ & $10$ & $9.5$ & $-$  \\
\hline
2021-08-06  & onOff & E2       & $202$ &  $199.7-207.5$, $215.4-223.1$ & $12.5-11.2$    & $19.3$ & $36.1$ & $37.3$ & $39.2$ & $-$ \\
2021-08-06  & onOff & E2       & $210$ &  $207.7-215.5$, $223.4-231.1$ & $12.1-10.8$    & $46.5$ & $43.9$ & $43$ & $45.9$ & $-$ \\
2021-08-06  & onOff & E2       & $234^{(a)}$ &  $231.7-239.5$, $247.4-255.1$ & $10.8-9.8$ & $37.8$ & $41.6$ & $54$ & $74.8$ & $-$ \\
2021-08-06  & onOff & E2       & $242$ &  $239.7-247.5$, $255.4-263.1$ & $10.4-9.5$ & $52.3$ & $52.4$ & $64.8$ & $67.8$ & $-$ \\
\hline \hline
\end{tabular}
\tablefoot{
\tablefoottext{*}{The rms noise levels were measured separately for the lower outer (LO), lower inner (LI), upper inner (UI), and upper outer (UO) bands of the lower and upper side bands. }
\tablefoottext{a}{The 234\,GHz tuning observations are not presented in this paper due to low atmospheric transmission in the covered frequency range.}
}
\end{table*}

In addition to the IRAM 30-m observations we conducted observations with the Atacama Pathfinder EXperiment (APEX) 12-m submillimeter telescope at selected frequencies centered around 227, 291, and 344 GHz (project ids: M9530C\_107, M9515A\_108, M9523C\_109; PI: Menten). APEX is located on a 5100\,m altitude site in the Llano de Chajnantor, Chile \citep{gusten2006}.
The APEX data were obtained using several instruments, summarised here in chronological order of their use (see also Table~\ref{tab:obs-log-apex}). We used the dual-sideband, dual-polarization receiver nFLASH230\footnote{https://www.apex-telescope.org/ns/nflash/} with the Fast Fourier Transform Spectrometer \citep[FFTS,][]{klein2006-apex-FFTS} backend, providing a bandwidth of 32\,GHz and a spectral resolution of 61\,kHz, covering frequencies between $\sim$213 and $\sim$245\,GHz. The observations were carried out on 2021 July 20, 21, and 31 with a total observing time of $\sim$5.5\,hours. For the July 20 and 21 observations the wobbler switching mode was used with a single pointing toward the position of the source and a wobbler throw of 200$\arcsec$ in azimuth, and for the July 31 observation mapping was conducted with a reference position at the relative equatorial offset (6000$\arcsec$, 4000$\arcsec$).
We also used LAsMA\footnote{https://www.mpifr-bonn.mpg.de/5278286/lasma}, a 7-pixel, dual-sideband, single-polarization, heterodyne receiver, consisting of a hexagonal array of six pixels surrounding a central pixel. The backend consisted of Fast Fourier Transform Spectrometers (FFTSs) covering an intermediate frequency (IF) bandwidth of 4 -- 8\,GHz. The observing time with LAsMA was $\sim$8.5\,hours. The LAsMA observations were conducted using the on-the-fly mapping mode with reference position at (400$\arcsec$, 400$\arcsec$).
Finally, we used the dual-sideband, dual-polarization receiver SEPIA345\footnote{https://www.apex-telescope.org/ns/instruments/sepia/}. The upper and lower sidebands each cover the IF range 4 -- 12\,GHz, giving a total of 32\,GHz IF bandwidth, with each sideband recorded by two FFTS units. The observations were conducted with the position-switching mode with a reference position at (300$\arcsec$, 300$\arcsec$). The observing time was $\sim$1\,hour with SEPIA. The total observing time with APEX was $\sim$15\,hours. 
For all APEX observations, the antenna temperature, $T_{\rm A}^*$, was converted to main-beam temperature, $T_{\rm MB}$, using the relationship $T_{\rm MB} = T_{\rm A}^*$ $\eta_{\rm f}/\eta_{\rm mb}$ where $\eta_{\rm f}$ and $\eta_{\rm mb}$ are the 
forward and main-beam efficiencies, respectively, from the APEX website\footnote{https://www.apex-telescope.org/telescope/efficiency/}. The main parameters of the observations are listed in Table~\ref{tab:obs-log-apex}. The data obtained with APEX were also reduced using the GILDAS/CLASS software (see Figs.~\ref{fig:survey-219.2ghz-small-apex}--\ref{fig:survey-apex-sepia}).

\begin{table*}
\tiny
\caption{Information on the APEX~12-m observations.}              
\label{tab:obs-log-apex}      
\centering                                      
\begin{tabular}{cccccccrrrrc}          
\hline\hline                 
\multirow{1}{*}{Date}  & \multirow{1}{*}{Receiver}      & \multirow{1}{*}{Mode} & \multirow{1}{*}{Band} & \multirow{1}{*}{Tuning} & \multirow{1}{*}{Frequency range} & \multirow{1}{*}{Beam}  & \multicolumn{4}{c}{Measured rms noise level} & Size of otf  \\
\cdashline{8-11}
\multirow{2}{*}{yyyy-mm-dd}  &   & & & \multirow{2}{*}{(GHz)} & \multirow{2}{*}{(GHz)}  & \multirow{2}{*}{($\arcsec$)} & LO & LI & UI & UO & \multirow{2}{*}{($\arcsec$)} \\ 
& & & & & & & (mK) & (mK) & (mK) & (mK) & \\
\hline \hline                                
2021-07-20  & nFLASH & onOff & nFLASH230 & $219.2$  &  $213.2-221.1$, $229.3-237.2$   & $29.5-26.5$    & $20.3$ & $38.5$ & $21$ & $122.5$ & $-$ \\ 
2021-07-21  & nFLASH & onOff & nFLASH230 & $227.2$  &  $221.1-229$, $237.3-245.2$  & $28.4-25.6$    & $19.7$ & $67.9$ & $20.2$ & $67.8$ & $-$ \\
2021-07-31  & nFLASH & otf   & nFLASH230 & $227.2$  &  $213.2-221.1$, $229.3-237.2$ & $28.4-25.6$ & $101$ & $350$ & $138$ & $-^*$ & $600\arcsec$$\times$$600\arcsec$ \\
\hline                                 
2021-11-19 & LAsMA & otf & LAsMA345 & $344.2$  & $342.2-346.3$, $354.2-358.2$ & $18.3-17.5$ & \multicolumn{2}{c}{$30.4^{(a)}$} & \multicolumn{2}{c}{$59.3^{(b)}$} & $120\arcsec$$\times$$120\arcsec$ \\
2021-11-20  & LAsMA & otf & LAsMA345 & $344.2$  & $342.2-346.3$, $354.2-358.2$ & $18.3-17.5$ & \multicolumn{2}{c}{$31.1^{(a)}$} & \multicolumn{2}{c}{$105.1^{(b)}$} & $120\arcsec$$\times$$120\arcsec$ \\
\hline                               
2022-05-28  & SEPIA & onOff & SEPIA345 & $291.0$  & $272.9-280.8$, $289.2-297$  & $23.0-21.1$ & $69.9$ & $29.5$ & $38$ & $41.5$ & $-$ \\ 
\hline \hline    
\end{tabular}
\tablefoot{
\tablefoottext{*}{Due to high noise level and deep atmospheric lines the rms of the UO cannot be measured accurately.}
\tablefoottext{a}{Measured rms noise level for the whole LSB.}
\tablefoottext{b}{Measured rms noise level for the whole USB.}
}
\end{table*}

\section{Results and analysis}
\label{sec:results}

\subsection{Line detection and identification}
\label{subsub:line-detection-identification}
\begin{figure*}[htbp!]
\centering 
\vspace{-8.4cm}
\includegraphics[width=\textwidth]{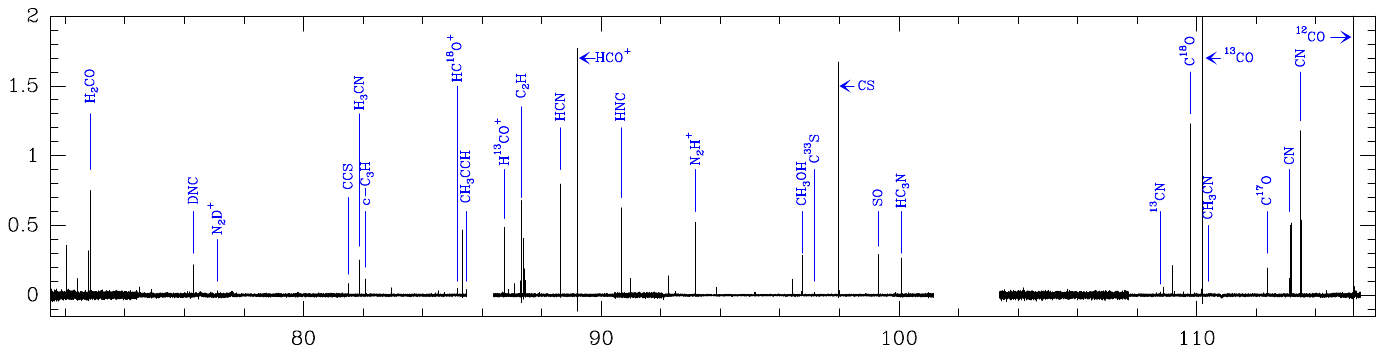} \\
\vspace{-10.3cm}
\includegraphics[width=\textwidth]{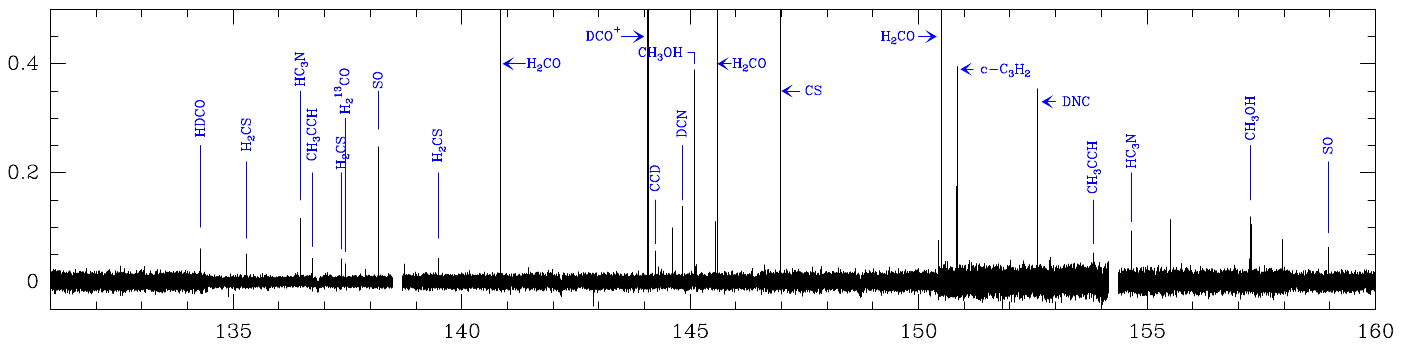} \\
\vspace{-10.3cm}
\includegraphics[width=\textwidth]{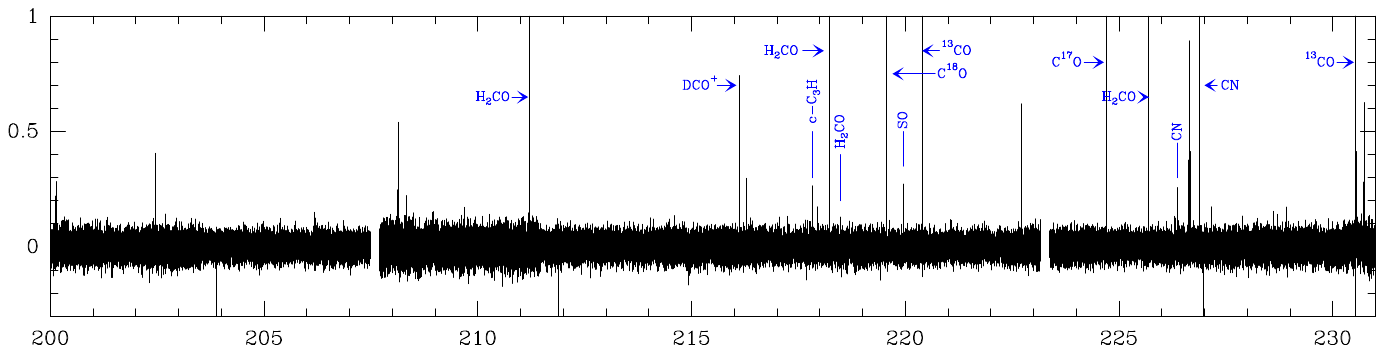} \\
\vspace{-10.3cm}
\includegraphics[width=\textwidth]{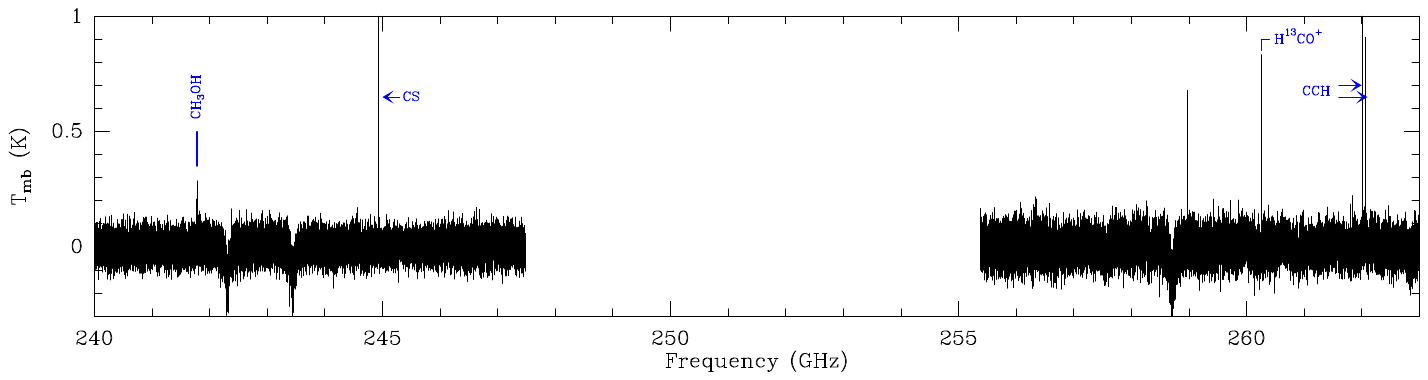} \\
\vspace{-2cm}
\caption{Complete spectrum of V1057~Cyg obtained with the IRAM\,30-m telescope, 
with selected lines labelled with the name of the molecular species. The complete list of identified lines is provided in Table~\ref{tab:appendix-iram-lines-list-fits} and zoom views of all lines are displayed in 
Fig.~\ref{fig:lines-example} and  Figs.~\ref{fig:survey-73ghz-small} -- ~\ref{fig:survey-242ghz-small}. The deep absorption features around 242--244\,GHz and 258--259\,GHz represent residual atmospheric absorption.}
\label{fig:composit-iram}
\end{figure*}

\begin{figure*}[htbp!]
\centering 
\vspace{-3.5cm}
\hspace{-1cm}
  \begin{minipage}[h]{0.33\textwidth}
    \includegraphics[width=2.2\textwidth]{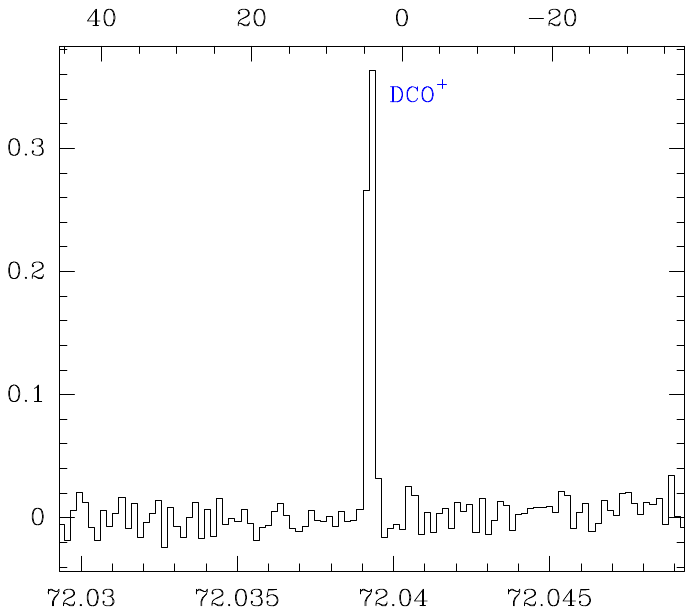}
  \end{minipage}
  \hspace{-0.4cm}
  \begin{minipage}[h]{0.33\textwidth}
    \includegraphics[width=2.2\textwidth]{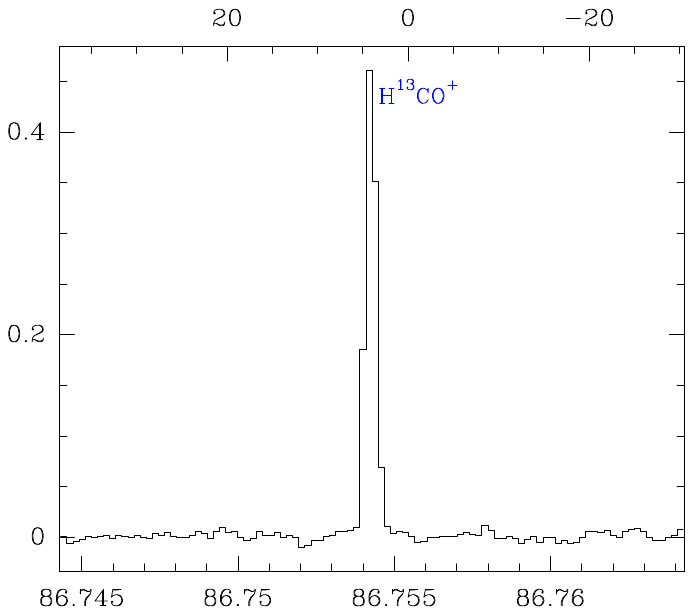}  
  \end{minipage}
  \hspace{-0.4cm}
  \begin{minipage}[h]{0.33\textwidth}
    \includegraphics[width=2.2\textwidth]{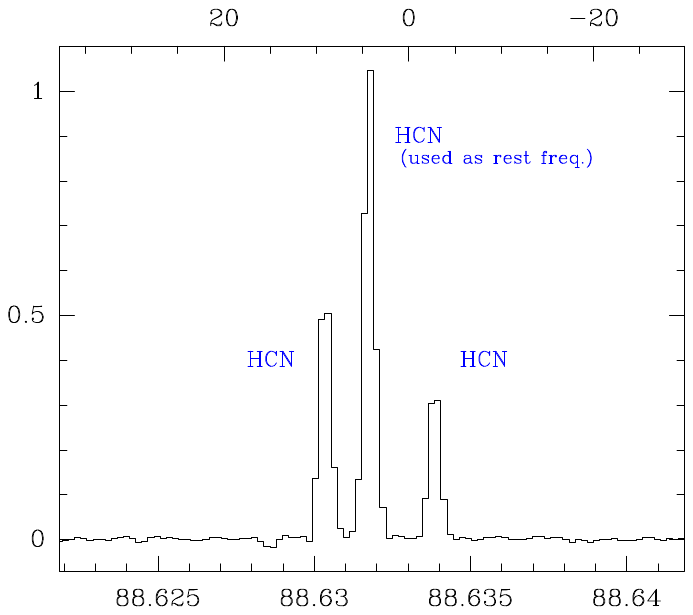}
  \end{minipage} \\
\vspace{-5cm}
\hspace{-0.8cm}
  \begin{minipage}[h]{0.33\textwidth}
    \includegraphics[width=2.2\textwidth]{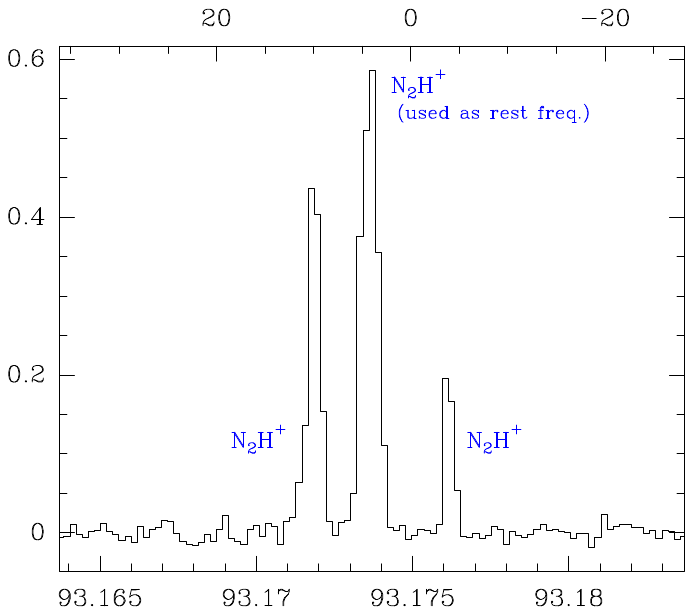}
  \end{minipage}
  \hspace{-0.4cm}
  \begin{minipage}[h]{0.33\textwidth}
    \includegraphics[width=2.2\textwidth]{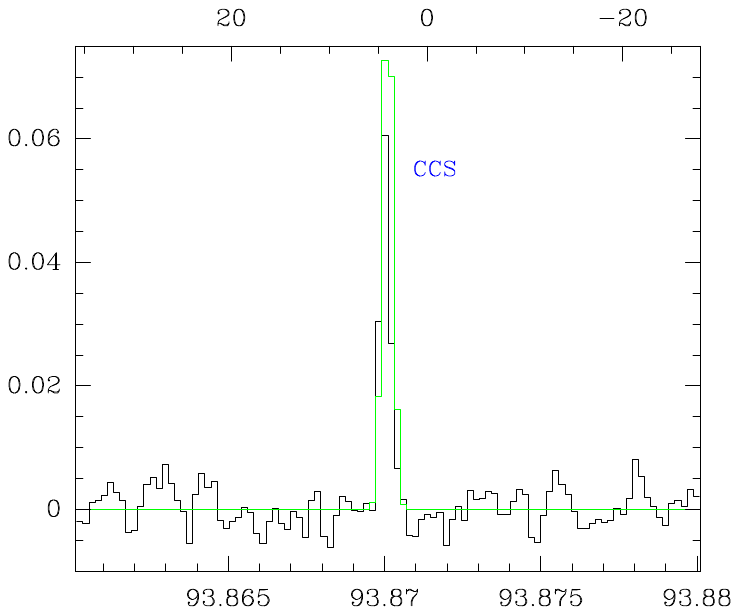}  
  \end{minipage}
  \hspace{-0.4cm}
  \begin{minipage}[h]{0.33\textwidth}
    \includegraphics[width=2.2\textwidth]{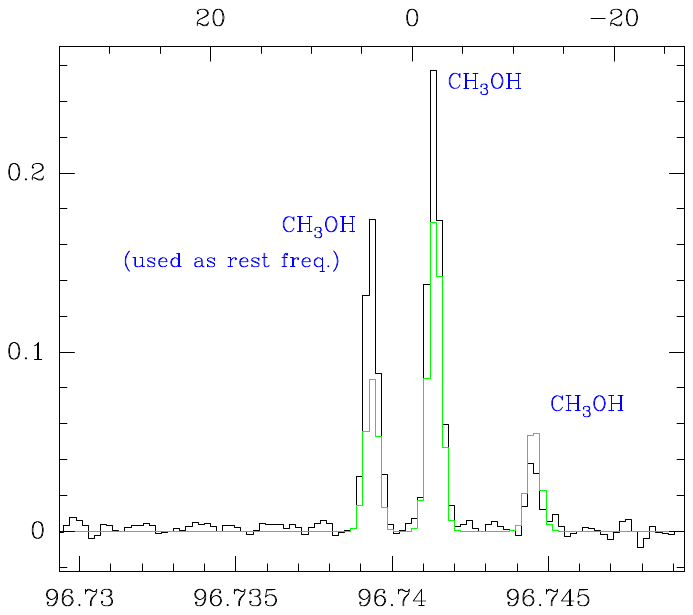}
  \end{minipage} \\
\vspace{-5cm}
\hspace{-0.8cm}
  \begin{minipage}[h]{0.33\textwidth}
    \includegraphics[width=2.2\textwidth]{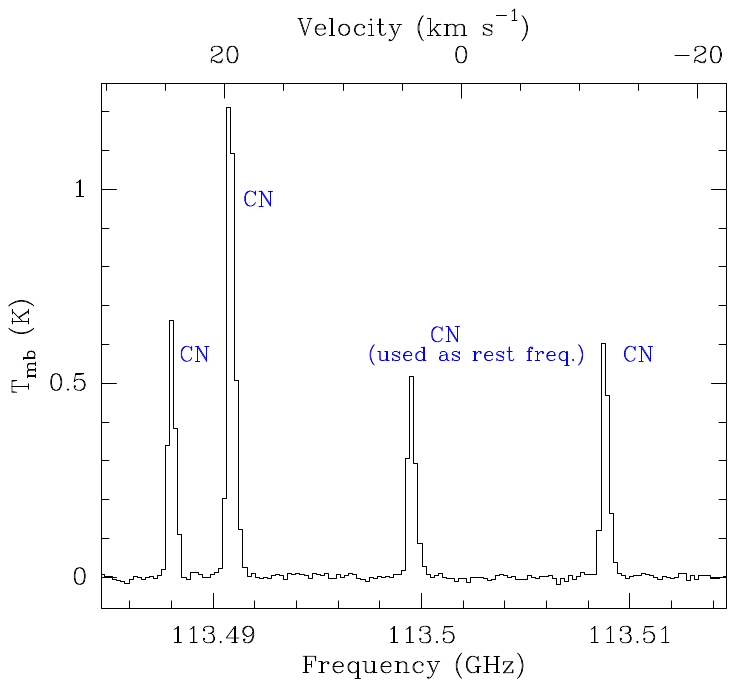}
  \end{minipage}
  \hspace{-0.4cm}
  \begin{minipage}[h]{0.33\textwidth}
    \includegraphics[width=2.2\textwidth]{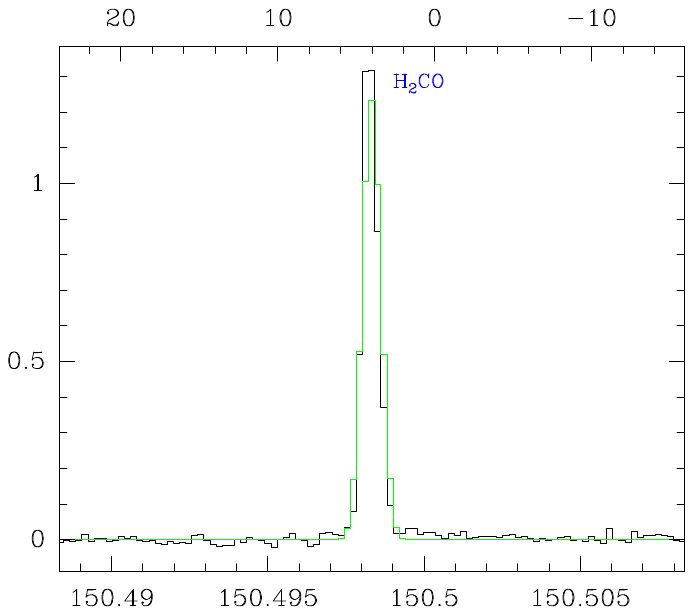}  
  \end{minipage}
  \hspace{-0.4cm}
  \begin{minipage}[h]{0.33\textwidth}
    \includegraphics[width=2.2\textwidth]{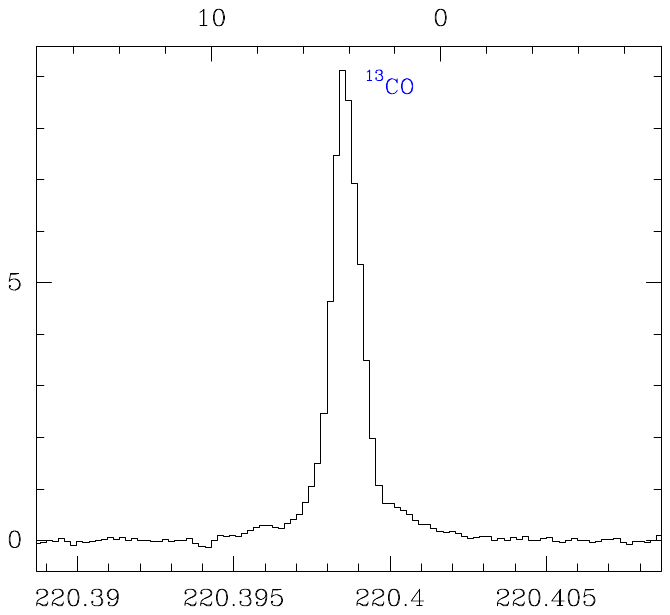}
  \end{minipage} \\
\vspace{-1.5cm} 
\caption{Selection of lines detected with the IRAM~30-m telescope, labelled with the molecular species name in blue; the other lines are shown in Figs.~\ref{fig:survey-73ghz-small}--\ref{fig:survey-242ghz-small}. LTE synthetic spectra are overlaid in green for selected species (see Sects.~\ref{subsec:weeds-modeling} and \ref{subsec:pop-diagrams}). In each panel, the bottom and top axes are labelled in frequency and velocity, respectively.}
\label{fig:lines-example}
\end{figure*}

Figure~\ref{fig:composit-iram} presents the complete spectrum obtained towards V1057~Cyg with the IRAM 30-m telescope, with frequencies of prominent lines marked with vertical lines. 
Line identifications were based on spectroscopic data from the Cologne Database for Molecular Spectroscopy\footnote{https://cdms.astro.uni-koeln.de} \citep[CDMS,][]{muller2001,muller2005} and the Jet Propulsion Laboratory catalog\footnote{https://spec.jpl.nasa.gov} \citep[JPL,][]{pickett1998-jpl}. 
A line is considered detected if its peak intensity is $\geq$3\,$\sigma$, where $\sigma$ is the rms noise level measured in the portion of the spectrum
surrounding the candidate line. We only report lines of molecules with at least 2 detected transitions within the coverage of our line survey.
In general, the detected lines are relatively narrow ($<$2.5\,km\,s$^{-1}$), and well isolated, thus their identification is straightforward compared to line-rich sources such as hot cores or hot corinos \citep[in high-mass or low-mass star-forming regions; e.g.,][]{belloche2013,bianchi2020,busch2024}.
Figure~\ref{fig:lines-example} shows examples of the detected line profiles for several molecules observed with the 30-m telescope. These examples illustrate the identification of molecules with specific transitions (e.g.,~DCO$^+$), hyperfine structure components (e.g.,~CN), multiple transitions within a narrow frequency range (e.g.,~CH$_3$OH), and the presence of broad wings indicating outflow activity (e.g.,~$^{13}$CO).
The complete list of identified lines is given in Table~\ref{tab:appendix-iram-lines-list-fits} with upper energy levels, Einstein coefficients, and the parameters derived from Gaussian fits (discussed in Sect.~\ref{subsec:line-characteristics}), while each detected line is displayed in a narrow frequency window in Figs.~\ref{fig:survey-73ghz-small}--\ref{fig:survey-242ghz-small}. 
There are several cases where the data suggest potential detections of molecules, however, these identifications remain uncertain due to low signal-to-noise ratios (with candidate lines detected $<$3$\sigma$) or because they are based on a single, weakly detected transition (below $<$3$\sigma$).
These occurrences are not included in the final line identification and fit results in Table~\ref{tab:appendix-iram-lines-list-fits}, but are instead marked with green labels in Figs.~\ref{fig:survey-73ghz-small}--\ref{fig:survey-242ghz-small} (e.g.,~some transitions of H$_2$CO, H$_2$CN, H$_5$CN, CH$_3$OH in Fig.~\ref{fig:survey-202ghz-small}).
The list of identified lines from the APEX observations is given in Table~\ref{tab:res-apex-fits} and the detected lines are 
displayed in Figs.~\ref{fig:survey-219.2ghz-small-apex}--\ref{fig:survey-apex-sepia}. 

We detect emission of molecules ranging from simple neutral molecular species to COMs. 
Identified species are listed in Table~\ref{tab:species-detected}, with the molecules organised from the simplest to the most complex.
In total, 35 molecular species are detected (including isotopologues) in the spectrum of V1057~Cyg. We detect $C$-, $N$-, $O$-, $S$-bearing molecules (e.g., CO, CN, CS, CCS, HCN, HNC, H$_2$CO, H$_2$CS), deuterated species (e.g., DCN, DNC), molecular ions (e.g., HCO$^+$, HCS$^+$), carbon-chain molecules (e.g., C$_2$H, HC$_3$N), a cyclic molecule (e.g., c-C$_3$H$_2$), COMs (e.g., CH$_3$OH), and in many cases some of their isotopologues (see Table~\ref{tab:appendix-iram-lines-list-fits}). 

The spectral energy distribution (SED) of V1057~Cyg implies a flared disk and envelope geometry \citep[e.g.,][]{kenyon-and-hartmann1991}, while an envelope with an estimated size of 7000\,au was confirmed from Spitzer/IRS observations \citep{green2006}, and later 50--670$\mu$m far infrared/submilllimeter wavelength imaging with  \textit{Herschel} indicated a reservoir of cold dust \citep{green2013}. From $^{13}$CO observations using the Plateau de Bure Interferometer (with a beam of 2.7$\arcsec \times$ 2.2$\arcsec$), \citet{feher2017} reported a rotating envelope around V1057~Cyg, with a radius of 5$\arcsec$ (3000\,au). 
Thus it is likely that our single-dish data, with a beam size larger than 9$\arcsec$, corresponding to $\sim$8100\,au at the 897\,pc distance to the target, traces emission both from the disk and the envelope around the source. For instance, CO and its isotopologues likely trace the disk and outflow, while CS (2--1), HCN (1--0), and the hyperfine structure components of CN (2--1) and C$_2$H (1--0) could sample the cold gas closer to the midplane \citep[seen in T~Tauri stars, e.g.,][]{dutrey2014}. 
We also detected other dense gas tracers, such as CS, CCS, HCO$^+$, including their isotopologues $^{13}$CS, C$^{33}$S, C$^{34}$S, and H$^{13}$CO$^+$, as well as N$_2$H$^+$ \citep[e.g.,][and references therein]{shirley2015,yamamoto2017-astrochem-book}.

\begin{table}
\caption{Overview of molecules and isotopic species detected detected towards V1057~Cyg in this survey.}         
\centering                                      
\begin{tabular}{l}          
\hline\hline 
\textit{Simple neutral molecules:} \\
CN, $^{13}$CN, CO, $^{13}$CO, C$^{17}$O, C$^{18}$O, SO, CS, $^{13}$CS, C$^{33}$S, \\
C$^{34}$S, HCN, HNC, H$_2$CO, H$_2$$^{13}$CO, H$_2$CS  \\
\hline
\textit{Deuterated species:} \\
CCD, DCN, DCO$^+$, DNC, N$_2$D$^+$, HDCO \\
\hline
\textit{Molecular ions:} \\
HCO$^+$, H$^{13}$CO$^+$ HCS$^+$, N$_2$H$^+$ \\
\hline
\textit{Carbon chain molecules:} \\
C$_2$H, CCS, C$_4$H, CH$_3$CCH, HC$_3$N \\
\hline
\textit{Cyclic molecules:} \\
c-C$_3$H$_2$ \\
\hline
\textit{Complex organic molecules:} \\
CH$_3$OH, CH$_3$CN \\
\hline
\textit{Other molecules:} \\
HNCO \\
\hline \hline
\end{tabular}
\label{tab:species-detected} 
\end{table}

For most molecular species, several transitions with different upper-level energies are detected,
which allows a more detailed analysis including simple radiative transfer modelling and the construction of population diagrams, presented in Sects.~\ref{subsec:weeds-modeling} and \ref{subsec:pop-diagrams}, respectively. 

%
\subsection{Line profiles, LSR velocities, and line widths}
\label{subsec:line-characteristics}
Based on single-dish observations, it can be challenging to distinguish emission from foreground/background clouds and from different physical structures within a source (e.g.,~disk, envelope) unless these components have different velocities.
In this survey, we only detect molecules with similar velocities, suggesting that the observed line emission is associated with V1057~Cyg and its environment.

Most of the detected lines show narrow single-peaked profiles that can be fitted by a single Gaussian. In the case of outflow tracers (e.g.,~$^{13}$CO) the line profiles are also single-peaked (except $^{12}$CO, see below), but show broader wings, suggesting outflow activity. 
For the spectrum at the source position, each line profile was fitted with a single Gaussian component using the built-in function in CLASS, and the results are presented in Tables~\ref{tab:appendix-iram-lines-list-fits} and \ref{tab:res-apex-fits}. 
The Gaussian fitting method does not yield good results for the $^{12}$CO transitions because they show self-absorption, suggesting that these lines are optically thick \citep[see examples for other FUors in][]{evans1994,kospal2017b,kospal2011b,cruz-saenz-de-miera2023}.
Therefore, these lines are only reported as detected in Tables~\ref{tab:appendix-iram-lines-list-fits} and \ref{tab:res-apex-fits}, while Gaussian fit results are not presented. 
The LSR velocities obtained from the Gaussian fits range from 2.45 to 4.80\,km\,s$^{-1}$ for the 30-m, and from 3.95 to 4.46\,km\,s$^{-1}$ for the APEX data.
Comparing to single-dish and interferometric observations from the past decade, we find that the previously reported velocities of, for example, $^{13}$CO (1--0) \citep[4.6 and 4.05\,km\,s$^{-1}$,][]{kospal2011b,feher2017}, C$^{18}$O (1--0) \citep[4.10\,km\,s$^{-1}$,][]{kospal2011b}, and NH$_3$ (1,1) \citep[4.35\,km\,s$^{-1}$,][]{szabo2023a} 
are generally within $\sim$0.5\,km\,s$^{-1}$ of the $\varv_{\rm lsr}$ values derived in this study, further indicating that the molecular species detected in our survey are associated with V1057~Cyg and its environment. 
Notably, we find that the measured velocities of dense gas tracers such as CS, CCS, H$^{13}$CO$^+$, N$_2$H$^+$ and HC$_3$N  coincide within $\sim$0.3\,km\,s$^{-1}$ (within the uncertainties of the Gaussian fits) with the systemic velocity of V1057~Cyg derived from the most recent ammonia observations \citep[4.35\,$\pm$\,0.02\,km\,s$^{-1}$,][]{szabo2023a}.

From the Gaussian fits, we find FWHM line widths between 0.67 and 3.43\,km\,s$^{-1}$ for the 30-m, and from 0.71 to 1.78\,km\,s$^{-1}$ for the APEX data.
In general, lines that show wings, indicating outflow activity, have FWHM line widths  broader than the other species (e.g.,~$^{12}$CO and its isotopologues).
The line widths of more than $>$50\% of the fitted lines cluster around 1.3--1.5\,km\,s$^{-1}$ for mainly the transitions of CH$_3$OH, HC$_3$N, and H$_2$CO. Narrower  widths are mainly found for CN and c-C$_3$H$_2$ lines, the former clustered around 1.1\,km\,s$^{-1}$ and the latter around 0.9 -- 1\,km\,s$^{-1}$.

We note that the transition of methanol (CH$_3$OH) at 84.5\,GHz is a well-known Class~{\footnotesize I} maser line \citep[e.g.,][and references therein]{batrla1988,breen2019-84ghz-methanol}. 
However, the line emission detected towards V1057~Cyg does not show clear
maser characteristics (such as strong, and narrow features in comparison to other transitions), thus not suggesting maser activity. 
The well-known Class~{\footnotesize I} CH$_3$OH maser line at 95.1\,GHz \citep{plambeck1990,chen2011-95ghz-methanol,yang2023} was not detected in our survey, which we discuss in Sect.~\ref{subsec:pop-diagrams}.

\subsection{Line mapping} 
\label{subsec:maps} 

Figures~\ref{fig:iram-maps} and ~\ref{fig:apex-maps-nflash} present integrated intensity maps for lines mapped in the otf mode with the IRAM~30-m and APEX~12-m telescopes, respectively.
In most cases (except $^{13}$CO 1--0 and 2--1, see Figs.~\ref{fig:iram-maps}i and \ref{fig:apex-maps-nflash}b), the molecular line emission does not peak on V1057~Cyg itself. 
The HCN (1--0) HNC (1--0), HC$_3$N (10--9), HCO$^+$ (1--0), and N$_2$H$^+$ (1--0) maps 
show a primary  
concentration of 
emission
offset from V1057~Cyg towards the north, while a secondary peak is distributed towards the southwest, 
connected with a ridge that traces a pc-scale filamentary structure (Fig.~\ref{fig:iram-maps}c,d,f,g,h).
The northern peak is offset by approximately 14.7$\arcsec$ ($\sim$0.06\,pc) to the north, while the southern peak is located approximately 1.8$\arcmin$ ($\sim$0.46\,pc) to the southwest of V1057~Cyg.
The offset of the northern concentration relative to V1057 Cyg corresponds to only half a beam width in the 3\,mm maps, but given the high signal-to-noise ratio of most maps, it is significant. It is unlikely the result of an optical depth effect because $^{13}$CO (1--0) does peak towards the source. Moreover, the offset is larger than the pointing error. Furthermore, this peak is also visible in the dust-based H$_2$ column density map \citep[][]{szabo2023a}.
The larger-scale ridge structure is well traced by the 1--0 and 2--1 transitions of $^{13}$CO and C$^{18}$O (see Figs.~\ref{fig:iram-maps}i,j and \ref{fig:apex-maps-nflash}b,c).
The ridge towards the southwest appears fainter compared to the northern peak seen in the maps of the 1--0 transitions of HCN, HCO$^+$, and N$_2$H$^+$.
Notably, the extent of the emission from some transitions (e.g.,~HC$_3$N 10--9, HNC 1--0) to the southwest seems comparable in size and intensity to that of the northern peak.
The integrated intensity maps of CCS (7--6), H$_2$CS (3--2), and HC$_3$N (12--11) have lower signal-to-noise ratios but appear to trace the same extensions toward the north and southwest.

\begin{figure*}[htbp!]
    \centering
    \vspace{-0.3cm}
    \hspace{-0.8cm}
    \subfigure{(a)\includegraphics[width=0.45\textwidth]{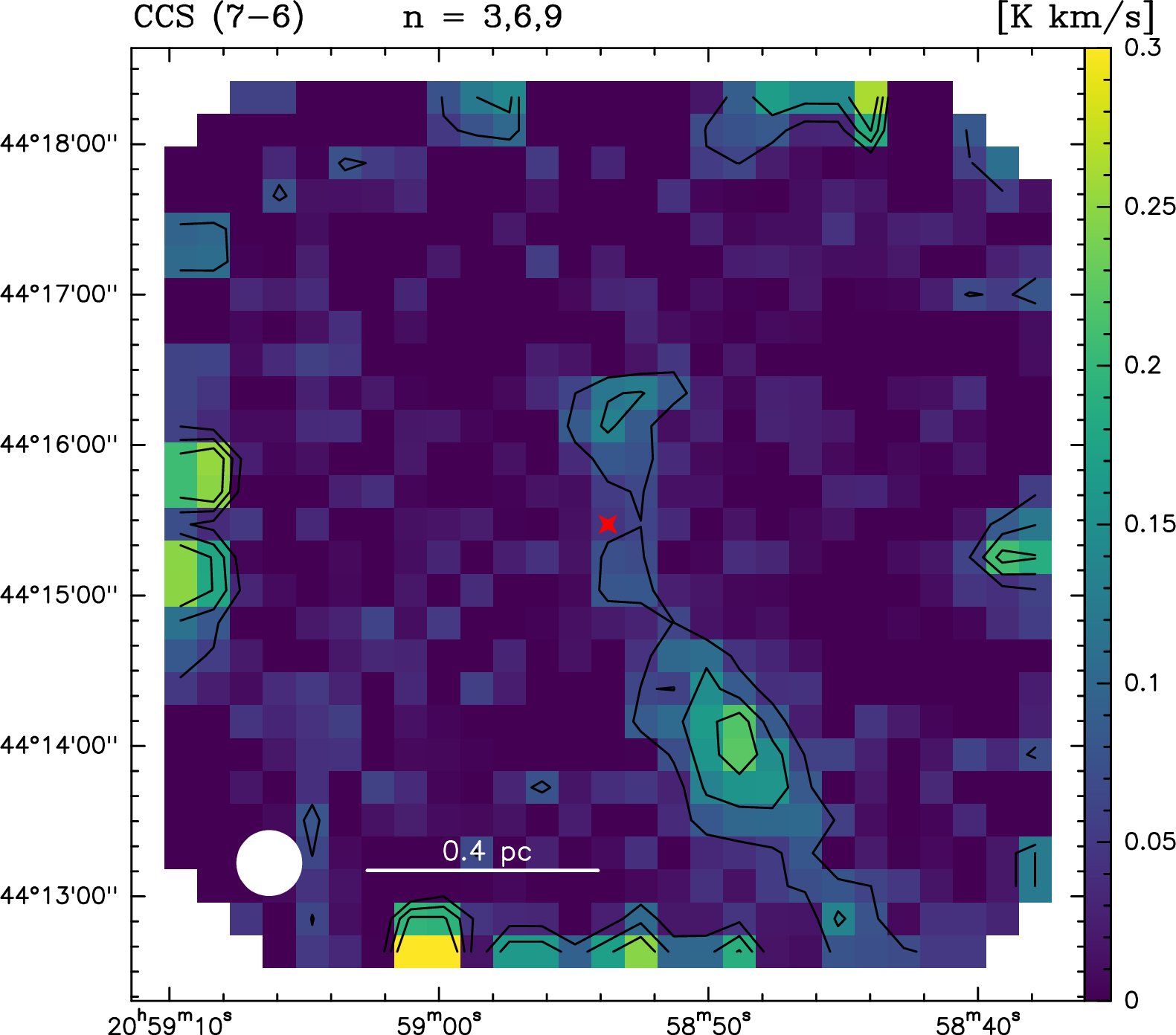}} 
    \hspace{0.3cm}
    \subfigure{(b)\includegraphics[width=0.45\textwidth]{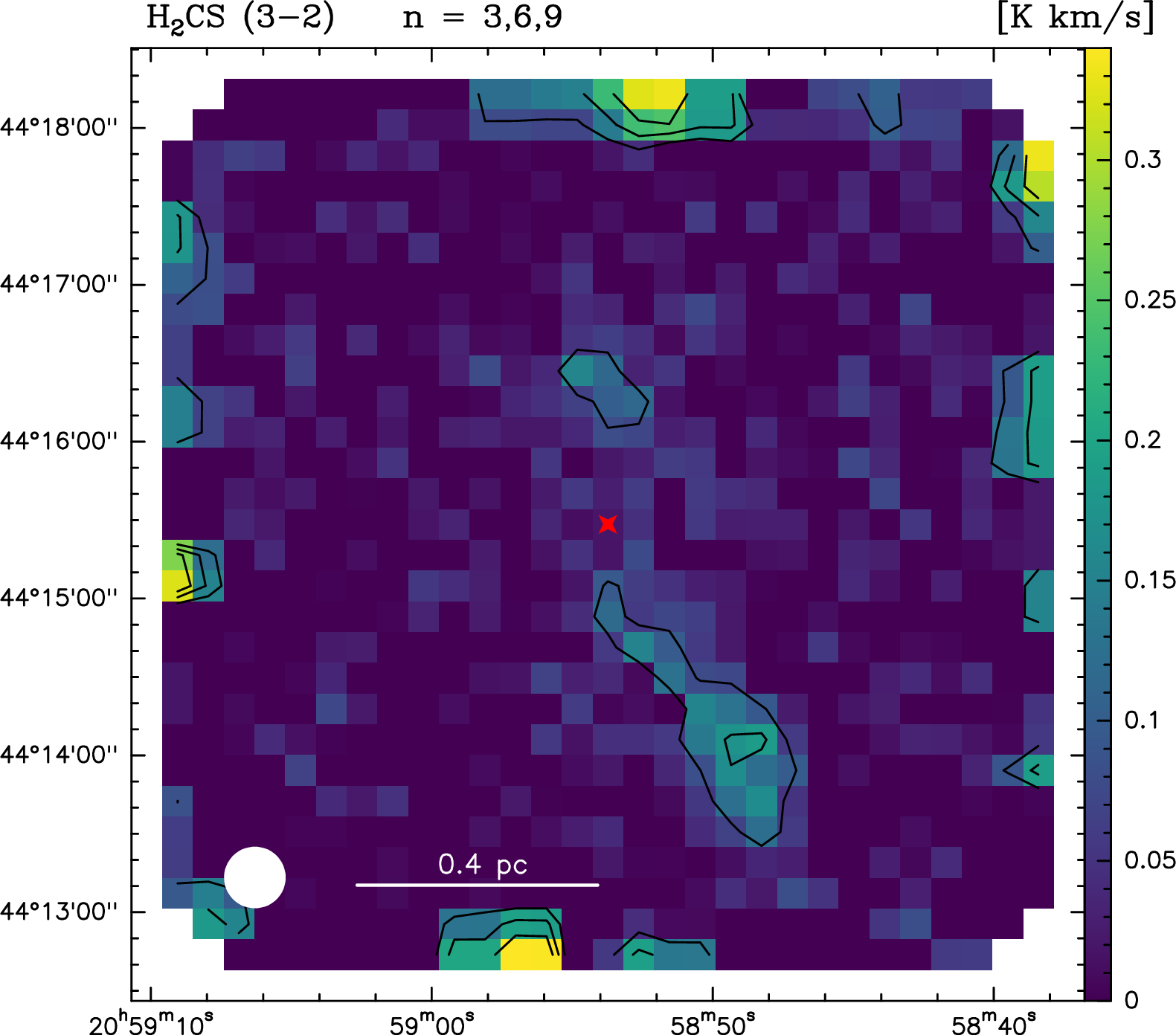}} \\
    \vspace{0.2cm}
    \hspace{-0.8cm}
    \subfigure{(c)\includegraphics[width=0.45\textwidth]{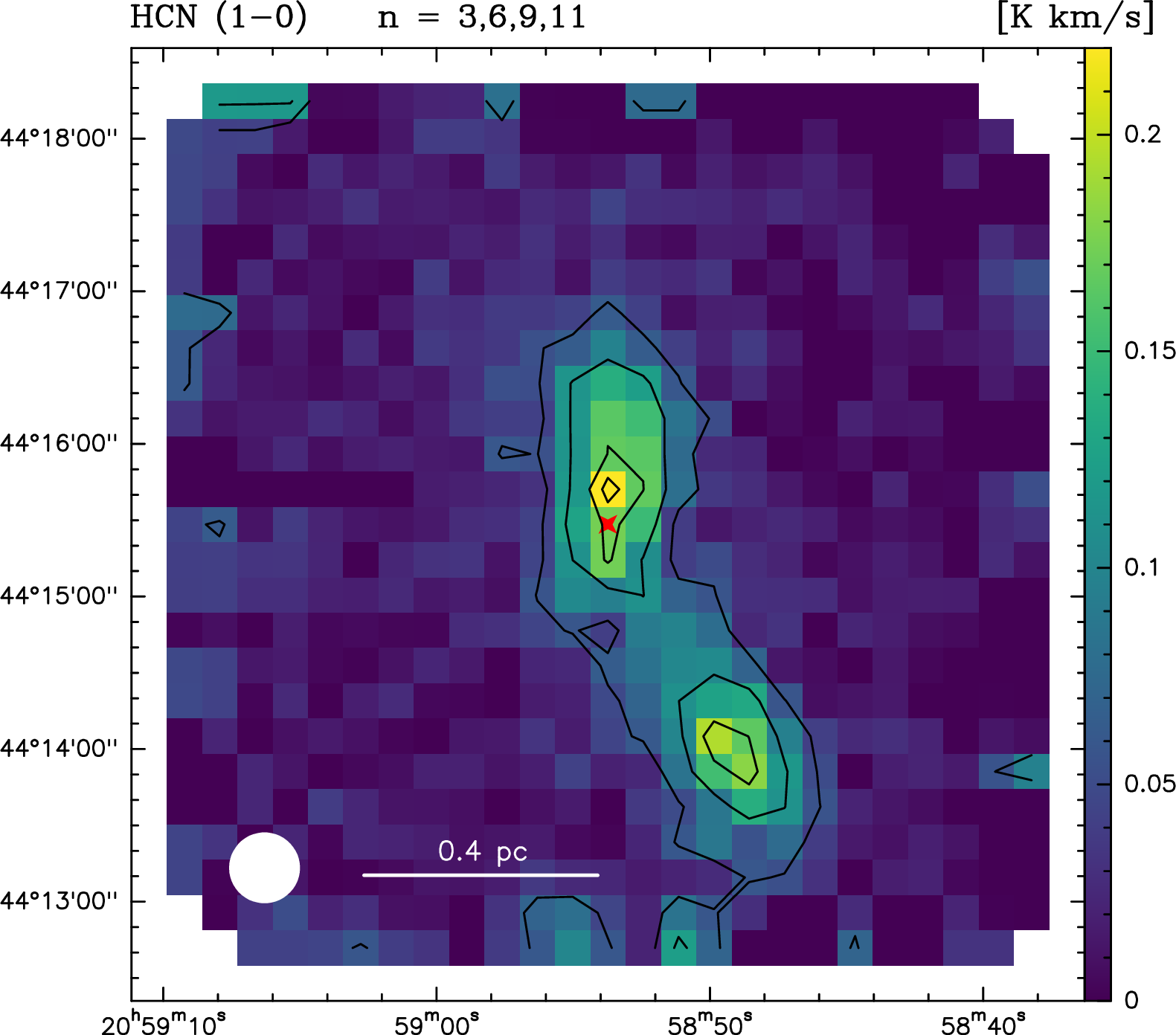}} 
    \hspace{0.3cm}
    \subfigure{(d)\includegraphics[width=0.45\textwidth]{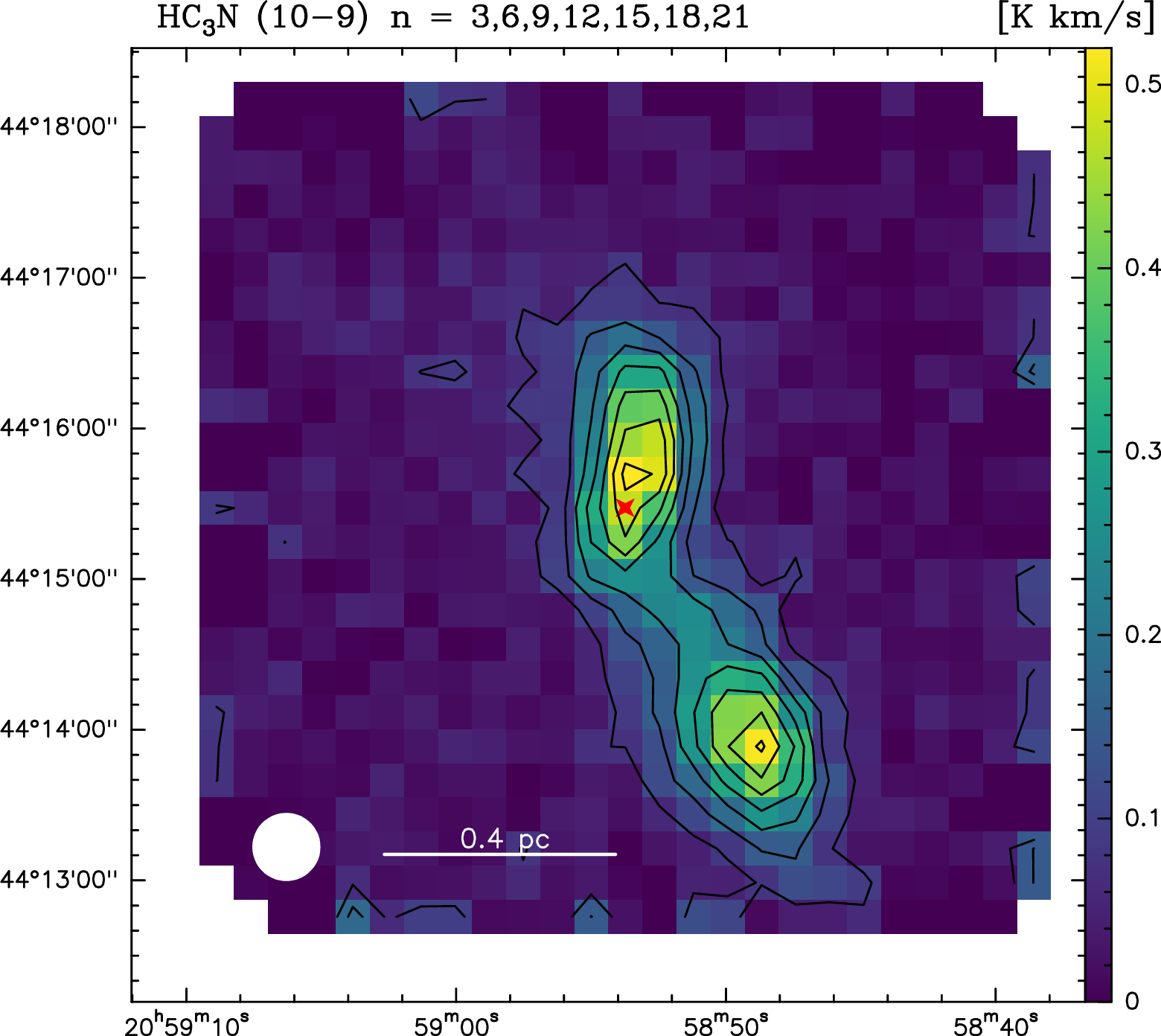}}
    \\
    \vspace{0.2cm}
    \hspace{-0.8cm}
    \subfigure{(e)\includegraphics[width=0.45\textwidth]{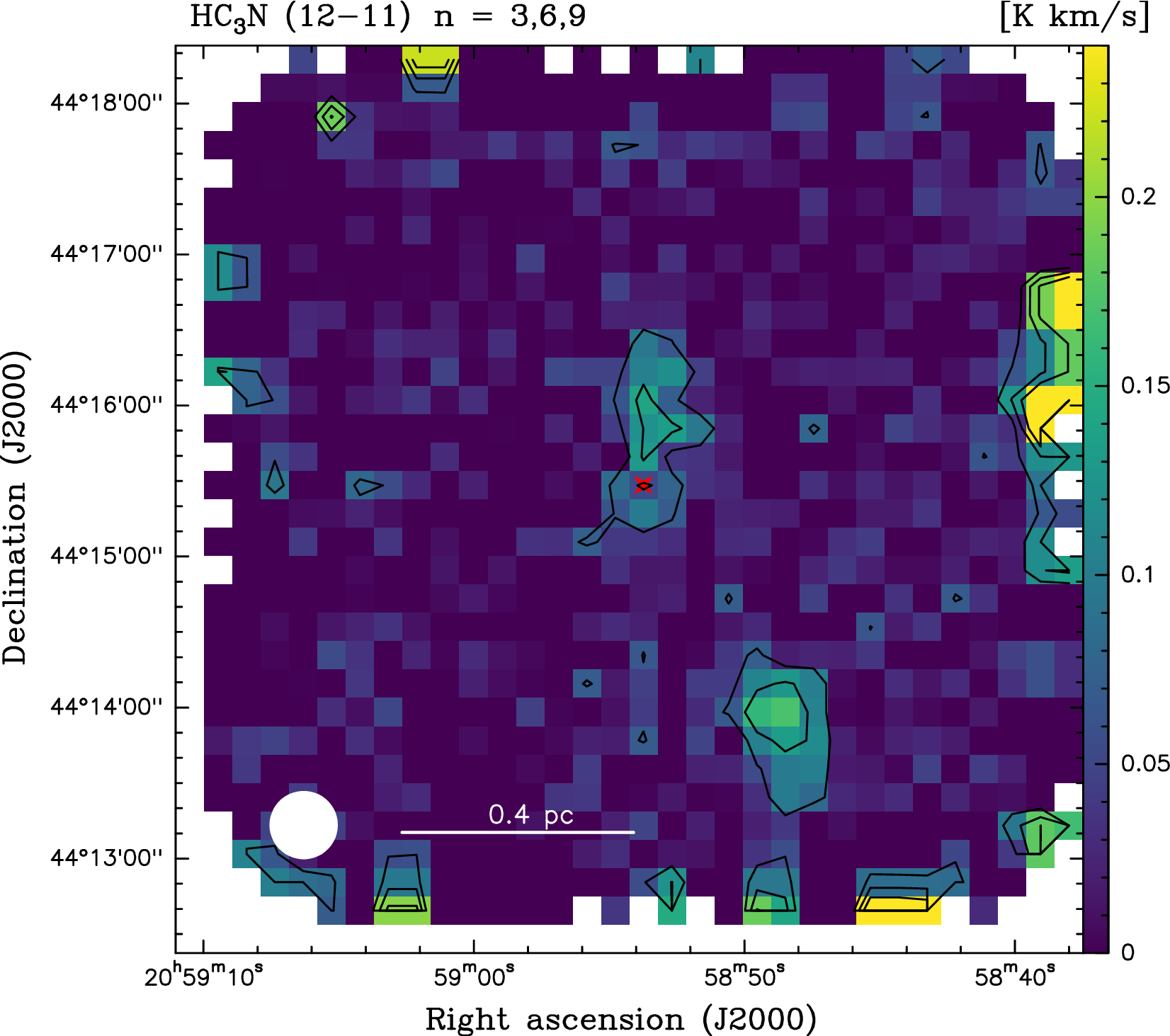}}
    \hspace{0.3cm}
    \subfigure{(f)\includegraphics[width=0.45\textwidth]{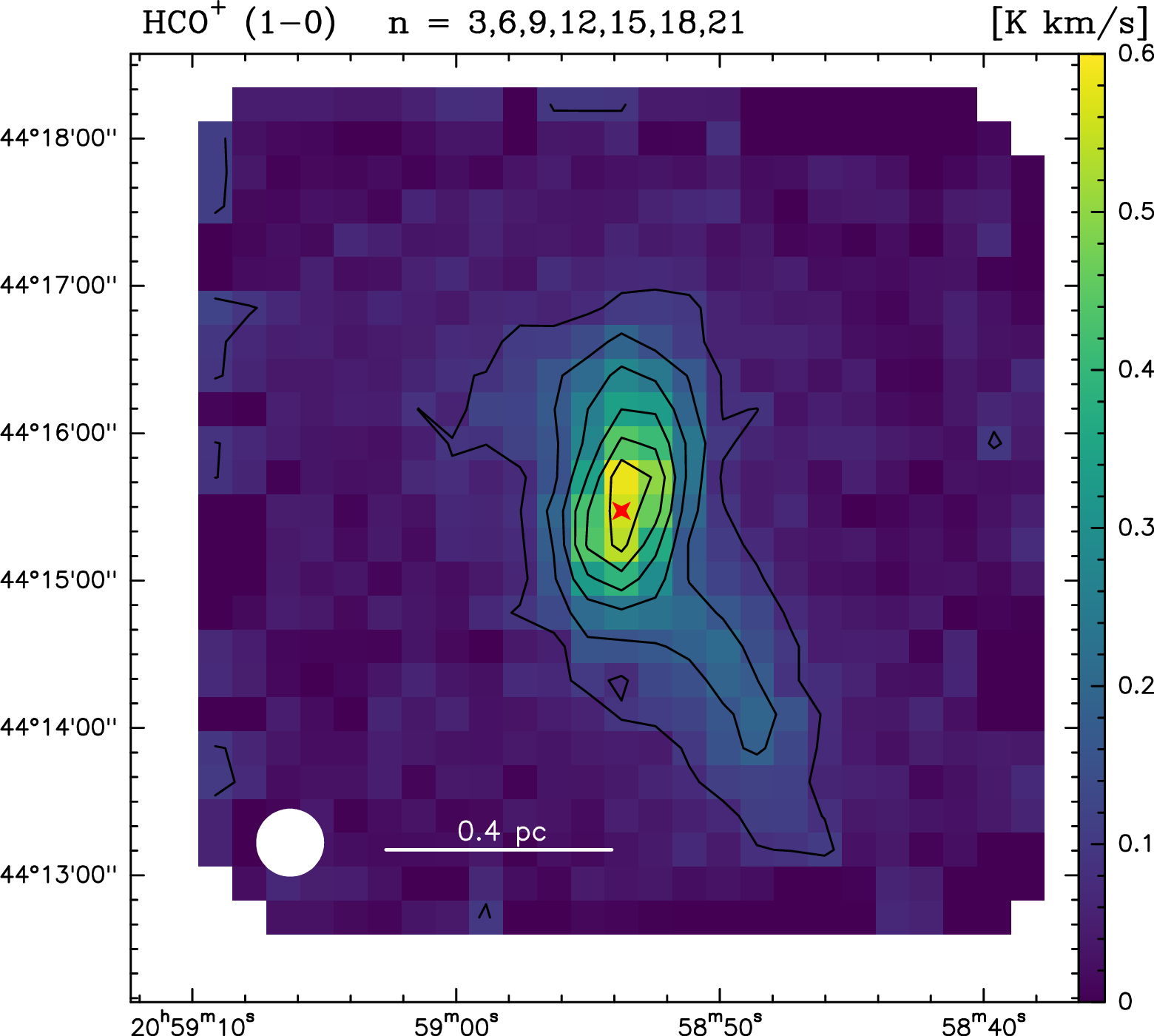}}
    \vspace{-0.1cm}
    \caption{Integrated intensity maps of molecular lines obtained with the IRAM~30-m telescope. The red cross marks the position of V1057~Cyg and the white circle shows the telescope beam size at the line frequency. The physical scale is presented near the beam size. Contour levels are $n\sigma_{int}$ (multiples of the noise level in each map), with the corresponding values of n given above each panel.}
    \label{fig:iram-maps}
\end{figure*}

\addtocounter{figure}{-1}
\begin{figure*}[htbp!]
    \centering
    \hspace{-0.8cm}
    \subfigure{(g)\includegraphics[width=0.45\textwidth]{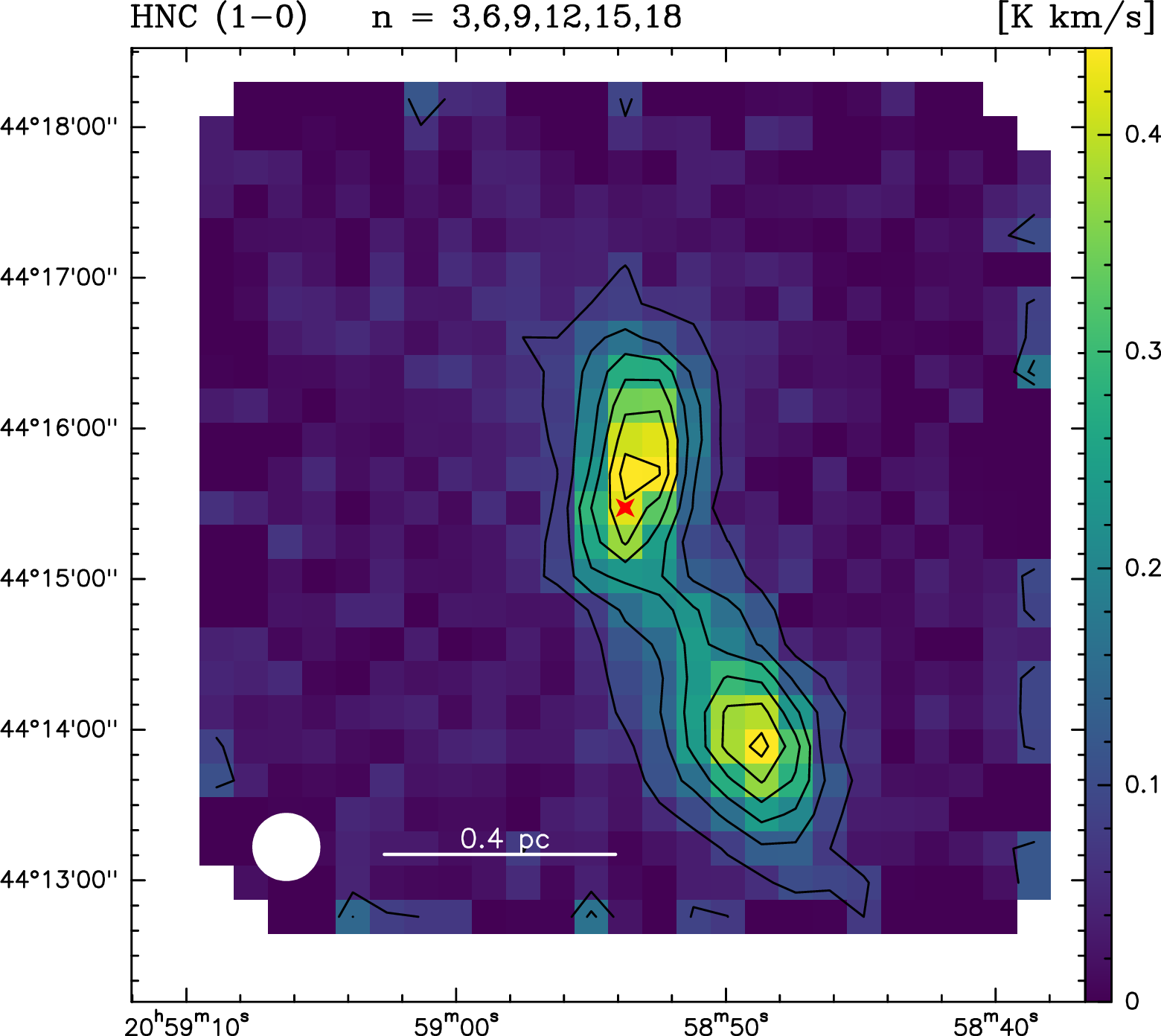}} 
    \hspace{0.3cm}
    \subfigure{(h)\includegraphics[width=0.45\textwidth]{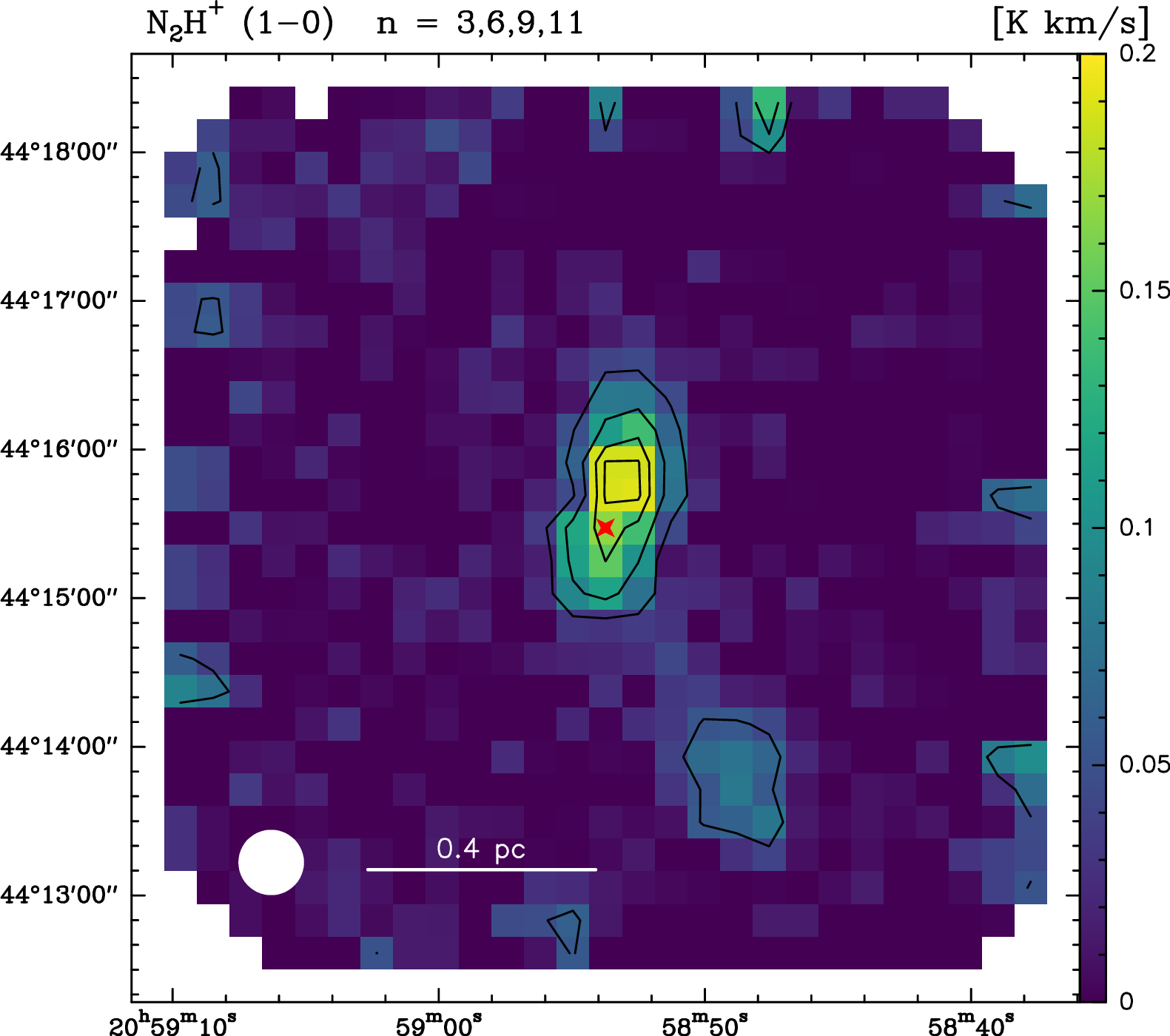}} \\
    \vspace{0.2cm}
    \hspace{-0.8cm}
    \subfigure{(i)\includegraphics[width=0.45\textwidth]{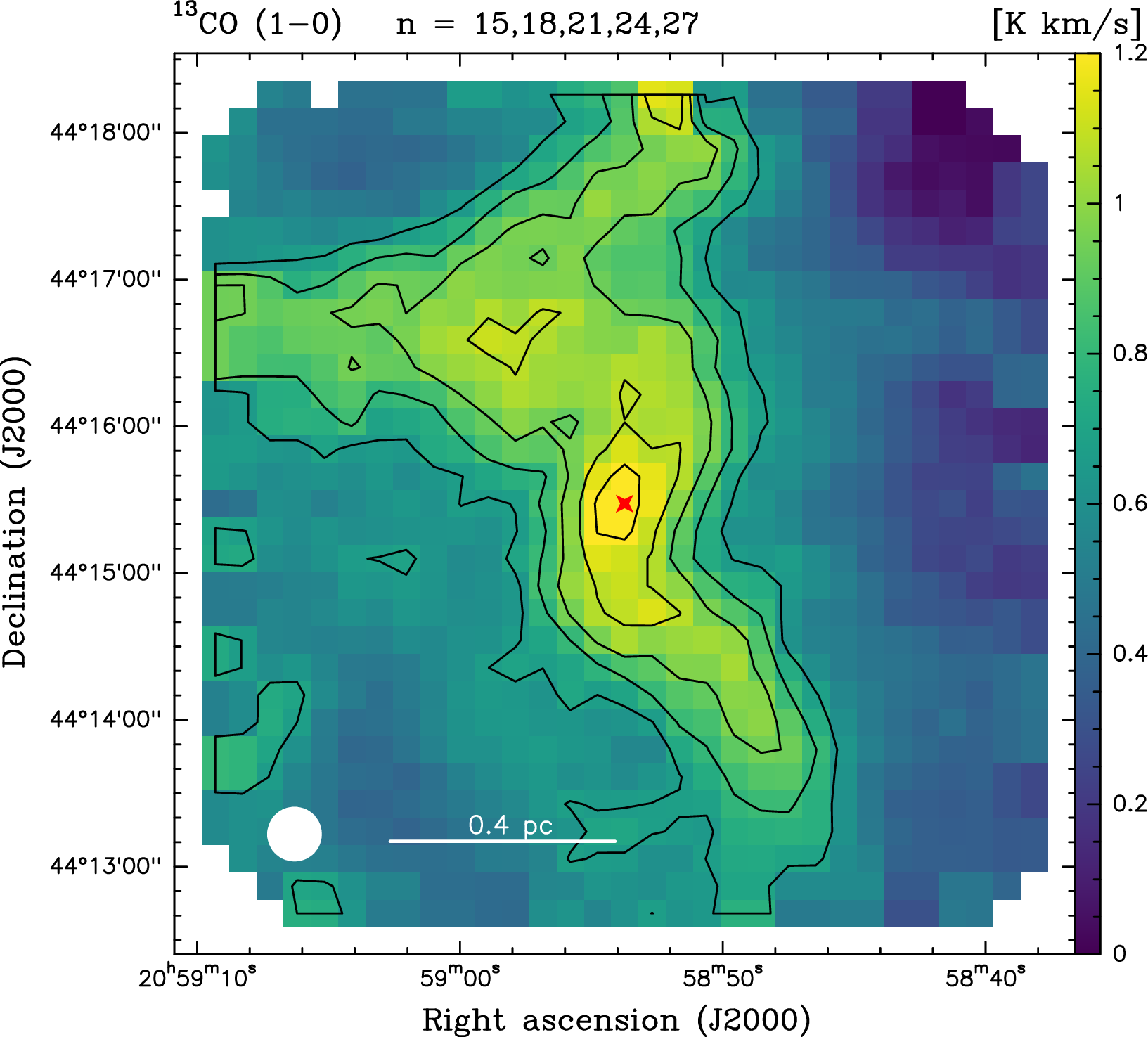}} 
    \hspace{0.3cm}
    \subfigure{(j)\includegraphics[width=0.45\textwidth]{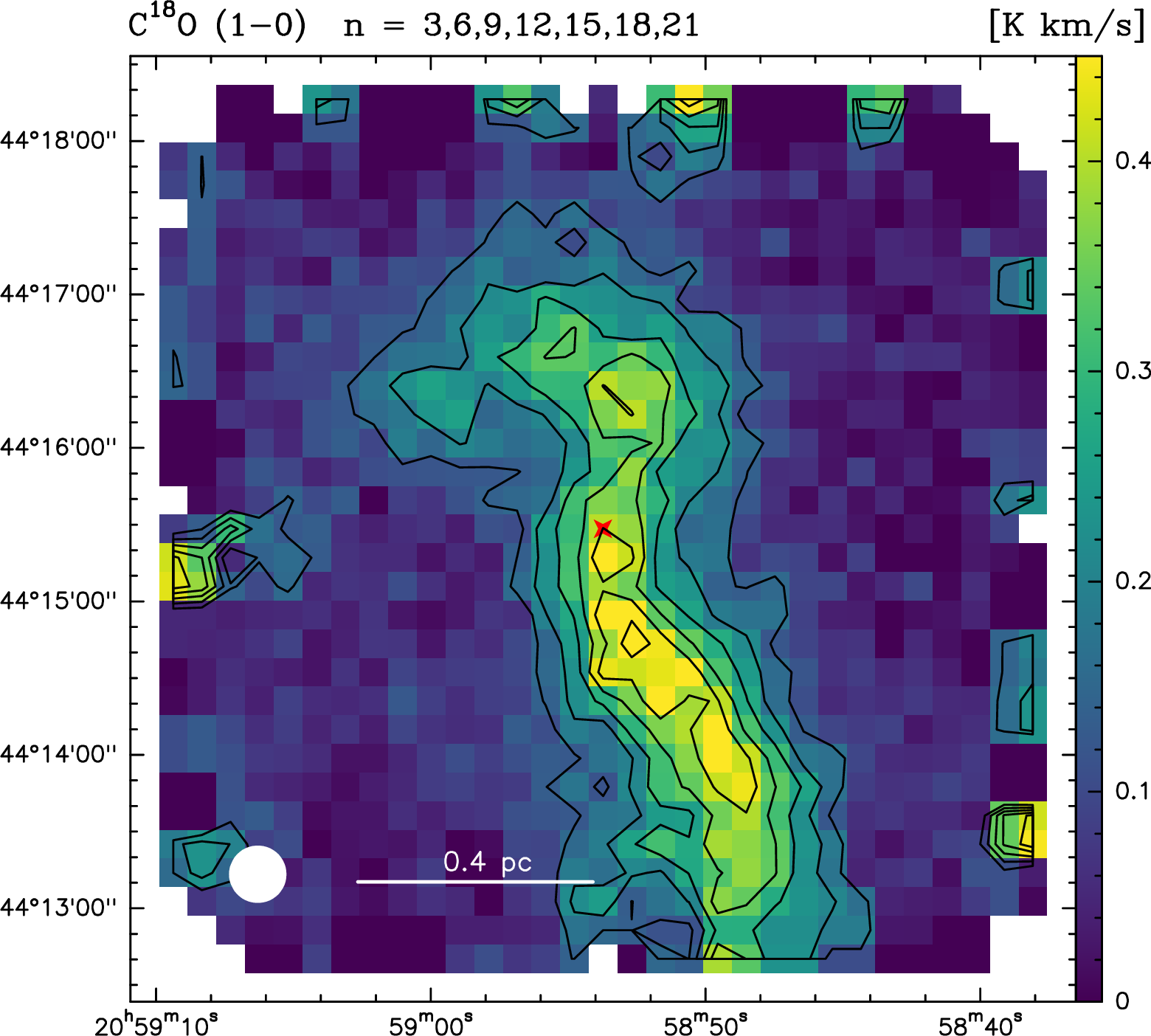}}
   \caption{Continued.}
\end{figure*}

\begin{figure*}[h]
\centering 
\vspace{-0.3cm}
\hspace{-0.9cm}
\subfigure{(a)}
\includegraphics[width=0.47\textwidth] {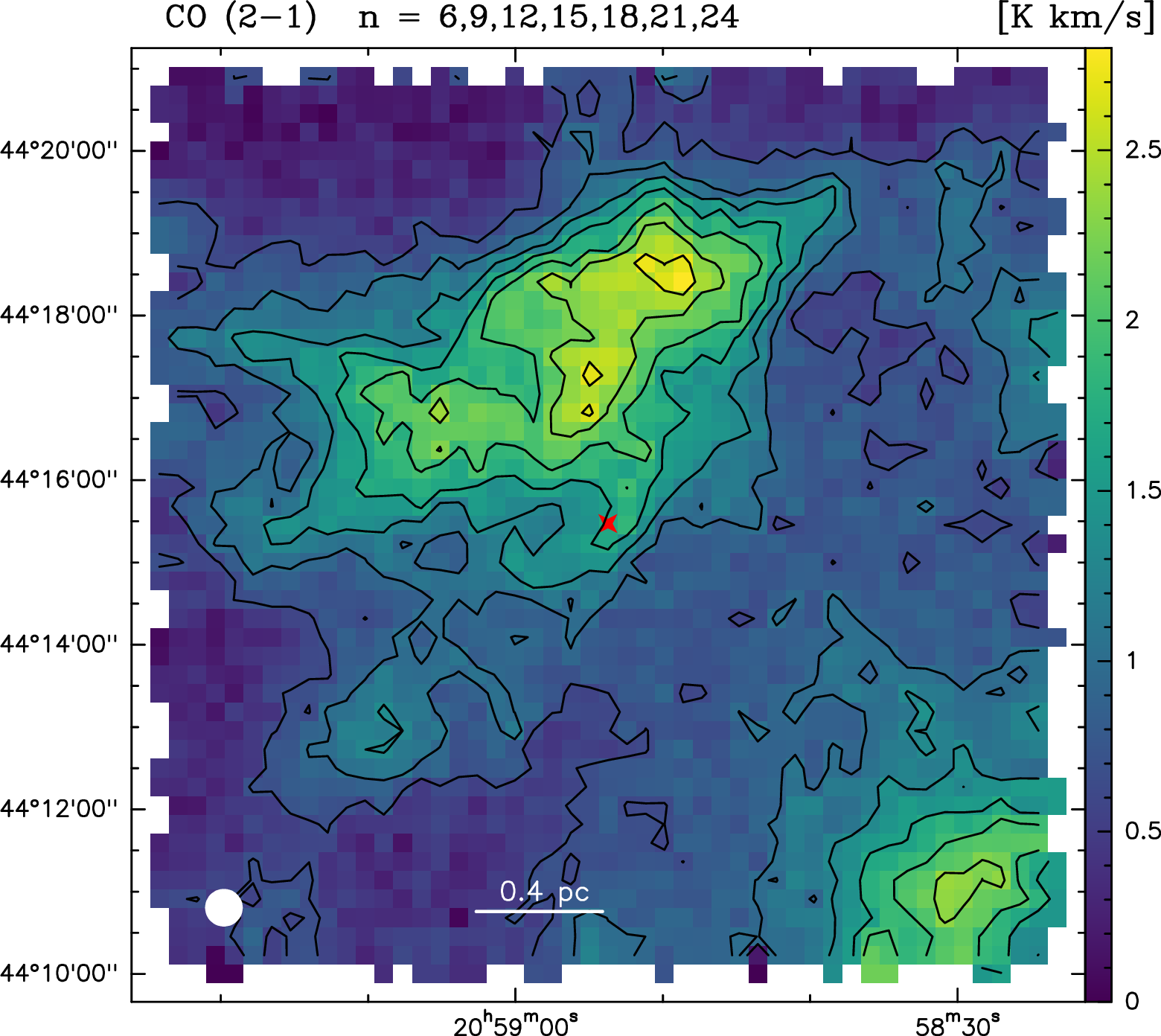}
\hspace{0.6cm}
\subfigure{(b)}
\includegraphics[width=0.47\textwidth]{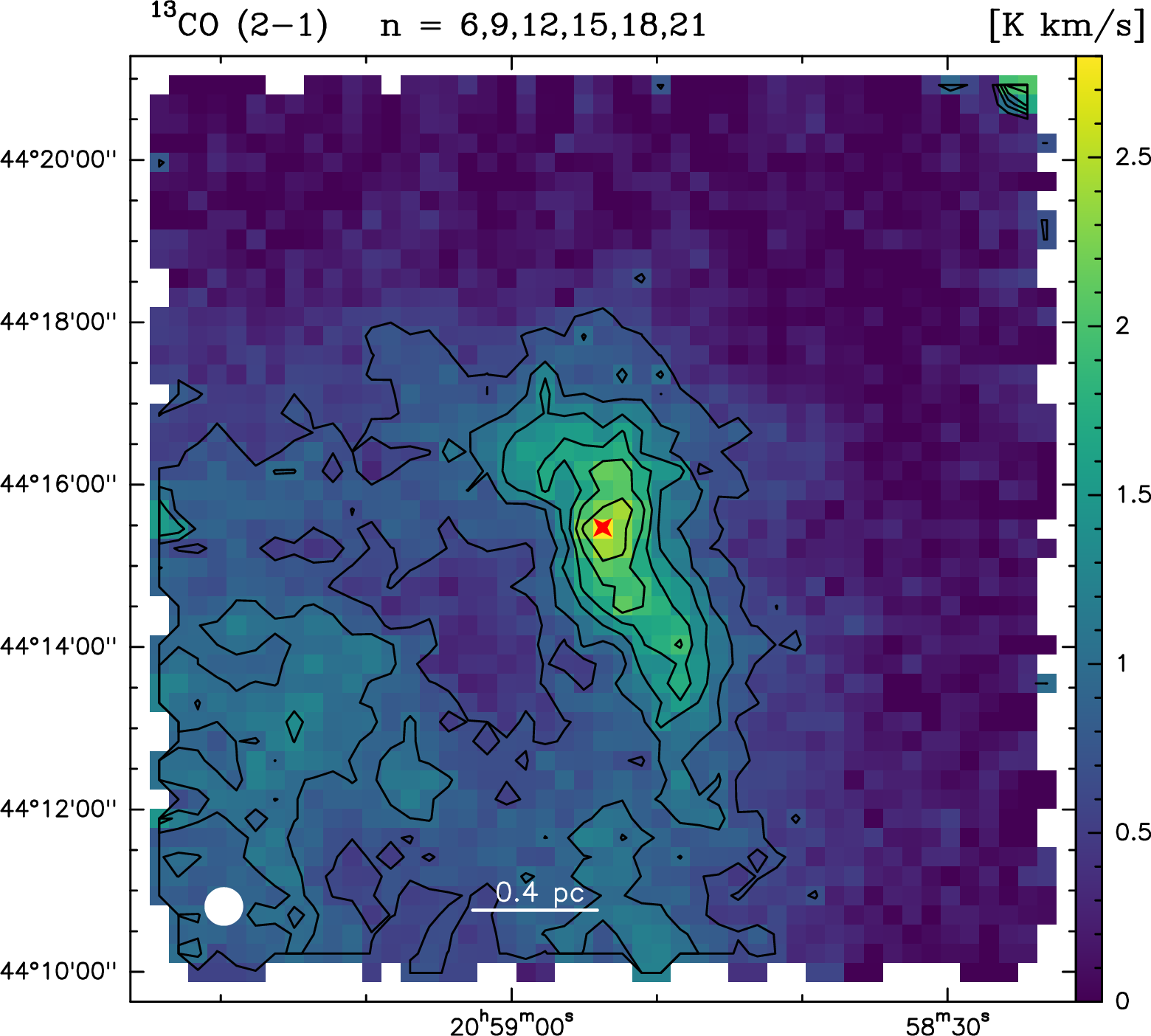} \\
\vspace{1cm}
\hspace{-0.9cm}
\subfigure{(c)}
\includegraphics[width=0.47\textwidth]{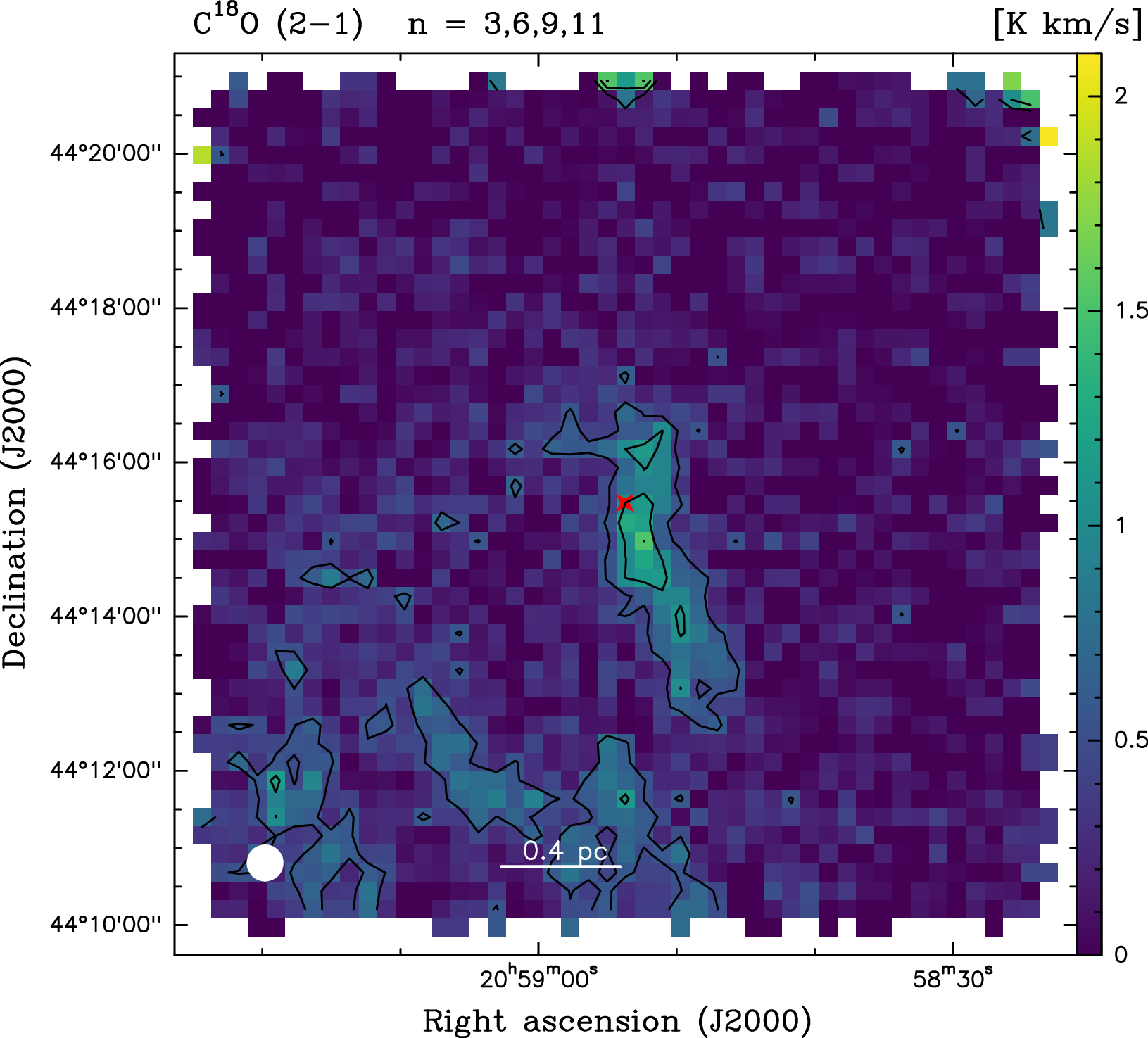}
\caption{Same as Fig.~\ref{fig:iram-maps}, but for the maps of (a) $^{12}$CO, (b) $^{13}$CO, and (c) C$^{18}$O (2--1) obtained with APEX.}
\label{fig:apex-maps-nflash}
\end{figure*}

\begin{figure}[!htbp]
\centering 
 \includegraphics[width=\columnwidth]{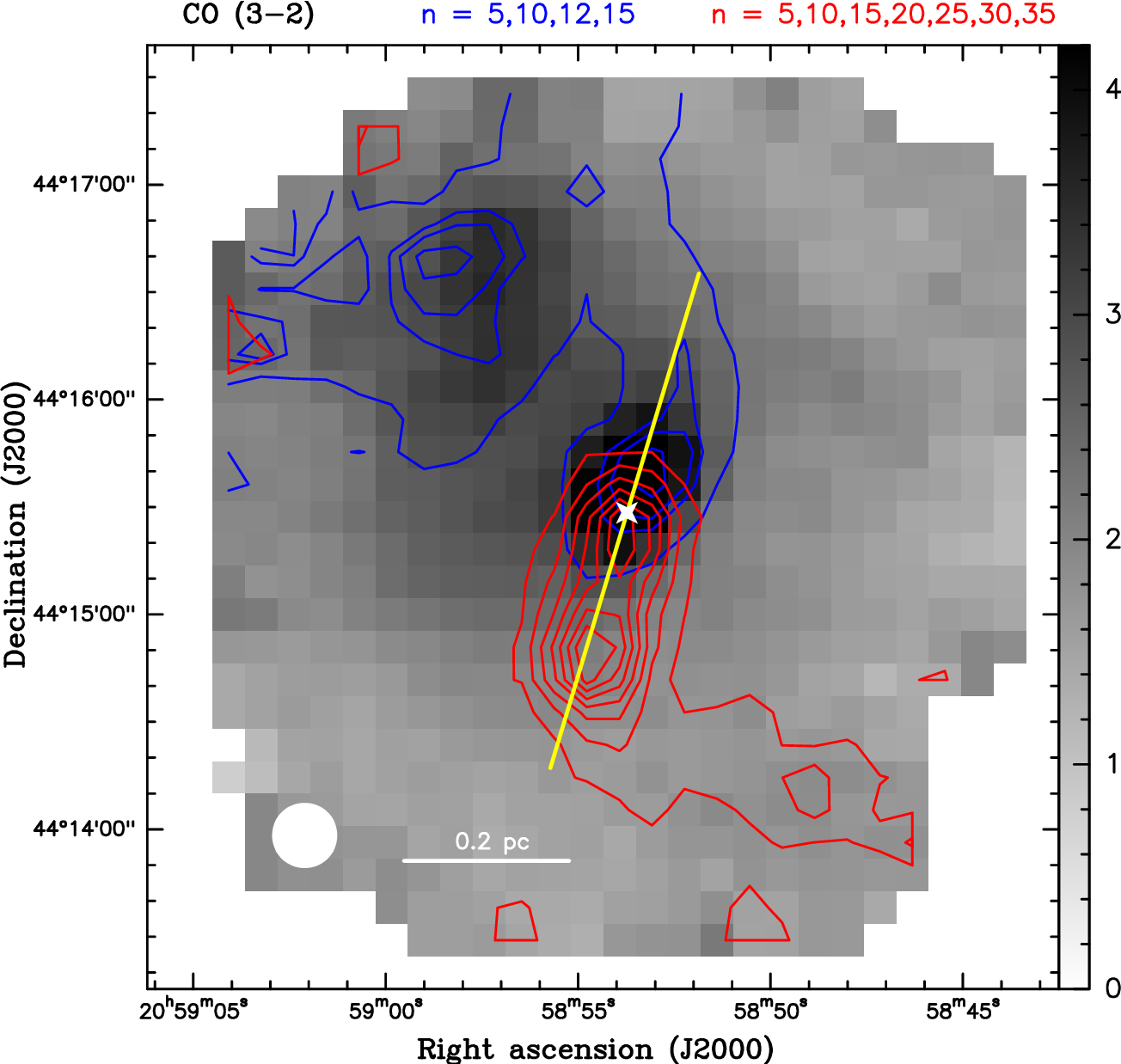}
\caption{Integrated intensity map of $^{12}$CO (3--2) (grey-scale image). The blue and red contours represent the blueshifted and redshifted wing emission integrated over the velocity ranges of $[-6,2]$\,km\,s$^{-1}$ and $[5.5,12]$\,km\,s$^{-1}$. The HPBW is shown in the bottom left corner in white. The position of V1057~Cyg is marked with a white cross. The yellow line shows the cut used for the position-velocity diagram shown in Fig.~\ref{fig:pv-diagram}. The position angle (PA) of the cut is $-17^{\circ}$ (East from North).}
\label{fig:apex-lasma-12COmaps-wing}
\end{figure}

\begin{figure}[!htbp]
\centering 
\hspace{0.2cm}
\includegraphics[width=\columnwidth]{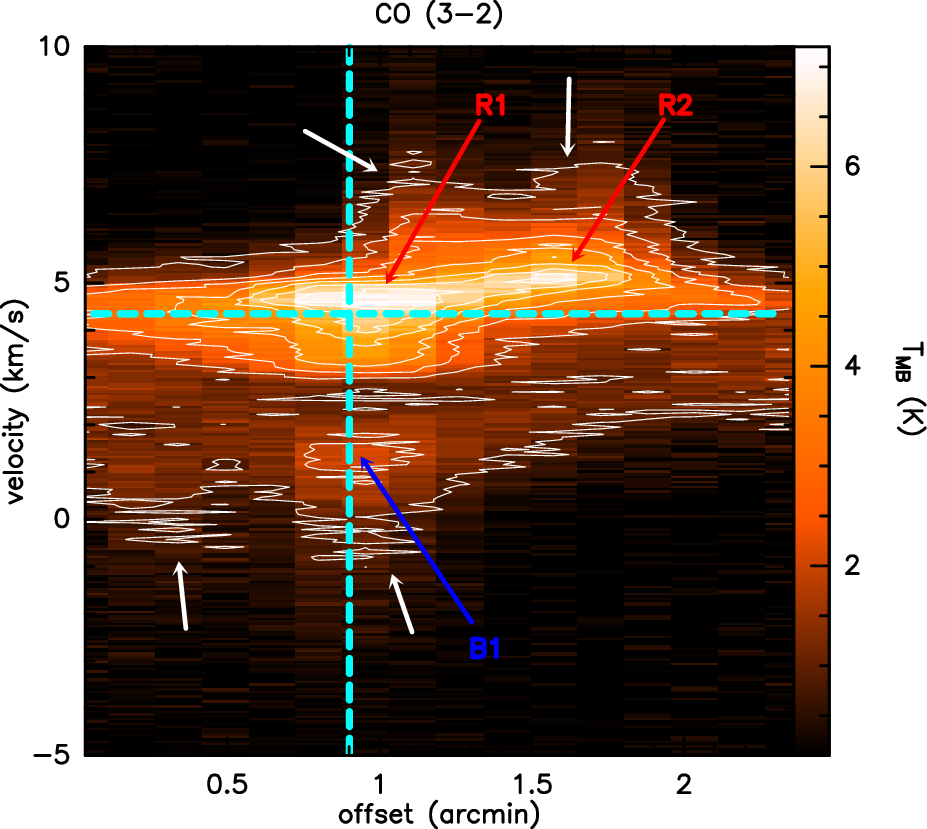}
\caption{Position-velocity (p-$\varv$) diagram along the yellow line shown in Fig.~\ref{fig:apex-lasma-12COmaps-wing}. The horizontal dashed cyan line corresponds to the systemic $\varv_{\rm lsr}$ velocity of 4.35\,km\,s$^{-1}$ \citep{szabo2023a}, while the vertical one corresponds to the source position. The contours start at 5$\sigma$ and end at 35$\sigma$ with 5$\sigma$ steps in between. 
Peaks of redshifted emission are labeled R1 and R2 and the peak of blueshifted emission is labeled B1. The white arrows highlight the presence of weak "fingers" that may trace episodic ejection events.}
\label{fig:pv-diagram}
\end{figure}

The molecular outflow of V1057~Cyg was first reported by \citet{evans1994} on the basis of the strong blueshifted wing emission detected in their $^{12}$CO 3--2 spectrum obtained with the Caltech Submillimeter Observatory 10.4 m telescope (CSO). These authors also mentioned the existence of maps towards this source as a private communication from S.~McMuldroch but we could not find any publication reporting them. 
The APEX spectrum has a similar line profile as the CSO spectrum, with a slightly more pronounced self-absorption, however, our more sensitive spectrum reveals the redshifted wing for the first time. 
The APEX map of the $^{12}$CO 3--2 emission, presented in Fig.~\ref{fig:apex-lasma-12COmaps-wing}, reveals a bipolar outflow that appears to originate from V1057~Cyg.
The outflow appears to be more collimated on the redshifted side, with discrete peaks to the south of the source position. 
Fig.~\ref{fig:pv-diagram} shows a $^{12}$CO (3--2) position-velocity (p-$\varv$) diagram taken along the cut shown in
Fig.~\ref{fig:apex-lasma-12COmaps-wing}.
The p-$\varv$ diagram shows several peaks 
(labelled R1, R2, and B1; R for red and B for blue), which are
more prominent in the redshifted emission. 
The highest velocity extensions are revealed by the first contour close to R1, which may mark an episodic ejection event \citep{plunkett2015}. Peak R1 appears close to the systemic velocity of the source, while peak R2 is located around 5\,km\,s$^{-1}$.
The fainter blueshifted peak, B1, appears at a velocity around 1\,km\,s$^{-1}$. 
The p-$\varv$ diagram also shows weak "fingers" (indicated by white arrows) revealed by the lowest contours of the $^{12}$CO emission. 
These results are further discussed in Sect.~\ref{subsec:outflow}. 

\subsection{Radiative transfer modelling}
\label{subsec:weeds-modeling}
The observed spectra were modeled using Weeds \citep[a GILDAS/CLASS extension for the analysis of millimeter and submillimeter data,][]{maret2011-weeds} to create synthetic spectra with the assumption of local thermodynamic equilibrium (LTE). The spectroscopic information required for the synthetic spectra was taken from the CDMS database \citep{muller2005,Endres2016}.
Multiple transitions are detected from molecules such as methanol (CH$_3$OH), formaldehyde (H$_2$CO), cyclopropenylidene (c-C$_3$H$_2$), carbon monosulfide (CS), thioethenylidene (CCS), cyanoacetylene (HC$_3$N), and thioformaldehyde (H$_2$CS) (see Tables~\ref{tab:res-LTE-model-res-pop-diag} and \ref{tab:appendix-iram-lines-list-fits}). 
These species were chosen because several of their lines show significant emission (i.e.,~$>3\sigma$) and their profiles do not show indication of outflow (i.e.,~extended wings, as seen in lines from $^{12}$CO and its isotopologues).
Only transitions detected with the IRAM~30-m telescope with a signal-to-noise ratio higher than 3$\sigma$ were used for the following analysis. The radiative transfer model presented here, along with the population diagrams in Sect.~\ref{subsec:pop-diagrams}, provides initial estimates of the temperatures and column densities of V1057~Cyg and its surrounding environment as observed in the single-dish data.

The purpose of applying a Weeds model to the chosen molecules is to improve the line identification and 
to investigate whether the spectrum is well-represented by an LTE model. 
The radiative transfer is computed by accounting for the line optical depth and the angular resolution of the observations. Weeds requires five input parameters for each molecule: the size of the emitting region ($\theta\rm{_s}$) in arcseconds, column density ($N_{\rm tot}$), temperature ($T_{\rm rot}$), line width ($\Delta \varv$ in km\,s$^{-1}$), and velocity offset ($\varv_{\rm off}$, with respect to the systemic velocity of the source). For each molecule, the line width was set to the average value of the line widths determined from the Gaussian fits of its detected transitions.
The column density and rotational temperature were set by eye, and subsequently adjusted until the best-fit was achieved. 
The population diagrams described in Sect.~\ref{subsec:pop-diagrams} were used to optimise the temperature, emission size, and, in turn, the column density, of the Weeds models to reach the best fits possible. 

We explored a range of source sizes between 1.5$\arcsec$ \citep[e.g.,][]{calahan2024b} and 35$\arcsec$ (larger than the largest beam size) and found the best fits with a source size of 25$\arcsec$ for HC$_3$N and 30$\arcsec$ for the rest of the molecular species. These results are consistent with the integrated intensity maps of the different species (see Fig.~\ref{fig:iram-maps}), and are further discussed in Sect.~\ref{subsec:pop-diagrams}.
\citet{calahan2024b} detected methanol emission towards V1057~Cyg using the NOEMA interferometer, which remained unresolved. They derived the column density and temperature of methanol with a similar LTE analysis, assuming a size of 1.5$\arcsec$ (Calahan priv.~comm.). Using the same parameters in our Weeds model of the single-dish data results in methanol lines that are much weaker than those detected with the 30-m telescope.
The 218.44\,GHz methanol line is detected in both the interferometric and the single-dish data.
The peak intensity of this line is 53\,mJy/beam in the NOEMA data (Calahan priv.~comm.) and 440 mJy/beam in the IRAM 30-m telescope spectrum, $\sim$8.3 times higher.
Since the CH$_3$OH emission in the NOEMA data was unresolved \citep{calahan2024b}, this difference implies that the single-dish observations trace extended emission that is not detected by the interferometer. With the single-dish telescope, we are likely capturing several components blended together, including disk, envelope, and surrounding cloud emission.

The best-fit column densities from our LTE modeling are used to derive abundances relative to H$_2$. 
We derived values for the H$_2$ column density by following the method used in \citet[][and references therein]{szabo2023a}. We used archival \textit{Herschel} data at 160, 250, and 350\,$\mu$m to fit a spectral energy distribution (SED) to the data and derive H$_2$ column densities at 25$\arcsec$ and 30$\arcsec$ resolution (corresponding to the sizes mentioned above) using the same method as used by \citet{szabo2023a}. 
However, unlike \citet{szabo2023a}, we did not use the 500\,$\mu$m data because of their relatively poor angular resolution (37$\arcsec$).
Based on the SED fit, we find H$_2$ column densities of 1.14$\times$10$^{22}$\,cm$^{-2}$ and 1.21$\times$10$^{22}$\,cm$^{-2}$ at the angular resolution of 30$\arcsec$ and 25$\arcsec$, respectively (beam centered on the optical stellar position), with typical uncertainties of about 10\%. The derived column densities and abundances are further discussed in Sect.~\ref{subsec:pop-diagrams} and listed in Table~\ref{tab:res-LTE-model-res-pop-diag}.
In a number of cases (especially for certain transitions of methanol) the LTE synthetic spectra do not match the observed spectra that well, whose excitation may deviate from LTE. 
Non-LTE modeling is beyond the scope of this work.

\subsection{Population diagrams and abundances}
\label{subsec:pop-diagrams}

\begin{figure*}[!htbp]
    \centering
    \hspace{-0.2cm}
    \subfigure(a){\includegraphics[width=0.3\textwidth]{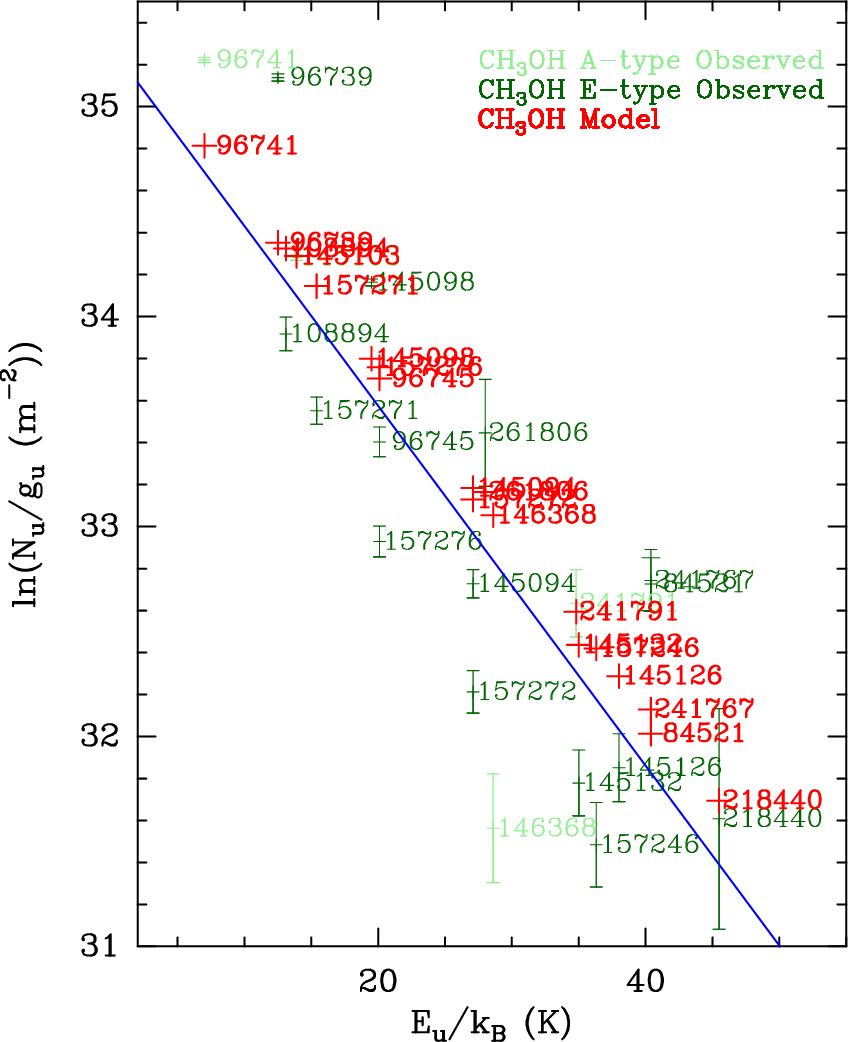}} 
    \hspace{0.1cm}
    \subfigure(b){\includegraphics[width=0.3\textwidth]{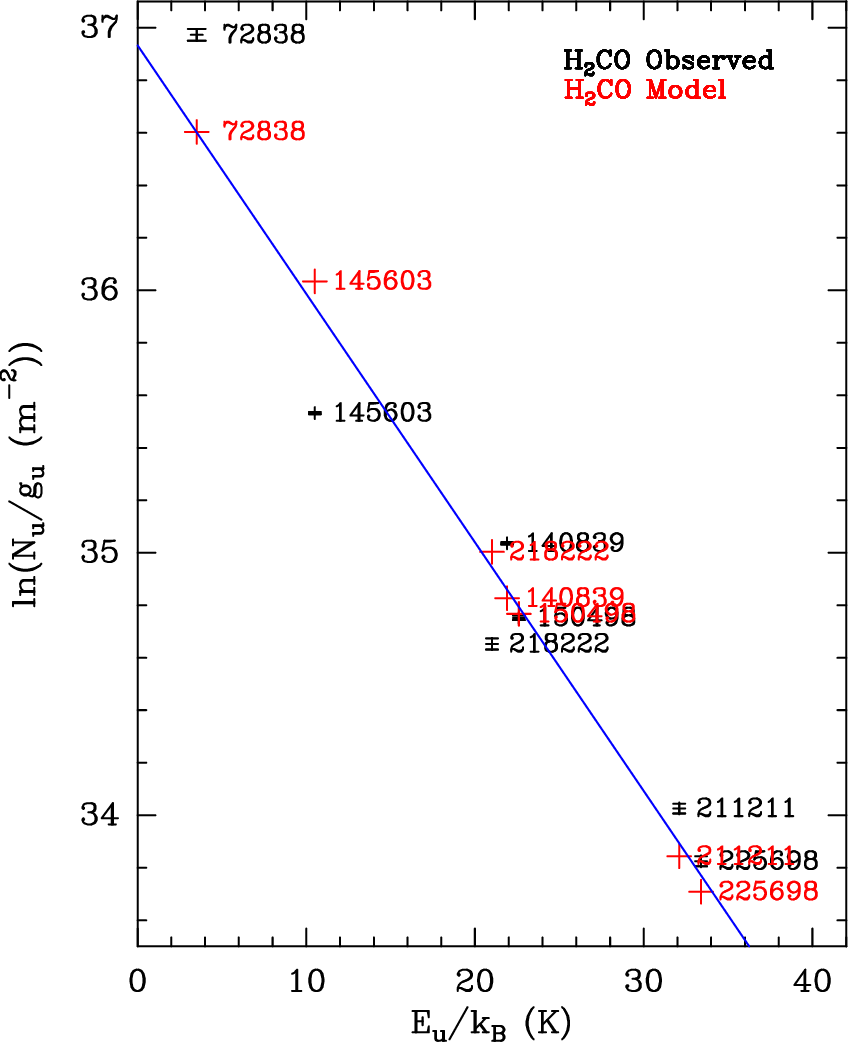}} 
    \hspace{0.1cm}
    \subfigure(c){\includegraphics[width=0.3\textwidth]{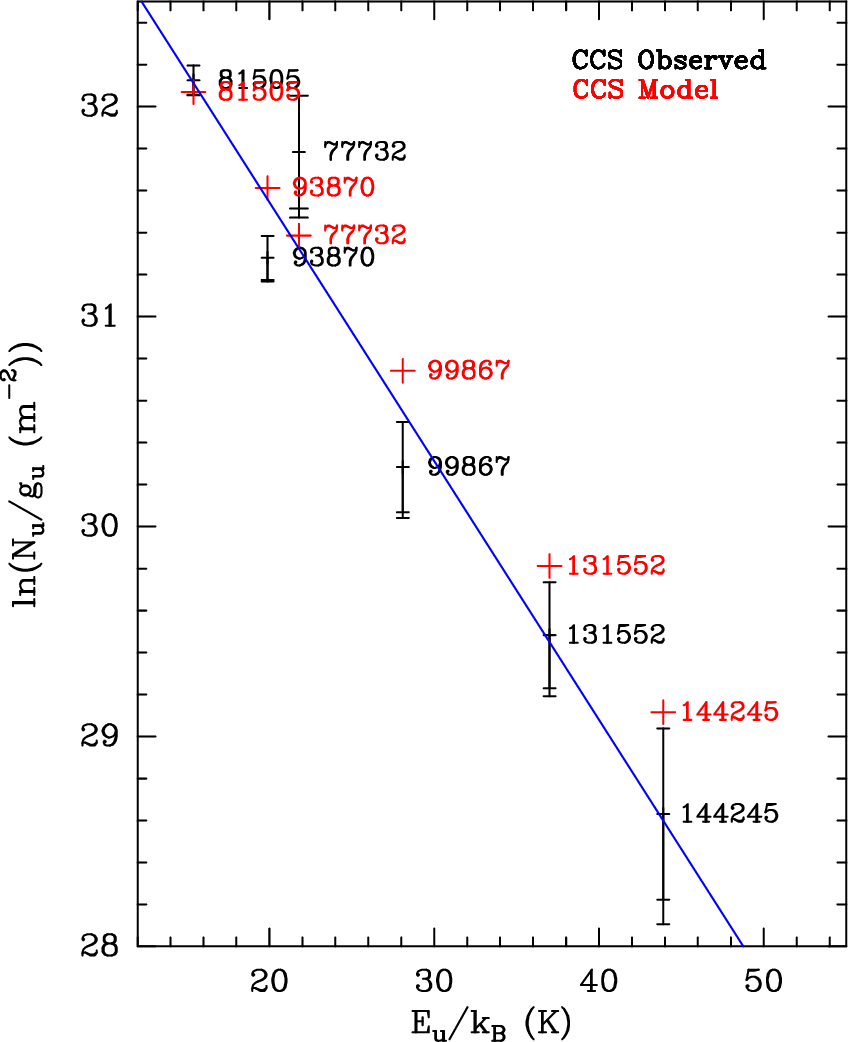}} 
    \\
    \vspace{0.2cm}
    \hspace{-0.2cm}
    \subfigure(d){\includegraphics[width=0.3\textwidth]{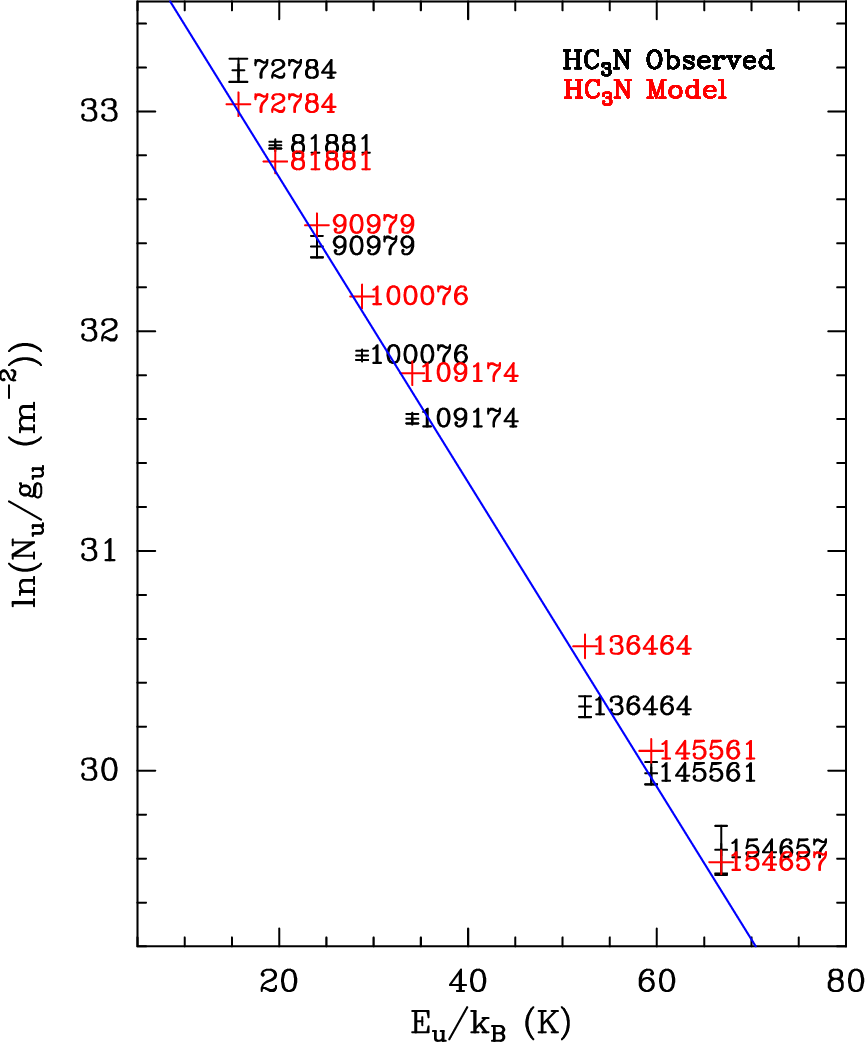}}
    \hspace{0.1cm}
    \subfigure(e){\includegraphics[width=0.3\textwidth]{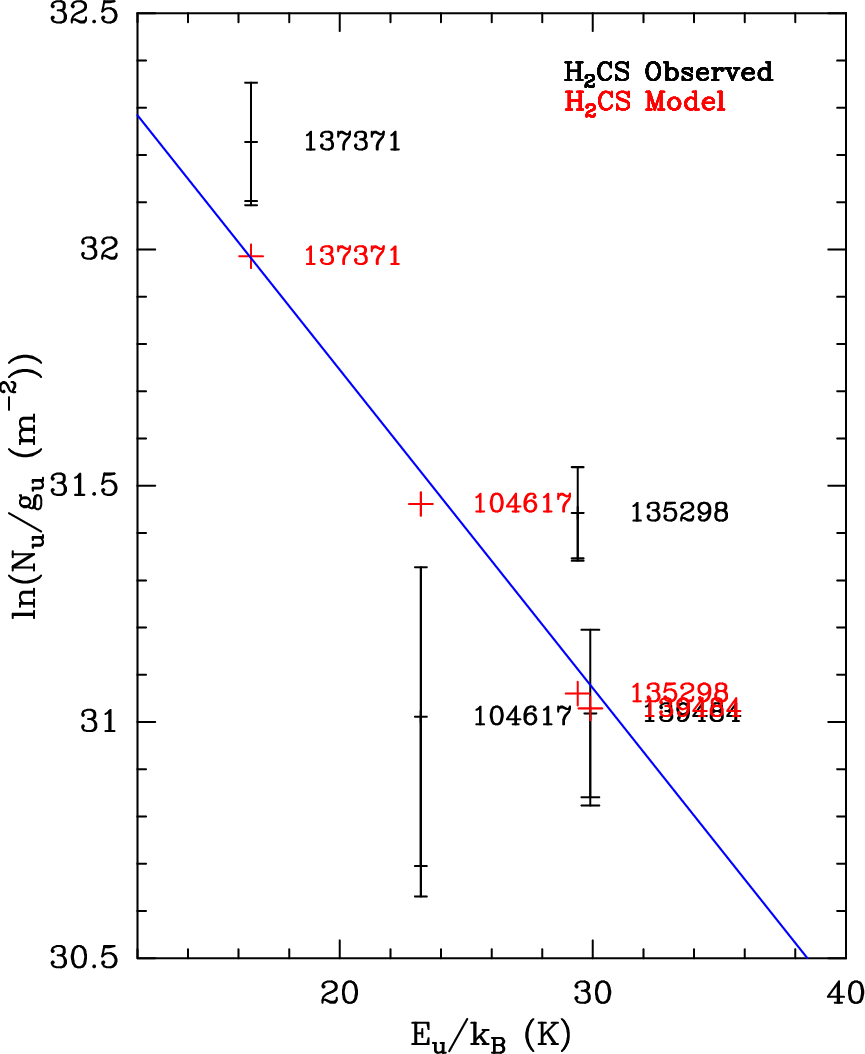}}
    \hspace{0.1cm}
    \subfigure(f){\includegraphics[width=0.3\textwidth]{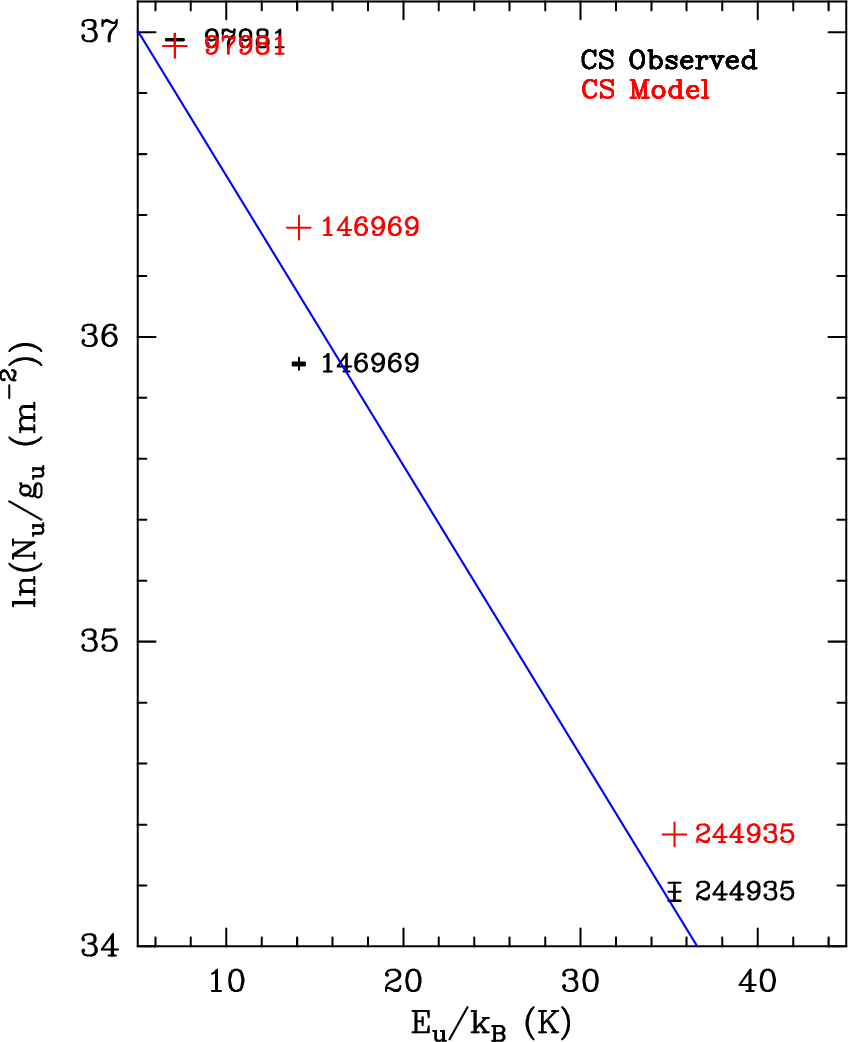}}\\
    \vspace{0.2cm}
    \hspace{-0.2cm}
    \subfigure(g){\includegraphics[width=0.3\textwidth]{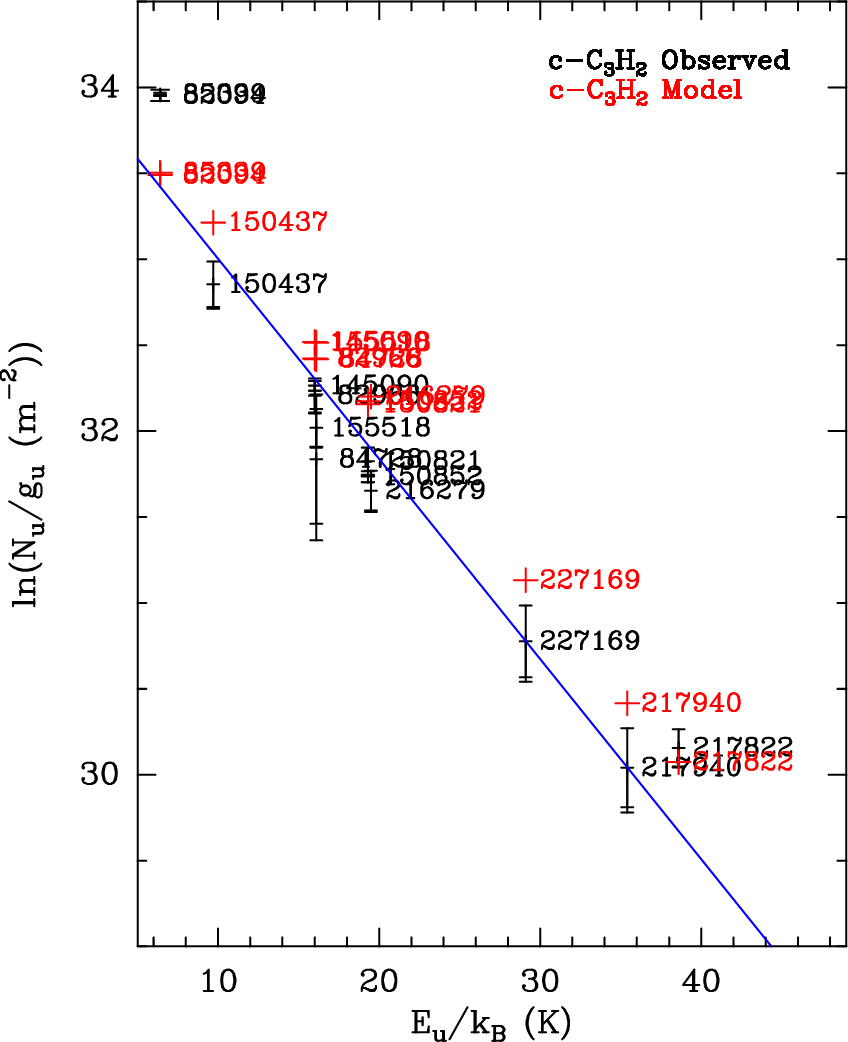}} \\
    \caption{Population diagrams of: (a) CH$_3$OH; (b) H$_2$CO; (c) CCS; (d) HC$_3$N, (e) H$_2$CS; (f) CS; (g) c-C$_3$H$_2$. The black points (green in the case of CH$_3$OH) represent the observations while the red points correspond to the Weeds synthetic spectra computed with the parameters given in Table~\ref{tab:res-LTE-model-res-pop-diag}. The blue line is a fit to the observed data points.}
    \label{fig:pop-diagrams}
\end{figure*}

The transitions used for the population diagrams are well isolated in the spectrum of V1057~Cyg, therefore it was not necessary to discard any transitions owing to contamination by other species \citep[a potential issue in line-rich sources; e.g.,][]{belloche2009,belloche2016}. 
The population diagrams were constructed using the integrated intensities of the observed profiles, and are presented in Fig.~\ref{fig:pop-diagrams}, while the fit results are given in Table~\ref{tab:res-LTE-model-res-pop-diag}.
The population diagrams presented in this paper account for beam dilution, using the sizes determined in Sect.~\ref{subsec:weeds-modeling}, including the changing beam size through the frequency coverage. Additionally, the diagrams are also corrected for optical depth, using the opacity correction factor $C_{\tau}=\frac{\tau}{1-e^{-\tau}}$ \citep[see][]{goldsmith1999} from the Weeds model.

In the case of methanol (Fig.~\ref{fig:pop-diagrams}a), we separated the $A$- and $E$-type transitions using two shades of green to test if there is a systematic difference, however, we found none. 
The population diagram is also a good way to see hints of maser activity of known maser species. The methanol emission in the 84.5\,GHz transition does not show deviation from the rest of the transitions in Fig.~\ref{fig:pop-diagrams}(a), so does not indicate maser activity. However, we note, that the best-fit model with Weeds underestimates the line strength. The other common maser transition at 95.1\,GHz was not detected in this survey.

In all cases, a linear trend is present without significant departure from the linearity that could indicate the effect of optically thick emission or sub-thermal excitation \citep[e.g.,][]{velilla-prieto2017-ik-tauri-line-survey}. 
The assumed emission size can affect the spread of data points in the population diagram due to the frequency-dependent beam size, while the opacities were found to be $\ll 1$ for all transitions, and thus do not affect the shape of the diagrams. We checked emission from CH$_3$OH, from which multiple lower-frequency transitions (below 120\,GHz, $\sim$3\,mm window) are detected, and we found that assuming emission sizes smaller than 30$\arcsec$ increases the spread of the data points. This leads to a larger gap between the low- and high-frequency transitions, which in turn compromises the fit to the data.

From the population diagram method, we find column densities ranging from 2.8$\times10^{13}$\,cm$^{-2}$ for CH$_3$OH, down to 1.4$\times10^{12}$\,cm$^{-2}$ for H$_2$CS, and rotational temperatures ranging from 8.1\,K to 14.8\,K.
The Weeds models were optimised on the basis of the population diagram results and the two methods agree within the errors of the population diagram method for both the column density and rotational temperature values.

\begin{table*}
\small
\caption{Parameters of our best-fit LTE model and population diagram fit of methanol, formaldehyde, thioethenylidene, cyanoacetylene, thioformaldehyde, carbon monosulfide, and cyclopropenylidene.}  
\centering                                      
\begin{tabular}{cccccccc}          
\hline\hline       
\multirow{3}{*}{Molecule} & \multicolumn{5}{c}{Parameters of the best-fit LTE model} & \multicolumn{2}{c}{Results of the population diagram method} \\
\cline{2-6} 
\cdashline{7-8} 
 & \multicolumn{1}{c}{Source size\tablefootmark{(a)}} & \multicolumn{1}{c}{$N_{\rm Weeds}$\tablefootmark{(b)}} & \multicolumn{1}{c}{$T_{\rm rot}$\tablefootmark{(c)}}  & \multicolumn{1}{c}{$\Delta \varv$\tablefootmark{(d)}} & \multicolumn{1}{c}{$N_{\rm X}/N_{\rm H_2}$\tablefootmark{(e)}} & \multicolumn{1}{c}{$N_{\rm pop}$\tablefootmark{(f)}} & \multicolumn{1}{c}{$T_{\rm rot}$\tablefootmark{(g)}}  \\
& \multicolumn{1}{c}{($\arcsec$)} & \multicolumn{1}{c}{(cm$^{-2}$)} & \multicolumn{1}{c}{(K)} & \multicolumn{1}{c}{(km\,s$^{-1}$)} & & \multicolumn{1}{c}{(cm$^{-2}$)} & \multicolumn{1}{c}{(K)}  \\
\hline \hline                                
CH$_3$OH  & 30$\arcsec$ & $3.3\times10^{13}$ & $12.0$ & $1.42$ & $2.8\times10^{-09}$ & $2.8\times10^{13}\pm1.0\times10^{13}$    & $11.7\pm1.9$  \\ 
H$_2$CO   & 30$\arcsec$ & $2.0\times10^{13}$ & $9.7$ & $1.27$  & $1.7\times10^{-09}$ & $1.6\times10^{13}\pm4.2\times10^{12}$    & $10.6\pm1.3$ \\ 
CCS       & 30$\arcsec$ & $3.0\times10^{12}$ & $9.4$   & $1.14$  & $2.8\times10^{-10}$ & $3.3\times10^{12}\pm1.2\times10^{12}$    & $8.1\pm0.8$  \\ 
HC$_3$N   & 25$\arcsec$ & $5.0\times10^{12}$ & $14.5$ & $1.19$ & $3.5\times10^{-10}$ & $4.2\times10^{12}\pm5.7\times10^{11}$    & $14.4\pm0.7$  \\ 
H$_2$CS   & 30$\arcsec$ & $1.5\times10^{12}$ & $14.0$ & $1.29$ & $1.2\times10^{-10}$ & $1.4\times10^{12}\pm1.5\times10^{12}$    & $14.8\pm9.5$  \\ 
CS        & 30$\arcsec$ & $2.2\times10^{13}$ & $10.5$ & $1.33$ & $1.9\times10^{-09}$ & $1.7\times10^{13}\pm5.4\times10^{12}$    & $10.5\pm1.5$  \\
c-C$_3$H$_2$ & 30$\arcsec$ & $6.0\times10^{12}$ & $9.0$ & $1.09$ & $4.7\times10^{-10}$ & $5.5\times10^{12}\pm1.5\times10^{12}$    & $8.6\pm0.7$  \\
\hline \hline
\end{tabular}
\label{tab:res-LTE-model-res-pop-diag}    
\tablefoot{
\tablefoottext{a}{Emission diameter (FWHM).}
\tablefoottext{b}{Column density.}
\tablefoottext{c}{Temperature.}
\tablefoottext{d}{Average line width of all detected transitions.}
\tablefoottext{e}{Abundance values with respect to the H$_2$ column density using $N_{\rm tot}$ values from the best-fit LTE model.}
\tablefoottext{f}{Column density from the population diagram method.}
\tablefoottext{g}{Rotational temperature from the population diagram method.}}
\end{table*}

\section{Discussion}
\label{sec:discussion}

\subsection{Large- and small-scale structures in the environment of V1057~Cyg}
\label{disc:subsec-large-small-scale}

In Sect.~\ref{subsec:maps} we presented the integrated intensity maps of several molecular species, in most cases tracing lower energy transitions (Figs.~\ref{fig:iram-maps} and \ref{fig:apex-maps-nflash}). Here, we discuss the different structures traced in our single-dish observations in the environment of our target and compare our findings with those in the literature.

The observed molecular emission traces a ridge structure seen in most of the integrated intensity maps with two peaks offset from V1057~Cyg: one towards the north at $\sim$14.7$\arcsec$ and one towards the southwest at $\sim$1.8$\arcmin$. 
The interferometric $^{13}$CO (1--0) data presented by \citet{feher2017} show a structure to the north of V1057~Cyg referred to as clump B with a similar offset to that seen in the single-dish data. Their maps show the close vicinity of V1057~Cyg and do not cover the southwestern extension.
Extended 850\,$\mu$m emission distributed over large scales has been reported by \citet{sandell2001}, who used the Submillimter Common User Bolometer Array (SCUBA) on the James Clerk Maxwell Telescope. These authors found V1057~Cyg to be located in a narrow ridge, traced by the dust continuum emission, extending to the north and southwest of the star. Our single-dish data extend further to the southwest providing more information. The ridge to the southwest seen in our molecular line data is visible in \textit{Herschel} dust-based H$_2$ column density maps \citep{szabo2023a}.
The dust map, along with our integrated intensity maps, further suggests a large-scale morphology of a filamentary cloud, in which V1057~Cyg is located, as proposed by \citet{sandell2001}.

Ridge structures or filaments, similar to what is seen around V1057~Cyg are typically found in star forming regions \citep[e.g.,][for a recent review]{andre2010,hacar2023-review}. A famous example is the Taurus Molecular Cloud 1 (TMC-1), where numerous cloud cores are located along a ridge structure \citep[e.g.,][]{hirahara1992-tmc1,langer1995}. Naturally, the size of the TMC-1 filaments are much larger \citep[$\sim$10\,pc with cores around 0.1\,pc in size, e.g.,~L1495 in TMC-1,][]{Schmalzl2010} than the ridge we see around V1057~Cyg ($\sim$1--1.5\,pc) within our field of view. Ridges with similar sizes to that seen around V1057 Cyg are found in many low-mass star forming regions, with NGC~1333 being a prominent example \citep[e.g.,][]{hacar2017}.
We speculate that the condensation detected to the southwest of V1057~Cyg in high density tracers (see Sect.~\ref{subsec:maps}) represents a secondary, possibly protostellar, dense core in the vicinity of V1057 Cyg.
The spatial extent of the 850\,$\mu$m data of \citet{sandell2001} do not extend far enough to the south to localise the peak, however, the \textit{Herschel} derived H$_2$ column density map shows a peak at the position of the southwestern peak seen in our single-dish molecular line data \citep[see, e.g.,][and Fig.~C.1 in the latter]{green2013,szabo2023a}. 

Previously, the NH$_3$ (1,1) transition towards V1057~Cyg has been detected with a beam size of $\sim$37$\arcsec$ using the Effelsberg 100-m telescope \citep{szabo2023a}. In light of our 30-m single-dish map of, for example, the N$_2$H$^+$ (1--0) emission, it is plausible that the NH$_3$ spectrum is contaminated by the northern peak, located at $\sim$15$\arcsec$ offset of V1057~Cyg.
High-angular resolution observations of high-density tracers with interferometers might probe if V1057~Cyg is an isolated protostar or a member of a cluster similarly to the RNO~1B/1C FUor system \citep{quanz2007}.

\subsection{Temporal variations in the molecular outflow}
\label{subsec:outflow}
Both single-dish and interferometric observations of $^{12}$CO (and its isotopologues) have been used to identify  molecular outflows associated with FUors and FUor-like objects \citep[e.g.,][and references therein]{evans1994,kospal2011b,feher2017,kospal2017b,cruz-saenz-de-miera2023}. 
As presented in Sect.~\ref{subsec:maps}, the molecular outflow of V1057~Cyg was first reported by \citet{evans1994}, based only on the presence of blueshifted wing emission in a single $^{12}$CO 3--2 spectrum of the source observed with the Caltech Submillimeter Observatory 10.4-meter telescope (CSO). 
Our new $^{12}$CO (3--2) APEX map reveals a clear bipolar outflow morphology (Fig.~\ref{fig:apex-lasma-12COmaps-wing}), with
the redshifted component reported here for the first time.
We note that the CSO spectrum of \citet{evans1994}, which peaks at 8.7\,km\,s$^{-1}$, is shifted by about 4.5\,km\,s$^{-1}$ compared to the APEX one. This discrepancy is puzzling.
Although the Gaussian fit is unreliable for the $^{12}$CO data, it would still be closer to our $^{13}$CO and C$^{18}$O (1--0) $\varv_{\rm lsr}$ of 3.86\,km\,s$^{-1}$ and 4.09\,km\,s$^{-1}$, and also to those derived from interferometric data by \citet{feher2017} for $^{13}$CO and C$^{18}$O (1--0) with values of 4.05\,km\,s$^{-1}$ and 4.10\,km\,s$^{-1}$, respectively.
We suspect that the frequency axis of the spectrum published in \citet{evans1994} was not properly calibrated.

Figure~\ref{fig:apex-lasma-12COmaps-wing} reveals the outflow originating from V1057~Cyg, with the redshifted lobe pointing to the South-South-East and the blueshifted one to the North-North-West. 
We measured the length of the lobes, $R_{\rm lobe}$, and maximum velocities, $\varv'_{\rm max}$, to estimate the dynamical time, $t_{\rm d}$, of the outflow with the formula $t_{\rm d}=\frac{R_{\rm lobe}}{\varv'_{\rm max}\tan i}$, with $i$ the inclination of the outflow axis to the line of sight \citep[summarised in Table~\ref{tab:outflow-lobe-td},][]{beuther2002}. 
The outflow map shows bipolar morphology with no overlap between the redshifted and blueshifted emissions, implying an intermediate inclination angle. Therefore we estimated $t_{\rm d}$ assuming a range of inclinations between 10\,$^\circ$ and 80\,$^\circ$ and obtained values ranging from 5$\times$10$^{3}$\,yrs to 234$\times$10$^{3}$\,yrs.
Previous studies analysing the disk of V1057~Cyg derived an inclination of 62\,$^\circ$ \citep{liu2018,szabo2021}.
In the following we adopt the dynamical times estimated with 62\,$^\circ$, which is 15~000\,years for the blue, and 22~000\,years for the red lobe of the outflow of V1057~Cyg.
A survey of 20 FUors, also using the APEX~12-m telescope \citep{cruz-saenz-de-miera2023}, found dynamical times on the order of thousands of years, from 3$\times$10$^{3}$\,yrs to 130$\times$10$^{3}$\,yrs, suggesting that the detected outflows are most likely not related to the current outbursts in the sample. V1057~Cyg has been in outburst for about 50\,years \citep{herbig1977,herbig2003,clarke2005,szabo2021}, therefore the outflow cannot result from the ongoing outburst alone. 
High velocity gas may have been ejected in the recent outburst, however, this gas would be confined to small-scales, remaining unresolved in single-dish observations \citep[see also the results of][]{cruz-saenz-de-miera2023}.

\begin{table}
\caption{Properties of each lobe of the outflow of V1057~Cyg determined from the APEX $^{12}$CO 3--2 observations. 
}  
\centering                                      
\begin{tabular}{ccrccr}          
\hline \hline 
Lobe & $\varv_{\rm max}$$^{(a)}$ & $\varv'_{\rm max}$$^{(b)}$ & $R^{\rm proj}_{\rm lobe}$$^{(c)}$ & $i^{(d)}$ & $t_{\rm d}$$^{(a)}$  \\
& (km\,s$^{-1}$) & (km\,s$^{-1}$) & ($\arcsec$) & ($^{\circ}$) &  (10$^3$\,years) \\
\hline \hline
Blue   & $-6$ & $10.35$ & $70$ & $10$ & $163$ \\
Red    & $12$ & $7.65$  & $74$ & $10$ & $235$ \\
\hline
Blue   & & & & $45$ & $29$ \\
Red    & & & & $45$ & $41$ \\
\hline
Blue   & & & & $62$ & $15$ \\
Red    & & & & $62$ & $22$ \\
\hline
Blue   & & & & $80$ & $5$ \\
Red    & & & & $80$ & $7.3$ \\
\hline \hline
\end{tabular}
\label{tab:outflow-lobe-td} 
\tablefoot{
\tablefoottext{a}{Minimum (blueshifted) and maximum (redshifted) velocities.}
\tablefoottext{b}{Maximum velocities $\varv'_{\rm max}=\left|\varv_{\rm max}-\varv_{\rm sys}\right|$, where $\varv_{\rm sys}=4.35$\,km\,s$^{-1}$.}
\tablefoottext{c}{Projected length of a lobe along the outflow axis.}
\tablefoottext{d}{Input inclination for estimating the dynamical time of the outflow lobe.}
\tablefoottext{e}{Dynamical time of the outflow lobe.}
}
\end{table}

Signs of potential past ejection activity in V1057~Cyg have not yet been explored.
One way to search for historical mass ejection events is to inspect the kinematics of the outflow in a position-velocity (p-$\varv$) diagram, which can be used as a "fossil record" 
\citep[e.g.,][]{arce-goodman2001,zhou2024-episodic-accretion}.
The p-$\varv$ diagram displayed in Fig.~\ref{fig:apex-lasma-12COmaps-wing} shows 
discrete peaks and hints at structures similar to the turbulent jet $+$ jet bow shock model, which can be interpreted as episodic accretion/ejection events \citep[see, e.g.,][]{plunkett2015}. 
In this model, the episodic variation in the jet velocity produces an internal bow shock driving an internal shell in addition to the terminal shock. 
The $^{12}$CO emission traces cool (i.e.,~lower than 100\,K) swept-up material, providing a record of the timing history of mass-loss event(s). Our p-$\varv$ diagram shows blue- and redshifted peaks (R1, R2, and B1 in Fig.~\ref{fig:pv-diagram}), with velocity offsets from the systemic velocity of V1057~Cyg.
Moreover, the p-$\varv$ diagram shows weak "fingers", revealed by the lowest contour in the blue- and redshifted emission (indicated by white arrows in Fig.~\ref{fig:pv-diagram}). This is reminiscent of the results of \citet{plunkett2015}, who report a saw-like pattern along the extent of the $^{12}$CO emission of the Class\,{\footnotesize 0} source CARMA-7.
\cite{plunkett2015} proposed that the peaks may indicate shocks and the fingers (in their case the saw-like structure) may be evidence for a period where the faster ejecta overtakes the slower one. Furthermore, previous ejections could entrain the outflow material, therefore appearing as clumpy (see peaks R1, R2, and B1 in Fig.~\ref{fig:pv-diagram}), creating distinct shocks when a newer ejecta overtakes the previous, older one \citep{plunkett2015}. 
The clumpy structure of the $^{12}$CO emission of V1057~Cyg might indicate episodic ejection mechanisms, rather than a smooth, continuous/homogeneous outflow. The prominent peaks and weak fingers complemented with the estimated dynamical times further suggest past outburst activity of V1057~Cyg.

\subsection{Brief comparison to other FUors}
\label{subsec:overview-molecules}
Millimeter-wavelength observations can provide information on both the small- and large-scales in the environments of YSOs. The majority of previous studies of FUors, whether single-dish or interferometric observations, have mainly focused on $^{12}$CO and its rarer isotopologues of $^{13}$CO and C$^{18}$O to probe the molecular outflows \citep[e.g.,][and references therein]{evans1994,feher2017,ruiz-rodriguez2017-hbc494-alma,ruiz-rodriguez2017-v883ori,hales2020-v582aur-v900mon,wendeborn2020,cruz-saenz-de-miera2023,nogueira2023-hbc494-alma}. However, employing other molecular species has been increasing over the past decade using both single-dish and interferometric facilities. 
Despite being a famous member within its class and a classical FUor, V1057~Cyg had not yet been systematically searched for molecular emission in the millimeter domain, which similar to the optical and near-infrared in depth-studies \citep[see, e.g.,][]{herbig2003,szabo2021,szabo2022} could be valuable when classifying new FUors.
Prior to this work, molecular line observations towards our target mainly focused on $^{12}$CO (and its isotopic species), which has been observed since the 1970s, following the outburst. $^{12}$CO and $^{13}$CO 1--0 line emission was first reported by \citet{bechis1975}, followed by the detection of other transitions of $^{12}$CO and its isotopic species throughout the decades, conducted both with single-dishes and interferometers (see Sect.~\ref{subsec:line-characteristics}).
Detection of the simple molecule CN (3--2) was also reported towards V1057~Cyg with a peak $\varv_{\rm lsr}$ comparable to the $^{12}$CO observations \citep{stauber2007}.
The recent NOEMA observations by \cite{calahan2024b} show that V1057~Cyg stands out among a small sample of surveyed eruptive YSOs, with detections of several COMs (e.g.,~CH$_3$OH, CH$_3$CN, CH$_3$OCHO), sulfur-bearing molecules (e.g.,~SO, SO$_2$), and others.
In the following, we give a brief overview of the most relevant molecular line studies targeting FUors and briefly compare their results to ours. These include both single-dish and interferometric observations. For the relevant comparison, we mention RNO~1B/1C, L1551~IRS5, and V883~Ori. 
It is important to note that beam dilution is an important factor when comparing single-dish and interferometric studies, since it limits the capability of single-dish telescopes to detect compact emission.

For the overview, we constructed Table~\ref{tab:comparison}, which lists a selection of species ranging from simple molecules to COMs. The table indicates for each molecule the sources for which a detection has been reported.
The FUor binary RNO~1B/1C, is one of the earliest documented examples of molecular emission associated with FUors \citep[apart from CO,][]{evans1994}, which include detection of SiO (5--4), H$_2$CO (3--2), CH$_3$OH (4--3), C$^{18}$O (2--1), CS (7--6 and 2--1), and HCO$^+$ (4--3) emission \citep{McMuldroch1995}. In this case, the CO emission was found to trace the bipolar outflow and CS (2--1) to probe the dense cloud core around the binary. In all cases, they find relatively cold (below 35\,K) emission tracing the environment of RNO~1B/1C, which is similar to the cold molecular emission detected towards V1057~Cyg, however the rotational temperatures that we derived toward V1057~Cyg are lower than toward RNO~1B/1C. The column densities are about an order of magnitude lower for V1057~Cyg than in RNO~1B/1C. Similarly to RNO~1B/1C, we also detect further indication of dense material \citep[together with the previous NH$_3$ observations and compact morphology of dust emission,][]{szabo2023a} from a number of transitions of CS, CCS, HCO$^+$ (and their isotopologues) or N$_2$H$^+$ \citep[see also,][and references therein]{McMuldroch1995,shirley2015,yamamoto2017-astrochem-book}, traced on single-dish scales. 

\begin{table*}
\small
\caption{Brief overview (for visualisation purposes) of the relevant molecular emission reported towards RNO~1B/1C, L1551~IRS5, V883~Ori, and V1057~Cyg discussed in Sect~\ref{subsec:disc-comparison}. For the table L1551~IRS5 is abbreviated as L1551. Only the main isotopologues are listed below, in many cases the secondary isotopologues of the main ones are also detected.}         
\centering                                      
\begin{tabular}{llll}          
\hline \hline 
Species & Detected & Species & Detected \\
        & (source name/names) & & (source name/names) \\
\hline \hline 
CN & L1551, V1057~Cyg & HDO & L1551, V883~Ori, V1057~Cyg \\
CO & RNO~1B/1C, L1551, V883~Ori, V1057~Cyg & HDCO & L1551, V883~Ori, V1057~Cyg  \\
SO & L1551, V883~Ori, V1057~Cyg & NH$_2$D & L1551 \\
SiO & RNO~1B/1C, V883~Ori & N$_2$D$^+$ & L1551 \\
CS &  RNO~1B/1C, L1551, V883~Ori, V1057~Cyg & C$_2$DCCH & L1551 \\
NS &  L1551 & CCH & L1551~IRS5, V1057~Cyg \\
HCN & L1551, V883~Ori, V1057~Cyg & CCS & L1551, V1057~Cyg \\
HCO & L1551, V883~Ori, V1057~Cyg  & c-C$_3$H & L1551 \\
H$_2$S & L1551 & l-C$_3$H$_2$ & L1551 \\
HNC & L1551, V883~Ori, V1057~Cyg & c-C$_3$H$_2$ & L1551, V1057~Cyg \\
OCS & L1551, V883~Ori & HC$_3$N &  L1551, V1057~Cyg \\
SO$_2$ & L1551, V883~Ori  & CH$_3$CCH & L1551, V1057~Cyg \\
NH$_3$ & RNO~1B/1C, L1551, V1057~Cyg & HC$_5$N & L1551, V1057~Cyg \\
H$_2$CO & RNO~1B/1C, L1551, V883~Ori, V1057~Cyg & CH$_3$CN & L1551, V883~Ori, V1057~Cyg \\
H$_2$CS & L1551, V1057~Cyg & CH$_3$OH & RNO~1B/1C, L1551, V883~Ori, V1057~Cyg \\
CH$_3$ & L1551 & CH$_3$CHO & L1551, V883~Ori \\
HCO$^+$ & RNO~1B/1C, L1551, V883~Ori, V1057~Cyg & CH$_3$CCH & L1551,  V883~Ori, V1057~Cyg \\
HCS$^+$ & L1551, V883~Ori, V1057~Cyg & CH$_3$CH$_2$OH & L1551 \\
N$_2$H$^+$ & L1551, V1057~Cyg & CH$_3$OCH$_3$ & L1551, V883~Ori\\
N$_2$D$^+$ & L1551 & C$_2$H$_3$CN & V883~Ori \\
HOCO$^+$ & L1551 & HCOOCH & L1551 \\
CCD & L1551, V1057~Cyg & H$_2$CCO & L1551, V883~Ori \\
DCN & L1551, V883~Ori, V1057~Cyg & HNCO & L1551, V883~Ori, V1057~Cyg \\
DCO$^+$ & L1551, V883~Ori, V1057~Cyg &  \\
D$_2$CO & L1551 \\
DNC & L1551, V883~Ori, V1057~Cyg \\
\hline \hline
\end{tabular}
\label{tab:comparison} 
\tablefoot{The most relevant references for each source. RNO~1B/1C: \citet{evans1994,McMuldroch1995}.
L1551~IRS5: \citet[][and references therein]{fridlund2002,white2006-ch3oh-l1551,bianchi2020,mercimek2022,marchand2024}. 
V883~Ori: \citet[][and references therein]{ruiz-rodriguez2017-v883ori,lee2019-v883ori,ruiz-rodriguez2022,cruz-saenz-de-miera2023}; \citet{tobin2023}; \citet{lee2024,yamato2024-v883ori-alma}.
V1057~Cyg: \citet[][]{bechis1975,levreault1988,evans1994,stauber2007,kospal2011b,feher2017,calahan2024b}; and this work.
}
\end{table*}

Probably the most prominent example in the literature for molecular line emission is the FUor-like, Class\,{\footnotesize I} protostar, L1551~IRS5 \citep[e.g.,][]{adams1987,looney1997,connelley2018}, located in Taurus, at a distance of $\sim$140\,pc \citep{zucker2019-gaia}, with a luminosity of $L=28\,L_{\odot}$ \citep{butner1991}. This source might be one of the most line-rich sources associated with the class, and includes detection of molecular species ranging from simple to complex molecules \citep[see Table~\ref{tab:comparison}, and also e.g.,][and references therein]{fridlund2002,white2006-ch3oh-l1551,mercimek2022}. The most notable result toward this source is the interferometric detection of a hot corino \citep[a region around a low-mass protostar bright in emission of COMs, e.g.,][]{yamamoto2017-astrochem-book} traced by CH$_3$OH (and its isotopologues), CH$_2$DOH, HCOOCH$_3$, and CH$_3$CH$_2$OH \citep{bianchi2020}. The most recent study on this source reports a survey, covering the 3\,mm and 2\,mm windows using the 30-m telescope, revealing a total of 75 species \citep[including isotopologues,][]{marchand2024}. V1057~Cyg is not as line-rich as L1551~IRS5 with a lower number of detected species and in particular only three COMs (CH$_3$OH, CH$_3$CCH, and CH$_3$CN) out of the 15 reported by \citet{marchand2024}.
The majority of species detected in the recent survey of L1551~IRS5 trace the cold envelope with temperatures $\leq$\,10\,K, however, the authors report molecules with $\geq$\,5 atoms and a few S-bearing species (C$_3$S, SO, SO$_2$) to trace warmer, $\geq$\,30\,K, regions. In V1057~Cyg, we do not detect molecular emission above 10\,K, implying that we only trace cold molecular emission towards this source. The column densities derived by the authors for L1551~IRS5 are within the same order of magnitude as the column densities we derived for V1057~Cyg, with the exception of c-C$_3$H$_2$, which has an order of magnitude lower column density in V1057~Cyg \citep{marchand2024}. However, we note that the emission detected by our single-dish observations is extended over 25--30$\arcsec$, i.e., $\sim$0.1\,pc at the distance of $\sim$900\,pc, and likely traces the disk, the envelope, and the surrounding cloud while the observations of L1551~IRS5 and V883~Ori trace smaller scales (see below).

Finally, we give a brief comparison to V883~Ori, a FUor transiting from the Class\,{\footnotesize I} to the Class\,{\footnotesize II} phase, located at a distance of $d=338$\,pc, with a bolometric luminosity of $L=\sim$220\,L$_{\odot}$ \citep[e.g.,][]{furlan2016,connelley2018,lee2019-v883ori}. 
This source has been the focus of a number of interferometric observations, revealing in particular a snow-line shift attributed to the outburst, a complex physical structure of its environment traced by high-density
(e.g.,~HCN, HCO$^+$, DCN) and shock tracers (e.g.,~CH$_3$OH, SiO, SO), a ring-like structure traced by HCO$^+$ emission, and a rich chemistry traced by COMs \citep[see Table~\ref{tab:comparison} and also e.g.,][]{ruiz-rodriguez2017-v883ori,lee2019-v883ori,ruiz-rodriguez2022,yamato2024-v883ori-alma,lee2024}. 
The most recent study of \citet{lee2024} focuses on the detection of simple molecular species and a small list of COMs using ALMA Band 6 data, a frequency regime partly covered by our single-dish observations. We find many molecular species detected in both V1057~Cyg and V883~Ori (see Table~\ref{tab:comparison}). 
The interferometic data traces different physical components of the environment of V883~Ori: HDCO and HNCO trace mainly the water sublimation front, HCN emission traces both the inner and outer disk, revealing an arm-like structure, while HNC and DNC trace a ring structure at the outer disk boundary \citep{lee2024}. With the single-dish data of V1057~Cyg, we are only able to trace the large scale emission (the ridge with the southern peak and the northern concentration, and the outflow, see also Sects.~\ref{subsec:maps}, \ref{disc:subsec-large-small-scale}, and \ref{subsec:outflow}). Between these two sources, we find V883~Ori to harbor more COMs, similarly to L1551~IRS5. However, for the more evolved nature of V1057~Cyg it still demonstrates a line-rich environment, motivating follow-up interferometric observations.
Furthermore, we do not detect water lines in this survey (despite transitions covered by the setup) compared to V883~Ori, however, we note that HDO 3$_{(1,2)}$--2$_{(2,1)}$ emission has been recently confirmed by sensitive interferometric NOEMA observations towards V1057~Cyg \citep{calahan2024b}.

\subsection{Chemical signatures of outbursts and outlook}
\label{subsec:disc-comparison} 

Episodic outbursts may strongly influence the disks around protostars, introducing changes to the thermal structure and inducing chemical changes \citep[e.g.,][]{fischer2023}, which may be used to identify past outburst events. 

\citet{rab2017} extended the radiation thermo-chemical disk code, PRODIMO \citep[PROtoplanetary DIsk MOdel,][]{woitke2009,woitke2016}, to treat envelope structures by feeding a representative Class\,{\footnotesize I} model. They used simulated, spatially resolved ALMA C$^{18}$O 2--1 observations and found this particular transition to be a good tool to probe post-outburst sources. In the model, the spatially resolved 2--1 transition emission exhibits distinct signatures, such as a gap due to the freeze out of the molecules \citep[see Fig.~4 of][]{rab2017}. 
In our line survey, we detected and mapped the C$^{18}$O 2--1 transition with APEX and the emission is resolved in the north-south direction, tracing the ridge, but our observations do not resolve the disk around V1057~Cyg (see Fig.~\ref{fig:apex-maps-nflash} and Table~\ref{tab:res-apex-fits}).
Interferometric observations of the C$^{18}$O 1-0 transition by \citet{feher2017} found a rotating envelope with a radius of 5$\arcsec$, or 3000\,au at $d=600$\,pc \citep[common distance used before \textit{Gaia},][]{bailerjones1,bailerjones2}, corresponding to $\sim$4500\,au at $\sim$900\,pc, however, with the single-dish observations of this transition with a beam size of 22.4$\arcsec$, we do not resolve the envelope either. 
The model of \citet{rab2017} accounts for both disk and envelope emission, and the outburst induced changes are seen in the radial distribution of the $^{12}$CO emission on scales of a few 1000\,au, thus much larger than the typical disk scales. 
Our single-dish observations have an angular resolution that corresponds to about $\sim$18~000\,au and, therefore, cannot resolve the $^{12}$CO structures predicted by the model which are at a few 1000\,au scales.

The other notable model is by \citet{molyarova2018}, who used the ANDES physical-chemical code \citep{akimin2013} to follow the impact of a FUor outburst on the chemical evolution and to identify molecular tracers on different timescales. Two main groups are specified: species with abundances sensitive to the ongoing outburst but returning quickly (i.e.,~a few years after the peak outburst) to the quiescent (pre-outburst) values; and those that take longer to return to pre-outburst or even remain overabundant for decades ($>$\,20\,yrs).
The most notable prediction is formaldehyde (H$_2$CO), whose abundance can grow by up to 4--6 magnitudes, and take 30--120\,years to deplete back onto dust grains, making it a good candidate tracer to, in theory, identify post-outburst sources.
They modeled the brightest H$_2$CO transition observed in disks, the 3$_{03}$--2$_{02}$ transition at 218.222\,GHz, with input parameters based on the FUor outburst of V346~Nor, which in 2010 seemed to finish abruptly its outburst \citep[and was thought to be the first FUor to do so, see][]{kraus2016}. However, the study of \citet{kospal2020} showed evidence for the source gradually brightening back to almost the same brightness level prior to the abrupt fading event, and concluded the fading was likely due to different mechanisms, including increase of the line-of-sight extinction, and appearance of warm material possibly refuelling the outburst.
We detect the H$_2$CO (3$_{03}$--2$_{02}$) transition toward V1057~Cyg in the 30-m data, however, it is fainter than other transitions of H$_2$CO detected in the survey. 
We find the abundances of H$_2$CO and CH$_3$OH (see Table~\ref{tab:res-LTE-model-res-pop-diag}) to be orders of magnitude lower than the abundances predicted by both the medium and grown dust models by \citet{molyarova2018}. However, our abundances correspond to material at the scale of $\sim$30$\arcsec$, while the model abundances are predicted at the much smaller scale of the disk.
In contrast, the HC$_3$N abundance is closer to the model predictions, although its abundance is lower than those of H$_2$CO and CH$_3$OH. 
We note that our HC$_3$N map shows extended emission, without a peak centered on V1057~Cyg, hence suggesting that the HC$_3$N line emission detected towards the target is not dominated by compact, unresolved emission. We do not have maps of H$_2$CO or CH$_3$OH emission to check whether the emission peaks on the target.
The CS abundance derived for V1057~Cyg is higher than the model predictions. 
The NH$_3$ abundance derived in an earlier study \citep{szabo2023a} is $\sim$1 order of magnitude lower than the abundance predicted by the model of \citet{molyarova2018}.

Considering that V1057~Cyg has been in outburst for over 50\,years, it is plausible that during the peak of the outburst, the abundance values were elevated due to the heat from the central source. The detection of multiple transitions of many molecular species and the lower abundances compared to the model predictions show that many species are still present and not yet frozen-out to dust surfaces. 
On the other hand, the single-dish data clearly trace large-scale structures (i.e.,~the ridge), and it is uncertain what fraction of the measured flux is related to a more compact disk around the source and what is related to the extended ridge. 

The available chemical models, together with line surveys such as the one presented here motivate follow-up studies both observational and theoretical. In the case of V1057~Cyg, high-angular resolution observations could resolve the smaller-scale structures (i.e.,~on scales of the disk and envelope system) and their connection to the larger scale emission traced by the ridge both to the north and south from the source position. Future spatially resolved observations can reveal similar structures predicted in the models due to the inside out freeze-out \citep{rab2017}, snow-line shifts (either water or other molecules) attributed to the outburst, or help distinguishing which species trace the different layers in the system, similarly to what has been found in V883~Ori \citep[e.g.,][]{tobin2023,yamato2024-v883ori-alma}. This may allow to probe the impact of the outburst on the physical and chemical structure of the disk and the close environment of V1057~Cyg \citep[e.g.,][]{molyarova2018,lee2024,andreu2023}. Further detection of indicators of the water snowline \citep[e.g.,][]{molyarova2018} and information on the D/H ratio in more evolved sources such as V1057~Cyg may add more information on Class\,{\footnotesize II} sources, currently absent from the statistics \citep[see Fig.~3 of][and references therein]{andreu2023}. 
Despite our single-dish data not being able to resolve the small scales, the line survey shows a richer line emission than might have been expected for a relatively more evolved, classical FUor like V1057~Cyg, and motivates higher-angular resolution studies to determine whether emission from molecules, such as, H$_2$CO and CH$_3$OH is coming from the large-scale ridge or a compact structure around the source. Furthermore, high-angular resolution observations of $^{12}$CO could reveal past episodic ejection events in greater detail.

\section{Conclusions}
\label{sec:conc}
In this paper, we present the first wide-band line survey in the millimeter wavelength regime of V1057~Cyg and its surrounding environment. This Class\,{\footnotesize II} source has been in outburst for about 50\,years, and has not yet shown signs of returning to quiescence \citep{szabo2021}. The line survey has an almost continuous frequency coverage from $\sim$72\,GHz to $\sim$263\,GHz (with spatial resolution between $\sim$36$\arcsec$ and $\sim$10$\arcsec$) with the IRAM~30-m telescope. Additionally, specific frequency ranges around 219, 227, 291, and 344\,GHz (with spatial resolution between $\sim$30$\arcsec$ and $\sim$19$\arcsec$) were observed with the APEX~12-m telescope. The line identification and analysis carried out on the data lead to the following conclusions: 

\begin{itemize}
    \item We identified (mostly) simple $C$-, $N$-, $O$-, $S$-bearing molecules (e.g.,~CN, CS, HCN), deuterated species (e.g.,~DCN), molecular ions (e.g.,~HCO$^+$), more simple $C$-chain molecules (e.g.,~C$_2$H, HC$_3$N), and complex organic molecules. V1057~Cyg displays lines from 35 molecular species (including isotopologues). 
    The different spatial distribution of emission in the integrated line intensity maps indicate that the various lines sample different regions in the environment of the source. 
    Most of the lines detected in this survey have been detected in disks around T~Tauri stars \citep[e.g.,][]{dutrey2014}, the closest in evolution to V1057~Cyg and other well studied FUors \citep[e.g.,][]{marchand2024,lee2024}.
    \item Most of the lines show narrow single-peaked line profiles that are well described by single Gaussian fits, with the exception of outflow tracers that show broader line wings, indicating outflow activity. The LSR velocities in all cases indicate that the line emission is associated with the circumstellar environment of V1057~Cyg. 
    \item Population diagrams of molecules detected in several transitions (CH$_3$OH, H$_2$CO, CCS, HC$_3$N, H$_2$CS, CS, c-C$_3$H$_2$) were constructed, indicating rotational temperatures ($T_{\rm rot}$) ranging from 8.1\,K to 14.8\,K. We derived column densities ($N_{\rm tot}$) of these molecules under the assumption of LTE with values ranging from 1.4$\times$10$^{12}$\,cm$^{-2}$ to 2.8$\times$10$^{13}$\,cm$^{-2}$. The derived column densities are $\sim$1 order of magnitude lower than those derived for younger, line rich FUors, but our single-dish survey probes larger scales.
    \item The emission of several molecular species was mapped with the IRAM~30-m or the APEX~12-m telescopes. The integrated intensity maps trace a parsec-scale ridge that runs from the Southwest to the North, going through the position of V1057~Cyg. This ridge coincides with the ridge traced earlier in dust continuum emission \citep{sandell2001,green2013,szabo2023a}, however, our molecular line maps cover a larger area towards the southwest. 
    Two clumps of molecular emission, clearly visible in the integrated intensity maps of HCN, HNC, N$_2$H$^+$ 1--0, and HC$_3$N 10--9, are detected along the ridge to the North and Southwest of V1057~Cyg, offset by $\sim$15$\arcsec$ ($\sim$0.06\,pc), and $\sim$108$\arcsec$ ($\sim$0.46\,pc) respectively from the source position. 
    Our results could indicate that V1057~Cyg is forming as part of a small cluster, similarly to several other FUors \citep[e.g.,][]{white2006-ch3oh-l1551,quanz2007}, and motivate future interferometric observations to reveal the protostellar content of the filamentary structures.  
    \item Our new $^{12}$CO 3--2 map reveals a bipolar outflow centered on V1057~Cyg.  We estimate a dynamical time for the outflow of tens of thousands of years (15\,000\,years for the blue and 22\,000\,years for the red lobe of the outflow), implying that the detected outflow does not result from the current outburst activity alone. The position-velocity diagram shows discrete peaks and hints at structures representing episodic ejection events, further indicating outburst activity in the past evolution of V1057~Cyg. 
\end{itemize}

Despite being at a distance of almost 900\,pc and being a more evolved (i.e.,~Class\,{\footnotesize II}) object, V1057~Cyg exhibits a line-rich environment in our single-dish observations. While the molecular emission in the environment of V1057~Cyg does not reveal clear evidence of the imprint of an outburst on the chemistry of the object on the scales traced by our single-dish data, the line-rich spectra motivate interferometric observations to spatially resolve the emission. Our results also motivate similar studies of a larger sample of FUors to study the range of astrochemical properties associated with this class of outbursting protostars, and to improve the current statistics regarding the outbursts’ effects on the chemical composition of their close environments, ultimately the future birthplace of planets.


\begin{acknowledgements}
We thank the anonymous referee for valuable feedback which has improved the quality of our manuscript.
Zs. M. Szabó acknowledges funding from a St. Leonards scholarship from the University of St. Andrews and C. J. C. acknowledges support from the STFC (grant ST/Y002229/1).
For the purpose of open access, the author has applied a Creative Commons Attribution (CC BY) licence to any Author Accepted Manuscript version arising.
Y.G. is supported by the Strategic Priority Research Program of the Chinese Academy of Sciences, Grant No. XDB0800301.
This work is based on observations carried out under project number 060-21 (PI: Karl M. Menten) with the IRAM 30m telescope. IRAM is supported by INSU/CNRS (France), MPG (Germany) and IGN (Spain).
This publication is based on data acquired with the Atacama Pathfinder Experiment (APEX) under programme ID M9530C\_107, M9515A\_108, and M9523C\_109. APEX has been a collaboration between the Max-Planck-Institut fur Radioastronomie, the European Southern Observatory, and the Onsala Space Observatory.
This work was also supported by the NKFIH excellence grant TKP2021-NKTA-64.
\end{acknowledgements}

\bibliographystyle{aa}
\bibliography{paper}{} 

\begin{appendix}
\section{Line detection}
In Tables~\ref{tab:appendix-iram-lines-list-fits} and \ref{tab:res-apex-fits}, we list the lines detected above 3$\sigma$, with their respective quantum numbers, upper energy levels, Einstein coefficients, and the Gaussian fit results for the IRAM~30-m and the APEX~12-m telescopes, respectively. The lines detected in the 30-m survey are presented in Figs.~\ref{fig:survey-73ghz-small}--\ref{fig:survey-242ghz-small}. In certain tunings contamination from the image band cause false signal. These occurrences are labelled in red indicating rest frequency and peak temperature in the image band. An artifact is labelled on Fig.~\ref{fig:survey-202ghz-small} around $\sim$222.7\,GHz, which does not show similarities to any of the line profiles detected (i.e.~extremely narrow). Lines are labeled in red and marked as unidentified ($\sim$15 lines in the 30-m spectrum) where we have a $\geq$3\,$\sigma$ detection but line identification does not match anything in the databases.
Lines detected in the other selected frequencies covered with the APEX telescope are shown in Figs.~\ref{fig:survey-219.2ghz-small-apex} and \ref{fig:survey-227.2ghz-small-apex}.

\onecolumn

\begin{small}
\begin{longtable}{ccccccccccc}
\caption{Lines detected toward V1057~Cyg with the IRAM~30-m telescope} \label{tab:appendix-iram-lines-list-fits} \\ 
\hline \hline
Rest frequency & \multirow{2}{*}{Molecule} & \multirow{2}{*}{Quantum numbers} & $E_{\rm up}^{(a)}$ & $log_{10}A_{\rm ij}^{(b)}$ & $\varv_{\rm LSR}$$^{(*)}$ & $\Delta \varv$$^{(*)}$ & $T_{\rm MB}$$^{(*)}$ & rms & Blended? \\
    (GHz) & & & (K) & (s$^{-1}$) & (km\,s$^{-1}$) & (km\,s$^{-1}$) & (K) & (mK) & (Y/N) \\
\hline
\endfirsthead
\caption{continued.}\\
\hline\hline
Rest frequency & \multirow{2}{*}{Molecule} & \multirow{2}{*}{Quantum numbers} & $E_{up}^{(a)}$ & $log_{10}A_{\rm ij}^{(b)}$ & $\varv_{\rm LSR}$$^{(*)}$ & $\Delta \varv$$^{(*)}$ & $T_{\rm MB}$$^{(*)}$ & rms & Blended?\\
    (GHz) & & & (K) & (s$^{-1}$) & (km\,s$^{-1}$) & (km\,s$^{-1}$) & (K) & (mK) & (Y/N) \\
\hline
\endhead
\hline
\endfoot
\hline     
    \multicolumn{9}{c}{Tuning: 73\,GHz} \\
    \hline
    $72.039$ & DCO$^+$  & $J=1-0, F=0-1$ & $3.5$ & $-4.65$ & $4.28$ $(0.01)$ & $1.12$ $(0.06)$ & $0.45$ $(0.01)$ & $10.50$ & N \\
    $72.413$ & DCN      & $J=1-0, F=1-1$   & $3.5$ & $-4.88$ & $4.21$ $(0.06)$ & $0.81$ $(0.10)$ & $0.12$ $(0.01)$ & $15.30$ & N \\
    $72.414$ & DCN      & $J=1-0, F=2-1$   & $3.5$ & $-4.88$ & $4.13$ $(0.12)$ & $1.01$ $(0.15)$ & $0.12$ $(0.01)$ & $15.30$ & N \\
    $72.783$  & HC$_3$N   & $J=8-7$   & $15.7$ & $-4.53$ & $4.24$ $(0.02)$ & $1.15$ $(0.04)$ & $0.34$ $(0.05)$ & $10.08$ & N \\
    $72.837$ & H$_2$CO   & $1(0,1)-0(0,0)$  & $3.5$ & $-5.08$ & $4.16$ $(0.01)$ & $0.81$ $(0.01)$ & $0.49$ $(0.01)$ & $10.68$ & N \\
    $76.117$ & C$_4$H & $N=8-7, J=17/2-15/2$ & $16.4$ & $-4.97$ & \multirow{2}{*}{$4.20$ $(0.16)$} & \multirow{2}{*}{$2.27$ $(0.41)$} & \multirow{2}{*}{$0.02$ $(0.01)$} & \multirow{2}{*}{$5.24$} & Y \\    
    $76.117$ & C$_4$H & $N=8-7, J=17/2-15/2$ & $16.4$ & $-4.97$  & & & & & Y \\    
    $76.156$ & C$_4$H & $N=8-7, J=15/2-13/2$ & $16.5$ & $-4.97$ & \multirow{2}{*}{$4.50$ $(0.14)$} & \multirow{2}{*}{$1.50$ $(0.35)$} & \multirow{2}{*}{$0.03$ $(0.01)$} & \multirow{2}{*}{$5.31$} & Y \\    
    $76.156$ & C$_4$H & $N=8-7, J=15/2-13/2$ & $16.5$ & $-4.97$ & & & & & Y \\    
    $76.305$ & DNC       & $J=1-0$  & $3.7$ & $-4.79$ & $4.32$ $(0.02)$ & $1.56$ $(0.05)$ & $0.22$ $(0.01)$ & $5.87$ & N \\
    $77.107$ & N$_2$D$^+$ & $J=1-0$  & $3.7$ & $-4.68$ & $4.68$ $(0.17)$ & $1.18$ $(0.45)$ & $0.03$ $(0.01)$ & $5.69$ & N \\
    $77.109$ & N$_2$D$^+$ & $J=1-0$  & $3.7$ & $-4.68$ & $4.49$ $(0.17)$ & $2.17$ $(0.34)$ & $0.03$ $(0.01)$ & $3.59  $ & N \\
    $77.731$ & CCS       & $N=6-5, J=6-5$ & $21.8$ & $-4.69$ & $4.07$ $(0.26)$ & $3.43$ $(0.64)$ & $0.03$ $(0.01)$ & $5.91$ & N \\ 	
    $86.670$ & HCO       & $1( 0, 1)- 0( 0, 0), J=3/2-1/2$ & $4.2$ & $-5.32$ & $4.45$ $(0.12)$ & $1.49$ $(0.28)$ & $0.02$ $(0.01)$ & $3.19$ & N \\
    $86.708$ & HCO       & $1( 0, 1)- 0( 0, 0), J=3/2-1/2$ & $4.2$ & $-5.33$ & $4.27$ $(0.21)$ & $1.96$ $(0.53)$ & $0.02$ $(0.01)$ & $4.32$ & N \\ 
    $86.754$ & H$^{13}$CO$^+$ & $J=1-0$ &  $4.2$ & $-4.41$ & $4.13$ $(0.01)$  &  $1.42$ $(0.01)$ & $0.48$ $(0.01)$ & $3.91$ & N \\
    $87.090$ & HN$^{13}$C & $J=1-0$ & $4.2$ & $-4.62$ & $4.37$ $(0.04)$ & $1.46$ $(0.09)$ & $0.08$ $(0.01)$ & $3.92$ & N \\
    $87.284$ & CCH   & $N=1-0, J=3/2-1/2$ & $4.2$ & $-6.58$ & $4.17$ $(0.02)$ & $1.47$ $(0.07)$ & $0.11$ $(0.01)$ & $3.65$ & N \\
    $87.316$ & CCH   & $N=1-0, J=3/2-1/2$ & $4.2$ & $-5.81$ & $4.18$ $(0.01)$ & $1.38$ $(0.02)$ & $0.75$ $(0.01)$ & $5.62$ & N \\
    $87.328$ & CCH   & $N=1-0, J=3/2-1/2$ & $4.2$ & $-5.89$ & $4.16 $ $(0.01)$ & $1.35$ $(0.02)$ & $0.42$ $(0.12)$ & $3.79$ & N \\
    $87.402$ & CCH   & $N=1-0, J=1/2-1/2$ & $4.2$ & $-5.89$ & $4.17$ $(0.04)$ & $1.32$ $(0.10)$ & $0.43$ $(0.02)$ & $33.94$ & N \\  
    $87.407$ & CCH   & $N=1-0, J=1/2-1/2$ & $4.2$ & $-5.81$ & $4.21$ $(0.20)$ & $1.29$ $(0.39)$ & $0.18$ $(0.09)$ & $72.50$ & N \\
    $87.446$ & CCH   & $N=1-0, J=1/2-1/2$ & $4.2$ & $-6.58$ & $4.15$ $(0.02)$ & $1.46$ $(0.05)$ & $0.11$ $(0.01)$ & $3.06$ & N \\
    $87.925$ & HNCO      & $4(0,4)-3(0,3)$  & $10.5$ & $-5.05$ & $3.53$ $(0.08)$ & $1.10$ $(0.14)$ & $0.03$ $(0.01)$ & $3.36$ & N \\
    $88.630$ & HCN       & $J=1-0$ & $4.3$ & $-4.61$ & $4.21$ $(0.17)$ & $1.41$ $(0.43)$ & $0.57$ $(0.15)$ & $3.48$ & N \\
    $88.631$ & HCN       & $J=1-0$ & $4.3$ & $-4.61$ & $4.26$ $(0.05)$ & $1.38$ $(0.13)$ & $1.07$ $(0.09)$ & $3.97$ & N \\
    $88.633$ & HCN       & $J=1-0$ & $4.3$ & $-4.61$ & $4.26$ $(0.28)$ & $1.27$ $(0.75)$ & $0.34$ $(0.18)$ & $98.23$ & N \\
    $89.188$ & HCO$^+$   & $J=1-0$  & $4.3$ & $-4.37$ & $4.25$ $(0.01)$ & $1.46$ $(0.01)$ & $2.26$ $(0.01)$ & $3.77$ & N \\
    $90.663$ & HNC       & $J=1-0$  & $4.4$ & $-4.57$ & $4.30$ $(0.01)$ & $1.45$ $(0.01)$ & $1.59$ $(0.01)$ & $10.08$ & N \\
    $90.979$ & HC$_3$N   & $J=10-9$ & $24$ & $-4.23$ & $4.36$ $(0.02)$ & $1.37$ $(0.05)$ & $0.26$ $(0.01)$ & $9.21$ & N \\
    $92.494$ & $^{13}$CS & $J=2-1$  &  $6.7$ & $-4.85$ & $3.89$ $(0.33)$ & $2.26$ $(0.87)$ & $0.02$ $(0.01)$ & $9.76$ & N \\
    $93.171$ & N$_2$H$^+$ & $J=1-0$ & $4.5$ & $-4.44$ & $4.11$ $(0.14)$ & $1.45$ $(0.36)$ & $0.47$ $(0.10)$ & $24.52$ & N \\
    $93.173$ & N$_2$H$^+$ & $J=1-0$ & $4.5$ & $-4.44$ & $4.25$ $(0.09)$ & $2.07$ $(0.20)$ & $0.60$ $(0.98)$ & $9.06$ & N \\
    $93.176$ & N$_2$H$^+$ & $J=1-0$ & $4.5$ & $-4.44$ & $3.97$ $(0.20)$ & $0.77$ $(0.70)$ & $0.25$ $(0.99)$ & $9.21$ & N \\
    $93.870$ & CCS        & $N=7-6, J=8-7$ & $19.9$ & $-4.42$ & $4.44$ $(0.12)$ & $0.96$ $(0.21)$ & $0.05$ $(0.01)$ & $9.65$ & N \\
    \hline     
    \multicolumn{9}{c}{Tuning: 80\,GHz} \\
    \hline
    $81.505$ & CCS       & $N=6-5, J=7-6$ & $15.4$ & $-4.61$ & $4.16$ $(0.02)$ & $1.34$ $(0.06)$ & $0.08$ $(0.01)$ & $3.29$ & N \\
    $81.881$ & HC$_3$N   & $J=9-8$ & $19.6$ & $-4.37$ & $4.26$ $(0.01)$ & $1.53$ $(0.01)$ & $0.27$ $(0.01)$ & $2.59$ & N \\
    $82.093$ & c-C$_3$H$_2$ & $2(0,2)-1(1,1)$ & $6.4$ & $-4.72$ & $4.25$ $(0.01)$ & $1.48$ $(0.04)$ & $0.12$ $(0.01)$ & $2.99$ & N \\
    $82.966$ & c-C$_3$H$_2$ & $3(1,2)-3(0,3)$ & $16$ & $-5.00$ & $4.26$ $(0.03)$ & $1.30$ $(0.08)$ & $0.05$ $(0.01)$ & $2.21$ & Y \\
    $84.521$ & CH$_3$OH & $5_{-1}-4_0, E$ & $40.4$ & $-5.71$ & $4.31$ $(0.07)$ & $1.60$ $(0.18)$ & $0.03$ $(0.01)$ & $3.57$ & N \\ 
    $84.727$ & c-C$_3$H$_2$ & $3(2,2)-3(1,3)$ & $16.1$ & $-4.98$ & $4.17$ $(0.12)$ & $1.15$ $(0.18)$ & $0.02$ $(0.01)$ & $3.22$ & N \\ 
    $85.162$ & HC$^{18}$O$^+$ & $J=1-0$ & $4.1$ & $-4.43$ & $4.16$ $(0.04)$ & $1.48$ $(0.11)$ & $0.05$ $(0.01)$ & $2.91$ & N \\
    $85.338$ & c-C$_3$H$_2$ & $2(1,2)-1(0,1)$ & $6.4$ & $-4.63$ & $4.22$ $(0.01)$ & $1.31$ $(0.02)$ & $0.47$ $(0.01)$ & $5.38$ & N\\
    $85.347$ & HCS$^+$ & $J=2-1$ & $6.1$ & $-4.95$ & $4.20$ $(0.07)$ & $1.22$ $(0.19)$ & $0.03$ $(0.01)$ & $3.72$ & N \\
    $85.455$ & CH$_3$CCH & $5(1)-4(1)$ & $19.5$ & $-5.71$ & $4.11$ $(0.12)$ & $1.20$ $(0.31)$ & $0.03$ $(0.01)$ & $4.07$ & N \\
    $85.457$ & CH$_3$CCH & $5(0)-4(0)$ & $12.3$ & $-5.69$ & $4.09$ $(0.09)$ & $1.37$ $(0.21)$ & $0.04$ $(0.01)$ & $4.07$ & N \\
    $93.870$ & CCS     & $N=7-6, J=8-7$ & $19.9$ & $-4.42$ & $4.31$ $(0.03)$ & $1.18$ $(0.06)$ & $0.06$ $(0.01)$ & $3.26$ & N \\
    $95.150$ & C$_4$H & $N=10-9$ & $25.1$ & $-4.67$ & \multirow{2}{*}{$4.36$ $(0.07)$} & \multirow{2}{*}{$1.32$ $(0.16)$} & \multirow{2}{*}{$0.02$ $(0.01)$} & \multirow{2}{*}{$3.04$} & Y \\ 
    $95.150$ & C$_4$H & $N=10-9$ &  $25.1$ & $-4.67$ &  & & & & Y \\
    $95.188$ & C$_4$H & $N=10-9$ & $25.1$ & $-4.68$ & \multirow{2}{*}{$4.34$ $(0.09)$} & \multirow{2}{*}{$1.46$ $(0.30)$} & \multirow{2}{*}{$0.02$ $(0.01)$} & \multirow{2}{*}{$3.14$} & Y \\
    $95.188$ & C$_4$H & $N=10-9$ & $25.1$ & $-4.67$ & & & & & Y \\
    $96.412$ & C$^{34}$S & $J=2-1$ & $6.9$ & $-4.79$ & $4.21$ $(0.01)$ & $1.41$ $(0.04)$ & $0.11$ $(0.01)$ & $3.15$ & N \\
    $96.739$ & CH$_3$OH & $2(1,2)-1(1,1)$ $E$ & $12.5$ & $-4.59$ & $4.08$ $(0.13)$ & $1.43$ $(0.31)$ & $0.17$ $(0.03)$ & $7.20$ & N \\
    $96.741$ & CH$_3$OH & $2(0,2)-1(0,1)$ $A$ & $7.0$ & $-5.46$ & $4.08$ $(0.06)$ & $1.45$ $(0.15)$ & $0.25$ $(0.02)$ & $2.95$ & N \\ 
    $96.744$ & CH$_3$OH & $2(-0,2)-1(-0,1)$ $E$ & $20.1$ & $-5.46$ & $4.07$ $(0.60)$ & $0.97$ $(1.57)$ & $0.04$ $(0.05)$ & $30.24$ & N \\ 
    $97.172$ & C$^{33}$S & $J=2-1$ & $7.0$ & $-4.78$ & $4.56$ $(0.14)$ & $1.93$ $(0.30)$ & $0.02$ $(0.01)$ & $2.93$ & N \\
    $97.980$ & CS & $J=2-1$ & $7.1$ & $-4.77$ & $4.25$ $(0.01)$ & $1.39$ $(0.01)$ & $1.61$ $(0.01)$ & $4.37$ & N \\
    $99.299$ & SO & $J=2-1$ & $9.2$ & $-4.94$ & $4.17$ $(0.01)$ & $1.23$ $(0.02)$ & $0.27$ $(0.01)$ & $4.14$ & N \\
    $99.866$ & CCS & $N=8-7, J=7-6$ & $28.1$ & $-4.35$ & $4.33$ $(0.08)$ & $1.08$ $(0.24)$ & $0.02$ $(0.01)$ & $3.28$ & N \\
    $100.076$ & HC$_3$N & $J=11-10$  & $28.8$ & $-4.11$ & $4.28$ $(0.01)$ & $1.14$ $(0.02)$ & $0.25$ $(0.01)$ & $4.02$ & N \\
      \hline
    \multicolumn{9}{c}{Tuning: 90\,GHz otf (spectrum towards the source position, i.e.~(0$\arcsec$,0$\arcsec$))} \\
    \hline
    $88.630$ & HCN & $J=1-0$ & $4.3$ & $-4.61$ & $4.06$ $(0.17)$ & $1.46$ $(0.40)$ & $0.10$ $(0.02)$ & $10.36$ & N \\
    $88.631$ & HCN & $J=1-0$ & $4.3$ & $-4.62$ & $3.32$ $(0.07)$ & $1.19$ $(0.20)$ & $0.15$ $(0.02)$ & $11.34$ & N \\
    $88.633$ & HCN & $J=1-0$ & $4.3$ & $-4.61$ & $4.35$ $(0.28)$ & $0.75$ $(0.77)$ & $0.08$ $(0.03)$ & $20.04$ & N \\
    $89.188$ & HCO$^+$ & $J=1-0$ & $4.3$ & $-4.37$ & $4.18$ $(0.03)$ & $1.53$ $(0.07)$ & $0.23$ $(0.01)$ & $10.60$ & N \\
    $90.663$ & HNC & $J=1-0$ & $4.4$ & $-4.57$ & $4.12$ $(0.04)$ & $1.51$ $(0.11)$ & $0.19$ $(0.01)$ & $11.58$ & N \\
    $90.979$ & HC$_3$N & $J=10-9$ & $24.0$ & $-4.23$ & $4.29$ $(0.08)$ & $0.65$ $(0.10)$ & $0.09$ $(0.01)$ & $12.14$ & N \\
    $93.171$ & N$_2$H$^+$ & $J=1-0$ & $4.5$ & $-4.44$ & $4.05$ $(0.17)$ & $1.42$ $(0.38)$ & $0.03$ $(0.01)$ & $6.85$ & N \\
    $93.173$ & N$_2$H$^+$ & $J=1-0$ & $4.5$ & $-4.44$ & $4.29$ $(0.20)$ & $2.38$ $(0.38)$ & $0.03$ $(0.01)$ & $8.45$ & N \\
    $93.870$ & CCS & $N=7-6, J=8-7$ & $19.9$ & $-4.42$ & $4.29$ $(0.17)$ & $1.10$ $(0.53)$ & $0.02$ $(0.01)$ & $7.73$ & N \\
    $109.782$ & C$^{18}$O & $J=1-0$ & $5.3$ & $-7.02$ & $4.09$ $(0.02)$ & $1.07$ $(0.04)$ & $0.35$ $(0.01)$ & $11.89$ & N \\
    $110.201$ & $^{13}$CO & $J=1-0$ & $5.3$ & $-7.19$ & $3.86$ $(0.01)$ & $1.80$ $(0.01)$ & $2.36$ $(0.02)$ & $13.47$ & N \\
    \hline     
    \multicolumn{9}{c}{Tuning: 94\,GHz} \\
    \hline
    $92.494$ & $^{13}$CS & $J=2-1$ & $6.7$ & $-4.85$ & $4.12$ $(0.08)$ & $1.26$ $(0.19)$ & $0.04$ $(0.01)$ & $5.21$ & N \\
    $93.171$ & N$_2$H$^+$ & $J=1-0$ &  $4.5$ & $-4.44$ & $4.09$ $(0.09)$ & $1.36$ $(0.25)$ & $0.55$ $(0.08)$ & $6.29$ & N \\
    $93.173$ & N$_2$H$^+$ & $J=1-0$ &  $4.5$ & $-4.44$ &$4.22$ $(0.07)$ & $1.92$ $(0.16)$ & $0.71$ $(0.01)$ & $6.29$ & N \\
    $93.176$ & N$_2$H$^+$ & $J=1-0$ & $4.5$ & $-4.44$ & $3.91$ $(0.21)$ & $1.03$ $(0.46)$ & $0.27$ $(0.01)$ & $6.29$ & N \\
    $93.870$ & CCS & $N=7-6, J=8-7$ & $19.9$ & $-4.42$ & $4.36$ $(0.02$) & $0.86$ $(0.14)$ & $0.08$ $(0.01)$ & $4.53$ & N \\ 
    $95.150$ & C$_4$H &$N=10-9$ &  $25.1$ & $-5.41$ &  \multirow{2}{*}{$4.32$ $(0.08)$} & \multirow{2}{*}{$0.97$ $(0.14)$} & \multirow{2}{*}{$0.03$ $(0.01)$} & \multirow{2}{*}{$4.49$} & Y \\ 
    $95.150$ & C$_4$H & $N=10-9$ &  $25.1$ & $-5.41$ & & & & & Y \\
    $96.412$ & C$^{34}$S & $J=2-1$ &  $6.9$ & $-5.41$ & $4.14$ $(0.01)$ & $1.41$ $(0.03)$ & $0.11$ $(0.01)$ & $2.46$ & N \\
    $96.739$ & CH$_3$OH & $2(1,2)-1(1,1)$ $E$ &  $12.5$ & $-5.59$ & $4.04$ $(0.10)$ & $1.48$ $(0.24)$ & $0.18$ $(0.01)$ & $5.83$ & N \\
    $96.741$ & CH$_3$OH & $2(0,2)-1(0,1)$ $A$ &  $7.0$ & $-5.46$ &$4.06$ $(0.05)$ & $1.45$ $(0.12)$ & $0.27$ $(0.01$) & $2.42$ & N \\
    $96.744$ & CH$_3$OH & $2(-0,2)-1(-0,1)$ $E$ &  $20.1$ & $-5.46$ & $4.10$ $(0.49)$ & $1.24$ $(1.02)$ & $0.04$ $(0.01)$ & $2.25$ & N \\
    $97.172$ & C$^{33}$S & $J=2-1$ &  $7.0$ & $-4.78$ & $4.80$ $(0.09)$ & $1.33$ $(0.21)$ & $0.02$ $(0.01)$ & $2.95$ & N \\
    $97.980$ & CS & $J=2-1$ &  $7.1$ & $-4.77$ & $4.17$ $(0.01)$ & $1.49$ $(0.01)$ & $1.52$ $(0.01)$ & $2.68$ & N \\
    $99.299$ & SO   & $J=3-2$ &  $9.2$ & $-4.94$ & $4.04$ $(0.01)$ & $1.45$ $(0.01)$ & $0.27$ $(0.01)$ & $2.93$ & N \\
    $108.651$ & $^{13}$CN & $N=1-0, J=1/2-1/2$ &  $5.2$ & $-5.01$ & $4.16$ $(0.11)$ & $1.24$ $(0.25)$ & $0.02$ $(0.01)$ & $3.72$ & N \\
    $108.780$ & $^{13}$CN & $N=1-0, J=3/2-1/2$ &  $5.2$ & $-4.97$ & $4.13$ $(0.06)$ & $1.01$ $(0.13)$ & $0.03$ $(0.01)$ & $3.39$ & N \\
    $108.893$ & CH$_3$OH & $0(0,0)-1(-1,1)$ $E$ &  $13.1$ & $-4.83$ & $4.05$ $(0.03)$ & $1.44$ $(0.07)$ & $0.05$ $(0.01)$ & $2.82$ & N \\
    $109.173$ & HC$_3$N & $J=12-11$ & $34.1$ & $-3.99$ & $2.45$ $(0.01)$ & $1.13$ $(0.01)$ & $0.21$ $(0.01)$ & $2.47$ & N \\
    $109.252$ & SO  & $2(3)-1(2)$ & $21.1$ & $-3.97$ & $3.68$ $(0.05)$ & $1.53$ $(0.15)$ & $0.03$ $(0.01)$ & $2.73$ & N \\
    $109.782$ & C$^{18}$O & $J=1-0$ & $5.3$ & $-7.20$ &  $4.14$ $(0.01)$ & $0.99$ $(0.01)$ & $1.12$ $(0.01)$ & $2.84$ & N \\
    $109.905$ & HNCO & $5(0,5)-4(0,4)$ & $15.8$ & $-4.75$ & $4.19$ $(0.10)$ & $1.61$ $(0.22)$ & $0.02$ $(0.01)$ & $3.34$ & N \\
    $110.153$ & NH$_2$D & $1(1,1)0-1(0,1)0$ &  $21.3$ & $-4.78$ & $4.46$ $(0.07)$ & $1.97$ $(0.28)$ & $0.04$ $(0.01)$ & $3.16$ & N \\
    $110.201$ & $^{13}$CO & $J=1-0$ &  $5.3$ & $-7.19$ & $3.02$ $(0.01)$ & $1.51$ $(0.01)$ & $5.66$ $(0.01)$ & $2.48$ & N \\
    $110.381$ & CH$_3$CN & $6(1)-5(1)$ &  $25.7$ & $-3.96$ & $4.15$ $(0.14)$ & $1.41$ $(0.31)$ & $0.02$ $(0.01)$ & $2.99$ & N \\
    $110.383$ & CH$_3$CN & $6(0)-5(0)$ &  $18.5$ & $-3.95$ & $3.76$ $(0.17)$ & $1.66$ $(0.41)$ & $0.01$ $(0.01)$ & $3.02$ & N \\
    $112.358$ & C$^{17}$O & $J=1-0$ & $5.4$ & $-7.17$ & $3.72$ $(0.03)$ & $1.29$ $(0.08)$ & $0.21$ $(0.01)$ & $4.81$ & N \\
    $112.360$ & C$^{17}$O &$J=1-0$ &  $5.4$ & $-7.17$ & $4.20$ $(0.12)$ & $1.13$ $(0.31)$ & $0.12$ $(0.01)$ & $5.07$ & N \\
    $113.123$ & CN & $N=1-0, J=1/2-1/2$ & $5.4$ & $-5.88$ & $4.26$ $(0.02)$ & $1.01$ $(0.05)$ & $0.15$ $(0.01)$ & $5.93$ & N \\
    $113.144$ & CN & $N=1-0, J=1/2-1/2$ & $5.4$ & $-5.97$ & $4.28$ $(0.01)$ & $1.15$ $(0.05)$ & $0.56$ $(0.01)$ & $5.42$ & N \\
    $113.170$ & CN & $N=1-0, J=1/2-1/2$ & $5.4$ & $-5.28$ & $4.21$ $(0.01)$ & $1.16$ $(0.01)$ & $0.56$ $(0.01)$ & $5.35$ & N \\
    $113.191$ & CN & $N=1-0, J=1/2-1/2$ & $5.4$ & $-5.17$ & $4.22$ $(0.01)$ & $1.13$ $(0.01)$ & $0.63$ $(0.01)$ & $4.68$ & N \\
    $113.488$ & CN & $N=1-0, J=3/2-1/2$ & $5.4$ & $-5.17$ & $4.24$ $(0.10)$ & $1.09$ $(0.24)$ & $0.65$ $(0.14)$ & $56.14$ & N \\
    $113.490$ & CN & $N=1-0, J=3/2-1/2$ & $5.4$ & $-4.92$ & $4.30$ $(0.04)$ & $1.05$ $(0.07)$ & $1.26$ $(0.12)$ & $84.52$ & N \\
    $113.499$ & CN & $N=1-0, J=3/2-1/2$ & $5.4$ & $-4.97$ & $4.27$ $(0.15)$ & $1.04$ $(0.32)$ & $0.49$ $(0.22)$ & $160$ & N \\
    $113.508$ & CN & $N=1-0, J=3/2-1/2$ & $5.4$ & $-5.28$ & $4.29$ $(0.04)$ & $0.99$ $(0.10)$ & $0.59$ $(0.07)$ & $56.98$ & N \\
    $113.520$ & CN & $N=1-0, J=3/2-1/2$ & $5.4$ & $-5.88$ & $4.31$ $(0.22)$ & $0.92$ $(0.59)$ & $0.13$ $(0.08)$ & $65.04$ & N \\
   $115.271$ & CO & $J=1-0$ & $5.5$ & $-7.14$ & $\dots$ & $\dots$ & $\dots$ & $\dots$ & $\dots$ \\ 
    \hline     
    \multicolumn{9}{c}{Tuning: 133\,GHz} \\
    \hline
    $131.551$ & CCS & $N=10-9, J=11-10$ & $37.0$ & $-3.97$ & $3.67$ $(0.18)$ & $1.54$ $(0.49)$ &  $0.03$ $(0.03)$ & $6.62$ & N \\
    $134.284$ & HDCO & $2(1,1)-1(1,0)$ & $17.6$ & $-4.33$ & $3.90$ $0.05)$ & $0.99$ $(0.12)$ & $0.07$ $(0.01)$ & $7.15$ & N \\
    $135.298$ & H$_2$CS & $4(1,4)-3(1,3)$ & $29.9$ & $-4.48$ & $3.93$ $(0.04)$ & $1.53$ $(0.15)$ & $0.05$ $(0.01)$ & $4.82$ & N \\
    $136.464$ & HC$_3$N & $J=15-14$ & $52.4$ & $-3.70$ & $4.08$ $(0.01)$ & $1.17$ $(0.04)$ & $0.12$ $(0.01)$ & $3.92$ & N \\
    $136.725$ & CH$_3$CCH & $8(1)-7(1)$ & $36.8$ & $-5.07$ & $4.12$ $(0.09)$ & $0.95$ $(0.19)$ & $0.04$ $(0.01)$ & $5.24$ & N \\
    $136.728$ & CH$_3$CCH & $8(0)-7(0)$ & $29.5$ & $-5.05$ & $3.95$ $(0.11)$ & $1.68$ $(0.31)$ & $0.04$ $(0.01)$ & $5.76$ & N \\
    $137.371$ & H$_2$CS & $4(0,4)-3(0,3)$ & $16.5$ & $-4.43$ &  $4.28$ $(0.06)$ & $1.33$ $(0.18)$ & $0.04$ $(0.01)$ & $4.29$ & N \\
    $137.449$ & H$_2$$^{13}$CO & $2(1,2)-1(1,1)$ & $21.7$ & $-4.30$ & $3.95$ $(0.08)$ & $1.67$ $(0.25)$ & $0.03$ $(0.01)$ & $4.29$ & N \\
    $138.178$ & SO & $4(3)-3(2)$ & $15.9$ & $-4.49$ & $3.96$ $(0.01)$ & $1.38$ $(0.02)$ & $0.24$ $(0.01)$ & $4.78$ & N \\
    $146.635$ & H$_2^{13}$CO & $2(1,1)-1(1,0)$ & $22.4$ & $-4.22$ & $4.11$ $(0.10)$ & $1.24$ $(0.20)$ & $0.03$ $(0.01)$ & $7.04$ & N \\
    $146.969$ & CS & $J=3-2$ & $7$ & $-4.21$ & $4.18$ $(0.01)$ & $1.26$ $(0.01)$ & $1.70$ $(0.01)$ & $6.87$ & N \\
    $150.436$ & c-C$_3$H$_2$ & $2(2,0)-1(1,1)$ & $9.7$ & $-4.22$ & $4.20$ $(0.04)$ & $1.03$ $(0.11)$ & $0.08$ $(0.01)$ & $8.78$ & N \\
    $150.498$ & H$_2$CO & $2(1,1)-1(1,0)$ & $22.6$ & $-4.18$ & $4.12$ $(0.01)$ & $1.18$ $(0.01)$ & $1.41$ $(0.01)$ & $10.81$ & N \\
    $150.820$ & c-C$_3$H$_2$ & $4(0,4)-3(1,3)$ & $19.3$ & $-3.74$ & $4.19$ $(0.02)$ & $0.93$ $(0.05)$ & $0.17$ $(0.01)$ & $10.44$ & N \\	
    $150.851$ & c-C$_3$H$_2$ & $4(1,4)-3(0,3)$ & $19.3$ & $-3.74$ & $4.23$ $(0.01)$ & $0.97$ $(0.02)$ & $0.41$ $(0.01)$ & $9.89$ & N \\
    $152.609$ & DNC & $J=2-1$ & $11.0$ & $-3.81$ & $4.32$ $(0.01)$ & $0.97$ $(0.03)$ & $0.34$ $(0.01)$ & $9.61$ & N \\
    $153.814$ & CH$_3$CCH & $9(1)-8(1)$ &  $44.1$ & $-4.95$ & $4.28$ $(0.16)$ & $1.36$ $(0.31)$ & $0.04$ $(0.01)$ & $9.94$ & N \\
    $153.817$ & CH$_3$CCH & $9(0)-8(0)$ & $36.9$ & $-4.94$ & $4.12$ $(0.09)$ & $1.01$ $(0.21)$ & $0.05$ $(0.01)$ & $9.14$ & N \\
    \hline     
    \multicolumn{9}{c}{Tuning: 141\,GHz} \\
    \hline
    $138.739$ & $^{13}$CS & $J=3-2$ & $13.3$ & $-4.29$ & $4.34$ $(0.09)$ & $1.18$ $(0.20)$ & $0.04$ $(0.0)$ & $7.71$ & N \\
    $139.483$ & H$_2$CS & $4(1,3)-3(1,2)$ & $29.9$ & $-4.44$ & $4.36$ $(0.07)$ & $1.18$ $(0.16)$ & $0.05$ $(0.01)$ & $7.11$ & N \\
    $140.839$ & H$_2$CO & $2(1,2)-1(1,1)$ & $21.9$ & $-4.27$ & $4.08$ $(0.01)$ & $1.17$ $(0.01)$ & $1.70$ $(0.01)$ & $7.76$ & N \\
    $144.077$ & DCO$^{+}$ & $J=2-1$ & $10.4$ & $-3.67$ & $4.21$ $(0.01)$ & $1.01$ $(0.01)$ & $0.71$ $(0.01)$ & $4.00$ & N \\
    $144.241$ & CCD & $N=2-1, J=5/2-3/2$ & $10.4$ & $-5.08$ & $4.17$ $(0.09)$ & $1.35$ $(0.45)$ & $0.04$ $(0.01)$ & $4.45$ & N \\
    $144.242$ & CCD & $N=2-1, J=5/2-3/2$ & $10.4$ & $-5.20$ & $4.06$ $(0.06)$ & $1.14$ $(0.18)$ & $0.05$ $(0.01)$ & $4.41$ & N \\
    $144.244$ & CCS & $N,J=11,12-10,11$ & $43.9$ & $-3.85$ &$4.20$ $(0.21)$ & $0.97$ $(0.49)$ & $0.02$ $(0.01)$ & $4.17$ & N \\
    $144.296$ & CCD & $N=2-1, J=3/2-1/2$ &  $10.4$ & $-5.17$ & $3.73$ $(0.17)$ & $2.75$ $(0.43)$ & $0.03$ $(0.01)$ & $4.18$ & N \\ 
    $144.376$ & CCD & $N=2-1, J=3/2-3/2$ & $10.4$ & $-5.42$ & $4.38$ $(0.10)$ & $0.91$ $(0.25)$ & $0.02$ $(0.01)$ & $4.59$ & N \\
    $144.617$ & C$^{34}$S & $3-2$ &  $13.9$ & $-4.23$ & $4.14$ $(0.02)$ & $1.33$ $(0.05)$ & $0.10$ $(0.01)$ & $4.20$ & N \\
    $144.826$ & DCN & $J=2-1, F=2-2$ & $10.4$ & $-4.50$ & \multirow{2}{*}{$3.93$ $(0.05)$} & \multirow{2}{*}{$1.23$ $(0.14)$} & \multirow{2}{*}{$0.03$ $(0.01)$} & \multirow{2}{*}{$3.52$} & Y \\
    $144.826$ & DCN & $J=2-1, F=1-0$ & $10.4$ & $-4.15$ & & & & & Y \\
    $144.828$& DCN & $J=2-1, F=2-1$ & $10.4s$ & $-3.89$ & $4.05$ $(0.02)$ & $1.15$ $(0.05)$ & $0.13$ $(0.01)$ & $3.367$ & N \\
    $145.089$ & c-C$_3$H$_2$ & $3(1,2)-2(2,1)$ & $16.0$ & $-4.16$ & $4.18$ $(0.13)$ & $0.95$ $(0.70)$ & $0.21$ $(0.06)$ & $70.95$ & N \\
    $145.093$ & CH$_3$OH & $3(-0,3)-2(-0,2)$ $E$ & $27.1$ & $-4.91$ & $3.99$ $(0.50)$ & $1.07$ $(1.00)$ & $0.06$ $(0.07)$ & $57.56$ & N \\
    $145.097$ & CH$_3$OH & $3(1,3)-2(1,2)$ $E$ & $19.5$ & $-4.95$ & $4.00$ $(0.10)$ & $1.23$ $(0.23)$ & $0.27$ $(0.06)$ & $33.63$ & N \\
    $145.103$ & CH$_3$OH & $3(0,3)-2(0,2)$ $A$ & $13.9$ & $-4.91$ &  $4.01$ $(0.05)$ & $1.07$ $(0.09)$ & $0.34$ $(0.06)$ & $46.30$ & N \\
    $145.126$ & CH$_3$OH & $3(-2,2)-2(-2,1)$ $E$ & $36.2$ & $-5.16$ & \multirow{2}{*}{$3.84$ $(0.12)$} & \multirow{2}{*}{$1.54$ $(0.41)$} & \multirow{2}{*}{$0.03$ $(0.01)$} & \multirow{2}{*}{$4.42$} & Y \\
    $145.126$ & CH$_3$OH & $3(2,1)-2(2,0)$ $E$ & $39.8$ & $-5.16$ & & & & & Y \\
    $145.131$ & CH$_3$OH & $3(-1,2)-2(-1,1)$ $E$ & $35.0$ & $-4.95$ & $3.96$ $(0.11)$ & $1.40$ $(0.40)$ & $0.03$ $(0.01)$ & $6.31$ & N \\ 
    $145.560$ & HC$_3$N & $J=16-15$ & $59.4$ & $-3.61$ & $4.05$ $(0.02)$ & $1.10$ $(0.04)$ & $0.11$ $(0.01)$ & $4.35$ & N \\
    $145.602$ & H$_2$CO & $2(0,2)-1(0,1)$ & $10.5$ & $-4.10$ & $4.08$ $(0.01)$ & $1.23$ $(0.01)$ & $1.16$ $(0.01)$ & $4.19$ & N \\
    $146.368$ & CH$_3$OH & $3(1,2)-2(1,1)$ $A$ & $28.6$ & $-4.96$ &  $3.93$ $(0.10)$ & $1.27$ $(0.22)$ & $0.02$ $(0.01)$ & $4.83$ & N \\
    $154.657$ & HC$_3$N & $J=17-16$ & $66.8$ & $-3.53$ & $4.07$ $(0.03)$ & $1.25$ $(0.09)$ & $0.10$ $(0.01)$ & $7.04$ & N \\
    $155.518$ & c-C$_3$H$_2$ & $3(2,2)-2(1,1)$ & $16.1$ & $-3.95$ & $4.19$ $(0.02)$ & $0.95$ $(0.06)$ & $0.11$ $(0.01)$ & $7.28$ & N \\
    $157.246$ & CH$_3$OH & $4(-0,4)-4(1,4)$ $E$ & $36.3$ & $-4.67$ & $4.25$ $(0.10)$ & $1.46$ $(0.23)$ & $0.04$ $(0.01)$ & $7.28$ & N \\
    $157.270$ & CH$_3$OH & $1(-0,1)-1(1,1)$ $E$ & $15.4$ & $-4.65$ & $3.99$ $(0.06)$ & $1.21$ $(0.14)$ & $0.12$ $(0.01)$ & $6.85$ & N \\
    $157.272$ & CH$_3$OH & $3(-0,3)-3(1,3)$ $E$ & $27.1$ & $-4.66$ & $4.00$ $(0.16)$ & $1.46$ $(0.63)$ & $0.07$ $(0.02)$ & $6.91$ & N \\
    $157.276$ & CH$_3$OH & $2(-0,2)-2(1,2)$ $E$ & $20.1$ & $-4.66$ & $3.98$ $(0.08)$ & $1.26$ $(0.21)$ & $0.10$ $(0.02)$ & $6.57$ & N \\
    $158.971$ & SO & $3(4)-2(3)$ & $28.7$ & $-4.37$ & $4.12$ $(0.06)$ & $1.80$ $(0.17)$ & $0.07$ $(0.01)$ & $7.01$ & N \\
    \hline
    \multicolumn{9}{c}{Tuning: 202\,GHz} \\
    \hline
    $206.176$ & SO & $4(5)-3(4)$ & $38.6$ & $-3.99$ & $4.11$ $(0.10)$ & $1.54$ $(0.26)$ & $0.13$ $(0.03)$ & $26.15$ & N \\
    $216.112$ & DCO$^+$ & $J=3-2$ & $20.7$ & $-3.11$ & $4.11$ $(0.01)$ & $0.86$ $(0.02)$ & $0.77$ $(0.03)$ & $27.50$ & N \\
    $216.278$ & c-C$_3$H$_2$ & $3(3,0)-2(2,1)$ & $19.5$ & $-3.59$ & $4.12$ $(0.03)$ & $0.97$ $(0.08)$ & $0.27$ $(0.03)$ & $23.81$ & N \\
    $216.372$ & CCD & $N=3-2,$ $J=7/2-5/2$ & $20.8$ & $-3.52$ & \multirow{3}{*}{$3.57$ $(0.10)$} & \multirow{3}{*}{$1.65$ $(0.24)$} & \multirow{3}{*}{$0.11$ $(0.03)$} & \multirow{3}{*}{$28.43$} & Y \\
    $216.373$ & CCD & $N=3-2,$ $J=7/2-5/2$ & $20.8$ & $-3.57$ & &  &  &  & Y \\
    $216.373$ & CCD & $N=3-2,$ $J=7/2-5/2$ & $20.8$ & $-3.55$ & &  &  &  & Y \\
    $217.238$ & DCN & $J=3-2$ & $20.9$ & $-3.33$ & $4.15$ $(0.08)$ & $1.09$ $(0.20)$ & $0.12$ $(0.03)$ & $24.58$ & N \\
    $217.822$ & c-C$_3$H$_2$ & $6(1,6)-5(0,5)$ & $38.6$ & $-3.26$ & $4.13$ $(0.03)$ &  $1.10$ $(0.08)$ & $0.27$ $(0.03)$ & $28.04$ & N \\
    $217.940$ & c-C$_3$H$_2$ & $5(1,4)-4(2,3)$ & $35.4$ & $-3.39$ & $4.15$ $(0.05)$ &  $0.91$ $(0.10)$ & $0.17$ $(0.03)$ & $25.74$ & N \\
    $218.222$ & H$_2$CO & $3(0,3)-2(0,2)$ & $21.0$ & $-3.54$ & $4.02$ $(0.01)$ & $1.16$ $(0.02)$ & $1.31$ $(0.03)$ & $24.59$ & N \\
    $218.440$ & CH$_3$OH & $4(-2,3)-3(-1,2)$ $E$ & $45.5$ & $-3.32$ & $3.91$ $(0.10)$ & $0.98$ $(0.20)$ & $0.09$ $(0.03)$ & $26.27$ & N \\
    $219.560$ & C$^{18}$O & $J=2-1$ & $15.8$ & $-6.22$ & $4.07$ $(0.01)$ & $1.09$ $(0.01)$ & $3.69$ $(0.05)$ & $32.17$ & N \\
    $219.560$ & H$_2$CN & $3(0,3)-2(0,2)$ & $21.1$ & $-5.05$ & $3.77$ $(0.14)$ & $0.90$ $(0.42)$ & $0.10$ $(0.04)$ & $33.09$ & N \\
    $219.949$ & SO & $6(5)-5(4)$ & $35.0$ & $-3.87$ & $4.14$ $(0.04)$ & $1.23$ $(0.09)$ & $0.29$ $(0.04)$ & $33.54$ & N \\
    $220.398$ & $^{13}$CO & $J=2-1$ & $15.9$ & $-6.21$ & $4.13$ $(0.01)$ & $1.55$ $(0.01)$ & $8.75$ $(0.11)$ & $30.68$ & N \\
    \hline
    \multicolumn{9}{c}{Tuning: 210\,GHz} \\
    \hline
    $211.211$ & H$_2$CO & $3(1,3)-2(1,2)$ & $32.1$ & $-3.64$ & $4.13$ $(0.01)$ & $1.07$ $(0.02)$ & $2.048$ $(0.06)$ & $44.13$ & N \\
    $212.928$ & H$_2$CN & $3(1,3)-2(1,2)$ & $32.3$ & $-3.68$ & $3.92$ $(0.15)$ & $1.29$ $(0.36)$ & $0.09$ $(0.04)$ & $34.86$ & N \\
    $224.714$ & C$^{17}$O & $J=2-1$ & $16.2$ & $-6.34$ & $4.77$ $(0.01)$ & $1.47$ $(0.04)$ & $1.04$ $(0.04)$ & $29.31$ & N \\
    $225.697$ & H$_2$CO & $3(1,2)-2(1,1)$ & $33.4$ & $-3.55$ & $4.09$ $(0.01)$ & $1.12$ $(0.01)$ & $1.78$ $(0.04)$ & $29.47$ & N \\
    $226.298$ & CN & $N=2-1, J=3/2-3/2$ & $16.3$ & $-5.08$ & $4.53$ $(0.14)$ & $1.00$ $(0.23)$ & $0.09$ $(0.04)$ & $31.07$ & N \\
    $226.303$ & CN & $N=2-1, J=3/2-3/2$ & $16.3$ & $-5.37$ & $4.03$ $(0.06)$ & $0.48$ $(0.17)$ & $0.13$ $(0.05)$ & $34.72$ & N \\
    $226.314$ & CN & $N=2-1, J=3/2-3/2$ & $16.3$ & $-5.00$ & $4.09$ $(0.07)$ & $0.76$ $(0.17)$ & $0.13$ $(0.04)$ & $36.27$ & N \\
    $226.359$ & CN & $N=2-1, J=3/2-3/2$ & $16.3$ & $-4.79$ & $4.11$ $(0.04)$ & $1.06$ $(0.09)$ & $0.26$ $(0.04)$ & $29.72$ & N \\
    $226.632$ & CN & $N=2-1, J=3/2-1/2$ & $16.3$ & $-4.37$ & $4.11$ $(0.02)$ & $0.92$ $(0.05)$ & $0.40$ $(0.04)$ & $31.92$ & N \\
    $226.659$ & CN & $N=2-1, J=3/2-1/2$ & $16.3$ & $-4.02$ & $4.13$ $(0.02)$ & $1.02$ $(0.06)$ & $0.91$ $(0.09)$ & $71.64$ & N \\
    $226.663$ & CN & $N=2-1, J=3/2-1/2$ & $16.3$ & $-4.07$ & $4.10$ $(0.15)$ & $0.96$ $(0.33)$ & $0.27$ $(0.13)$ & $70.84$ & N \\
    $226.679$ & CN & $N=2-1, J=3/2-1/2$ & $16.3$ & $-4.27$ & $4.12$ $(0.05)$ & $0.93$ $(0.11)$ & $0.43$ $(0.11)$ & $85.94$ & N \\
    $226.874$ & CN & $N=2-1, J=5/2-3/2$ & $16.3$ & $-4.06$ & $3.59$ $(0.02)$ & $1.72$ $(0.06)$ & $1.63$ $(0.09)$ & $73.74$ & N \\
    $226.887$ & CN & $N=2-1, J=5/2-3/2$ & $16.3$ & $-4.56$ & $4.25$ $(0.26)$ & $0.66$ $(0.45)$ & $0.26$ $(0.36)$ & $284$ & N \\
    $226.892$ & CN & $N=2-1, J=5/2-3/2$ & $16.3$ & $-4.74$ & $4.14$ $(0.27)$ & $0.79$ $(0.47)$ & $0.28$ $(0.36)$ & $285$ & N \\
    $227.169$ & c-C$_3$H$_2$ & $4(3,2)-3(2,1)$ & $29.1$ & $-3.51$ & $4.20$ $(0.05)$ & $0.86$ $(0.13)$ & $0.18$ $(0.04)$ & $29.02$ & N \\
    $228.910$ & DNC & $J=3-2$ & $22.0$ & $-3.25$ & $4.18$ $(0.08)$ & $1.05$ $(0.18)$ & $0.16$ $(0.05)$ & $41.83$ & N \\
    $230.538$ & CO & $J=2-1$ & $16.6$ & $-6.16$ & \dots & \dots & \dots & \dots & \dots \\
    \hline
    \multicolumn{9}{c}{Tuning: 242\,GHz} \\
    \hline
    $241.767$ & CH$_3$OH & $5(1,5)-4(1,4)$ $E$ & $40.4$ & $-4.23$ & $4.22$ $(0.14)$ & $1.42$ $(0.47)$ & $0.16$ $(0.08)$ & $61.14$ & N \\
    $241.791$ & CH$_3$OH & $5(0,5)-4(0,4)$ $A$ & $34.8$ & $-4.21$ & $4.08$ $(0.08)$ & $1.01$ $(0.18)$ & $0.19$ $(0.08)$ & $57.21$ & N \\
    $244.935$ & CS & $J=5-4$ & $35.3$ & $-3.52$ & $4.12$ $(0.01)$ & $1.15$ $(0.02)$ & $1.23$ $(0.05)$ & $37.76$ & N \\
    $258.255$ & SO & $6(6)-5(5)$ & $56.5$ & $-3.66$ & $4.24$ $(0.09)$ & $1.26$ $(0.22)$ & $0.18$ $(0.06)$ & $45.20$ & N \\ 
    $260.255$ & H$^{13}$CO$^+$ & $J=3-2$ & $25.0$ & $-2.87$ & $4.05$ $(0.02)$ & $0.93$ $(0.05)$ & $0.84$ $(0.08)$ & $63.97$ & N \\
    $261.805$ & CH$_3$OH & $2(-1,1)-1(-0,1)$ $E$ & $28.0$ & $-4.25$ & $4.50$ $(0.11)$ & $1.17$ $(0.28)$ & $0.20$ $(0.08)$ & $58.26$ & N \\
    $261.843$ & SO & $7(6)-6(5)$ & $47.6$ & $-3.64$ & $4.06$ $(0.09)$ & $1.20$ $(0.21)$ & $0.22$ $(0.08)$ & $61.15$ & N \\
    $262.004$ & CCH & $N=3-2, J=7/2-5/2$ & $25.1$ & $-4.27$ & $4.06$ $(0.03)$ & $1.04$ $(0.07)$ & $1.20$ $(0.13)$ & $56.72$ & N \\
    $262.006$ & CCH & $N=3-2, J=7/2-5/2$ & $25.1$ & $-4.29$ & $4.12$ $(0.06)$ & $1.13$ $(0.15)$ & $0.85$ $(0.16)$ & $53.22$ & N \\
    $262.064$ & CCH & $N=3-2, J=5/2-3/2$ & $25.2$ & $-4.31$ & $4.12$ $(0.03)$ & $1.06$ $(0.08)$ & $0.88$ $(0.11)$ & $61.20$ & N \\
    $262.067$ & CCH & $N=3-2, J=7/2-5/2$ & $25.2$ & $-4.34$ & $4.09$ $(0.06)$ & $0.98$ $(0.14)$ & $0.67$ $(0.13)$ & $58.27$ & N \\
    \hline
\end{longtable}
\tablefoot{
\tablefoottext{a}{Upper energy level.}
\tablefoottext{b}{Einstein coefficient.}
\tablefoottext{*}{Parameters from the Gaussian fit results, where the formal errors of the fits are given in parentheses. The rms is a result from the baseline fitting.}
}
\end{small}

\begin{tiny}
\begin{longtable}{cccccccccc} 
\caption{\label{tab:res-apex-fits}Lines detected toward V1057~Cyg with the APEX 12-m telescope.}\\
\hline \hline
Rest frequency & \multirow{2}{*}{Molecule} & \multirow{2}{*}{Quantum numbers} & $E_{\rm up}^{(a)}$ & $log_{10}A_{\rm ij}^{(b)}$ & $\varv_{\rm LSR}$ & $\Delta \varv$ & $T_{\rm MB}$ & rms & Blended? \\
    (GHz) & & & (K) & (s$^{-1}$) & (km\,s$^{-1}$) & (km\,s$^{-1}$) & (K) & (mK) & (Y/N) \\
\hline
\endfirsthead
\caption{Continued.}\\
\hline\hline
Rest frequency & \multirow{2}{*}{Molecular species} & \multirow{2}{*}{Quantum numbers} & $E_{\rm up}^{(a)}$ & \multirow{2}{*}{$log_{10}A_{\rm ij}^{(b)}$} & $\varv_{\rm LSR}$ & $\Delta \varv$ & $T_{\rm MB}$ & rms & Blended? \\
    (GHz) & & & (K) & & (km\,s$^{-1}$) & (km\,s$^{-1}$) & (K) & (mK) & (Y/N) \\
\hline
\endhead
\hline
\endfoot
\hline
\multicolumn{10}{c}{Receiver: nFLASH230, Tuning: 219.2\,GHz} \\
\hline
$216.112$  & DCO$^+$ & $J=3-2$ & $20.7$ & $-3.11$ & $4.31$ $(0.01)$ & $0.76$ $(0.02)$ & $0.36$ $(0.02)$ & $18.45$ & N \\
$216.278$  & c-C$_3$H$_2$ & $3(3,0)-2(2,1)$ & $19.5$ & $-3.55$ & $4.30$ $(0.04)$ & $1.07$ $(0.09)$ & $0.13$ $(0.02)$ & $19.43$ & N \\
$218.222$  & H$_2$CO & $3(0,3)-2(0,2)$ & $21.0$ & $-3.54$ & $4.09$ $(0.01)$ & $1.26$ $(0.03)$ & $0.38$ $(0.02)$ & $20.38$ & N \\
$217.822$  & c-C$_3$H$_2$ & $6(0,6)-5(1,5)$ & $38.6$ & $-3.26$ & \multirow{2}{*}{$4.14$ $(0.04)$} & \multirow{2}{*}{$0.77$ $(0.10)$} & \multirow{2}{*}{$0.82$ $(0.02)$} & \multirow{2}{*}{$20.04$} & Y \\
$217.822$  & c-C$_3$H$_2$ & $6(1,6)-5(0,5)$ & $38.6$ & $-3.26$ & &  & & & Y \\
$219.560$  & C$^{18}$O & $J=2-1$ & $15.8$ & $-6.22$ & $4.15$ $(0.01)$ & $0.97$ $(0.01)$ & $1.70$ $(0.02)$ & $19.27$ & N \\
$219.949$  & SO & $6(5)-5(4)$ & $35.0$ & $-3.87$ & $4.12$ $(0.06)$ & $1.48$ $(0.14)$ & $0.08$ $(0.02)$ & $20.01$ & N \\
$220.398$  & $^{13}$CO & $J=2-1$ & $15.9$ & $-6.21$ & $4.25$ $(0.01)$ & $1.21$ $(0.01)$ & $4.77$ $(0.05)$ & $21.09$ & N \\
$230.538$ & CO & $J=2-1$ & $16.6$ & $-6.16$ & \dots & \dots & \dots & \dots & \dots  \\
$231.321$ & N$_2$D$^+$ & $J=3-2$ & $22.2$ & $-3.14$ & $4.45$ $(0.06)$ & $0.71$ $(0.19)$ & $0.08$ $(0.02)$ & $19.79$ & N \\
\hline
\multicolumn{10}{c}{Receiver: nFLASH230, Tuning: 227.2\,GHz, Mode: onOff} \\
\hline
$224.714$ & C$^{17}$O & $J=3-2$ & $16.2$ & $-6.19$ & $4.28$ $(0.01)$ & $1.51$ $(0.05)$ & $0.28$ $(0.02)$ & $18.25$ & N \\
$225.697$ & H$_2$CO & $3(1,2)-2(1,1)$ & $33.4$ & $-3.55$ & $4.14$ $(0.01)$ & $1.21$ $(0.02)$ & $0.49$ $(0.02)$ & $18.74$ & N \\
$226.359$ & CN & $N=2-1, J=3/2-3/2$ & $16.3$ & $-4.79$ & $4.24$ $(0.04)$ & $0.96$ $(0.12)$ & $0.10$ $(0.02)$ & $20.78$ & N \\
$226.632$ & CN & $N=2-1, J=3/2-1/2$ & $16.3$ & $-4.37$ & $4.26$ $(0.03)$ & $0.90$ $(0.08)$ & $0.11$ $(0.02)$ & $19.47$ & N \\
$226.659$ & CN & $N=2-1, J=3/2-1/2$ & $16.3$ & $-4.02$ & $4.31$ $(0.02)$ & $0.78$ $(0.03)$ & $0.31$ $(0.04)$ & $30.83$ & N \\
$226.663$ & CN & $N=2-1, J=3/2-1/2$ & $16.3$ & $-4.07$ & $4.24$ $(0.08)$ & $0.99$ $(0.20)$ & $0.10$ $(0.04)$ & $19.41$ & N \\
$226.679$ & CN & $N=2-1, J=3/2-1/2$ & $16.3$ & $-4.27$ & $4.33$ $(0.08)$ & $0.93$ $(0.08)$ & $0.12$ $(0.04)$ & $31.88$ & N \\
$226.874$ & CN & $N=2-1, J=5/2-3/2$ & $16.3$ & $-4.01$ & \multirow{2}{*}{$4.41$ $(0.01)$} & \multirow{2}{*}{$1.78$ $(0.04)$} & \multirow{2}{*}{$0.48$ $(0.03)$} & $18.40$ & Y \\
$226.875$ & CN & $N=2-1, J=5/2-3/2$ & $16.3$ & $-3.94$ & &  &  & $18.40$ & Y \\
$226.887$ & CN & $N=2-1, J=5/2-3/2$ & $16.3$ & $-4.56$ & $4.18$ $(0.04)$ & $0.93$ $(0.10)$ & $0.11$ $(0.04)$ & $20.60$ & N \\
$226.892$ & CN & $N=2-1, J=5/2-3/2$ & $16.3$ & $-4.74$ & $4.26$ $(0.04)$ & $0.87$ $(0.07)$ & $0.13$ $(0.03)$ & $20.97$ & N \\
$228.910$ & DNC & $J=3-2$ & $22.0$ & $-3.25$ & $4.40$ $(0.02)$ & $0.61$ $(0.05)$ & $0.16$ $(0.03)$ & $21.64$ & N \\
$241.767$ & CH$_3$OH & $5(1,5)-4(1,4)$ $E$  & $40.4$ & $-4.23$ &  $4.12$ $(0.05)$ & $1.04$ $(0.12)$ & $0.08$ $(0.03)$ & $26.21$ & N \\
$241.791$ & CH$_3$OH & $5(0,5)-4(0,4)$ $A$ & $34.8$ & $-4.21$ & $4.01$ $(0.03)$ & $1.35$ $(0.07)$ & $0.09$ $(0.02)$ & $14.66$ & N \\
$244.935$ & CS & $J=5-4$ & $35.3$ & $-3.52$ & $4.16$ $(0.01)$ & $1.21$ $(0.02)$ & $0.31$ $(0.02)$ & $16.86$ & N \\ 
\hline
\multicolumn{10}{c}{Receiver: nFLASH230, Tuning: 227.2\,GHz, Mode: otf} \\
\hline
$219.560$  & C$^{18}$O & $J=2-1$ & $15.8$ & $-6.22$ & $4.03$ $(0.03)$ & $0.81$ $(0.10)$ & $0.20$ $(0.05)$ & $35.45$ & N \\
$220.398$  & $^{13}$CO & $J=2-1$ & $15.9$ & $-6.21$ & $3.95$ $(0.01)$ & $1.36$ $(0.02)$ & $1.01$ $(0.05)$ & $34.67$ & N \\
$230.538$  & CO & $J=2-1$ & $16.6$ & $-6.16$ & $\dots$ & $\dots$ & $\dots$ & $\dots$ & $\dots$ \\
\hline
\multicolumn{10}{c}{Receiver: LAsMA345, Tuning: 344.2\,GHz, Observation date: 2021-11-19, Mode: otf} \\
\hline
$345.795$  & CO & $J=3-2$ & $33.2$ & $-5.60$ & $\dots$ & $\dots$ & $\dots$ & $\dots$ & $\dots$ \\
$356.734^{(*)}$  & HCO$^+$ & $4-3$ & $42.8$ & $-2.44$ & $4.05$ $(0.06)$ & $1.31$ $(0.12)$ & $2.65$ $(0.63)$ & $455$ & N \\
\hline
\multicolumn{10}{c}{Receiver: SEPIA345, Tuning: 291\,GHz} \\
\hline
$279.511$  & N$_2$H$^+$ & $J=3-2$ & $26.8$ & $-2.89$ & $4.46$ $(0.01)$ & $0.89$ $(0.02)$ & $0.47$ $(0.03)$ & $26.47$ & N \\
$290.623$  & H$_2$CO & $4(0,4)-3(0,3)$ & $34.9$ & $-3.16$ & $4.09$ $(0.04)$ & $1.36$ $(0.10)$ & $0.79$ $(0.02)$ & $18.83$ & N \\
$293.912$  & CS & $J=6-5$ & $49.4$ & $-3.28$ & $4.36$ $(0.03)$ & $0.99$ $(0.08)$ & $0.14$ $(0.02)$ & $20.34$ & N \\
\end{longtable}
\tablefoot{
\tablefoottext{a}{Upper energy level.}
\tablefoottext{b}{Einstein coefficient.}
\tablefoottext{*}{The spectrum was spectrally smoothed by a factor of 2 using the \texttt{smooth} built-in function in CLASS, and presented as such throughout this paper.}
}
\end{tiny}

\begin{figure*}[h]
\centering 
\vspace{-4cm}
\hspace{-2cm}
  \begin{minipage}[h]{0.32\textwidth}
    \includegraphics[width=2.5\textwidth]{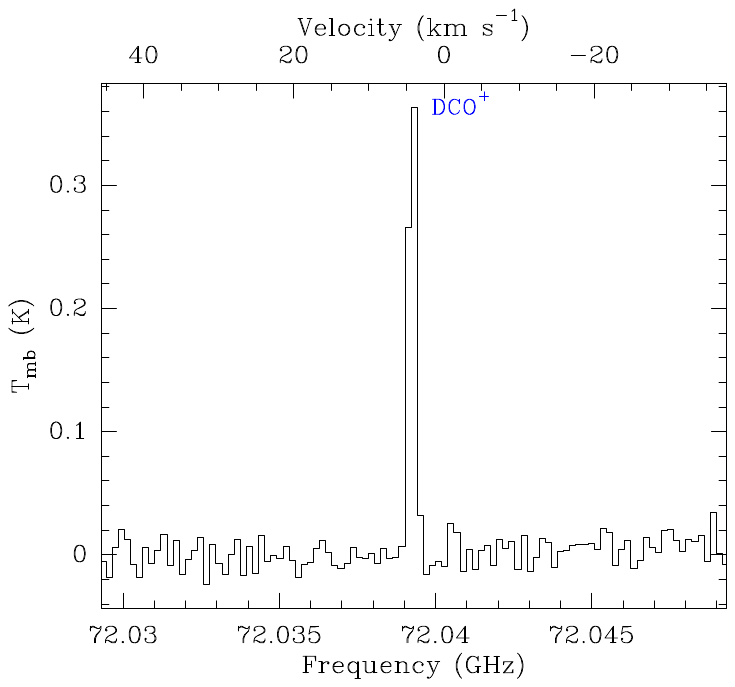}
  \end{minipage}
  \hspace{0.6cm}
  \begin{minipage}[h]{0.32\textwidth}
    \includegraphics[width=2.5\textwidth]{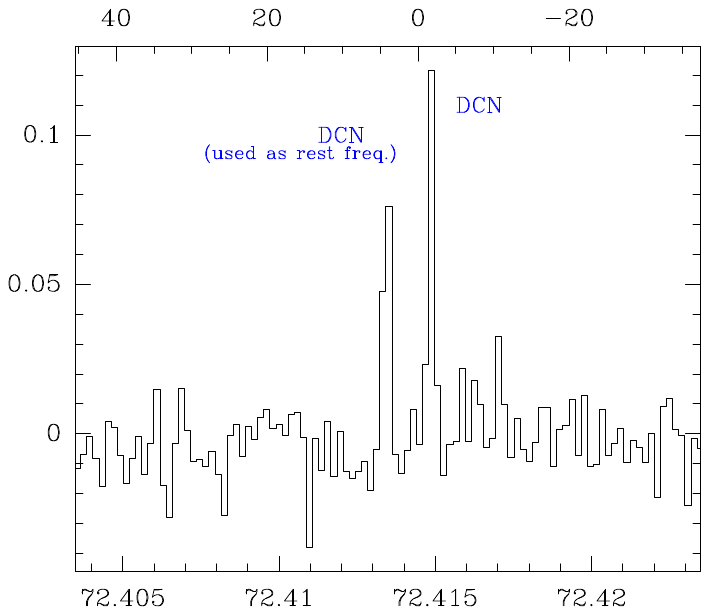}  
  \end{minipage}
  \hspace{0.6cm}
  \begin{minipage}[h]{0.32\textwidth}
    \includegraphics[width=2.5\textwidth]{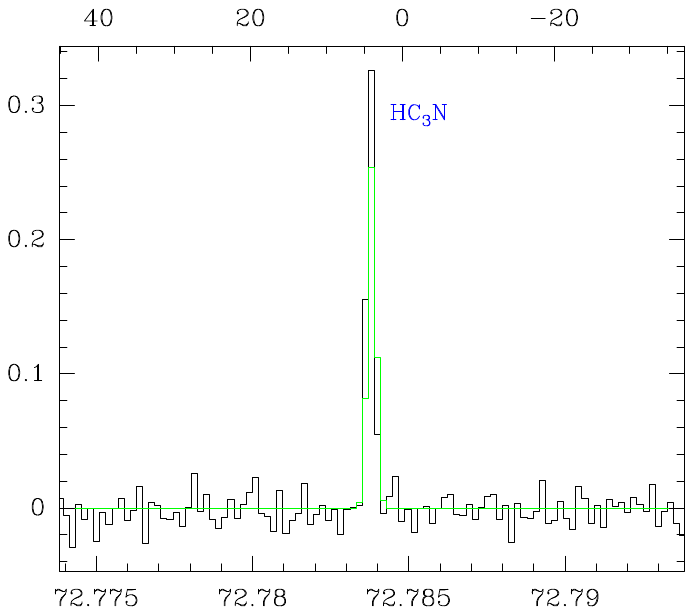}
  \end{minipage} \\
\vspace{-5.5cm}
\hspace{-2cm}
  \begin{minipage}[h]{0.32\textwidth}
    \includegraphics[width=2.5\textwidth]{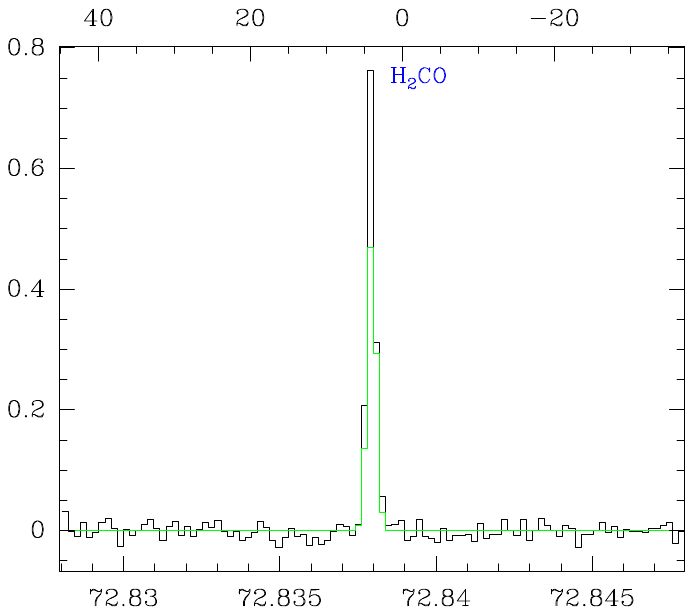}
  \end{minipage}
  \hspace{0.6cm}
  \begin{minipage}[h]{0.32\textwidth}
    \includegraphics[width=2.5\textwidth]{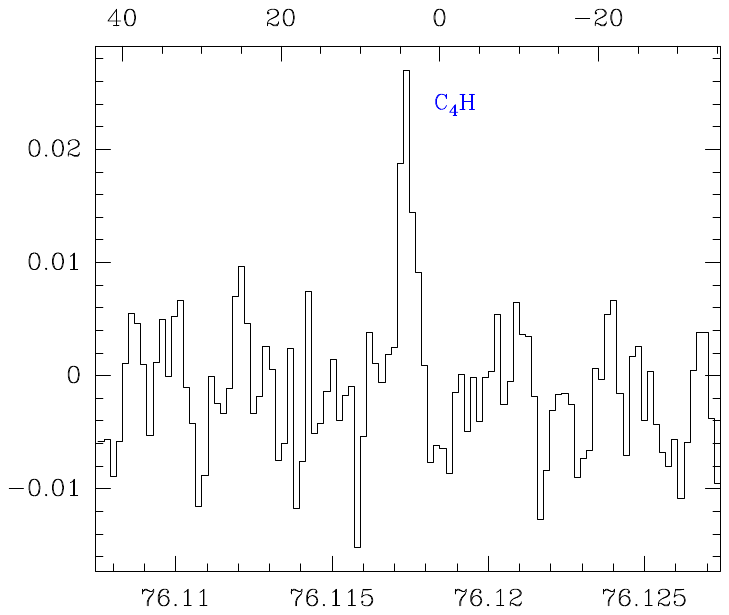}  
  \end{minipage}
  \hspace{0.6cm}
  \begin{minipage}[h]{0.32\textwidth}
    \includegraphics[width=2.5\textwidth]{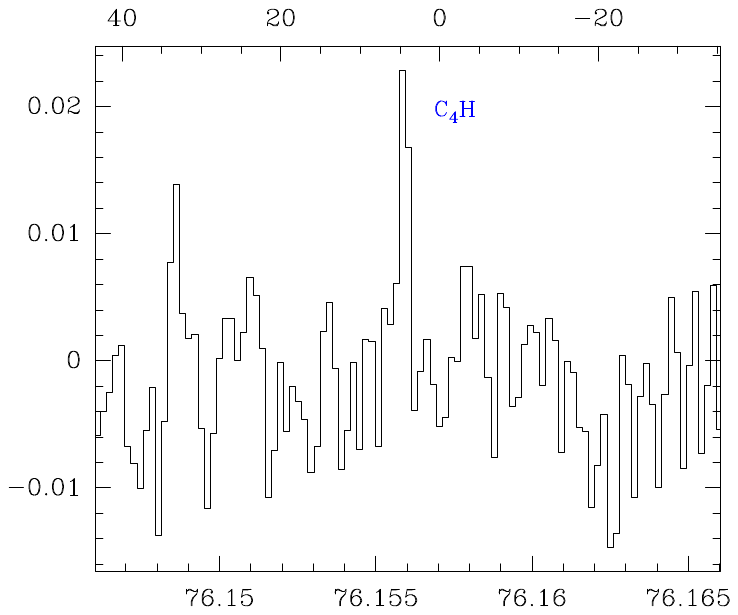}
  \end{minipage} \\
\vspace{-5.5cm}
\hspace{-2cm}
  \begin{minipage}[h]{0.32\textwidth}
    \includegraphics[width=2.5\textwidth]{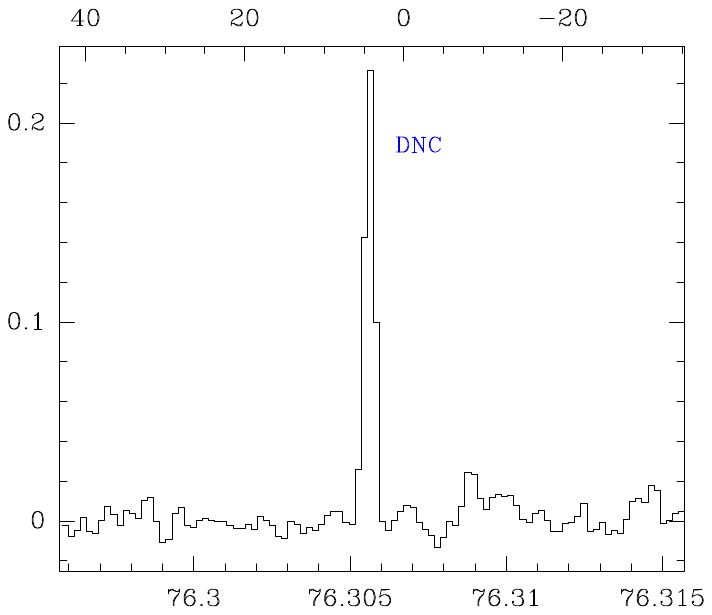}
  \end{minipage}
  \hspace{0.6cm}
  \begin{minipage}[h]{0.32\textwidth}
    \includegraphics[width=2.5\textwidth]{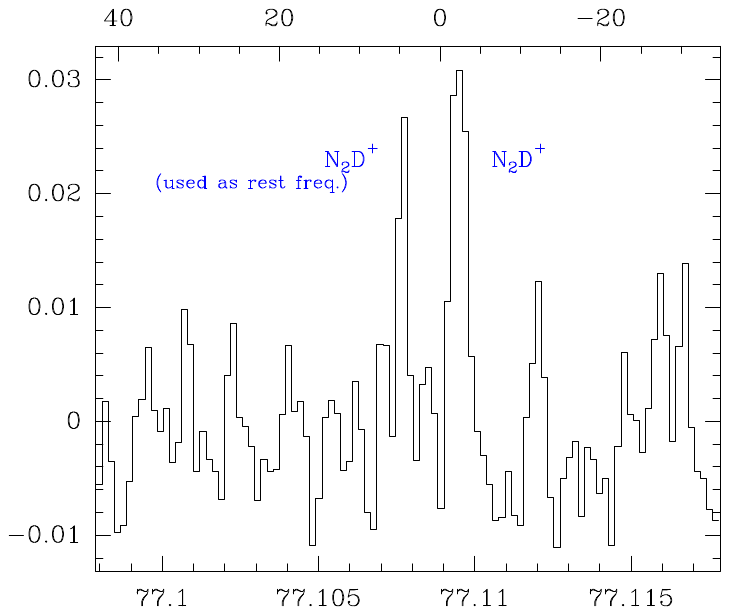}  
  \end{minipage}
  \hspace{0.6cm}
  \begin{minipage}[h]{0.32\textwidth}
    \includegraphics[width=2.5\textwidth]{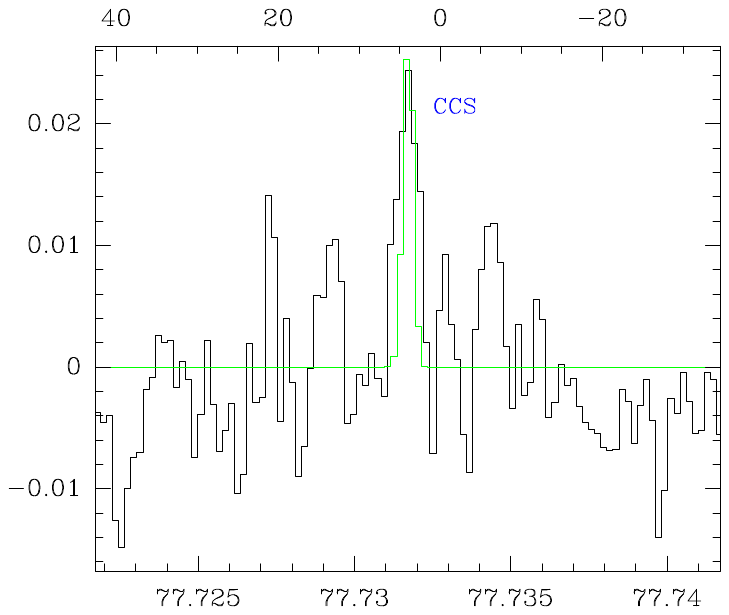}
  \end{minipage} \\
\vspace{-5.5cm}
\hspace{-2cm}
  \begin{minipage}[h]{0.32\textwidth}
    \includegraphics[width=2.5\textwidth]{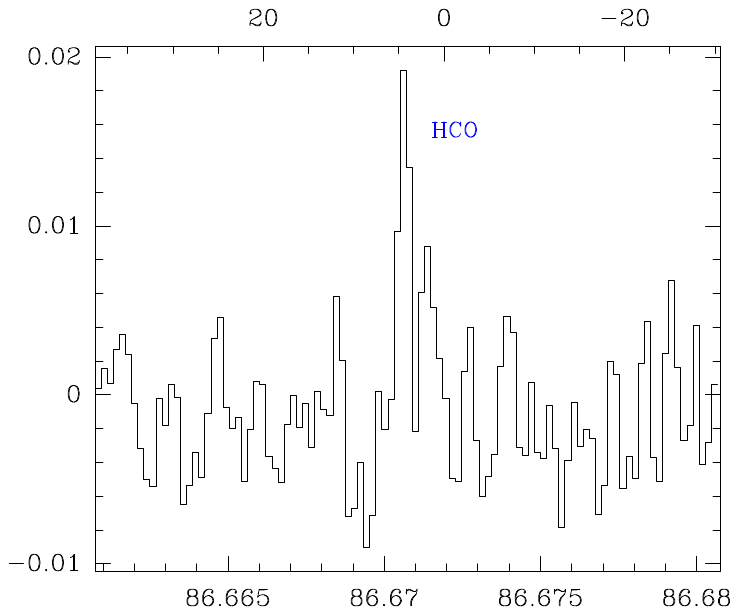}
  \end{minipage}
  \hspace{0.6cm}
  \begin{minipage}[h]{0.32\textwidth}
    \includegraphics[width=2.5\textwidth]{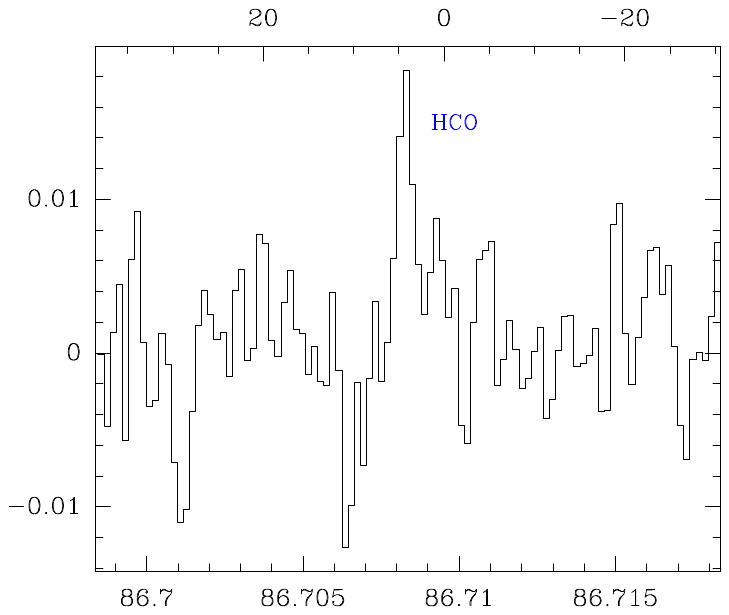}  
  \end{minipage}
  \hspace{0.6cm}
  \begin{minipage}[h]{0.32\textwidth}
    \includegraphics[width=2.5\textwidth]{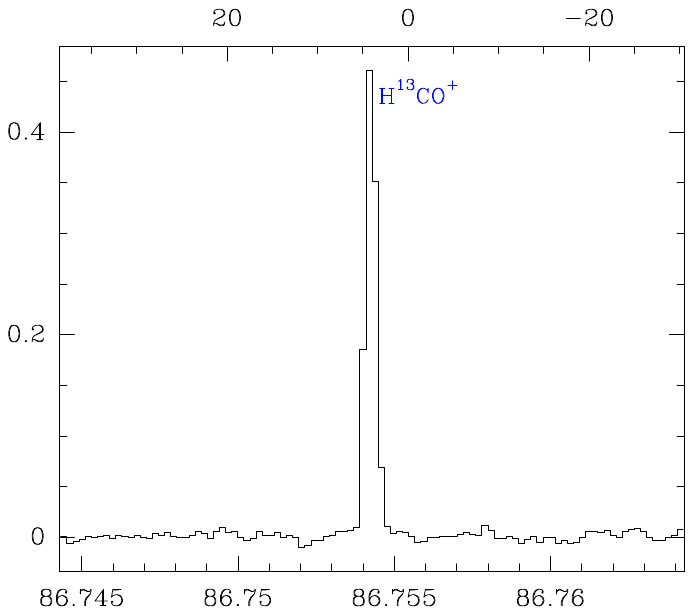}
  \end{minipage} 
 \vspace{-1.8cm} 
\caption{Spectra of lines detected towards V1057~Cyg with the IRAM~30-m telescope in the 73\,GHz tuning, displayed in main-beam temperature scale. Molecule names for identified lines are labelled in blue, the synthetic model (displayed in the population diagrams in Fig.~\ref{fig:pop-diagrams}) is overlaid in green, tentative detections are labelled in dark green, and unidentified lines are labelled in red. 
In certain panels, extra text describes the species or transition used as rest frequency to display the top axis.}
\label{fig:survey-73ghz-small}
\end{figure*}

\addtocounter{figure}{-1}
\begin{figure*}[h]
\centering 
\vspace{-4cm}
\hspace{-2cm}
  \begin{minipage}[h]{0.32\textwidth}
    \includegraphics[width=2.5\textwidth]{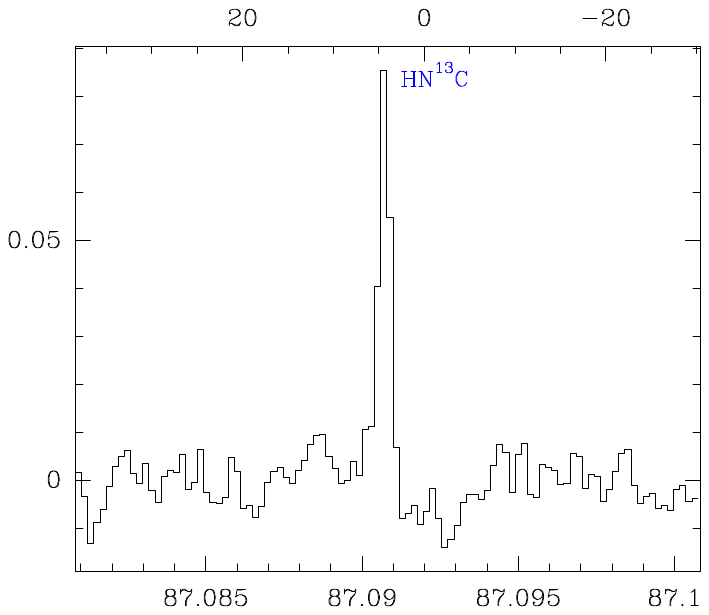}
  \end{minipage}
  \hspace{0.6cm}
  \begin{minipage}[h]{0.32\textwidth}
    \includegraphics[width=2.5\textwidth]{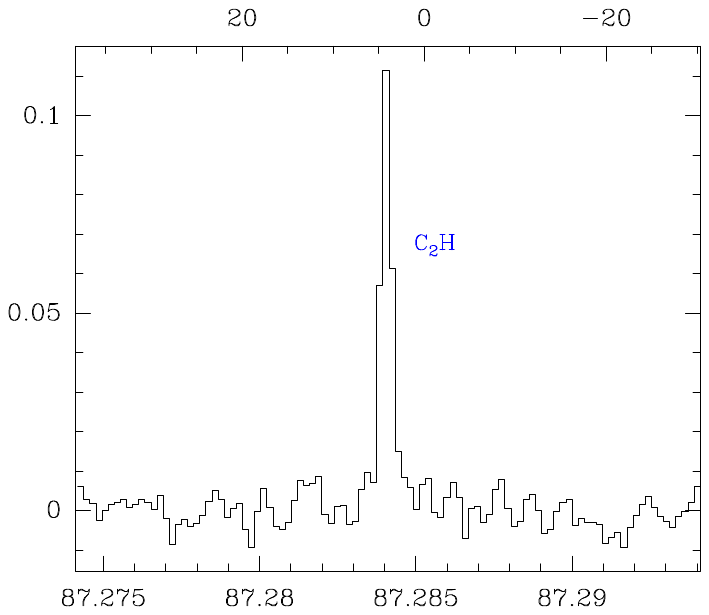}  
  \end{minipage}
  \hspace{0.6cm}
  \begin{minipage}[h]{0.32\textwidth}
    \includegraphics[width=2.5\textwidth]{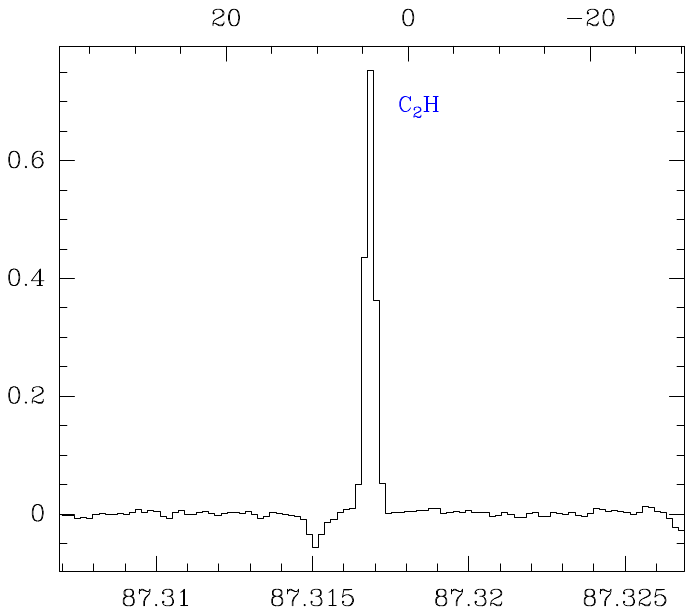}
  \end{minipage} \\
\vspace{-5.5cm}
\hspace{-2cm}
  \begin{minipage}[h]{0.32\textwidth}
    \includegraphics[width=2.5\textwidth]{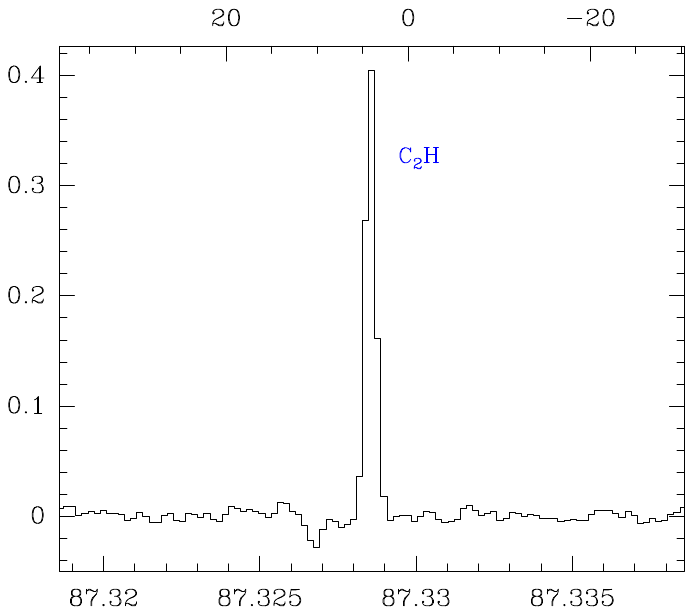}
  \end{minipage}
  \hspace{0.6cm}
  \begin{minipage}[h]{0.32\textwidth}
    \includegraphics[width=2.5\textwidth]{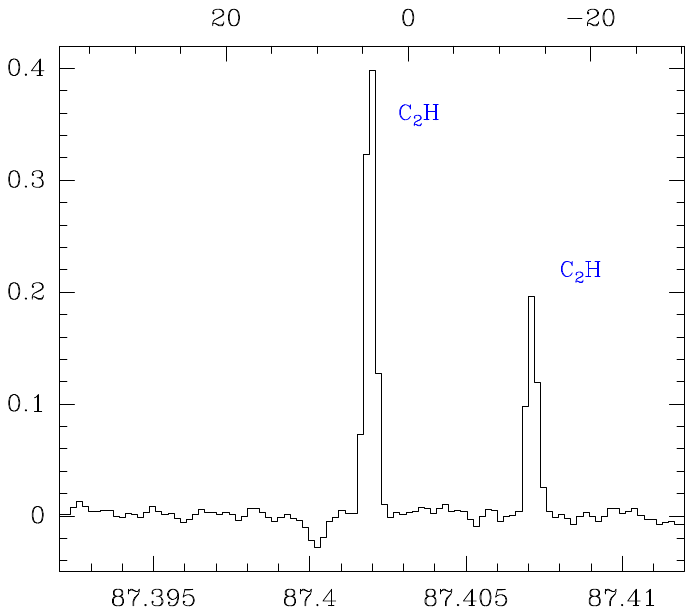}  
  \end{minipage}
  \hspace{0.6cm}
  \begin{minipage}[h]{0.32\textwidth}
    \includegraphics[width=2.5\textwidth]{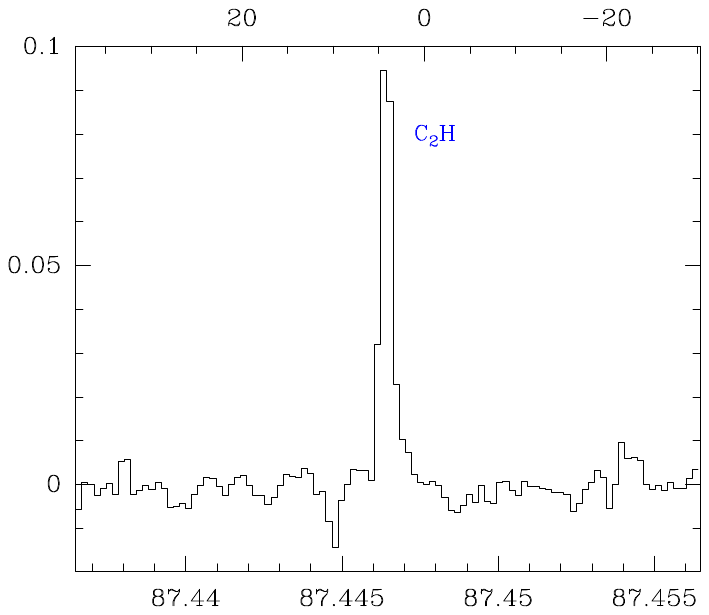}
  \end{minipage} \\
\vspace{-5.5cm}
\hspace{-2cm}
  \begin{minipage}[h]{0.32\textwidth}
    \includegraphics[width=2.5\textwidth]{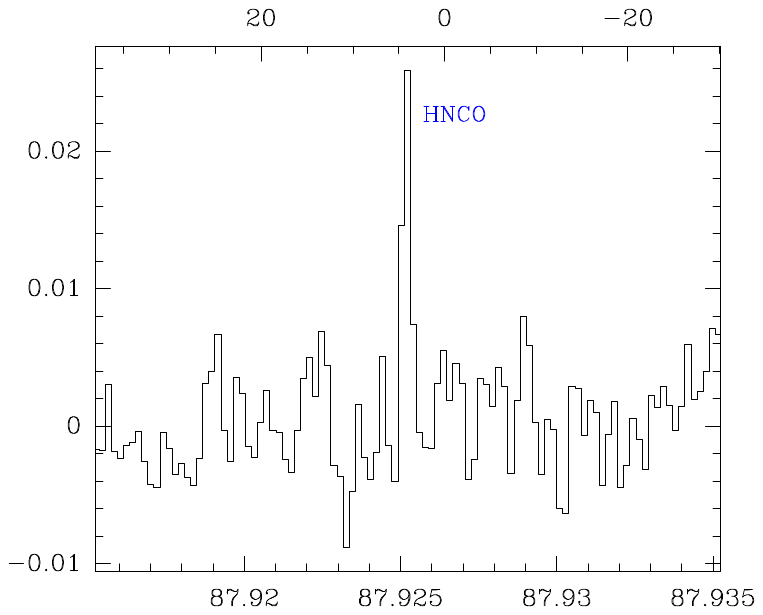}
  \end{minipage}
  \hspace{0.6cm}
  \begin{minipage}[h]{0.32\textwidth}
    \includegraphics[width=2.5\textwidth]{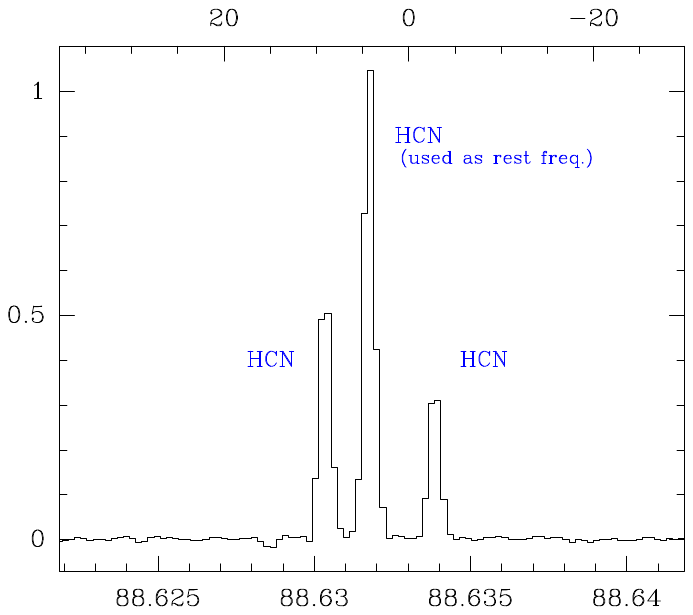}  
  \end{minipage}
  \hspace{0.6cm}
  \begin{minipage}[h]{0.32\textwidth}
    \includegraphics[width=2.5\textwidth]{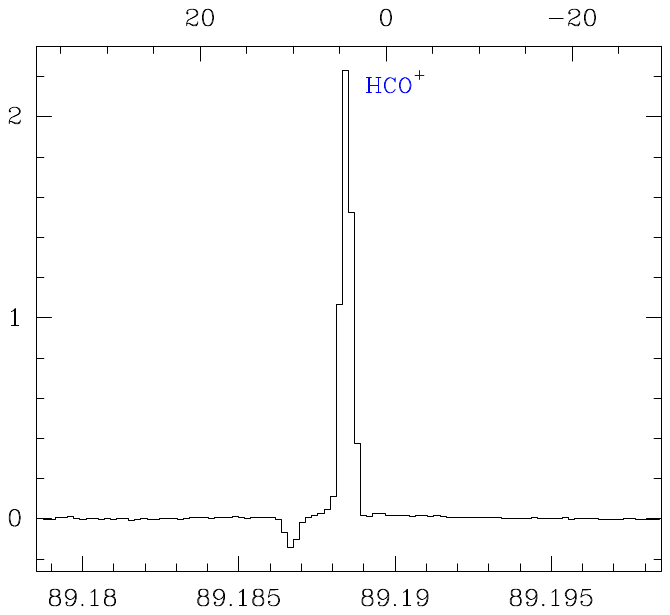}
  \end{minipage} \\
\vspace{-5.5cm}
\hspace{-2cm}
  \begin{minipage}[h]{0.32\textwidth}
    \includegraphics[width=2.5\textwidth]{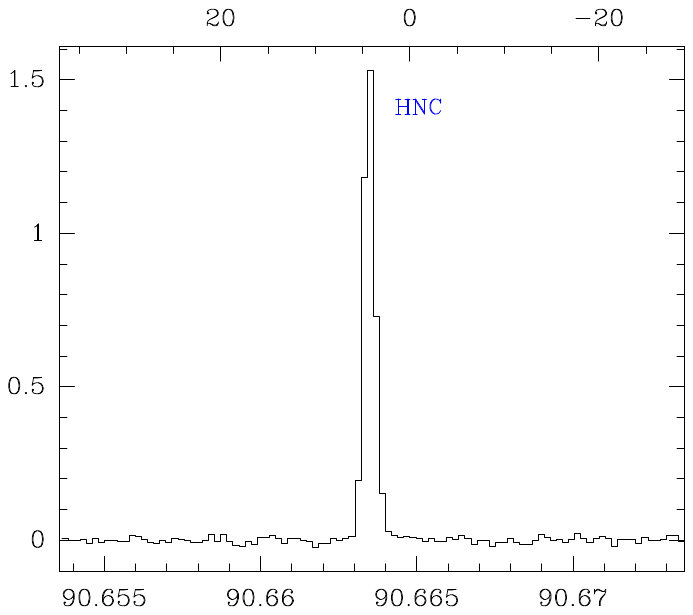}
  \end{minipage}
  \hspace{0.6cm}
  \begin{minipage}[h]{0.32\textwidth}
    \includegraphics[width=2.5\textwidth]{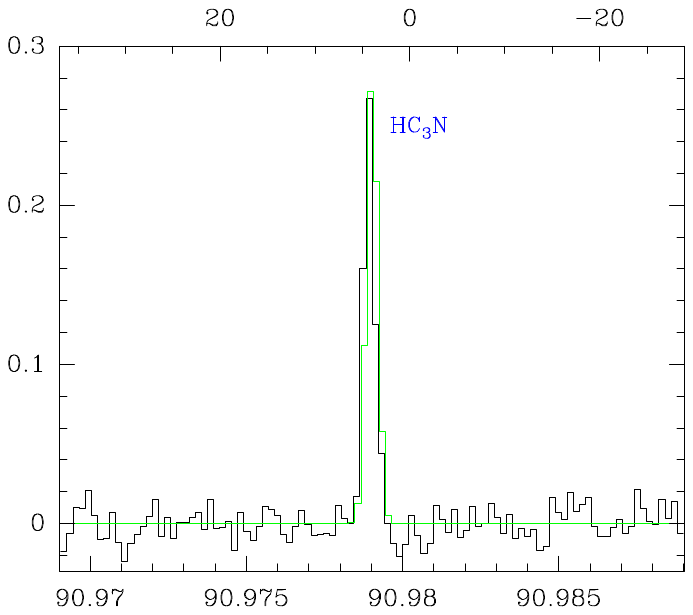}  
  \end{minipage}
  \hspace{0.6cm}
  \begin{minipage}[h]{0.32\textwidth}
    \includegraphics[width=2.5\textwidth]{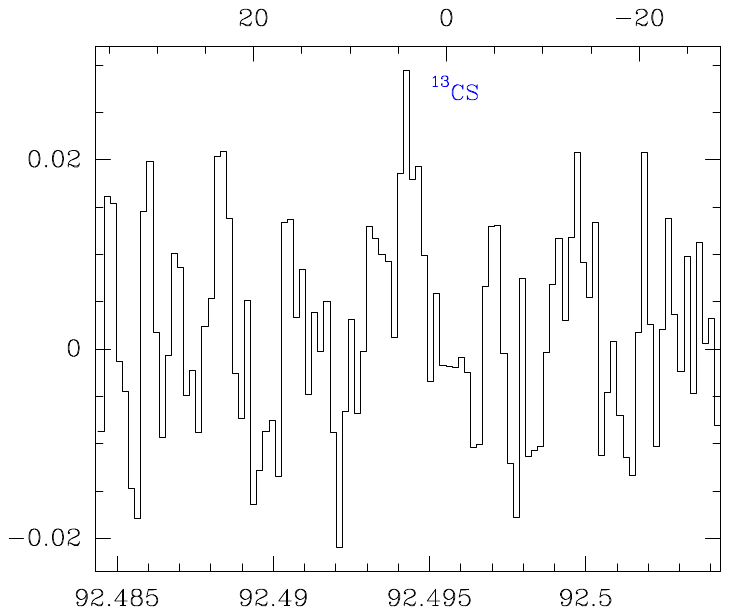}
  \end{minipage} 
 \vspace{-1cm} 
\caption{Continued. 
}
\end{figure*}

\addtocounter{figure}{-1}
\begin{figure*}[h]
\centering 
\vspace{-1cm}
\hspace{-2cm}
  \begin{minipage}[h]{0.32\textwidth}
    \includegraphics[width=2.5\textwidth]{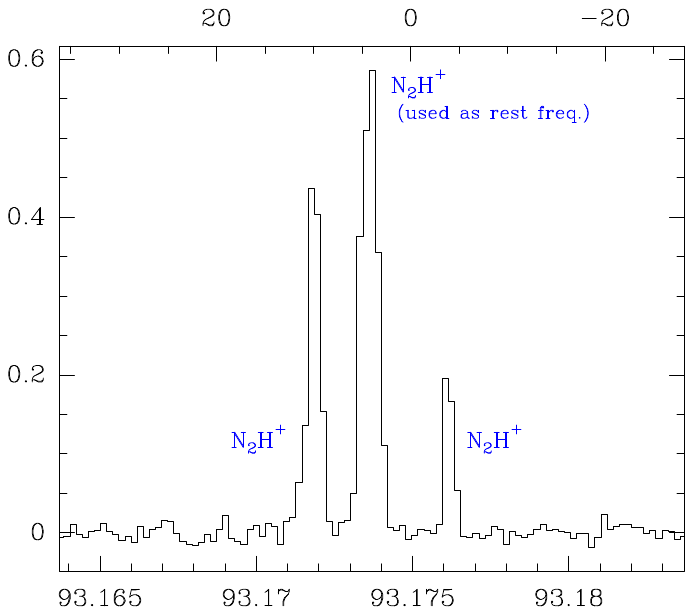}
  \end{minipage}
  \hspace{0.6cm}
  \begin{minipage}[h]{0.32\textwidth}
    \includegraphics[width=2.5\textwidth]{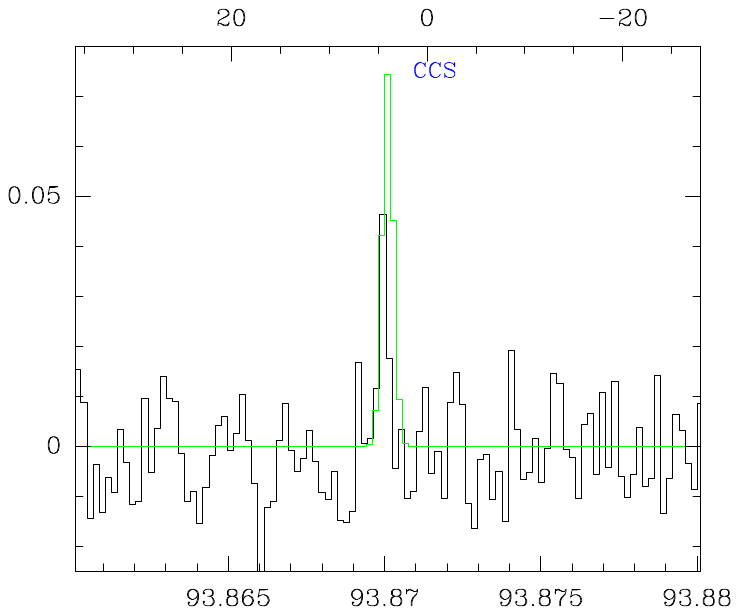}  
  \end{minipage}
  \hspace{0.6cm}
  \begin{minipage}[h]{0.32\textwidth}
    \includegraphics[width=2.5\textwidth]{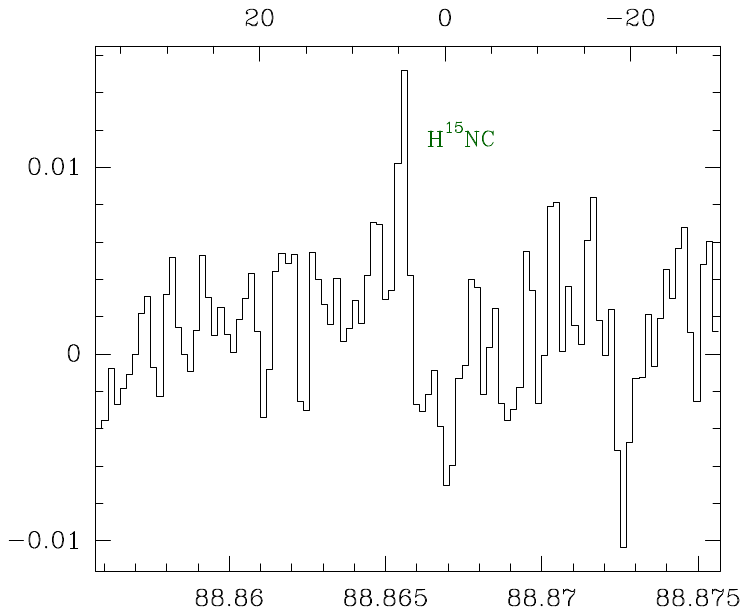}
  \end{minipage} \\
\vspace{-5.5cm}
\hspace{-2cm}
  \begin{minipage}[h]{0.32\textwidth}
    \includegraphics[width=2.5\textwidth]{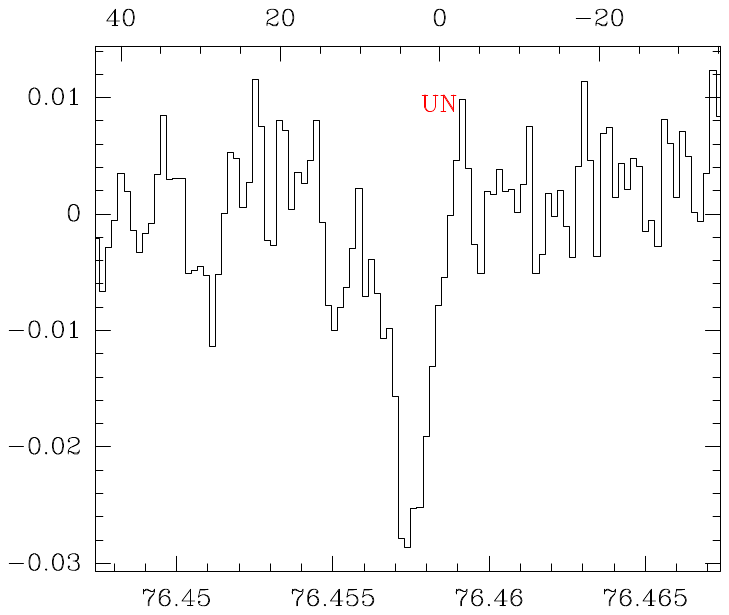}
  \end{minipage}
  \hspace{0.6cm}
  \begin{minipage}[h]{0.32\textwidth}
    \includegraphics[width=2.5\textwidth]{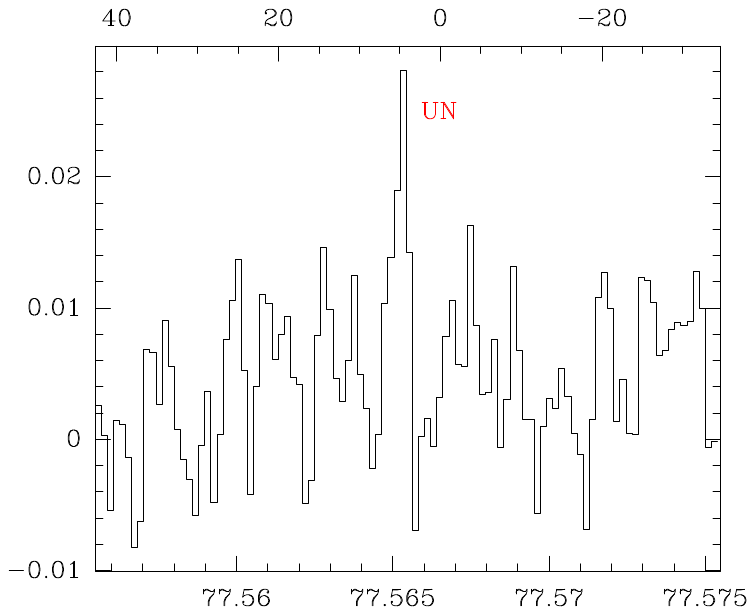}  
  \end{minipage}
  \hspace{0.6cm}
  \begin{minipage}[h]{0.32\textwidth}
    \includegraphics[width=2.5\textwidth]{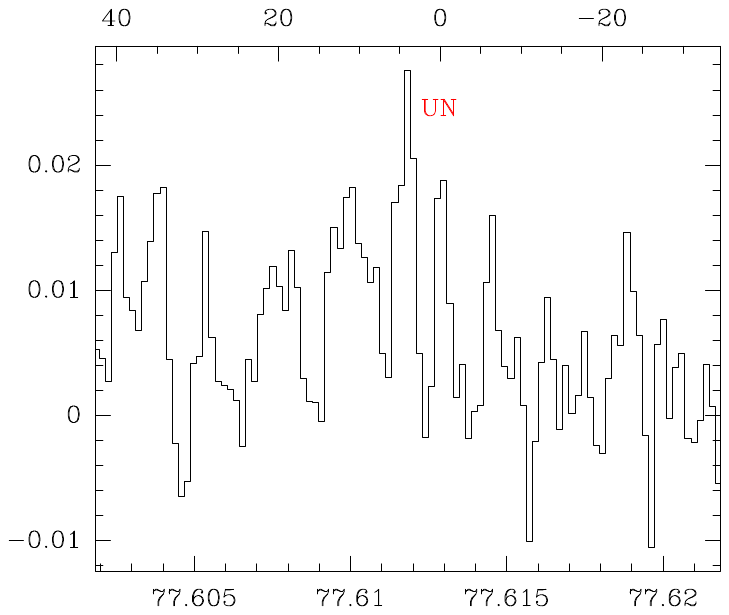}
  \end{minipage} \\
\vspace{-5.5cm}
\hspace{-2cm}
  \begin{minipage}[h]{0.32\textwidth}
    \includegraphics[width=2.5\textwidth]{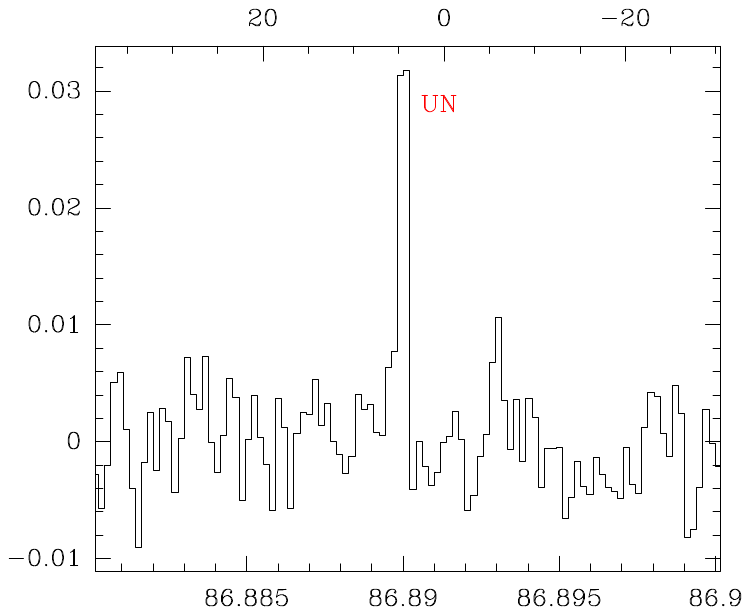}
  \end{minipage}
  \hspace{0.6cm}
  \begin{minipage}[h]{0.32\textwidth}
    \includegraphics[width=2.5\textwidth]{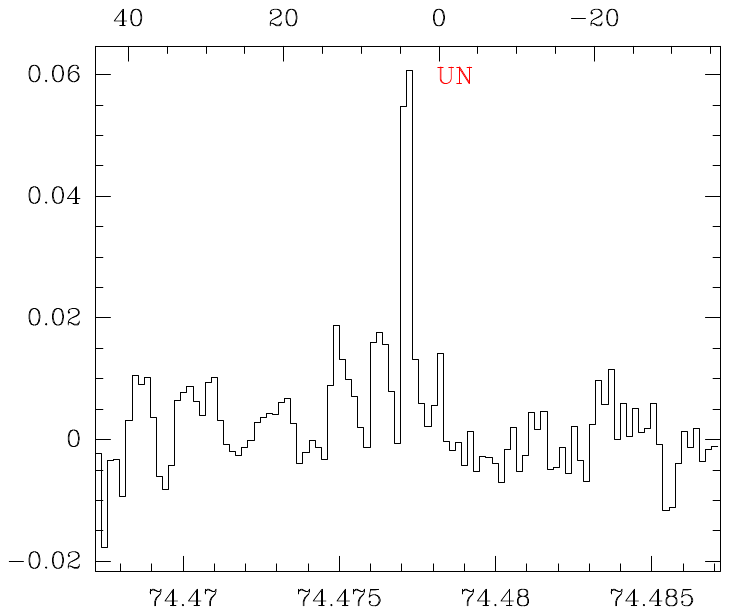}  
  \end{minipage}
 \vspace{1cm} 
\caption{Continued.}
\end{figure*}

\begin{figure*}[h]
\centering 
\vspace{-4cm}
\hspace{-2cm}
  \begin{minipage}[h]{0.32\textwidth}
    \includegraphics[width=2.5\textwidth]{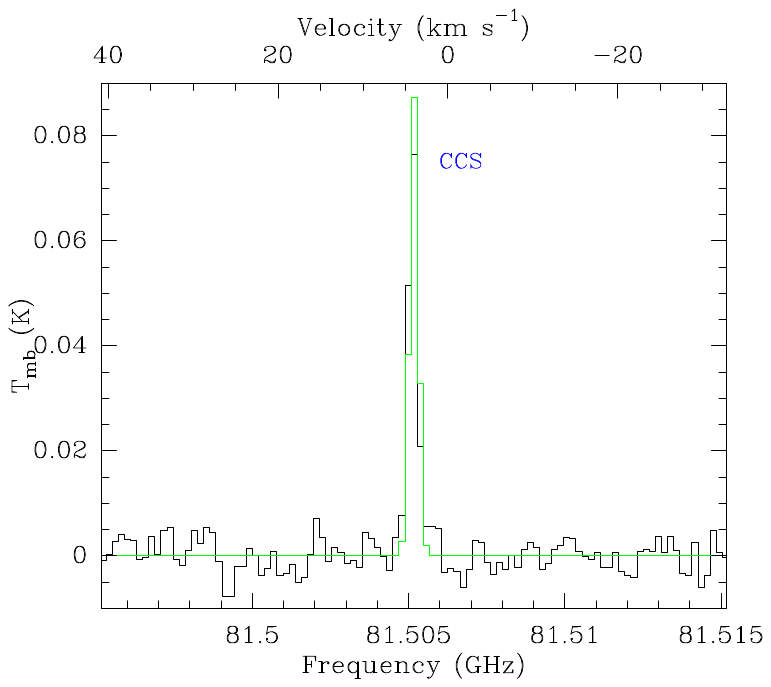}
  \end{minipage}
  \hspace{0.6cm}
  \begin{minipage}[h]{0.32\textwidth}
    \includegraphics[width=2.5\textwidth]{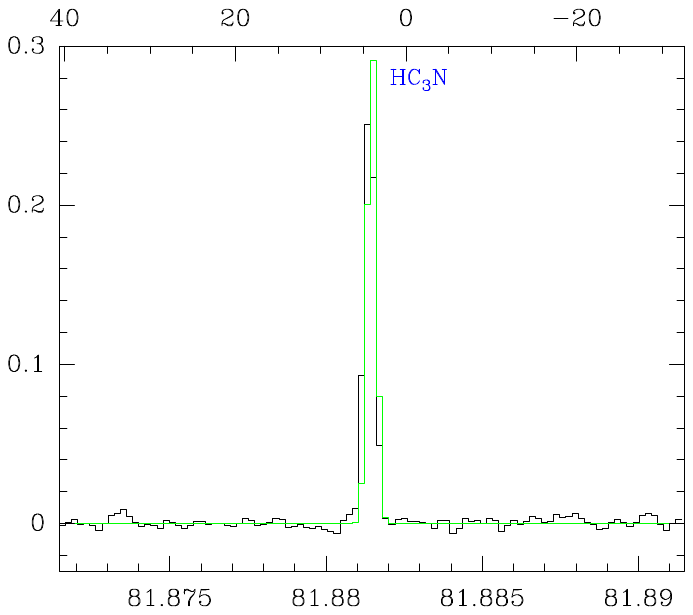}  
  \end{minipage}
  \hspace{0.6cm}
  \begin{minipage}[h]{0.32\textwidth}
    \includegraphics[width=2.5\textwidth]{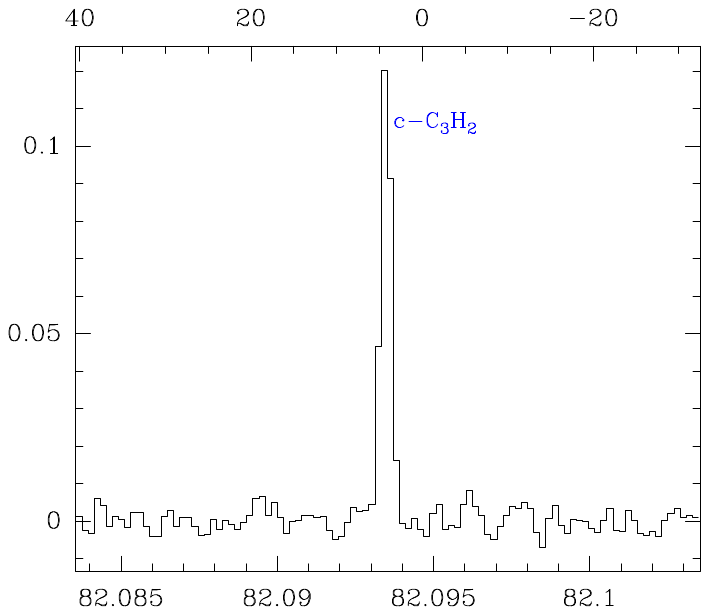}
  \end{minipage} \\
\vspace{-5.5cm}
\hspace{-2cm}
  \begin{minipage}[h]{0.32\textwidth}
    \includegraphics[width=2.5\textwidth]{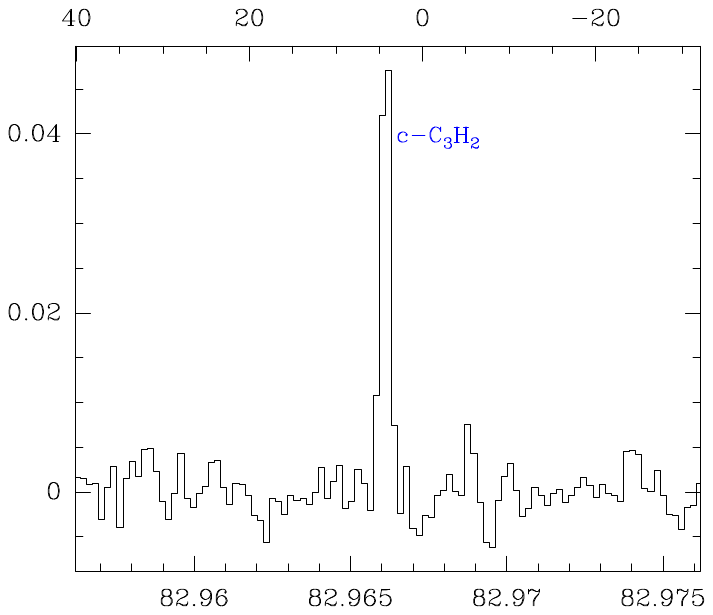}
  \end{minipage}
  \hspace{0.6cm}
  \begin{minipage}[h]{0.32\textwidth}
    \includegraphics[width=2.5\textwidth]{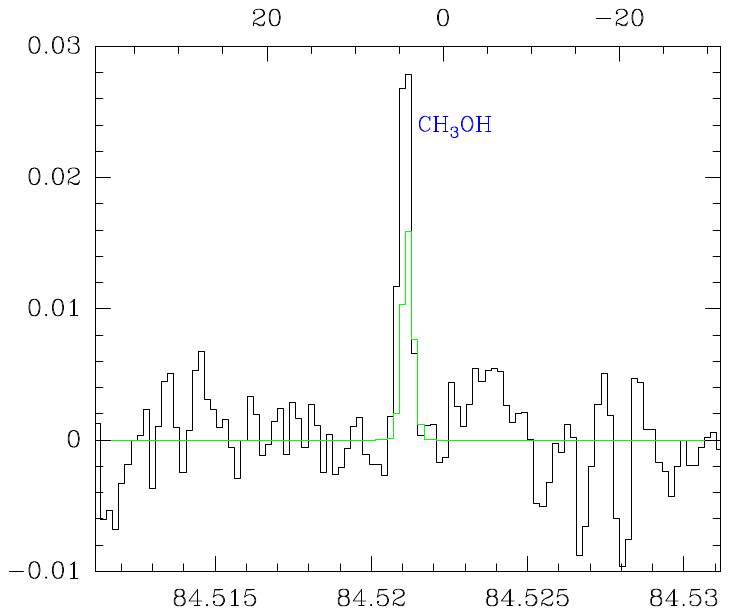}  
  \end{minipage}
  \hspace{0.6cm}
  \begin{minipage}[h]{0.32\textwidth}
    \includegraphics[width=2.5\textwidth]{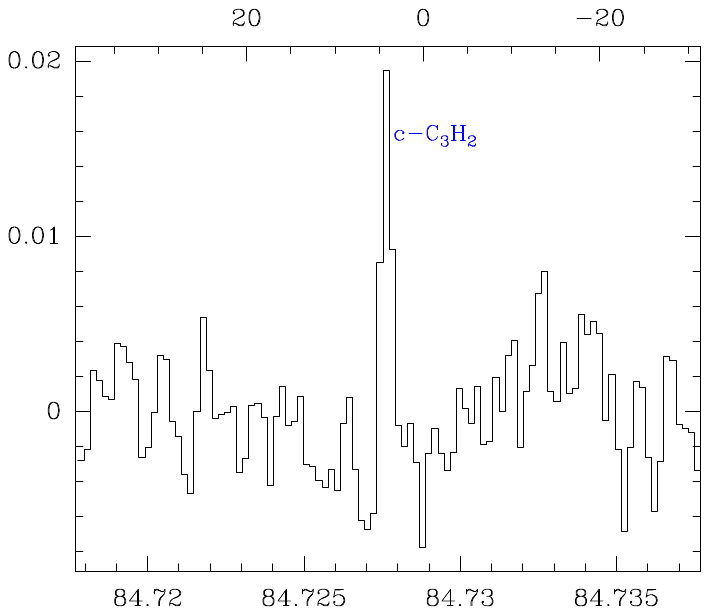}
  \end{minipage} \\
\vspace{-5.5cm}
\hspace{-2cm}
  \begin{minipage}[h]{0.32\textwidth}
    \includegraphics[width=2.5\textwidth]{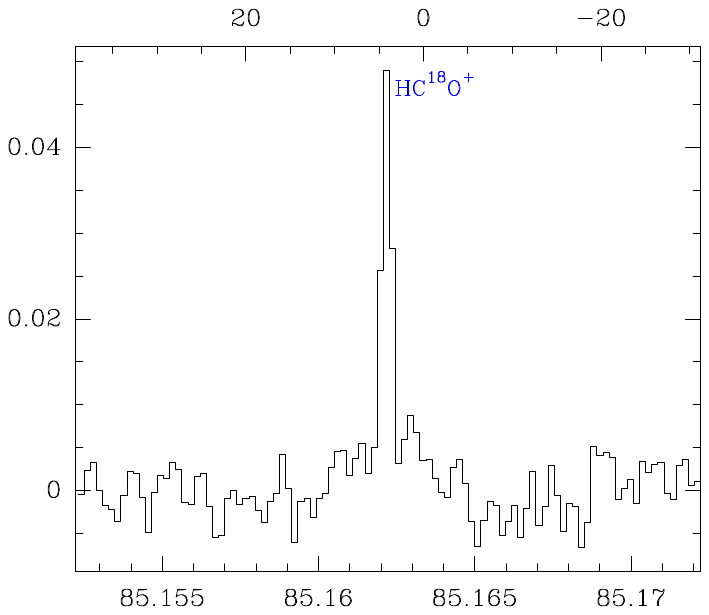}
  \end{minipage}
  \hspace{0.6cm}
  \begin{minipage}[h]{0.32\textwidth}
    \includegraphics[width=2.5\textwidth]{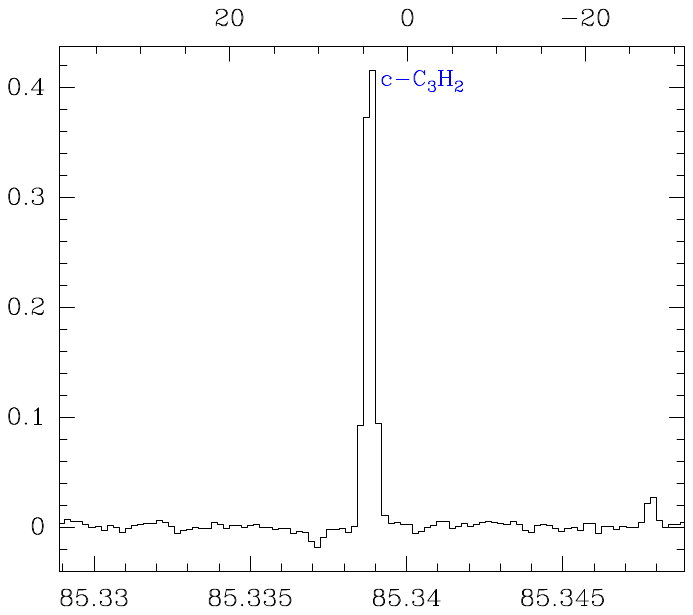}  
  \end{minipage}
  \hspace{0.6cm}
  \begin{minipage}[h]{0.32\textwidth}
    \includegraphics[width=2.5\textwidth]{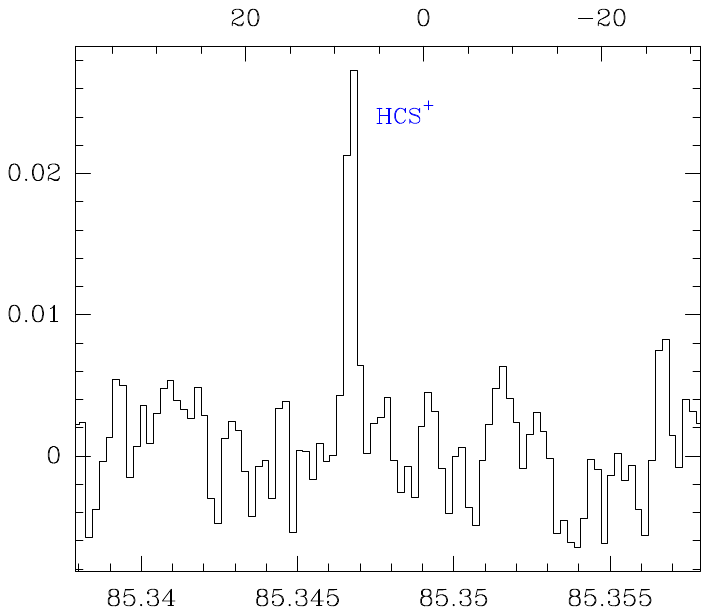}
  \end{minipage} \\
\vspace{-5.5cm}
\hspace{-2cm}
  \begin{minipage}[h]{0.32\textwidth}
    \includegraphics[width=2.5\textwidth]{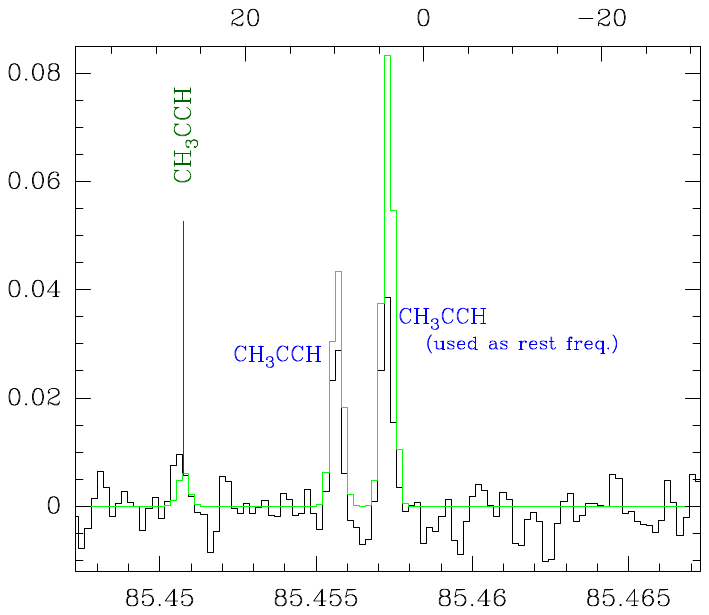}
  \end{minipage}
  \hspace{0.6cm}
  \begin{minipage}[h]{0.32\textwidth}
    \includegraphics[width=2.5\textwidth]{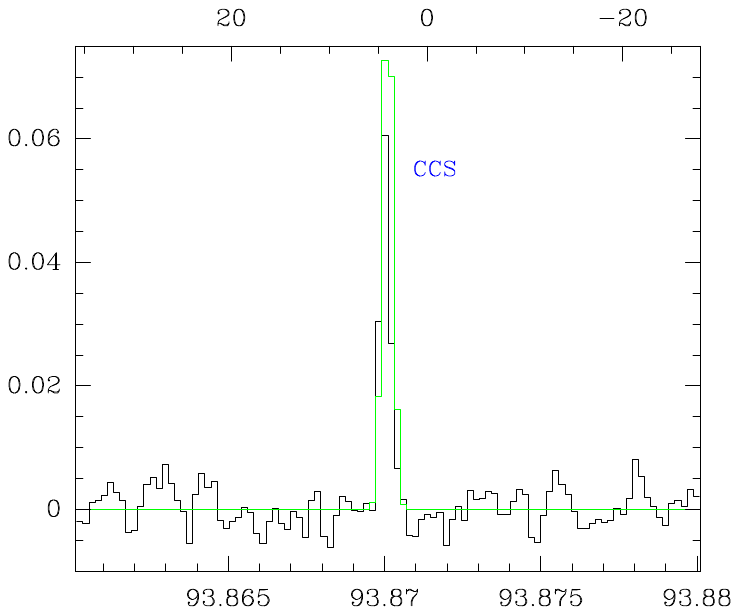}  
  \end{minipage}
  \hspace{0.6cm}
  \begin{minipage}[h]{0.32\textwidth}
    \includegraphics[width=2.5\textwidth]{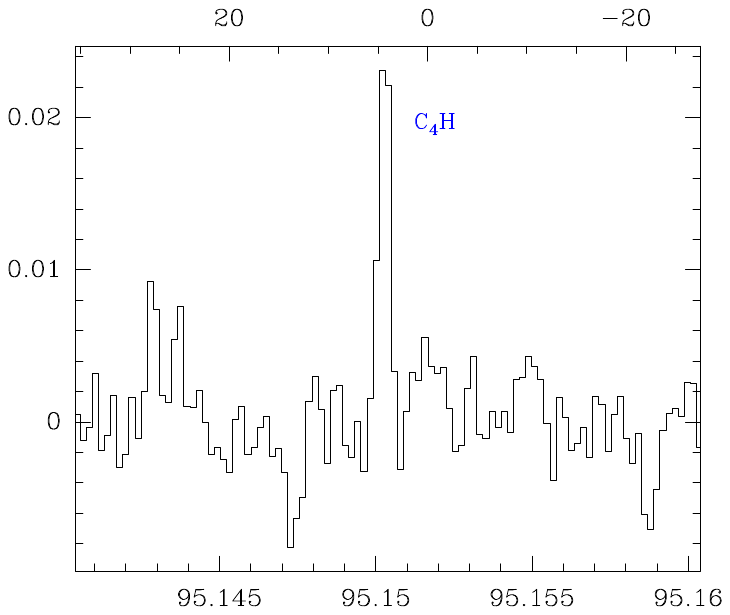}
  \end{minipage} 
 \vspace{-1.5cm} 
\caption{Same as Fig.~\ref{fig:survey-73ghz-small} but for the tuning at 80\,GHz.
}
\label{fig:survey-80ghz-small}
\end{figure*}

\addtocounter{figure}{-1}
\begin{figure*}[h]
\centering 
\vspace{-4cm}
\hspace{-2cm}
  \begin{minipage}[h]{0.32\textwidth}
    \includegraphics[width=2.5\textwidth]{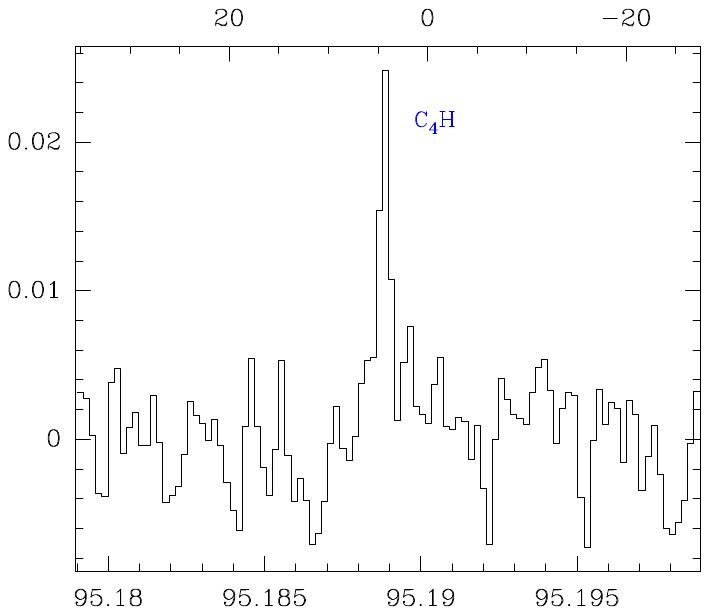}
  \end{minipage}
  \hspace{0.6cm}
  \begin{minipage}[h]{0.32\textwidth}
    \includegraphics[width=2.5\textwidth]{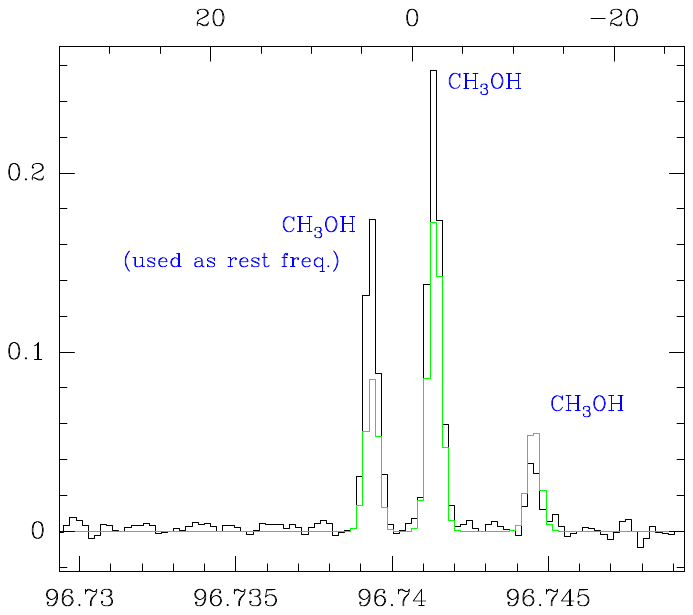}  
  \end{minipage}
  \hspace{0.6cm}
  \begin{minipage}[h]{0.32\textwidth}
    \includegraphics[width=2.5\textwidth]{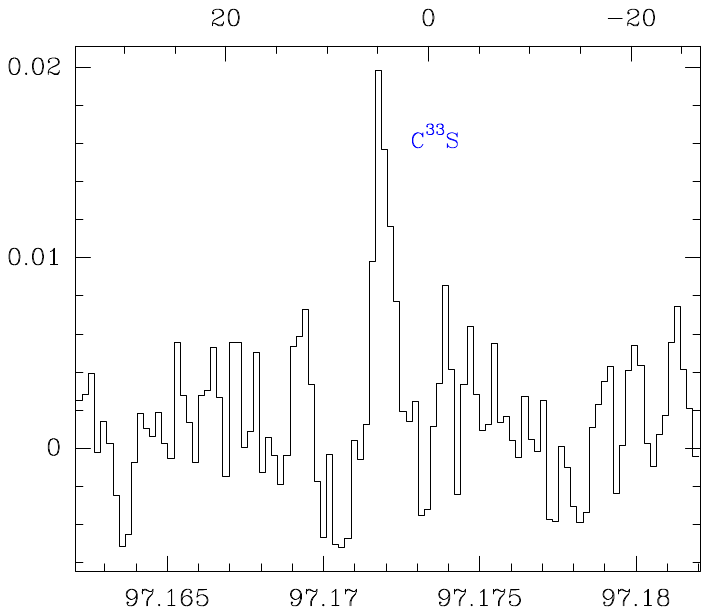}
  \end{minipage} \\
\vspace{-5.5cm}
\hspace{-2cm}
  \begin{minipage}[h]{0.32\textwidth}
    \includegraphics[width=2.5\textwidth]{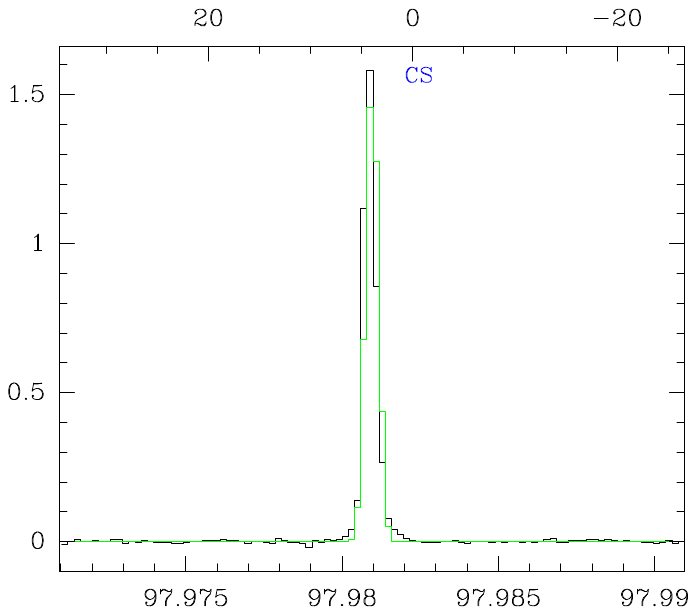}
  \end{minipage}
  \hspace{0.6cm}
  \begin{minipage}[h]{0.32\textwidth}
    \includegraphics[width=2.5\textwidth]{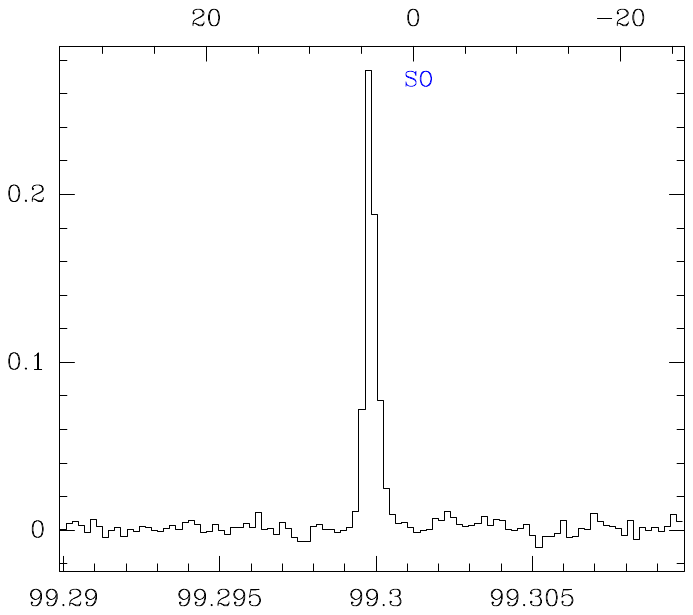}  
  \end{minipage}
  \hspace{0.6cm}
  \begin{minipage}[h]{0.32\textwidth}
    \includegraphics[width=2.5\textwidth]{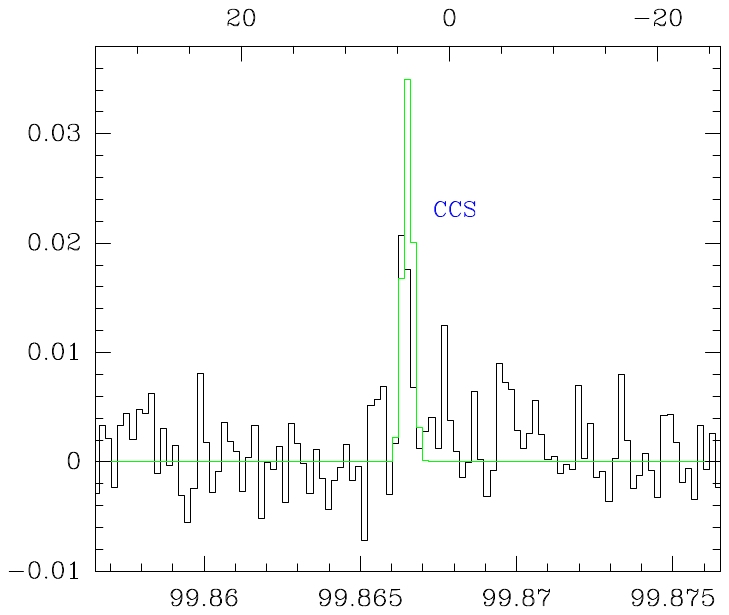}
  \end{minipage} \\
\vspace{-5.5cm}
\hspace{-2cm}
  \begin{minipage}[h]{0.32\textwidth}
    \includegraphics[width=2.5\textwidth]{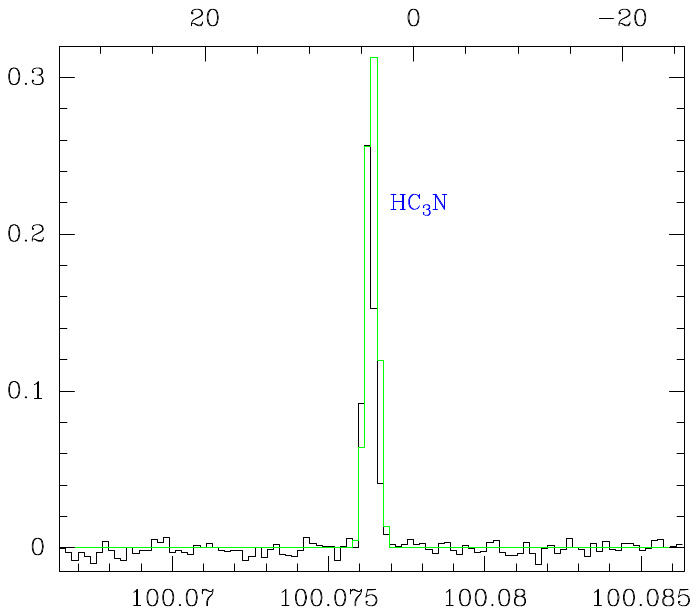}
  \end{minipage}
  \hspace{0.6cm}
  \begin{minipage}[h]{0.32\textwidth}
    \includegraphics[width=2.5\textwidth]{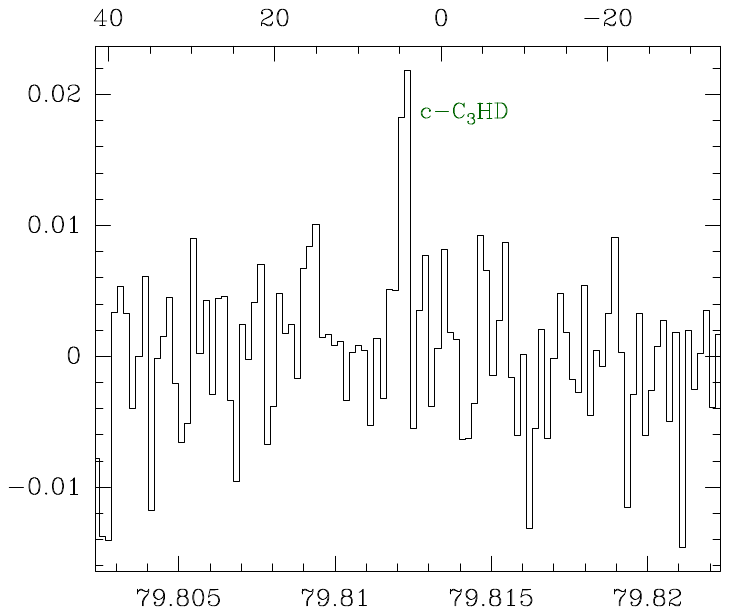}  
  \end{minipage}
  \hspace{0.6cm}
  \begin{minipage}[h]{0.32\textwidth}
    \includegraphics[width=2.5\textwidth]{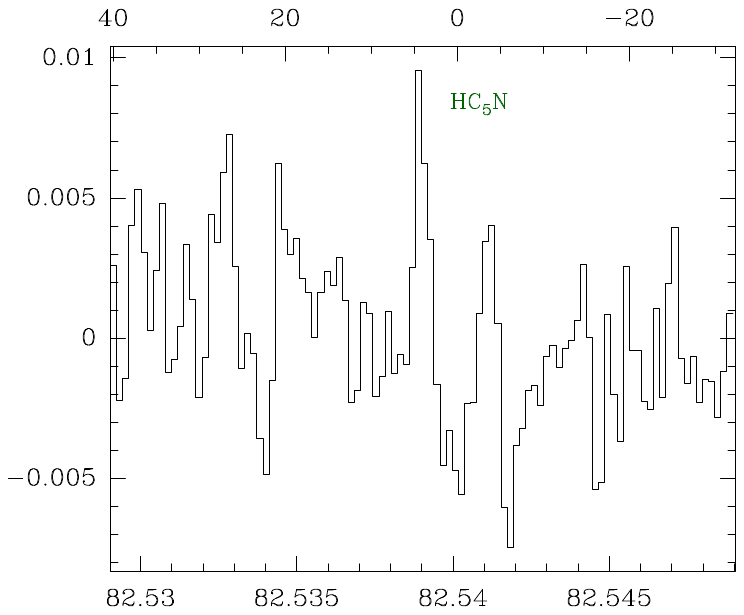}
  \end{minipage} \\
\vspace{-5.5cm}
\hspace{-2cm}
  \begin{minipage}[h]{0.32\textwidth}
    \includegraphics[width=2.5\textwidth]{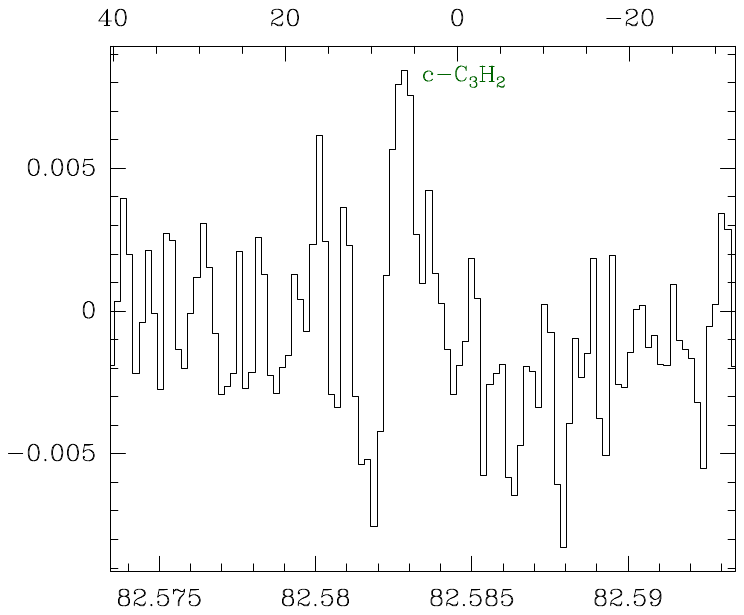}
  \end{minipage}
 \vspace{-1.5cm} 
\caption{Continued. 
}
\end{figure*}

\newpage
\begin{figure*}[h]
\centering 
\vspace{-4cm}
\hspace{-2cm}
  \begin{minipage}[h]{0.32\textwidth}
    \includegraphics[width=2.5\textwidth]{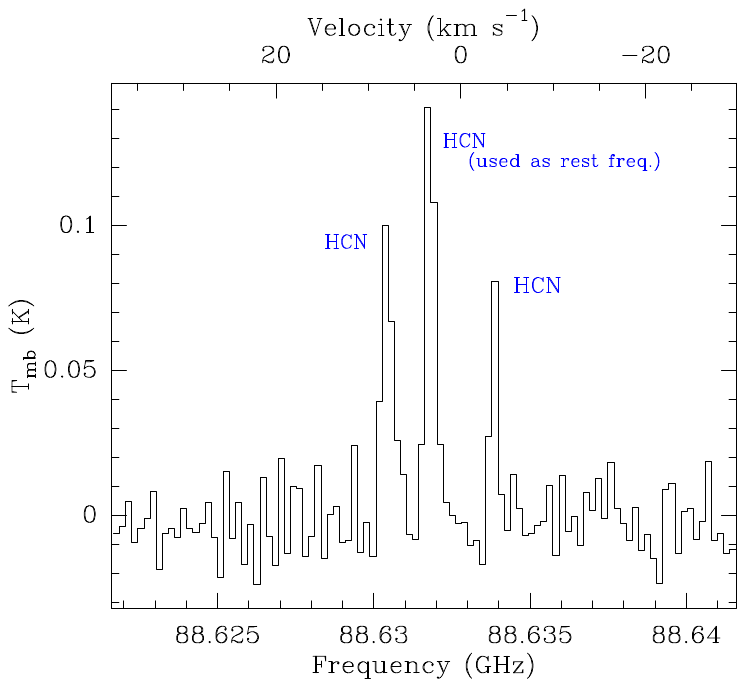}
  \end{minipage}
  \hspace{0.6cm}
  \begin{minipage}[h]{0.32\textwidth}
    \includegraphics[width=2.5\textwidth]{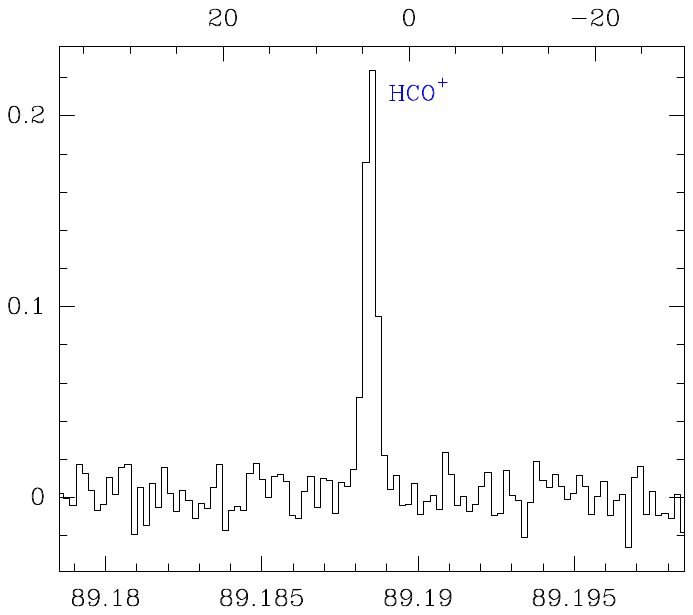}  
  \end{minipage}
  \hspace{0.6cm}
  \begin{minipage}[h]{0.32\textwidth}
    \includegraphics[width=2.5\textwidth]{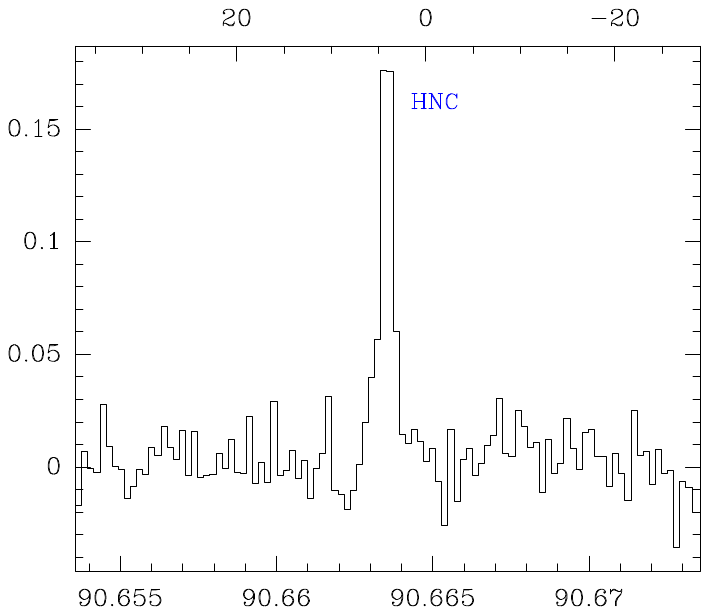}
  \end{minipage} \\
\vspace{-5.5cm}
\hspace{-2cm}
  \begin{minipage}[h]{0.32\textwidth}
    \includegraphics[width=2.5\textwidth]{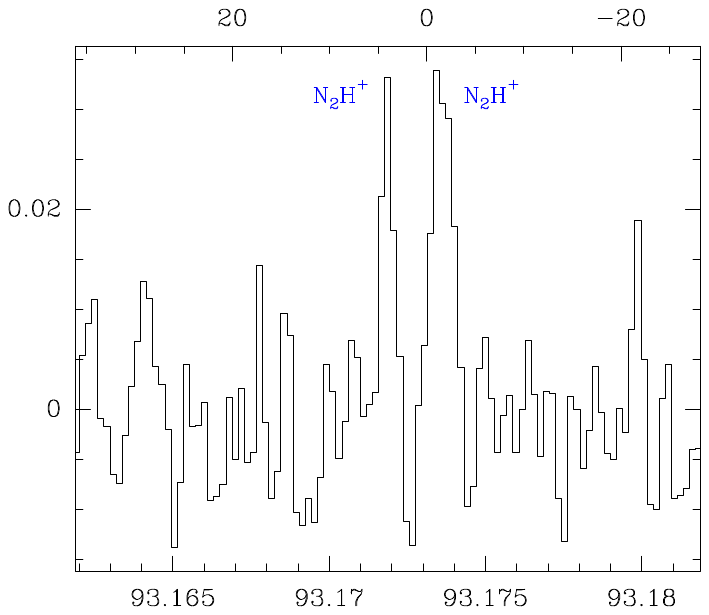}
  \end{minipage}
  \hspace{0.6cm}
  \begin{minipage}[h]{0.32\textwidth}
    \includegraphics[width=2.5\textwidth]{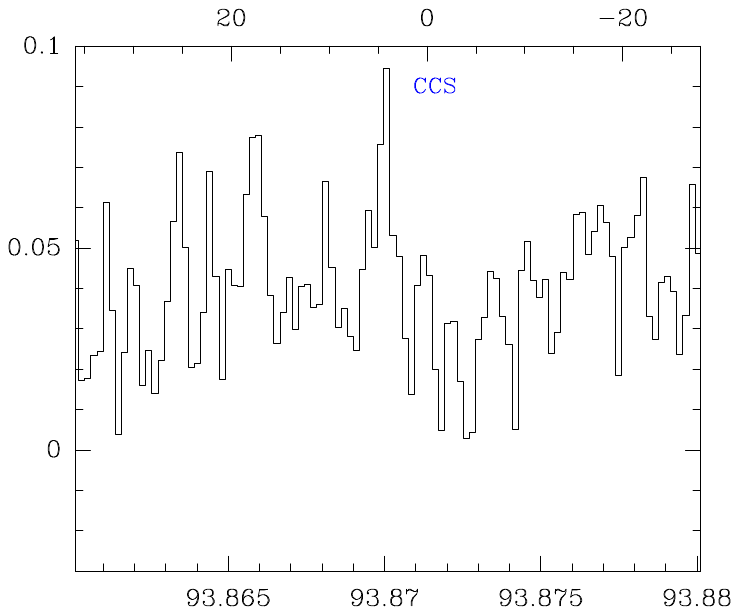}  
  \end{minipage}
  \hspace{0.6cm}
  \begin{minipage}[h]{0.32\textwidth}
    \includegraphics[width=2.5\textwidth]{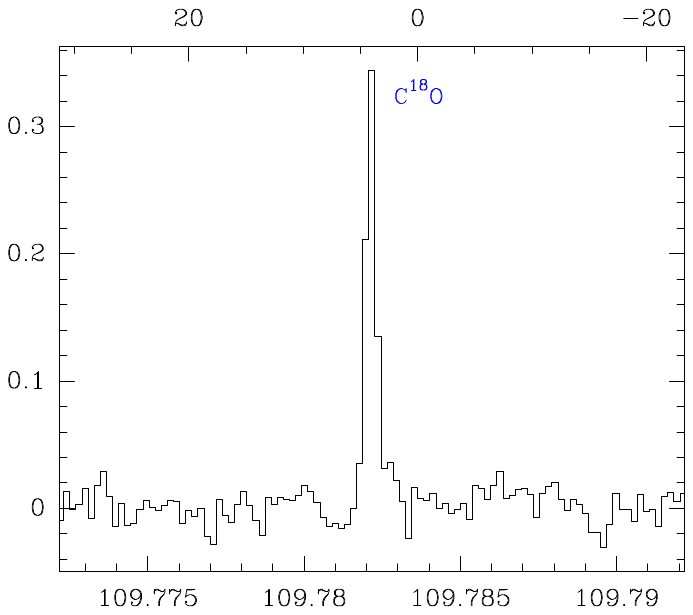}
  \end{minipage} \\
\vspace{-5.5cm}
\hspace{-2cm}
  \begin{minipage}[h]{0.32\textwidth}
    \includegraphics[width=2.5\textwidth]{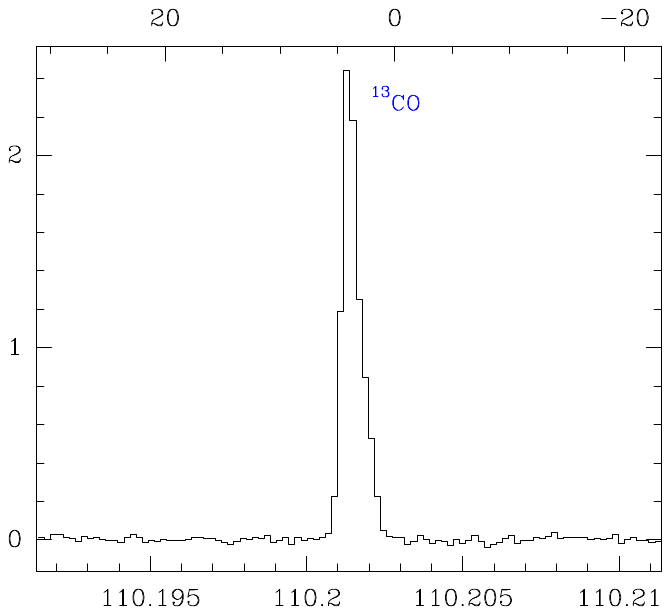}
  \end{minipage}
  \hspace{0.6cm}
  \begin{minipage}[h]{0.32\textwidth}
    \includegraphics[width=2.5\textwidth]{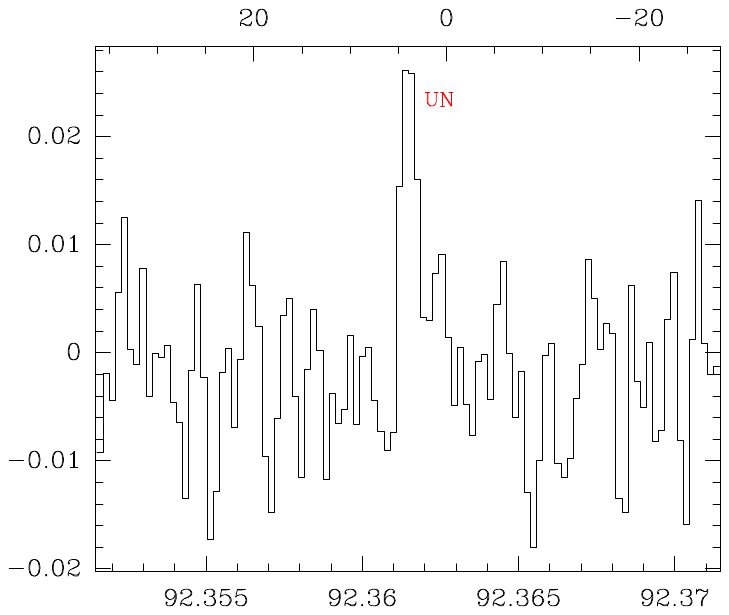}  
  \end{minipage}
  \hspace{0.6cm}
  \begin{minipage}[h]{0.32\textwidth}
    \includegraphics[width=2.5\textwidth]{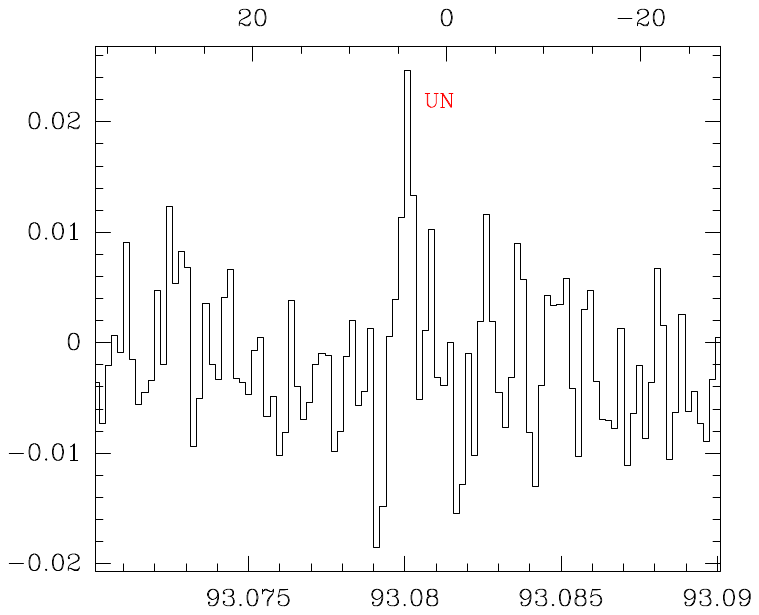}
  \end{minipage} \\
\vspace{-5.5cm}
\hspace{-2cm}
  \begin{minipage}[h]{0.32\textwidth}
    \includegraphics[width=2.5\textwidth]{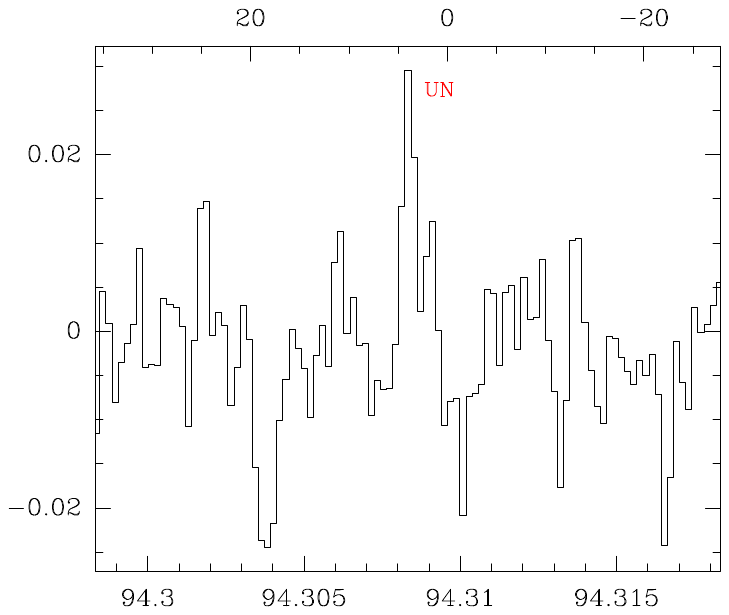}
  \end{minipage}
  \hspace{0.6cm}
  \begin{minipage}[h]{0.32\textwidth}
    \includegraphics[width=2.5\textwidth]{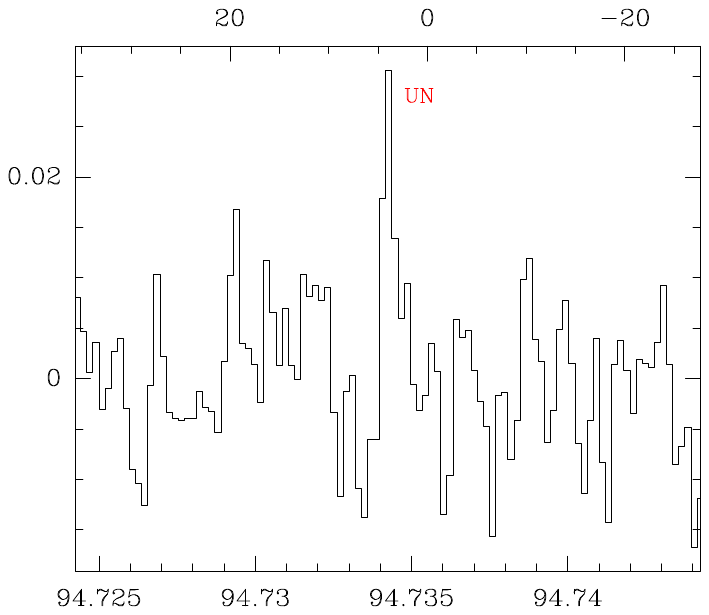}  
  \end{minipage}
  \hspace{0.6cm}
  \begin{minipage}[h]{0.32\textwidth}
    \includegraphics[width=2.5\textwidth]{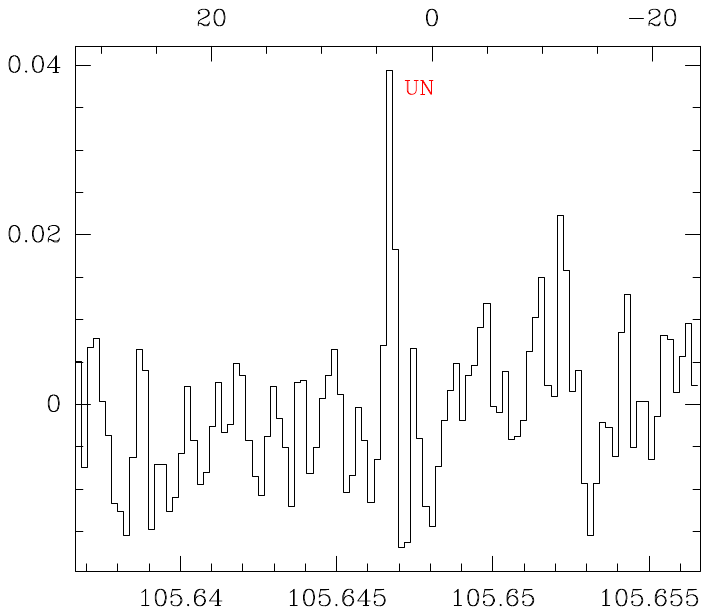}
  \end{minipage} 
 \vspace{-1.85cm} 
\caption{Same as Fig.~\ref{fig:survey-73ghz-small} but for the tuning at 90\,GHz in the otf mode at the source position.
}
\label{fig:survey-90ghz-small}
\end{figure*}

\addtocounter{figure}{-1}
\begin{figure*}[h]
\centering 
\vspace{2cm}
\hspace{-2cm}
  \begin{minipage}[h]{0.32\textwidth}
    \includegraphics[width=2.5\textwidth]{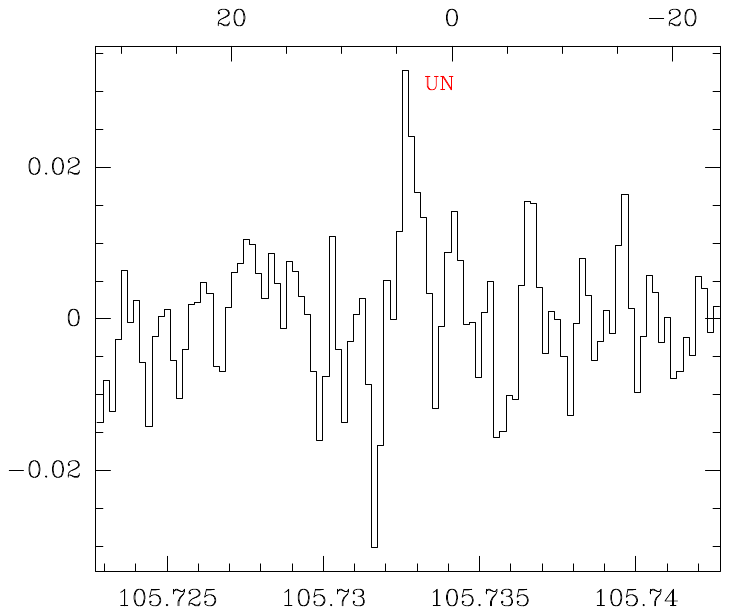}
  \end{minipage}
  \hspace{0.6cm}
  \begin{minipage}[h]{0.32\textwidth}
    \includegraphics[width=2.5\textwidth]{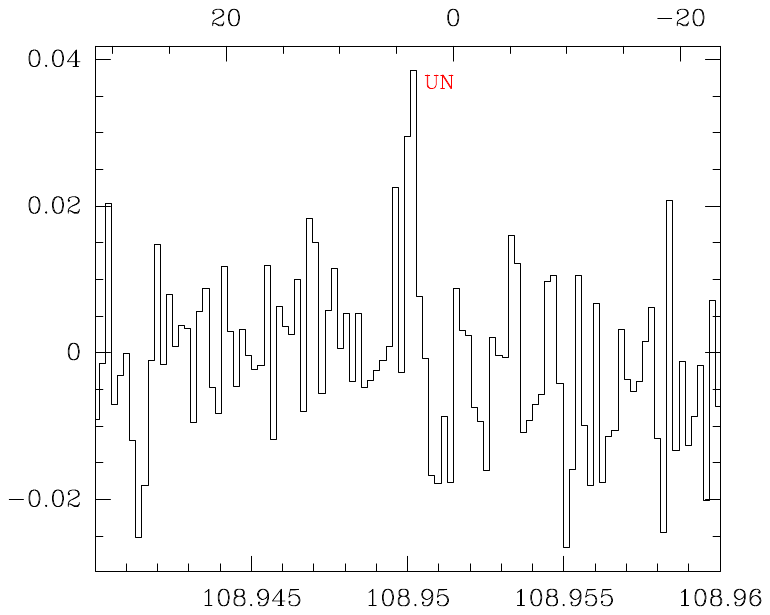}  
  \end{minipage}
  \hspace{0.6cm}
  \begin{minipage}[h]{0.32\textwidth}
    \includegraphics[width=2.5\textwidth]{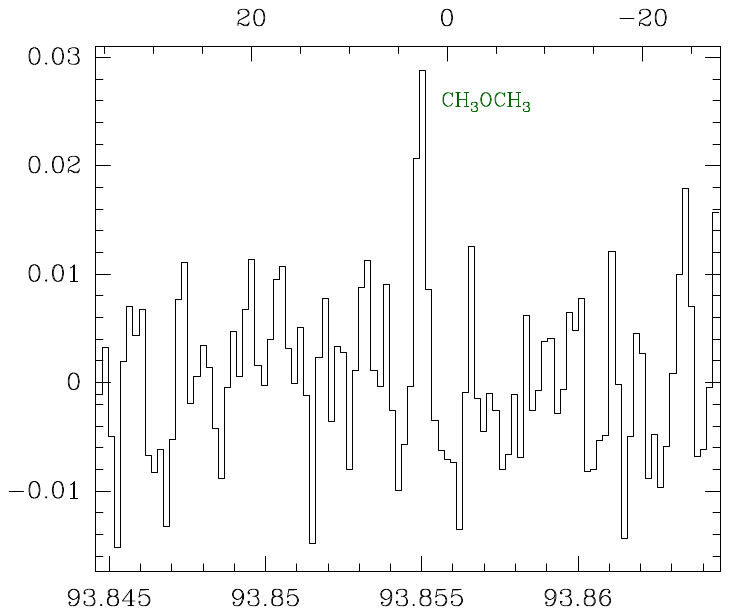}
  \end{minipage} \\
 \vspace{-1.5cm} 
\caption{Continued. 
}
\end{figure*}

\newpage
\begin{figure*}[h]
\centering 
\vspace{-4cm}
\hspace{-2cm}
  \begin{minipage}[h]{0.32\textwidth}
    \includegraphics[width=2.5\textwidth]{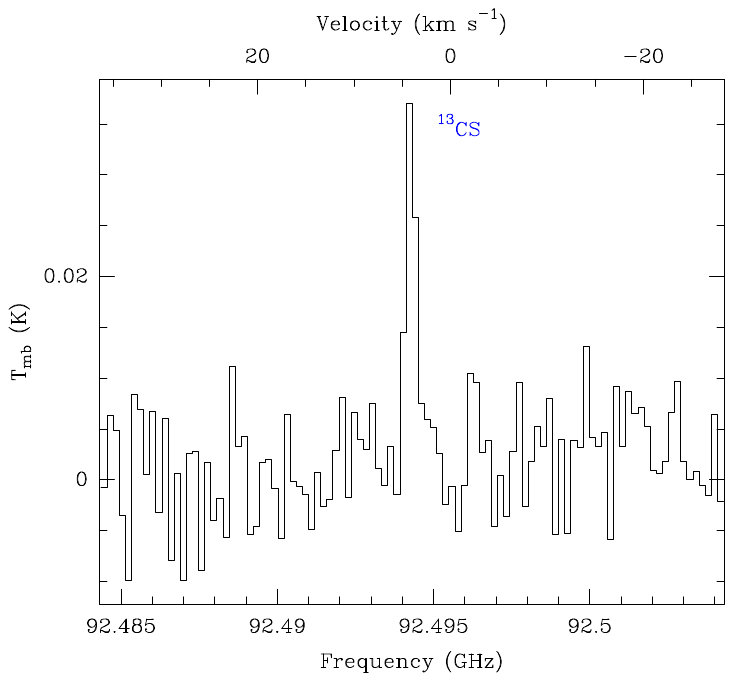}
  \end{minipage}
  \hspace{0.6cm}
  \begin{minipage}[h]{0.32\textwidth}
    \includegraphics[width=2.5\textwidth]{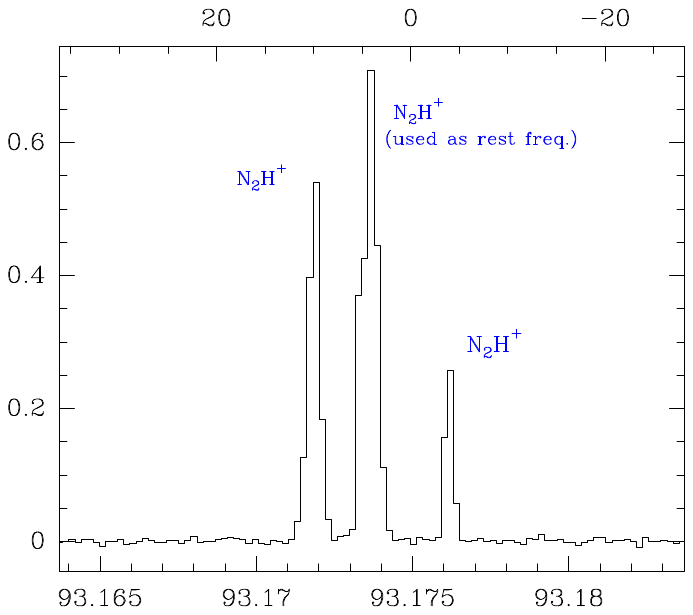}  
  \end{minipage}
  \hspace{0.6cm}
  \begin{minipage}[h]{0.32\textwidth}
    \includegraphics[width=2.5\textwidth]{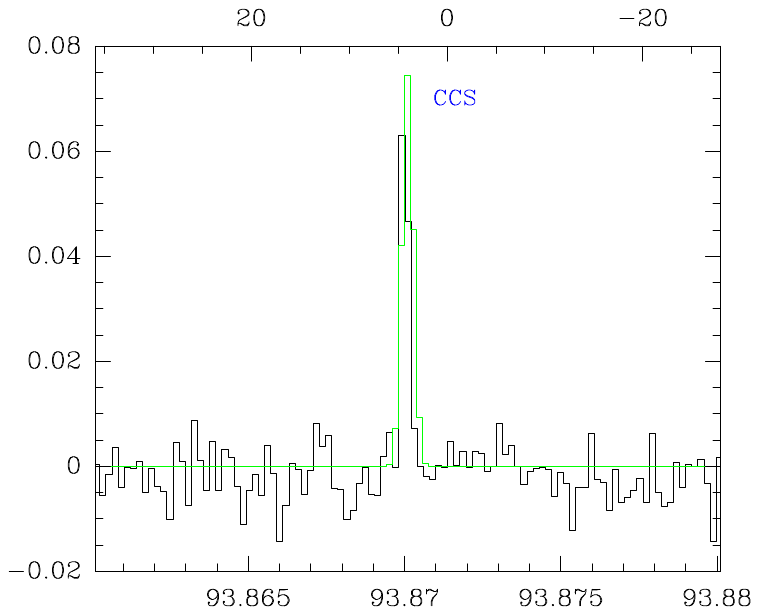}
  \end{minipage} \\
\vspace{-5.5cm}
\hspace{-2cm}
  \begin{minipage}[h]{0.32\textwidth}
    \includegraphics[width=2.5\textwidth]{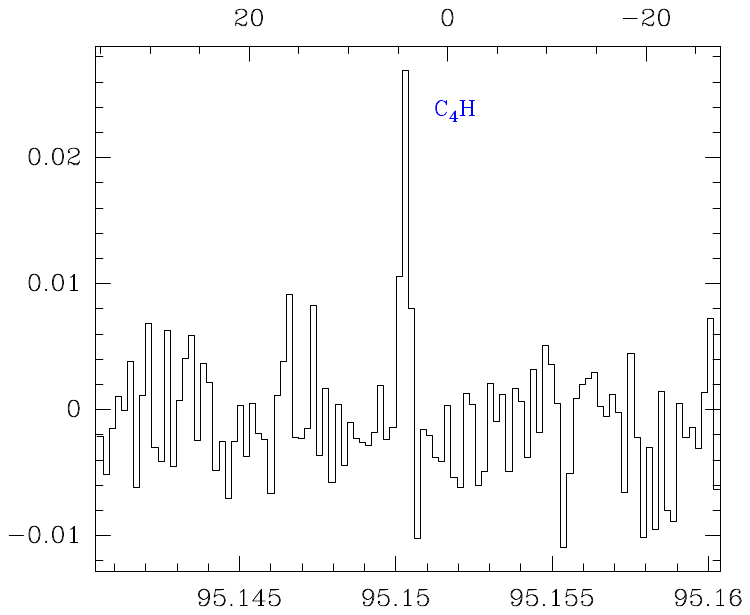}
  \end{minipage}
  \hspace{0.6cm}
  \begin{minipage}[h]{0.32\textwidth}
    \includegraphics[width=2.5\textwidth]{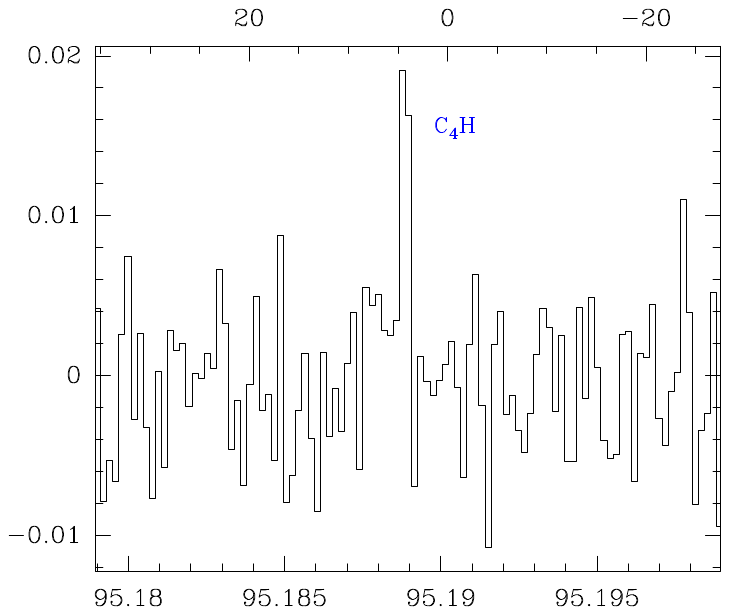}  
  \end{minipage}
  \hspace{0.6cm}
  \begin{minipage}[h]{0.32\textwidth}
    \includegraphics[width=2.5\textwidth]{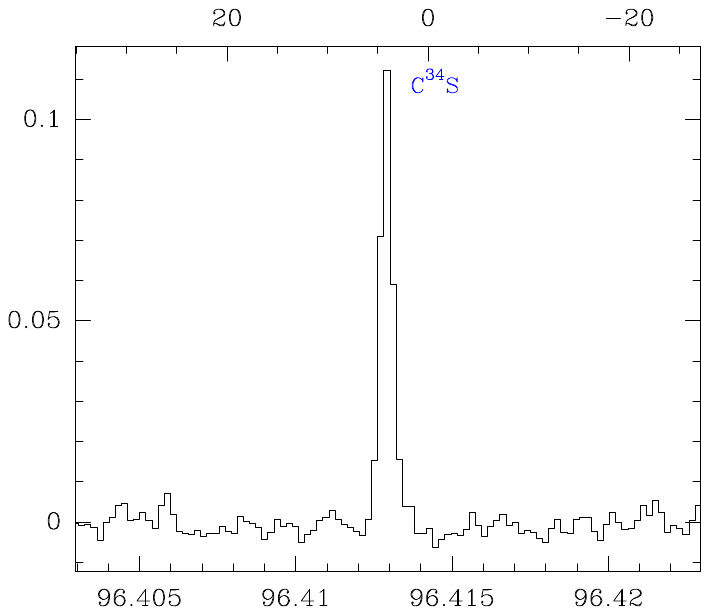}
  \end{minipage} \\
\vspace{-5.5cm}
\hspace{-2cm}
  \begin{minipage}[h]{0.32\textwidth}
    \includegraphics[width=2.5\textwidth]{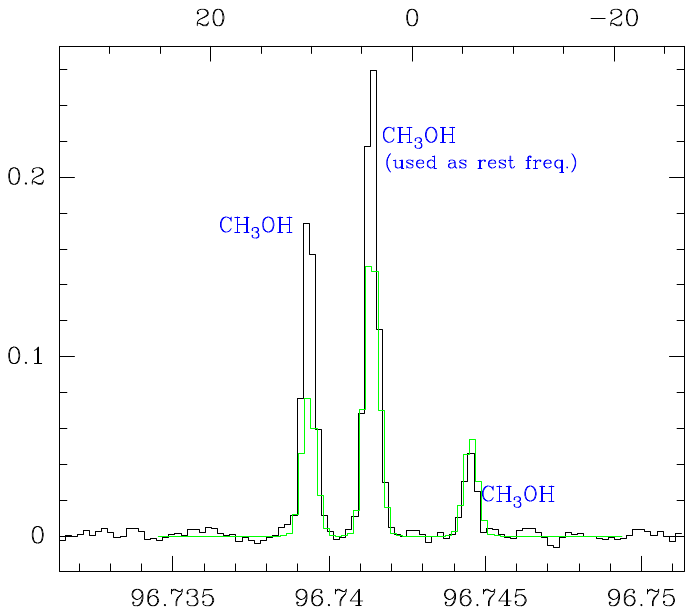}
  \end{minipage}
  \hspace{0.6cm}
  \begin{minipage}[h]{0.32\textwidth}
    \includegraphics[width=2.5\textwidth]{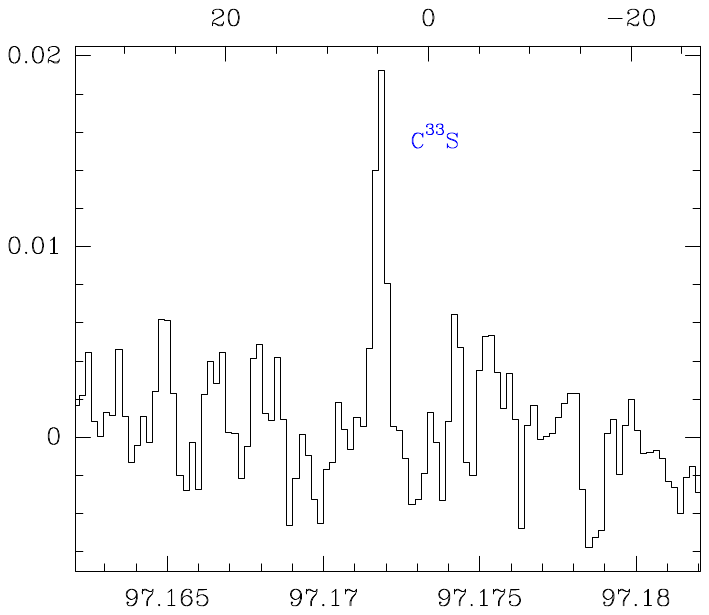}  
  \end{minipage}
  \hspace{0.6cm}
  \begin{minipage}[h]{0.32\textwidth}
    \includegraphics[width=2.5\textwidth]{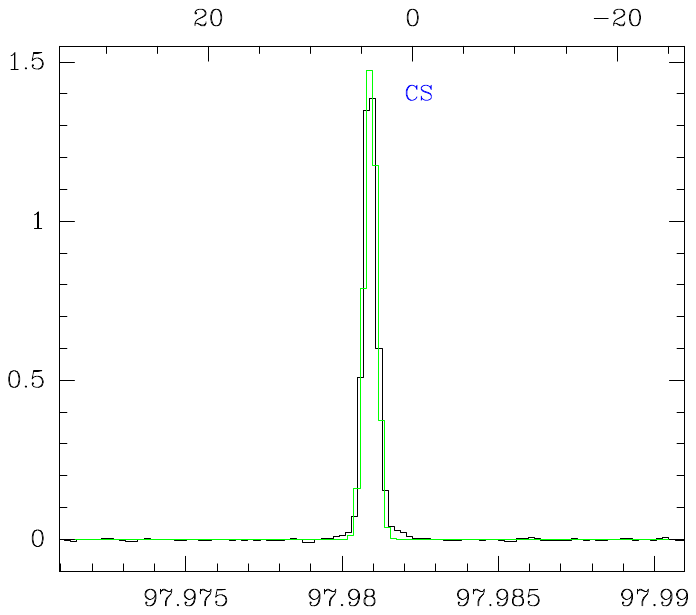}
  \end{minipage} \\
\vspace{-5.5cm}
\hspace{-2cm}
  \begin{minipage}[h]{0.32\textwidth}
    \includegraphics[width=2.5\textwidth]{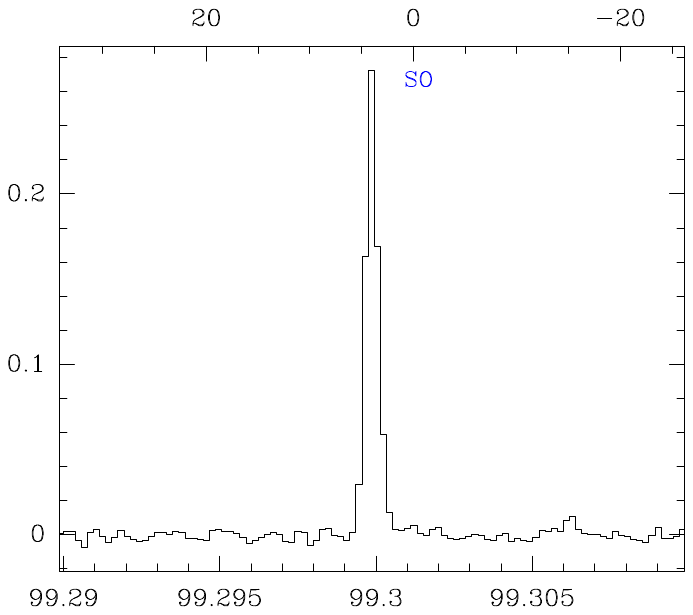}
  \end{minipage}
  \hspace{0.6cm}
  \begin{minipage}[h]{0.32\textwidth}
    \includegraphics[width=2.5\textwidth]{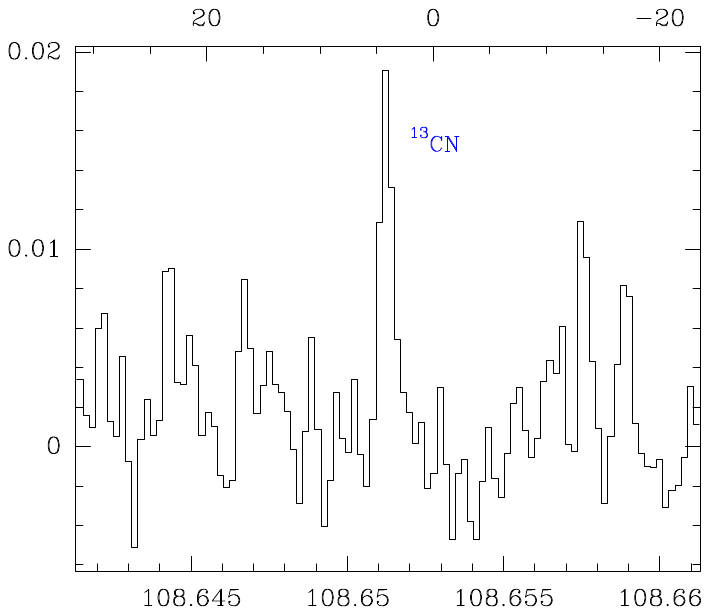}  
  \end{minipage}
  \hspace{0.6cm}
  \begin{minipage}[h]{0.32\textwidth}
    \includegraphics[width=2.5\textwidth]{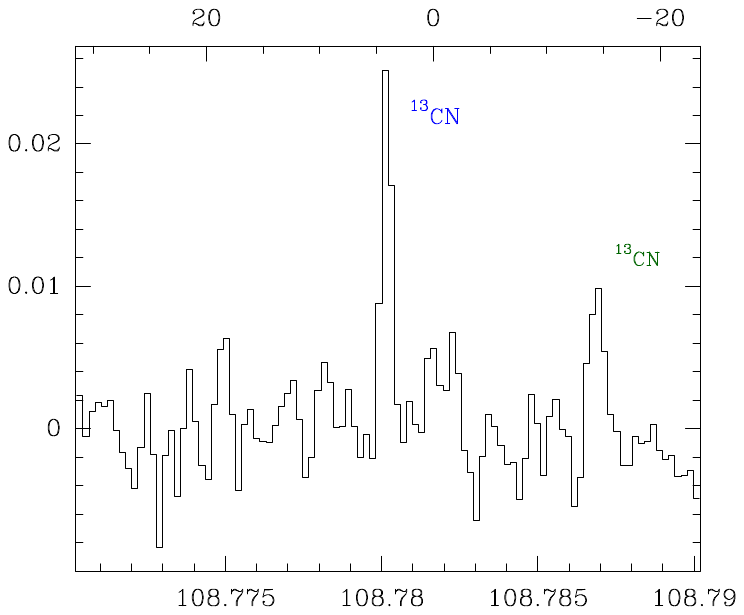}
  \end{minipage} 
 \vspace{-1.85cm} 
\caption{Same as Fig.~\ref{fig:survey-73ghz-small} but for the tuning at 94\,GHz. Possible contamination from the image band (indicating rest frequency and peak temperature in K in the image band) are labelled in red.
}
\label{fig:survey-94ghz-small}
\end{figure*}

\addtocounter{figure}{-1}
\begin{figure*}[h]
\centering 
\vspace{-4cm}
\hspace{-2cm}
  \begin{minipage}[h]{0.32\textwidth}
    \includegraphics[width=2.5\textwidth]{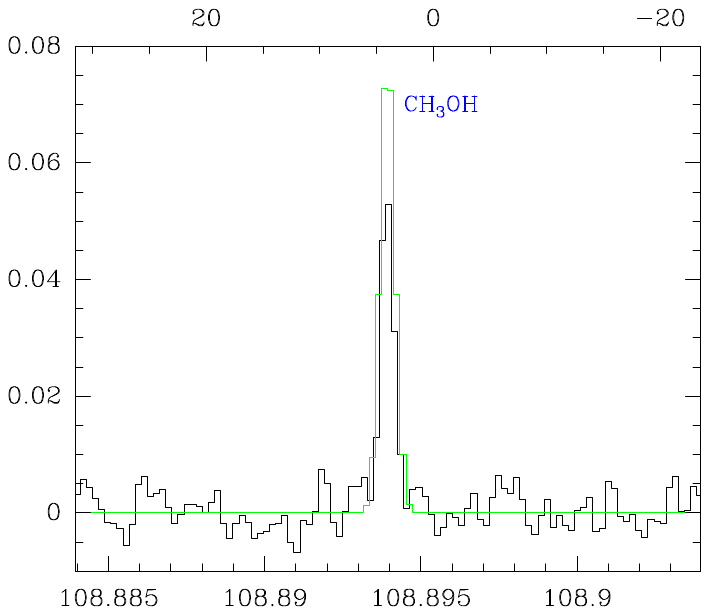}
  \end{minipage}
  \hspace{0.6cm}
  \begin{minipage}[h]{0.32\textwidth}
    \includegraphics[width=2.5\textwidth]{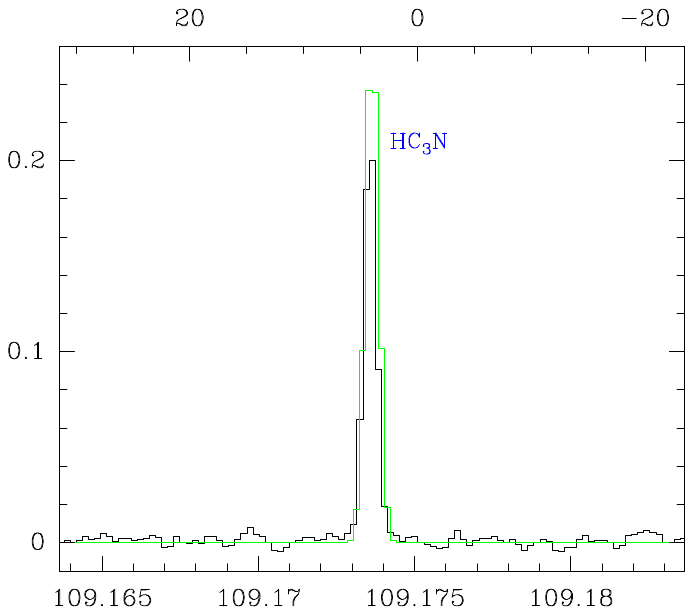}  
  \end{minipage}
  \hspace{0.6cm}
  \begin{minipage}[h]{0.32\textwidth}
    \includegraphics[width=2.5\textwidth]{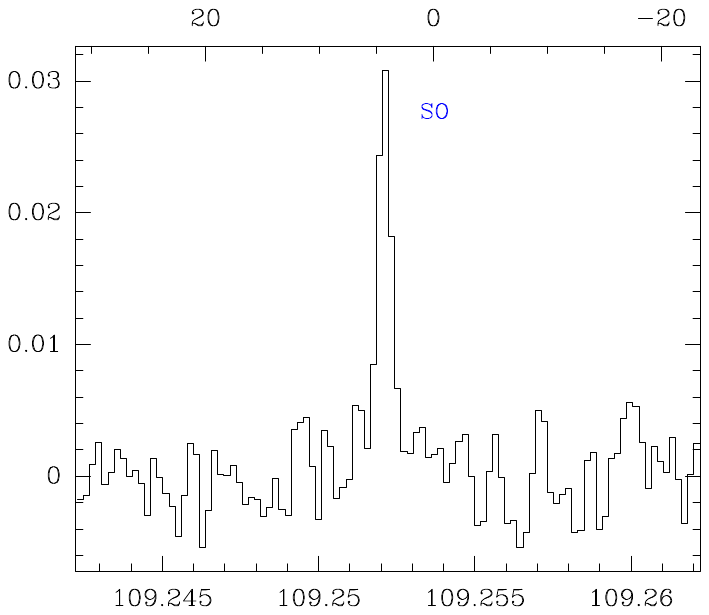}
  \end{minipage} \\
\vspace{-5.5cm}
\hspace{-2cm}
  \begin{minipage}[h]{0.32\textwidth}
    \includegraphics[width=2.5\textwidth]{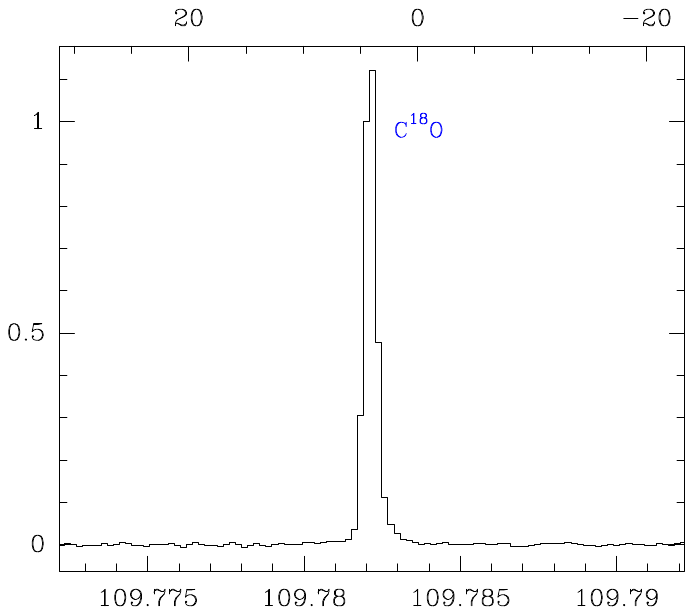}
  \end{minipage}
  \hspace{0.6cm}
  \begin{minipage}[h]{0.32\textwidth}
    \includegraphics[width=2.5\textwidth]{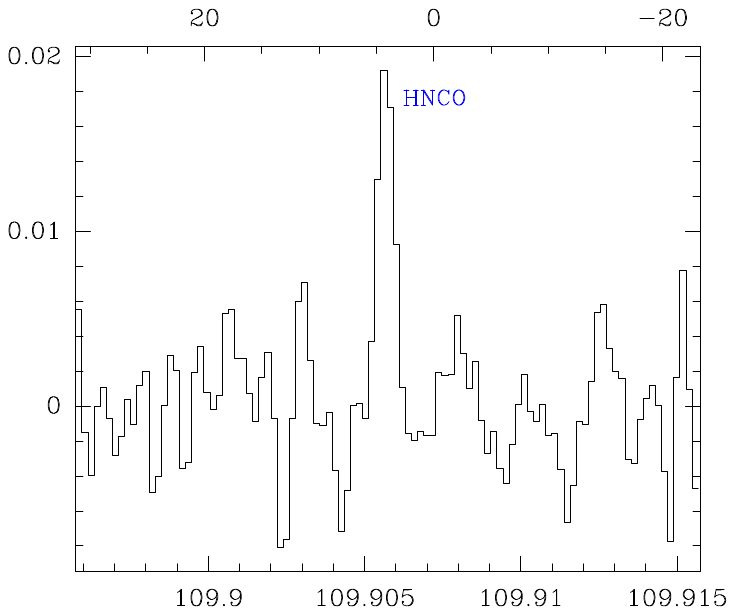}  
  \end{minipage}
  \hspace{0.6cm}
  \begin{minipage}[h]{0.32\textwidth}
    \includegraphics[width=2.5\textwidth]{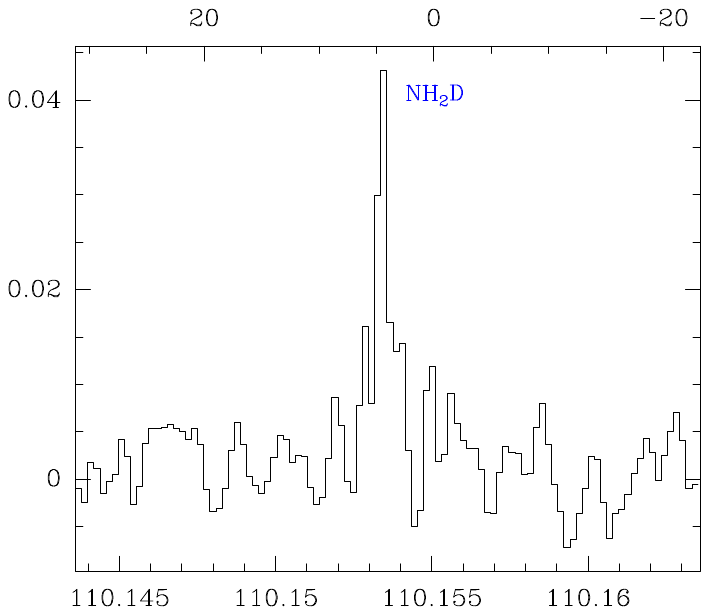}
  \end{minipage} \\
\vspace{-5.5cm}
\hspace{-2cm}
  \begin{minipage}[h]{0.32\textwidth}
    \includegraphics[width=2.5\textwidth]{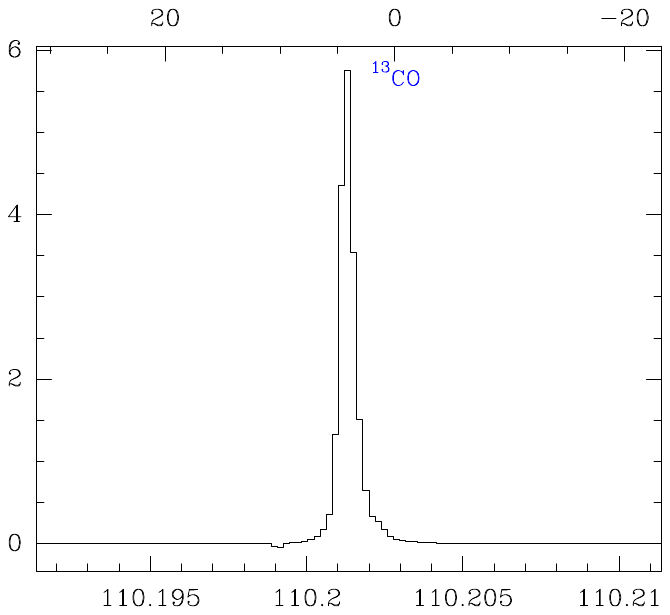}
  \end{minipage}
  \hspace{0.6cm}
  \begin{minipage}[h]{0.32\textwidth}
    \includegraphics[width=2.5\textwidth]{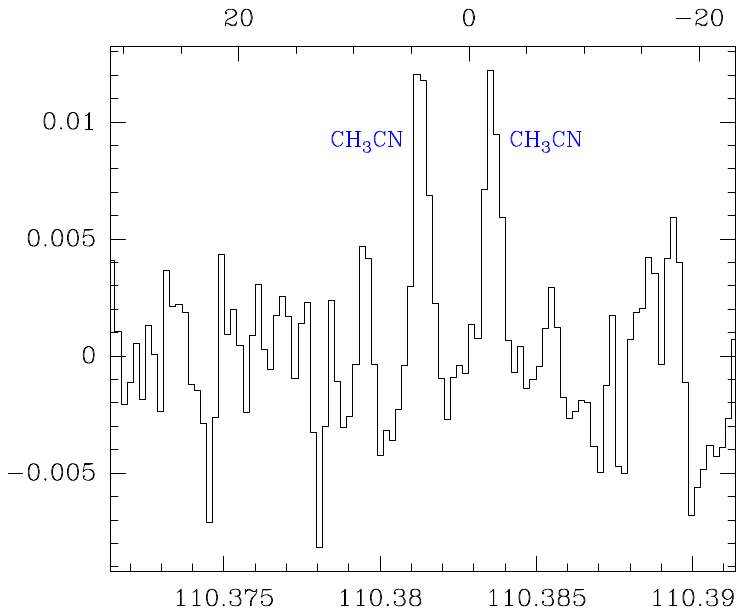}  
  \end{minipage}
  \hspace{0.6cm}
  \begin{minipage}[h]{0.32\textwidth}
    \includegraphics[width=2.5\textwidth]{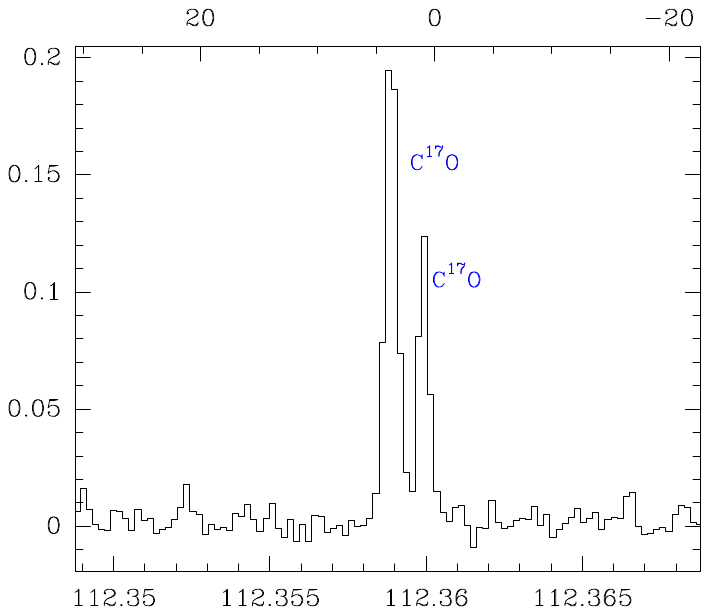}
  \end{minipage} \\
\vspace{-5.5cm}
\hspace{-2cm}
  \begin{minipage}[h]{0.32\textwidth}
    \includegraphics[width=2.5\textwidth]{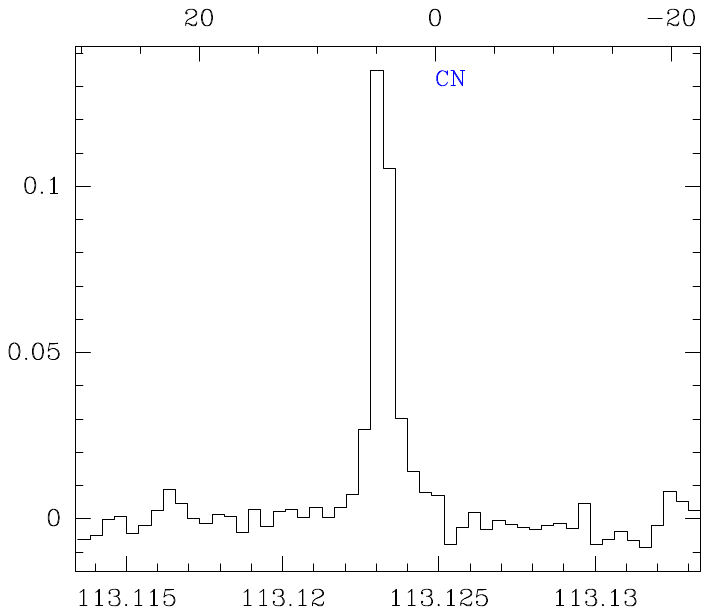}
  \end{minipage}
  \hspace{0.6cm}
  \begin{minipage}[h]{0.32\textwidth}
    \includegraphics[width=2.5\textwidth]{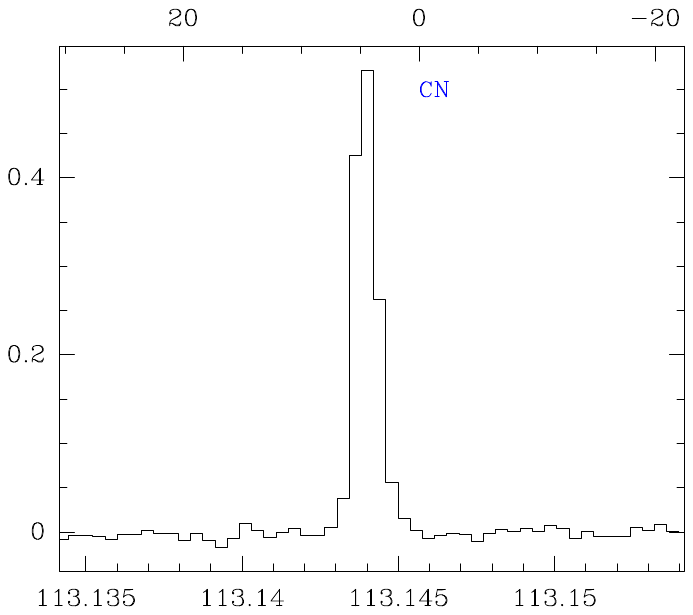}  
  \end{minipage}
  \hspace{0.6cm}
  \begin{minipage}[h]{0.32\textwidth}
    \includegraphics[width=2.5\textwidth]{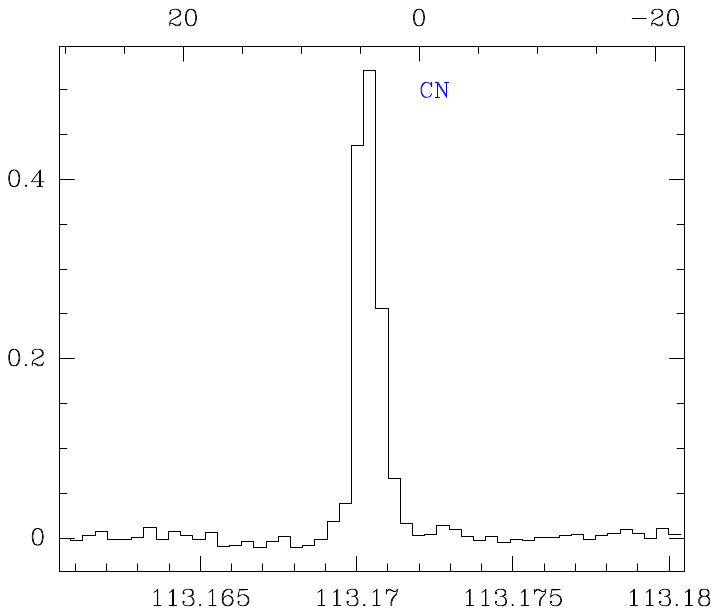}
  \end{minipage} 
 \vspace{-1.5cm} 
\caption{Continued. 
}
\end{figure*}

\addtocounter{figure}{-1}
\begin{figure*}[h]
\centering 
\vspace{-4cm}
\hspace{-2cm}
  \begin{minipage}[h]{0.32\textwidth}
    \includegraphics[width=2.5\textwidth]{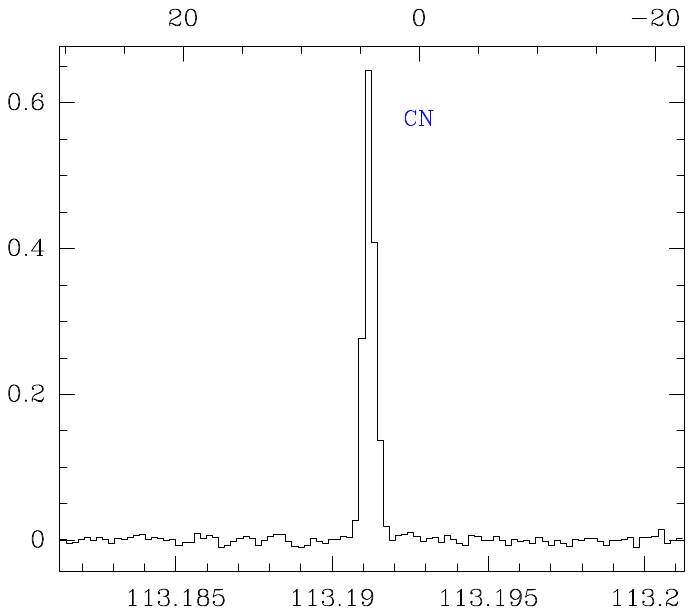}
  \end{minipage}
  \hspace{0.6cm}
  \begin{minipage}[h]{0.32\textwidth}
    \includegraphics[width=2.5\textwidth]{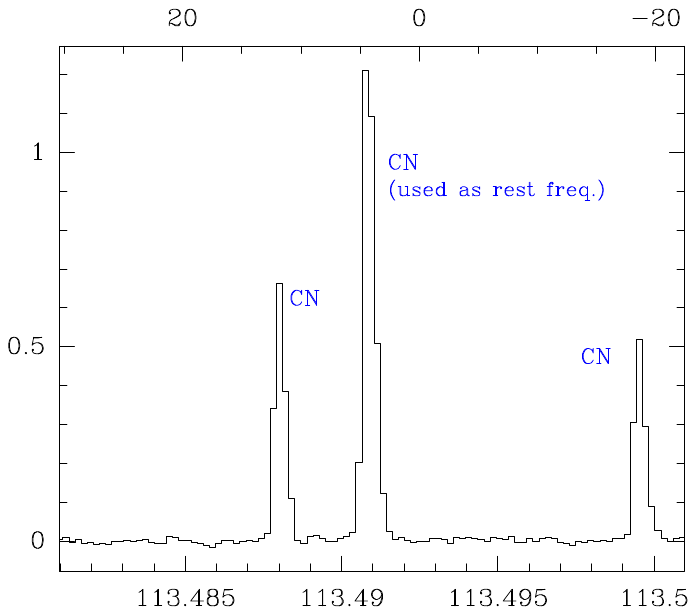}  
  \end{minipage}
  \hspace{0.6cm}
  \begin{minipage}[h]{0.32\textwidth}
    \includegraphics[width=2.5\textwidth]{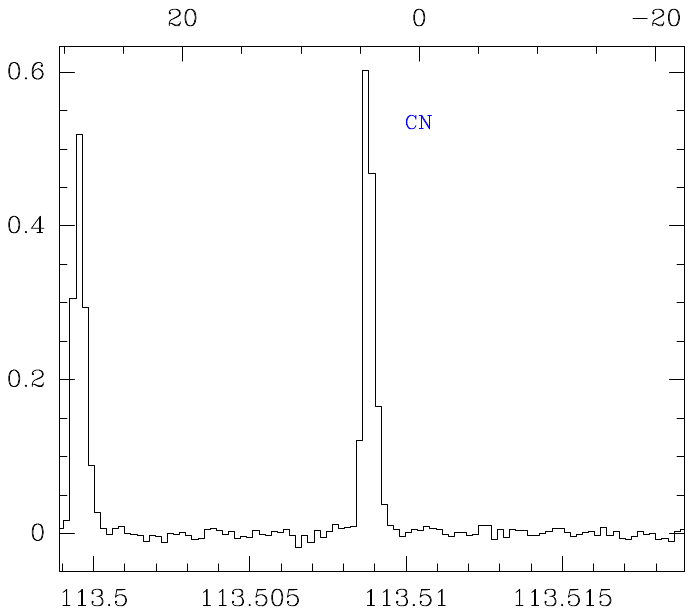}
  \end{minipage} \\
\vspace{-5.5cm}
\hspace{-2cm}
  \begin{minipage}[h]{0.32\textwidth}
    \includegraphics[width=2.5\textwidth]{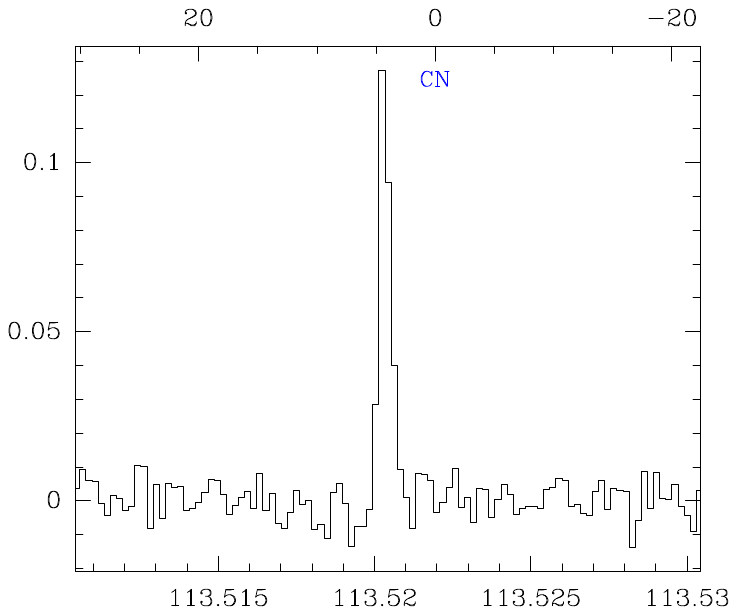}
  \end{minipage}
  \hspace{0.6cm}
  \begin{minipage}[h]{0.32\textwidth}
    \includegraphics[width=2.5\textwidth]{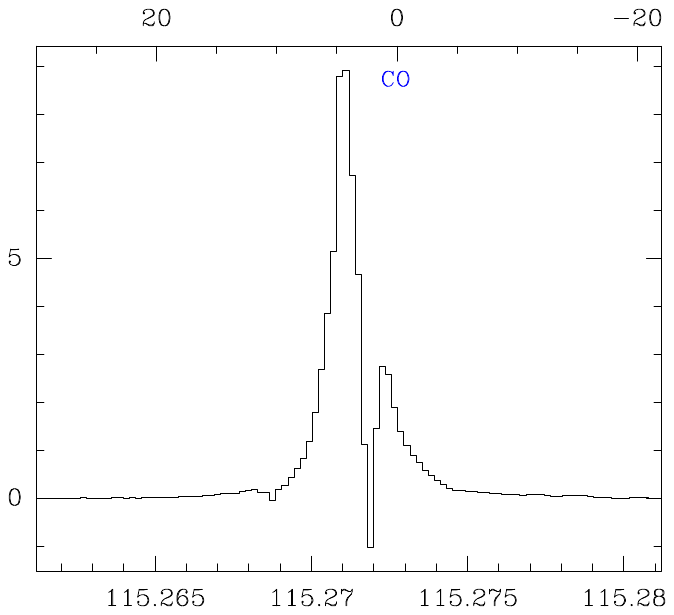}  
  \end{minipage}
  \hspace{0.6cm}
  \begin{minipage}[h]{0.32\textwidth}
    \includegraphics[width=2.5\textwidth]{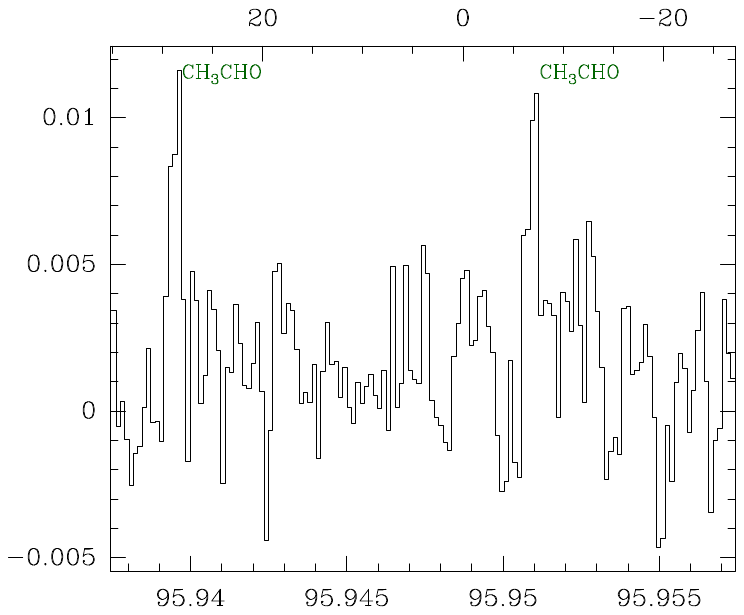}
  \end{minipage} \\
\vspace{-5.5cm}
\hspace{-2cm}
  \begin{minipage}[h]{0.32\textwidth}
    \includegraphics[width=2.5\textwidth]{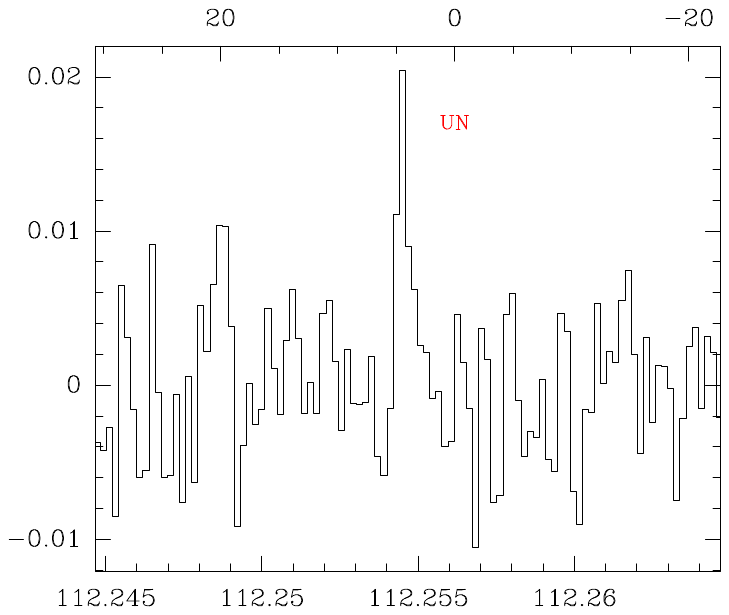}
  \end{minipage}
 \hspace{0.6cm}
 \begin{minipage}[h]{0.32\textwidth}
   \includegraphics[width=2.5\textwidth]{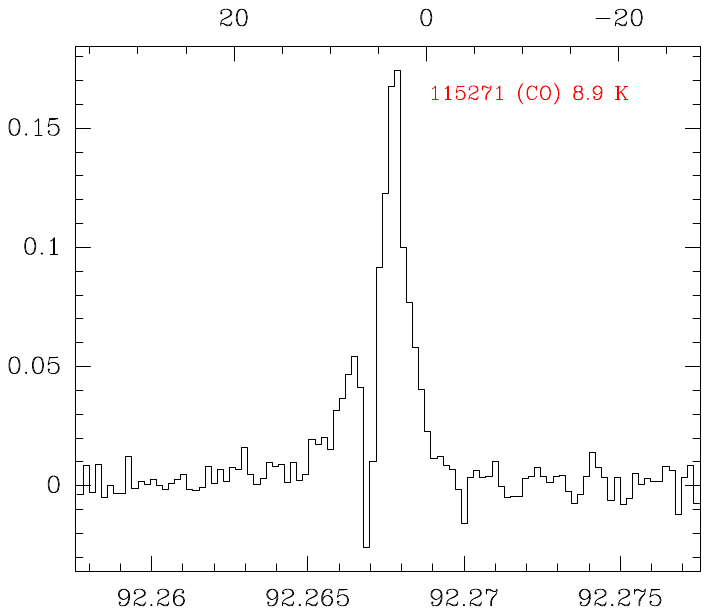}  
 \end{minipage}
 \hspace{0.6cm}
 \begin{minipage}[h]{0.32\textwidth}
   \includegraphics[width=2.5\textwidth]{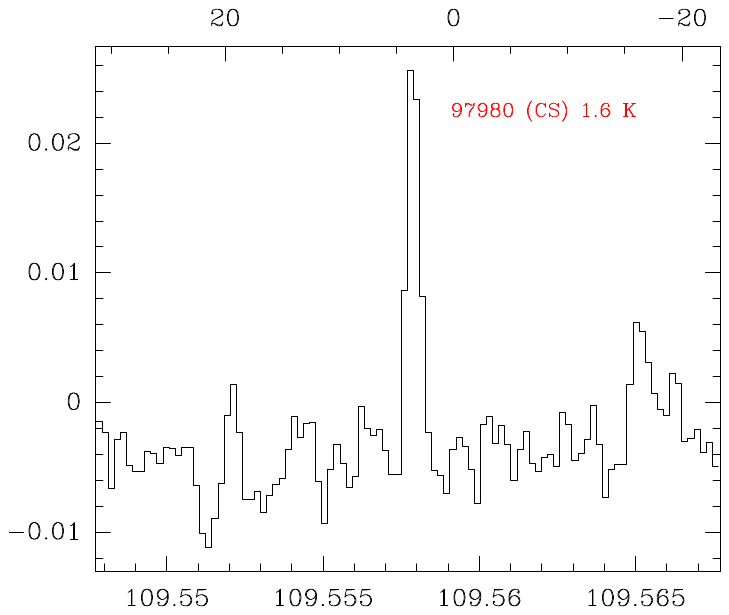}
 \end{minipage} \\
\vspace{-5.5cm}
\hspace{-2cm}
 \begin{minipage}[h]{0.32\textwidth}
   \includegraphics[width=2.5\textwidth]{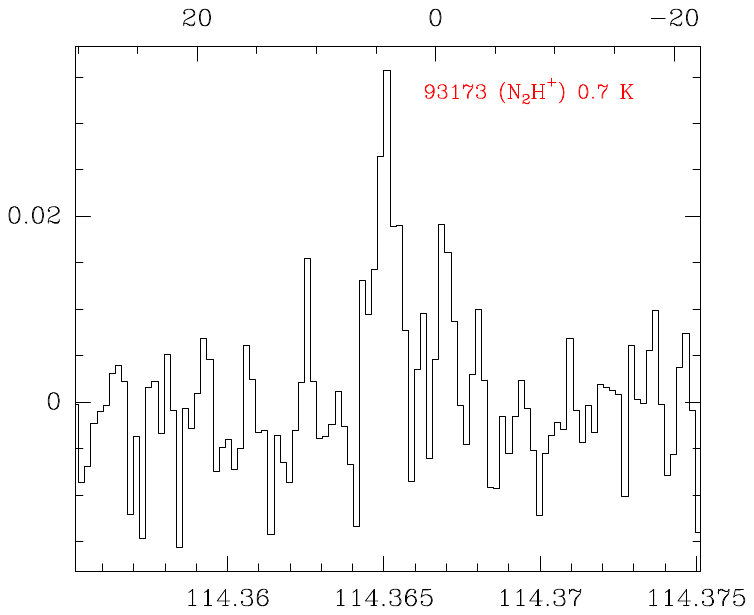}
 \end{minipage}
 \vspace{-1.5cm} 
\caption{Continued. 
}
\end{figure*}

\newpage
\begin{figure*}[h]
\centering 
\vspace{-4cm}
\hspace{-2cm}
  \begin{minipage}[h]{0.32\textwidth}
    \includegraphics[width=2.5\textwidth]{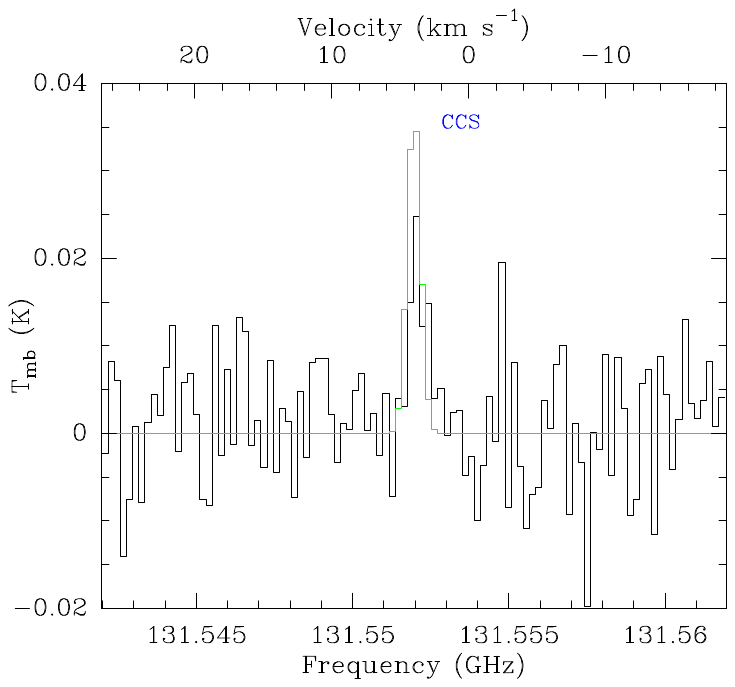}
  \end{minipage}
  \hspace{0.6cm}
  \begin{minipage}[h]{0.32\textwidth}
    \includegraphics[width=2.5\textwidth]{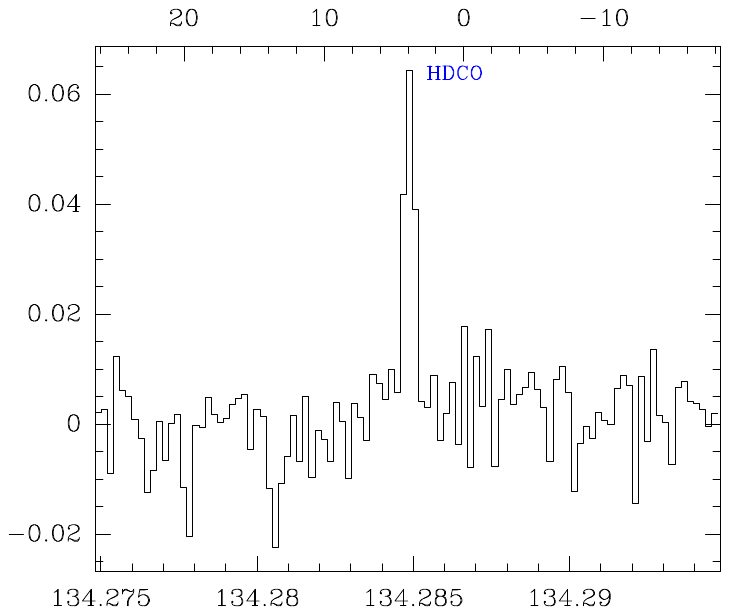}  
  \end{minipage}
  \hspace{0.6cm}
  \begin{minipage}[h]{0.32\textwidth}
    \includegraphics[width=2.5\textwidth]{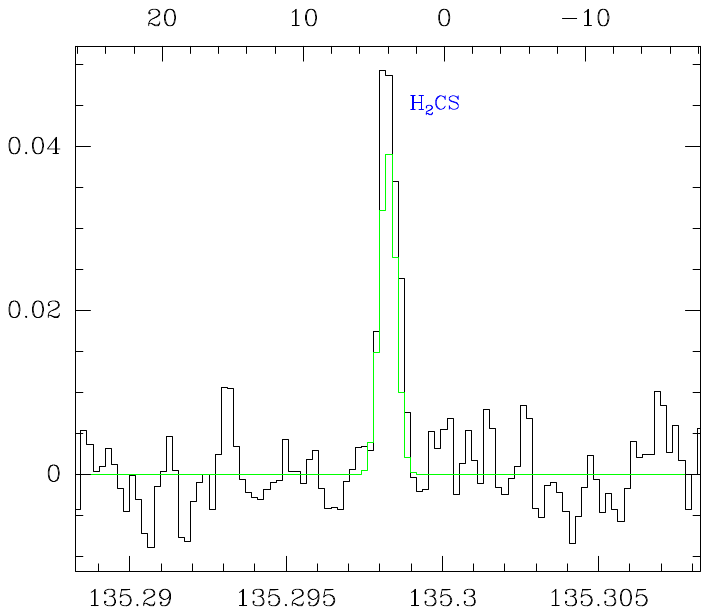}
  \end{minipage} \\
\vspace{-5.5cm}
\hspace{-2cm}
  \begin{minipage}[h]{0.32\textwidth}
    \includegraphics[width=2.5\textwidth]{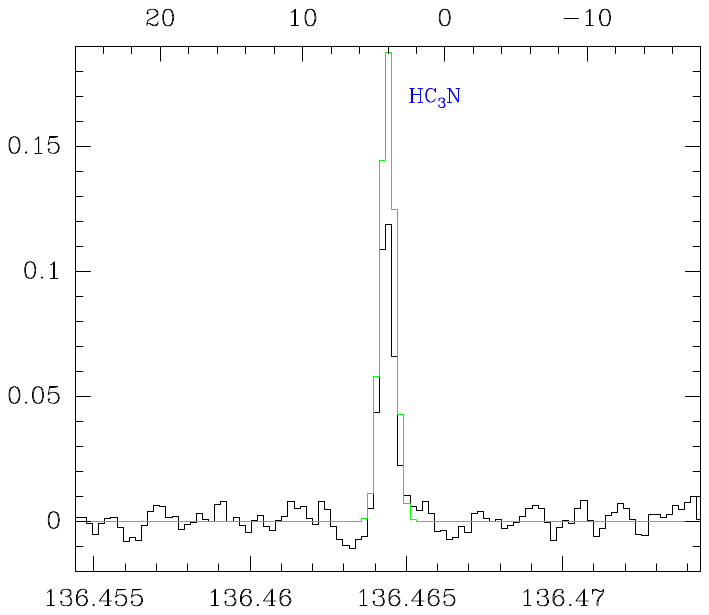}
  \end{minipage}
  \hspace{0.6cm}
  \begin{minipage}[h]{0.32\textwidth}
    \includegraphics[width=2.5\textwidth]{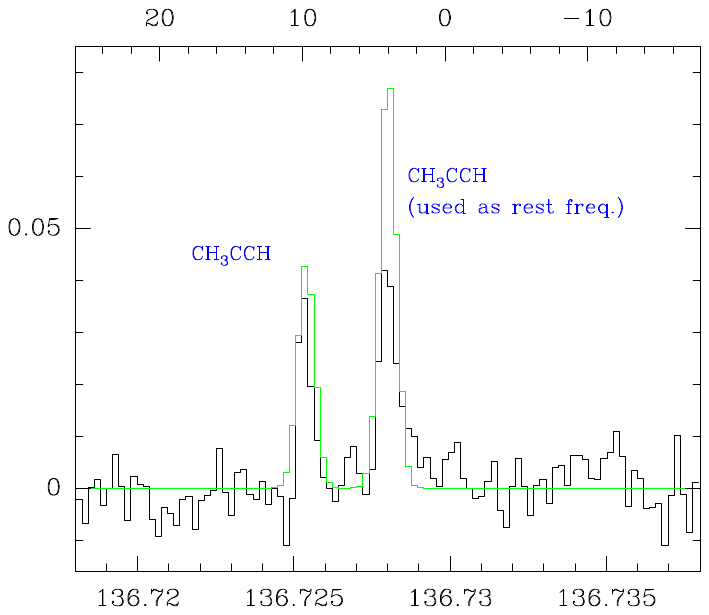}  
  \end{minipage}
  \hspace{0.6cm}
  \begin{minipage}[h]{0.32\textwidth}
    \includegraphics[width=2.5\textwidth]{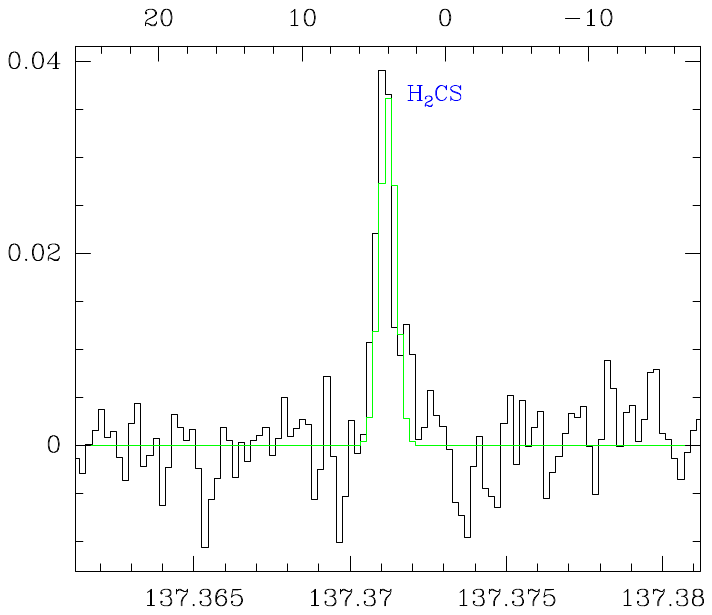}
  \end{minipage} \\
\vspace{-5.5cm}
\hspace{-2cm}
  \begin{minipage}[h]{0.32\textwidth}
    \includegraphics[width=2.5\textwidth]{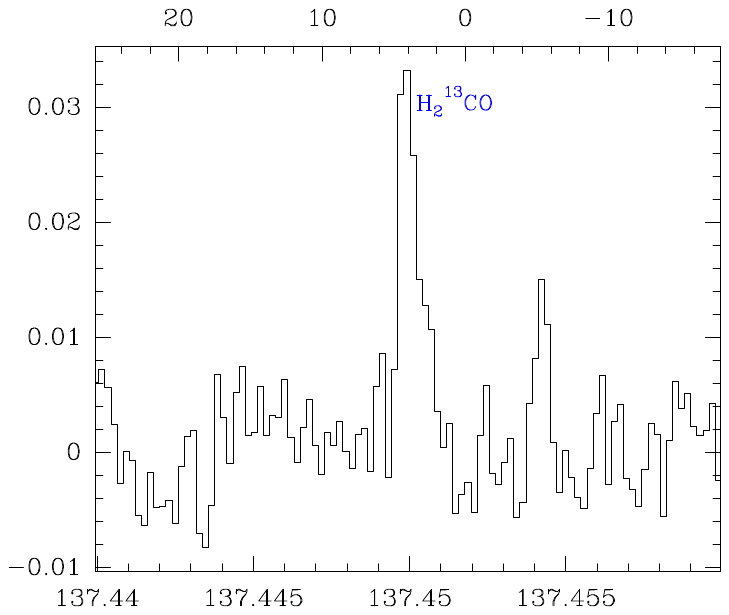}
  \end{minipage}
  \hspace{0.6cm}
  \begin{minipage}[h]{0.32\textwidth}
    \includegraphics[width=2.5\textwidth]{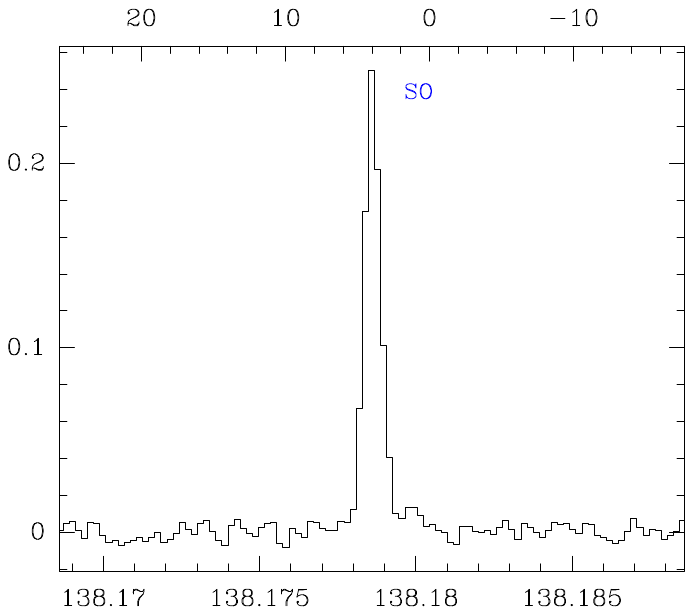}  
  \end{minipage}
  \hspace{0.6cm}
  \begin{minipage}[h]{0.32\textwidth}
    \includegraphics[width=2.5\textwidth]{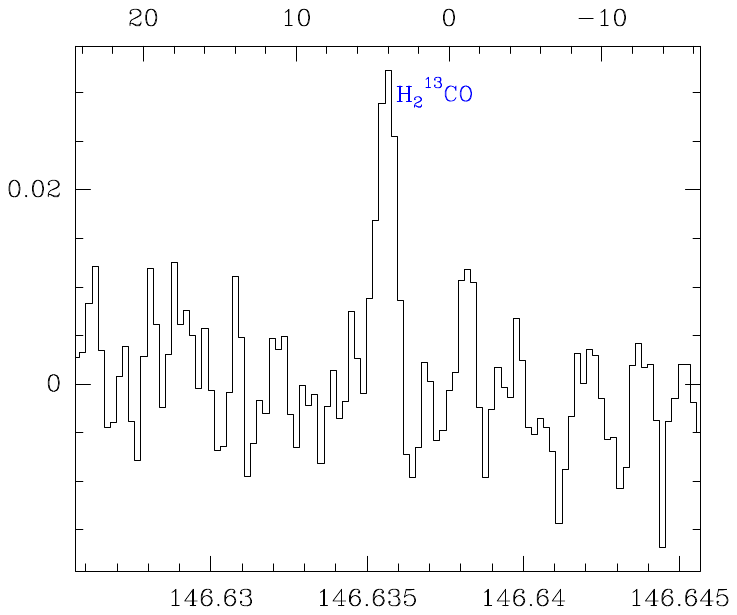}
  \end{minipage} \\
\vspace{-5.5cm}
\hspace{-2cm}
  \begin{minipage}[h]{0.32\textwidth}
    \includegraphics[width=2.5\textwidth]{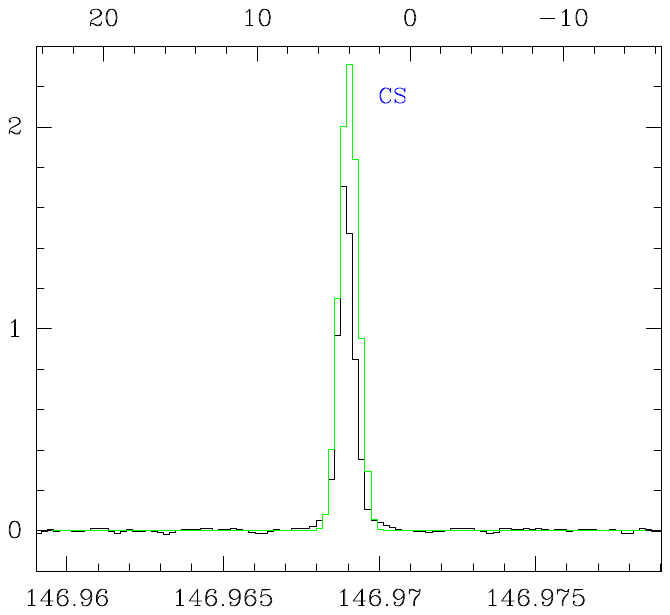}
  \end{minipage}
  \hspace{0.6cm}
  \begin{minipage}[h]{0.32\textwidth}
    \includegraphics[width=2.5\textwidth]{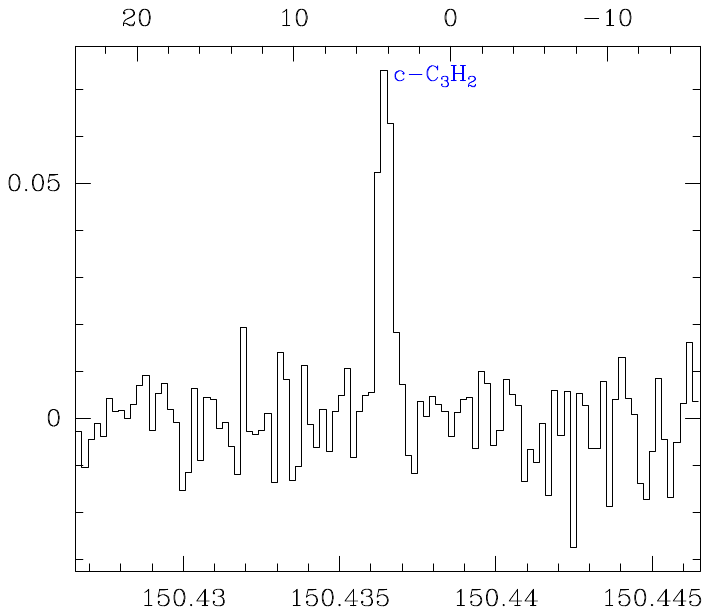}  
  \end{minipage}
  \hspace{0.6cm}
  \begin{minipage}[h]{0.32\textwidth}
    \includegraphics[width=2.5\textwidth]{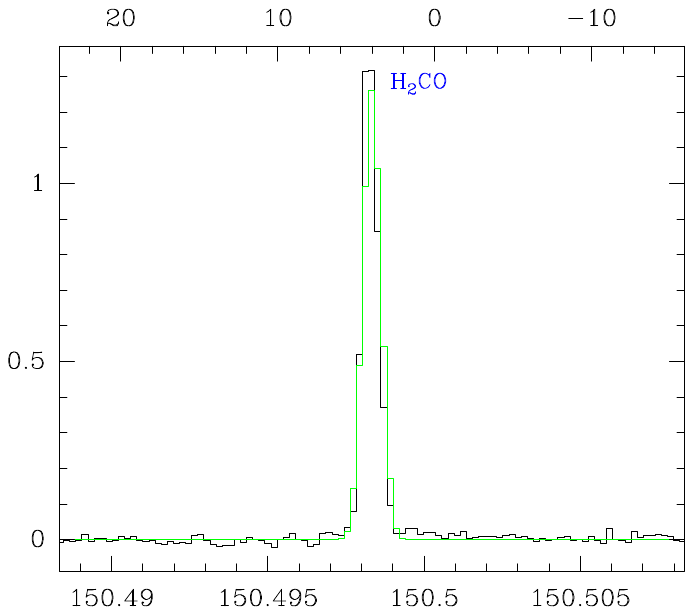}
  \end{minipage} 
 \vspace{-1.5cm} 
\caption{Same as Fig.~\ref{fig:survey-73ghz-small} but for the tuning at 133\,GHz.
}
\label{fig:survey-133ghz-small}
\end{figure*}

\addtocounter{figure}{-1}
\begin{figure*}[h]
\centering 
\vspace{-4cm}
\hspace{-2cm}
  \begin{minipage}[h]{0.32\textwidth}
    \includegraphics[width=2.5\textwidth]{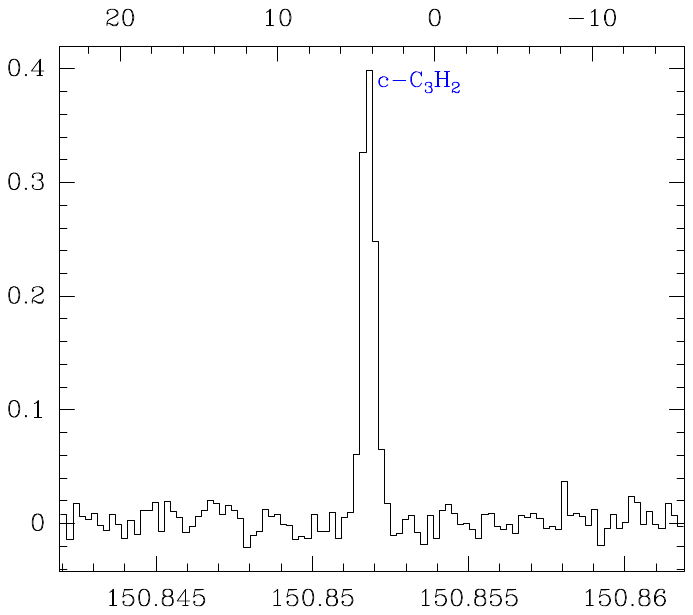}
  \end{minipage}
  \hspace{0.6cm}
  \begin{minipage}[h]{0.32\textwidth}
    \includegraphics[width=2.5\textwidth]{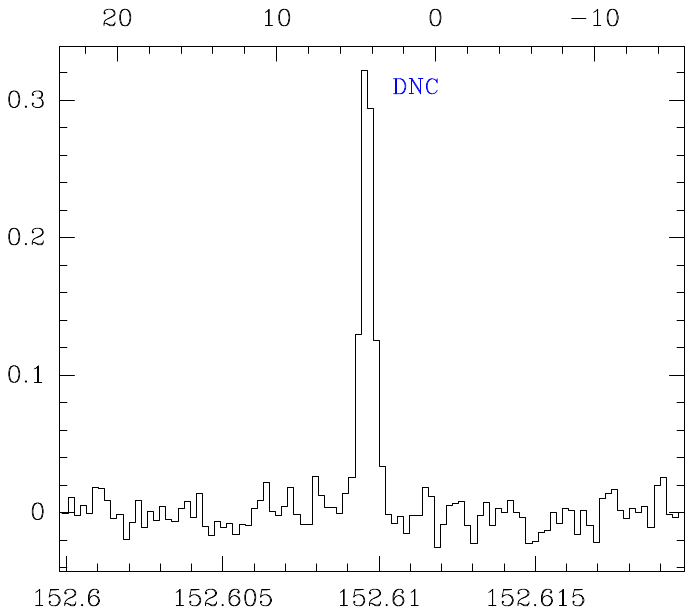}  
  \end{minipage}
  \hspace{0.6cm}
  \begin{minipage}[h]{0.32\textwidth}
    \includegraphics[width=2.5\textwidth]{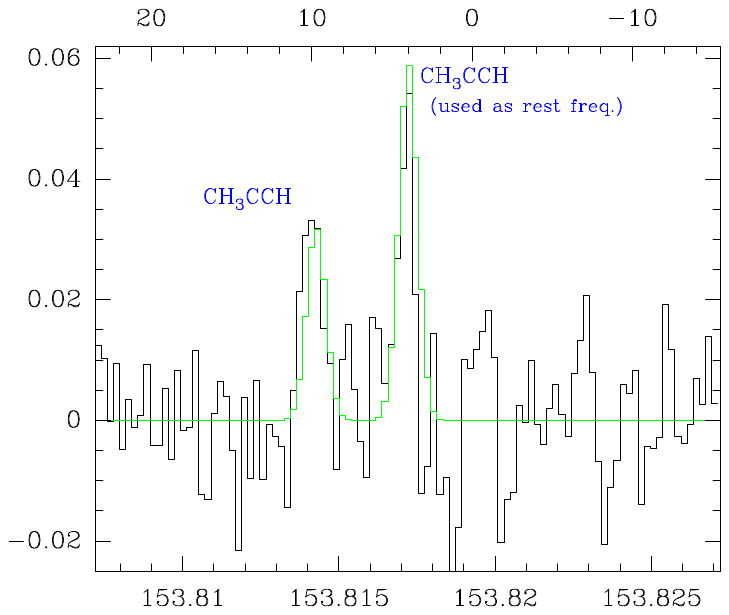}
  \end{minipage} \\
\vspace{-5.5cm}
\hspace{-2cm}
  \begin{minipage}[h]{0.32\textwidth}
    \includegraphics[width=2.5\textwidth]{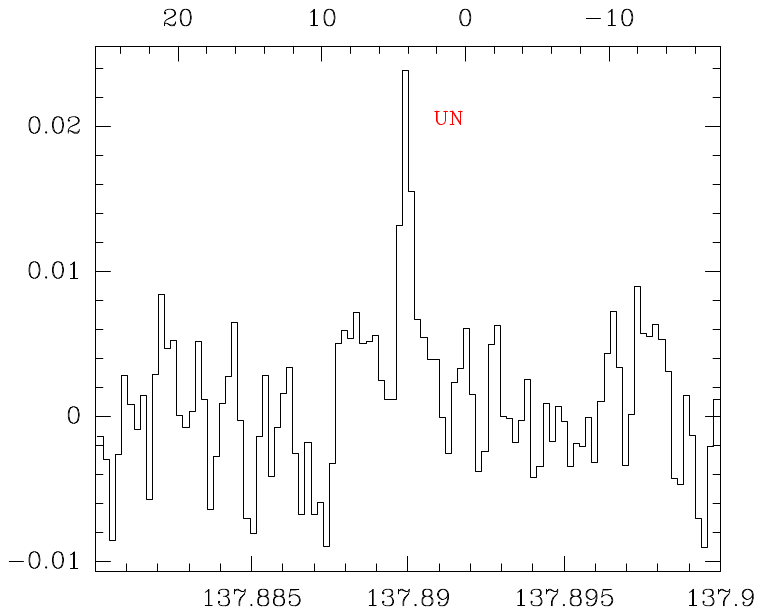}
  \end{minipage}
  \hspace{0.6cm}
  \begin{minipage}[h]{0.32\textwidth}
    \includegraphics[width=2.5\textwidth]{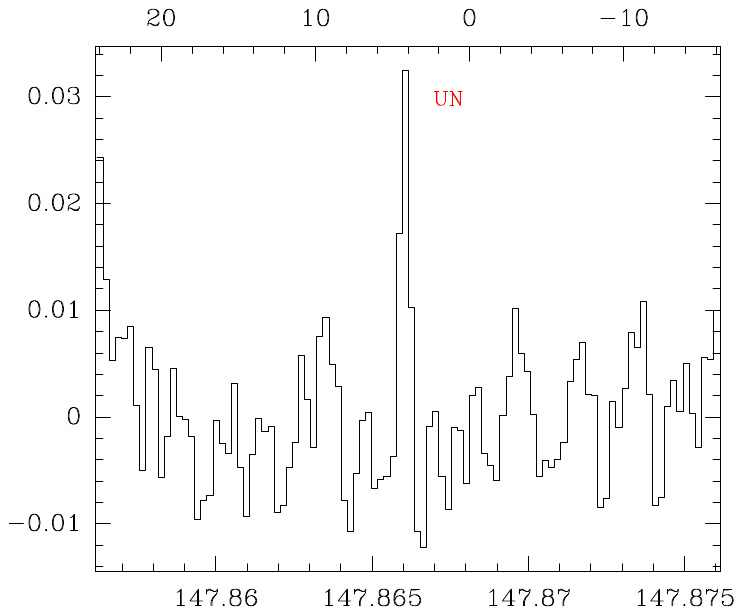}  
  \end{minipage}
  \hspace{0.6cm}
  \begin{minipage}[h]{0.32\textwidth}
    \includegraphics[width=2.5\textwidth]{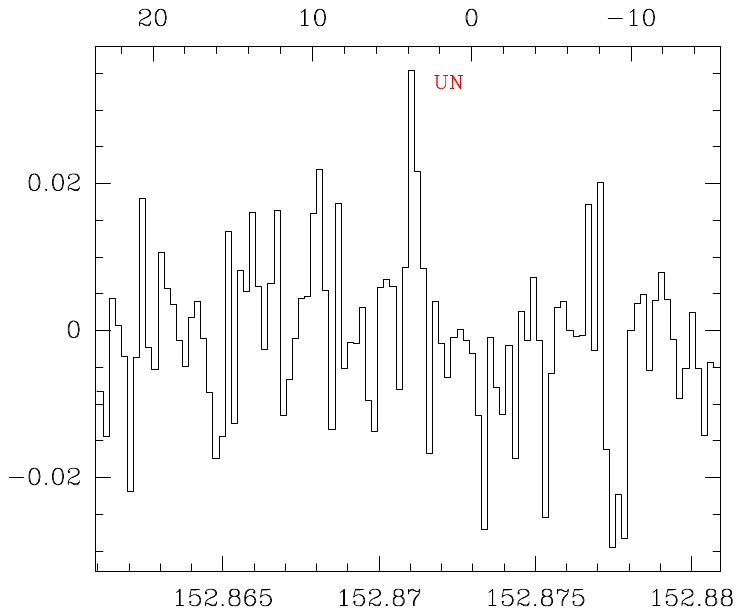}
  \end{minipage} \\
\vspace{-5.5cm}
\hspace{-2cm}
  \begin{minipage}[h]{0.32\textwidth}
    \includegraphics[width=2.5\textwidth]{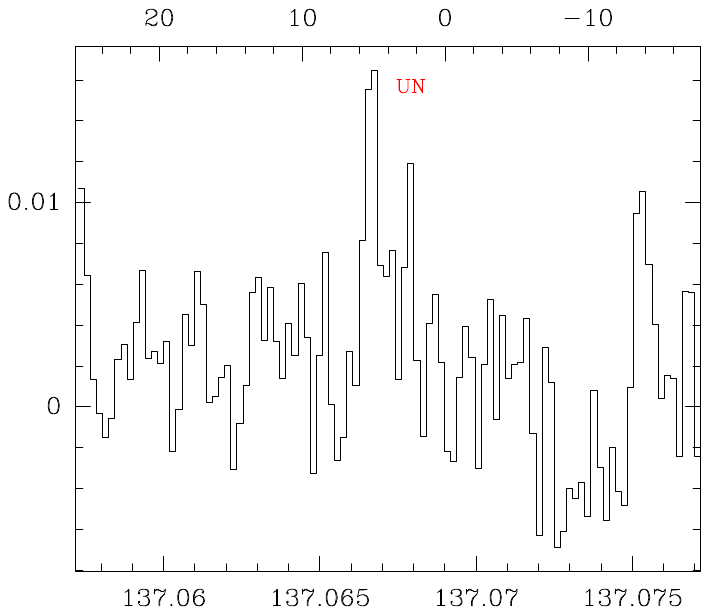}
  \end{minipage}
  \hspace{0.6cm}
  \begin{minipage}[h]{0.32\textwidth}
    \includegraphics[width=2.5\textwidth]{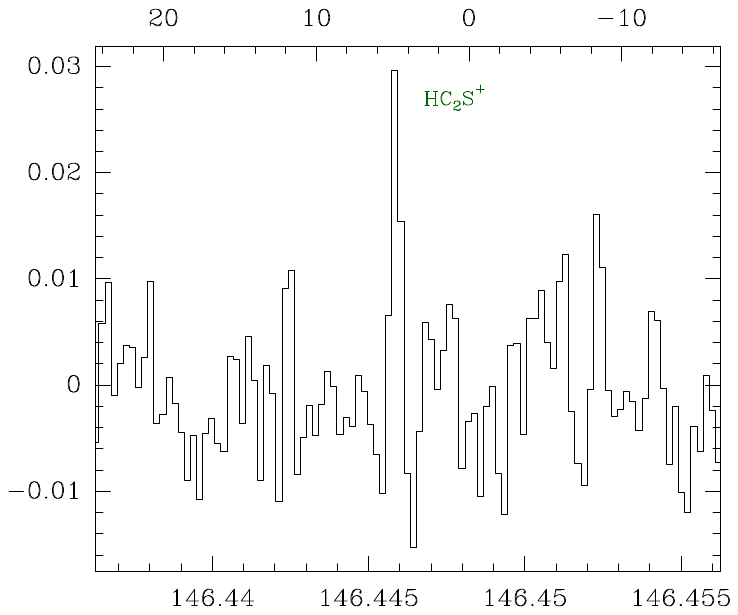}  
  \end{minipage}
  \hspace{0.6cm}
  \begin{minipage}[h]{0.32\textwidth}
    \includegraphics[width=2.5\textwidth]{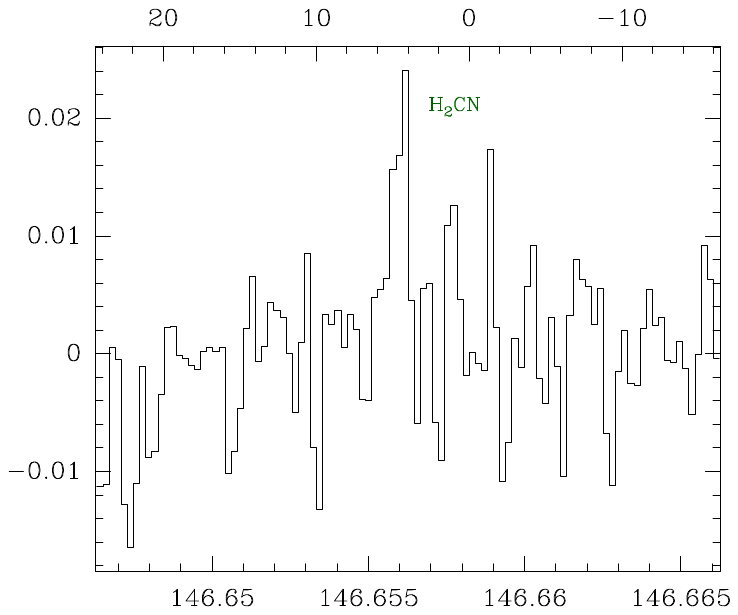}
  \end{minipage} \\
\vspace{-5.5cm}
\hspace{-2cm}
  \begin{minipage}[h]{0.32\textwidth}
    \includegraphics[width=2.5\textwidth]{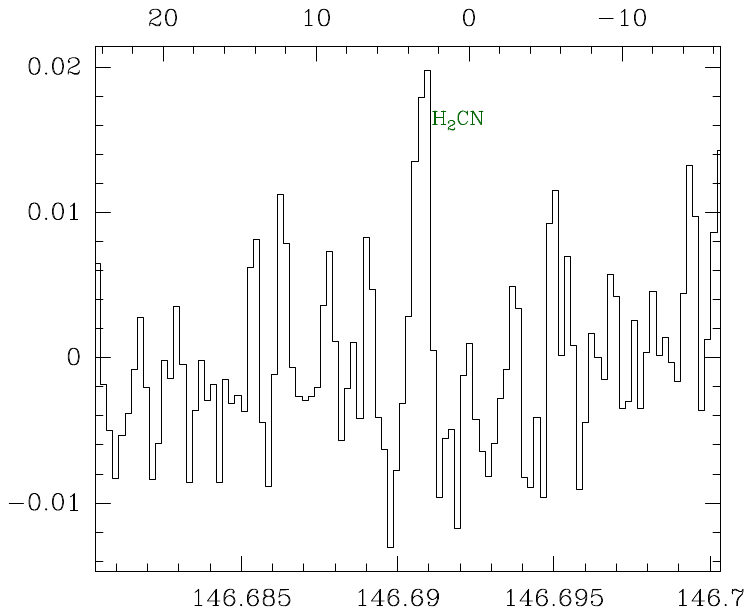}
  \end{minipage}
 \vspace{-1.5cm} 
\caption{Continued. 
}
\end{figure*}

\newpage
\begin{figure*}[h]
\centering 
\vspace{-4cm}
\hspace{-2cm}
  \begin{minipage}[h]{0.32\textwidth}
    \includegraphics[width=2.5\textwidth]{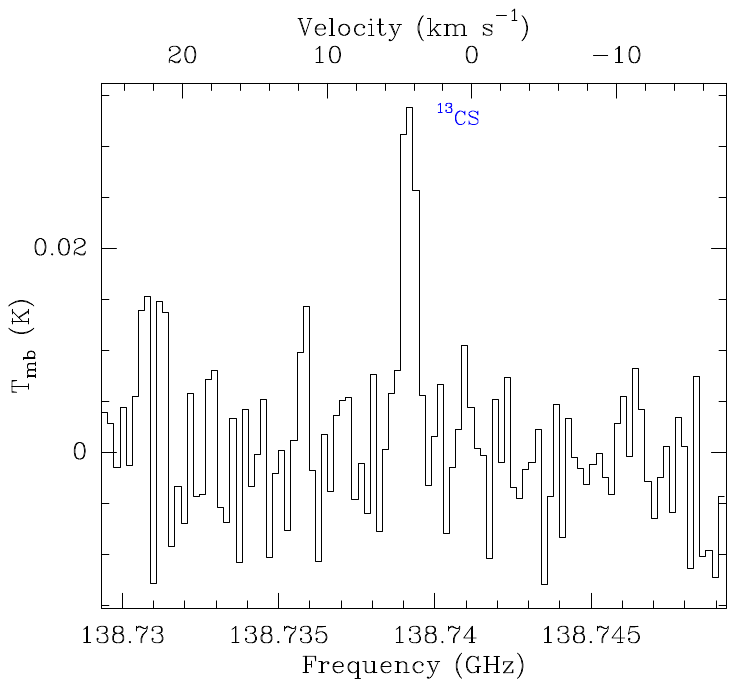}
  \end{minipage}
  \hspace{0.6cm}
  \begin{minipage}[h]{0.32\textwidth}
    \includegraphics[width=2.5\textwidth]{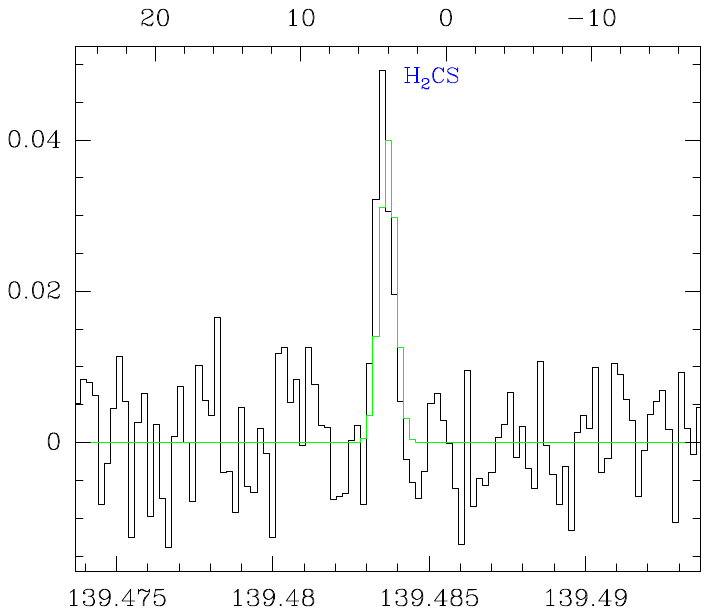}  
  \end{minipage}
  \hspace{0.6cm}
  \begin{minipage}[h]{0.32\textwidth}
    \includegraphics[width=2.5\textwidth]{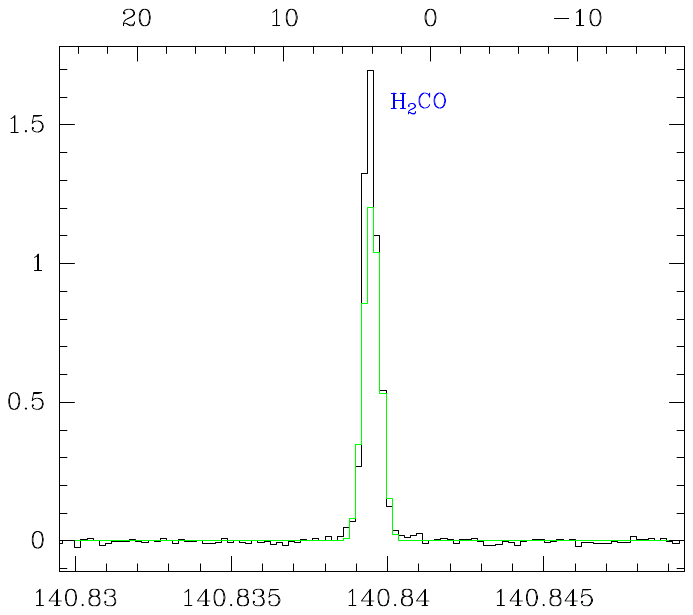}
  \end{minipage} \\
\vspace{-5.5cm}
\hspace{-2cm}
  \begin{minipage}[h]{0.32\textwidth}
    \includegraphics[width=2.5\textwidth]{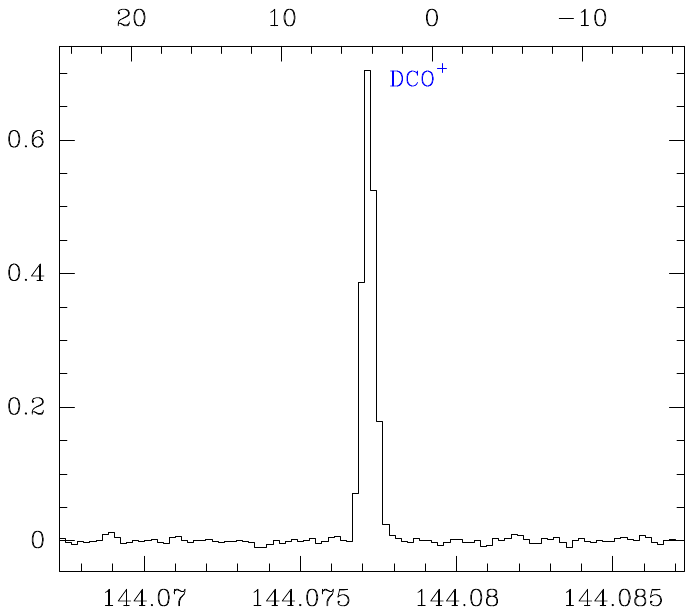}
  \end{minipage}
  \hspace{0.6cm}
  \begin{minipage}[h]{0.32\textwidth}
    \includegraphics[width=2.5\textwidth]{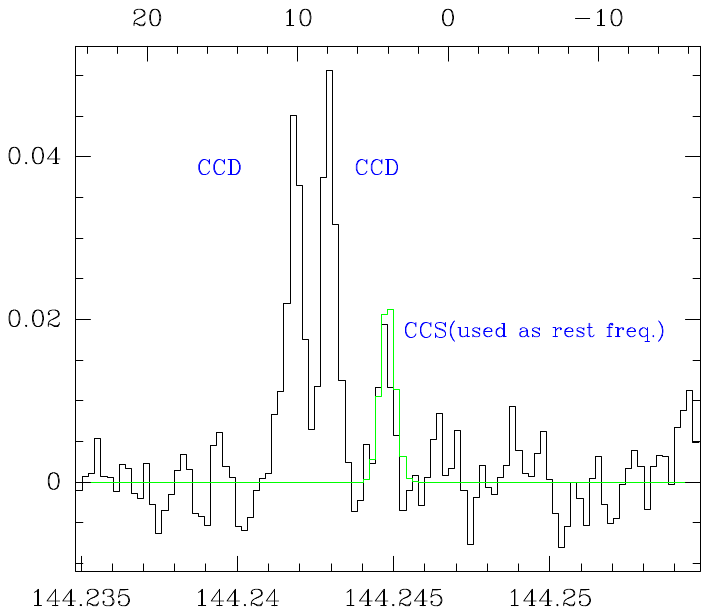}  
  \end{minipage}
  \hspace{0.6cm}
  \begin{minipage}[h]{0.32\textwidth}
    \includegraphics[width=2.5\textwidth]{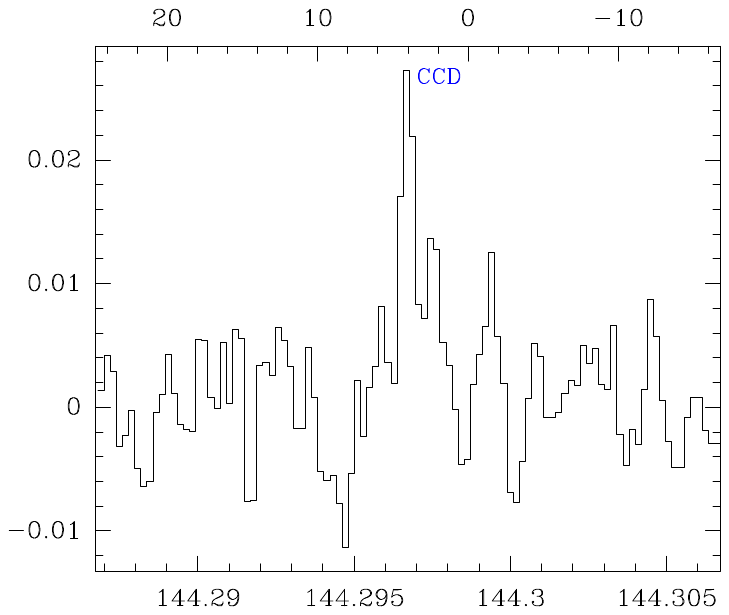}
  \end{minipage} \\
\vspace{-5.5cm}
\hspace{-2cm}
  \begin{minipage}[h]{0.32\textwidth}
    \includegraphics[width=2.5\textwidth]{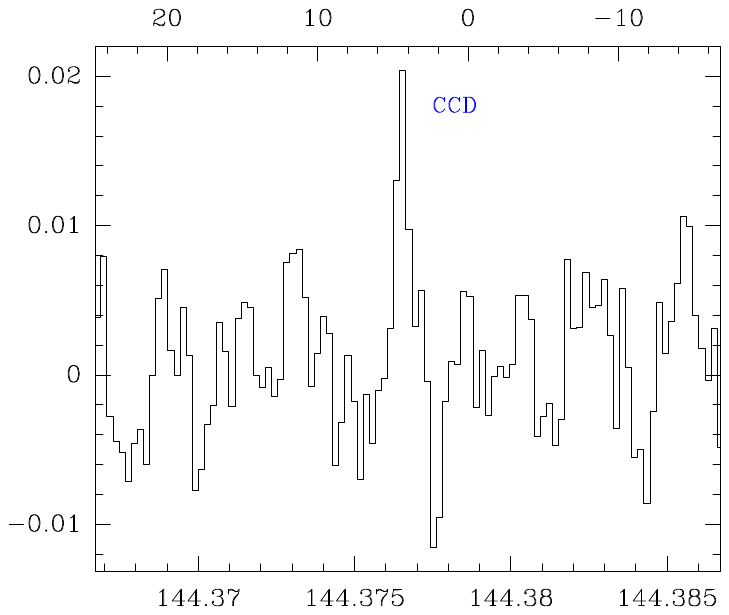}
  \end{minipage}
  \hspace{0.6cm}
  \begin{minipage}[h]{0.32\textwidth}
    \includegraphics[width=2.5\textwidth]{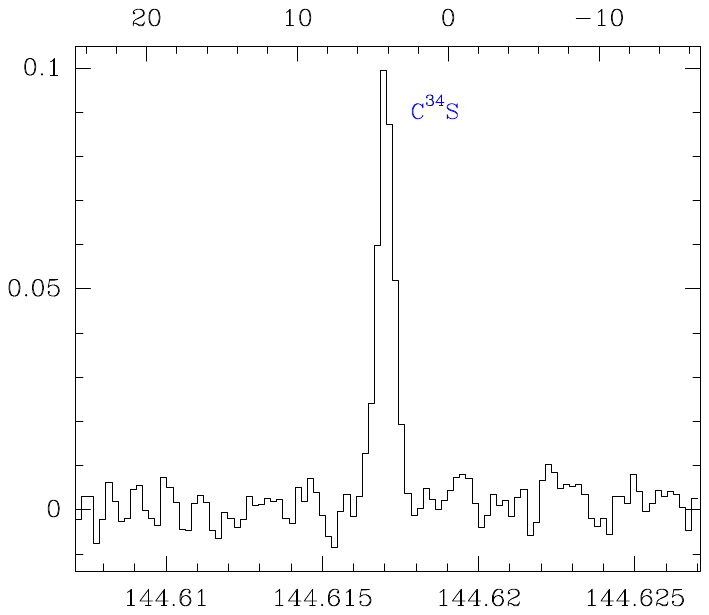}  
  \end{minipage}
  \hspace{0.6cm}
  \begin{minipage}[h]{0.32\textwidth}
    \includegraphics[width=2.5\textwidth]{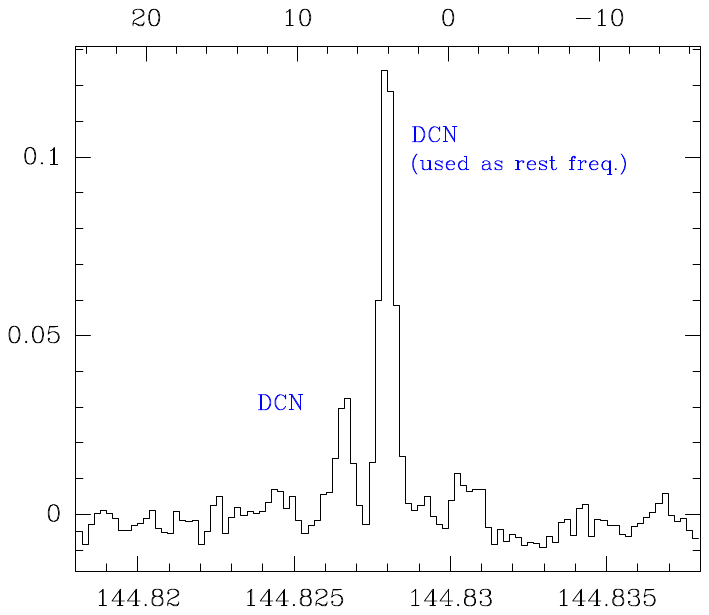}
  \end{minipage} \\
\vspace{-5.5cm}
\hspace{-2cm}
  \begin{minipage}[h]{0.32\textwidth}
    \includegraphics[width=2.5\textwidth]{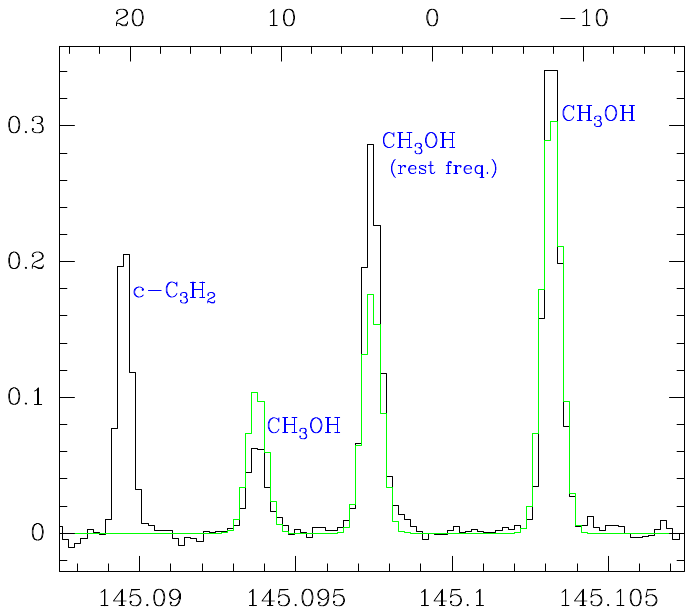}
  \end{minipage}
  \hspace{0.6cm}
  \begin{minipage}[h]{0.32\textwidth}
    \includegraphics[width=2.5\textwidth]{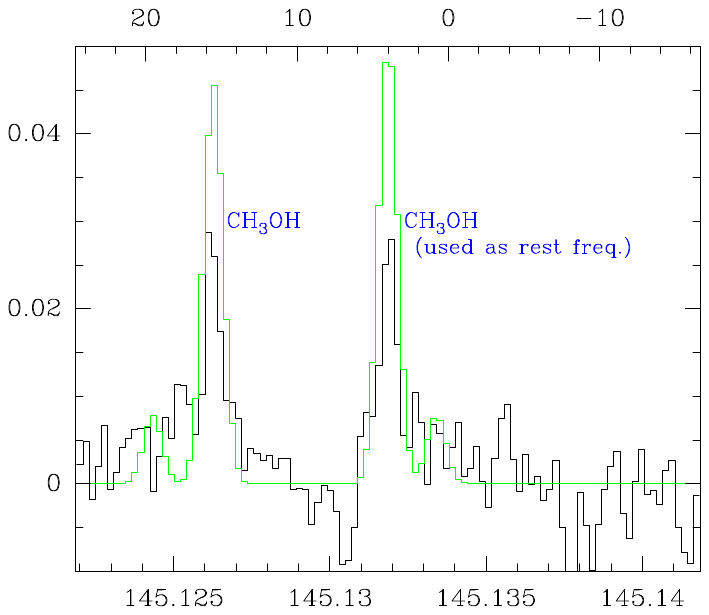}  
  \end{minipage}
  \hspace{0.6cm}
  \begin{minipage}[h]{0.32\textwidth}
    \includegraphics[width=2.5\textwidth]{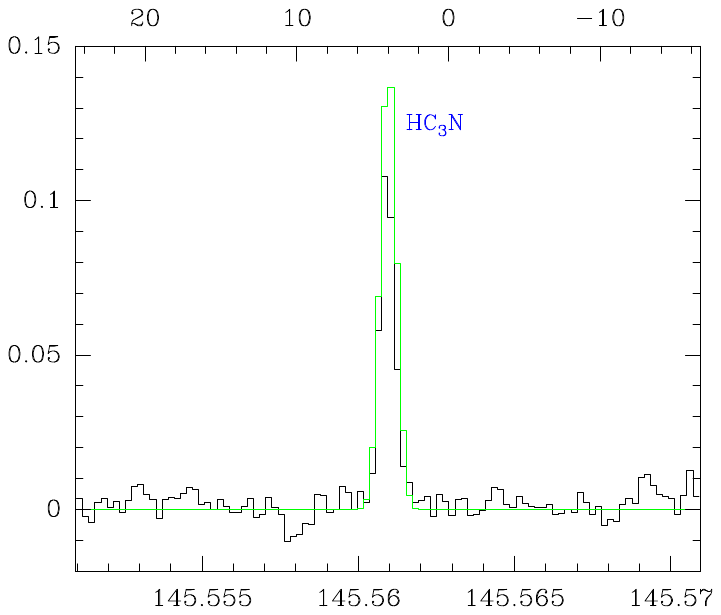}
  \end{minipage} 
 \vspace{-1.5cm} 
\caption{Same as Fig.~\ref{fig:survey-73ghz-small} but for the tuning at 141\,GHz.
}
\label{fig:survey-141ghz-small}
\end{figure*}

\addtocounter{figure}{-1}
\begin{figure*}[h]
\centering 
\vspace{-4cm}
\hspace{-2cm}
  \begin{minipage}[h]{0.32\textwidth}
    \includegraphics[width=2.5\textwidth]{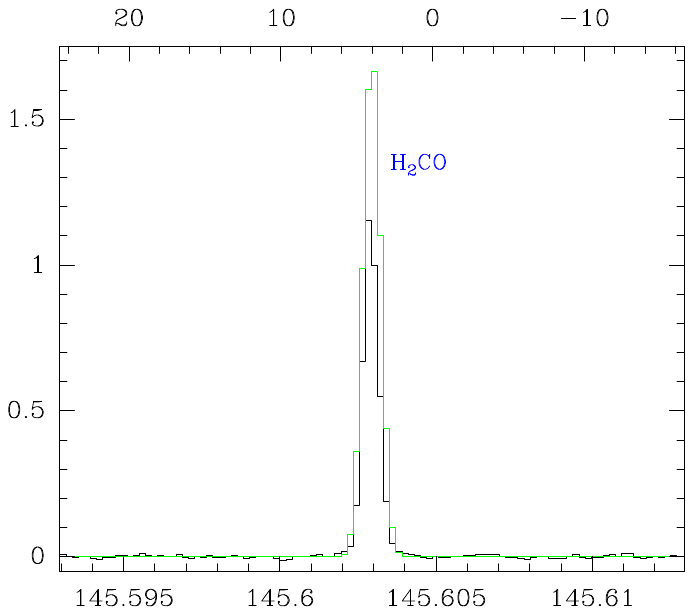}
  \end{minipage}
  \hspace{0.6cm}
  \begin{minipage}[h]{0.32\textwidth}
    \includegraphics[width=2.5\textwidth]{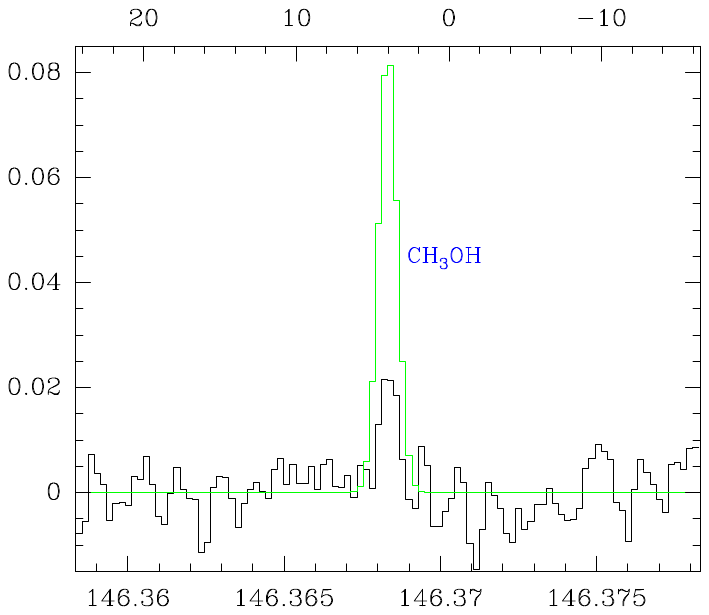}  
  \end{minipage}
  \hspace{0.6cm}
  \begin{minipage}[h]{0.32\textwidth}
    \includegraphics[width=2.5\textwidth]{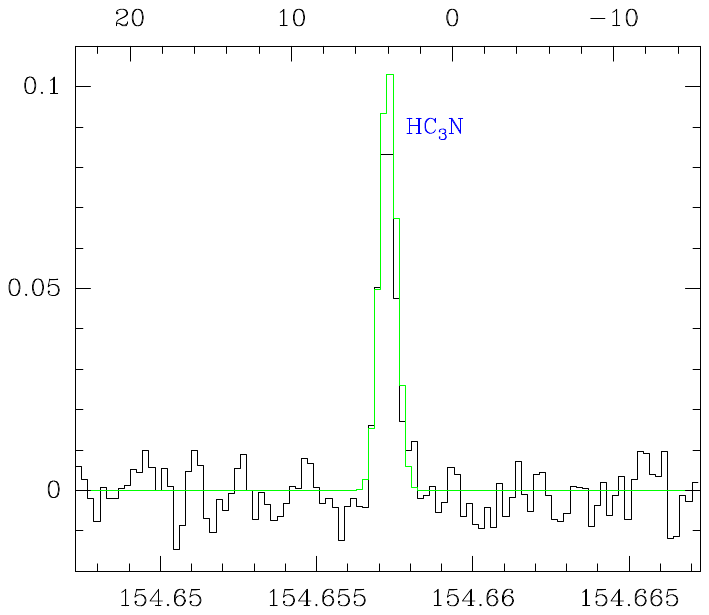}
  \end{minipage} \\
\vspace{-5.5cm}
\hspace{-2cm}
  \begin{minipage}[h]{0.32\textwidth}
    \includegraphics[width=2.5\textwidth]{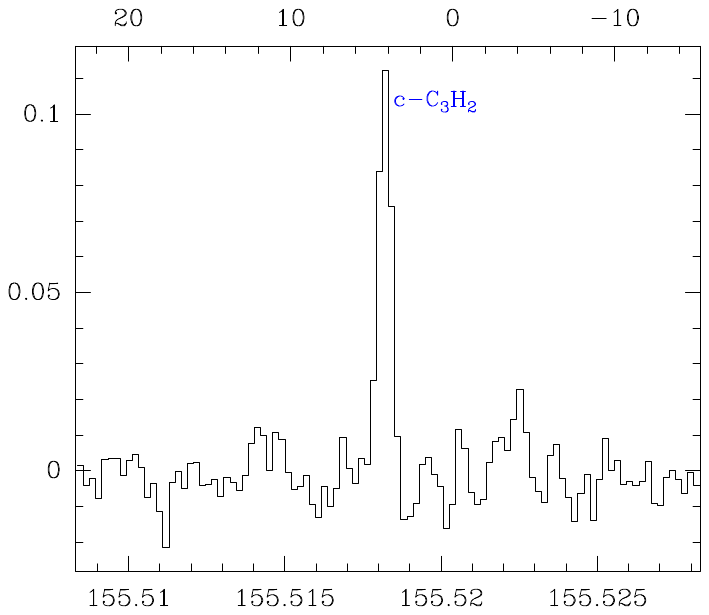}
  \end{minipage}
  \hspace{0.6cm}
  \begin{minipage}[h]{0.32\textwidth}
    \includegraphics[width=2.5\textwidth]{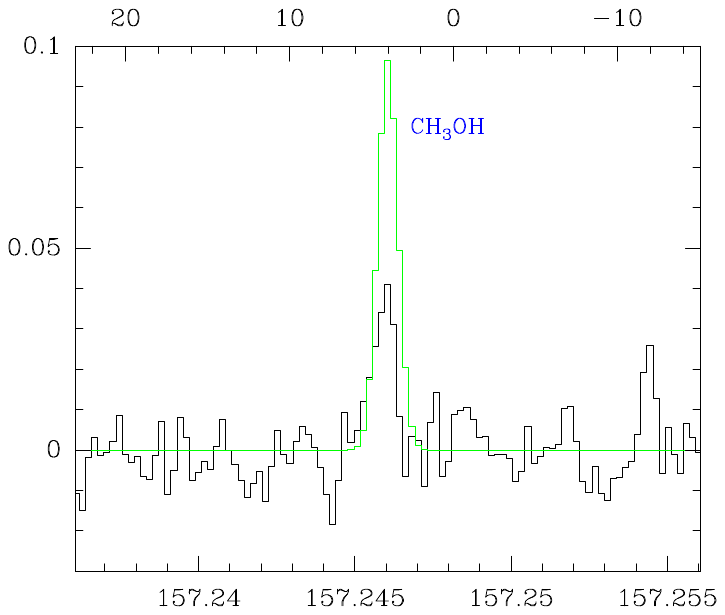}  
  \end{minipage}
  \hspace{0.6cm}
  \begin{minipage}[h]{0.32\textwidth}
    \includegraphics[width=2.5\textwidth]{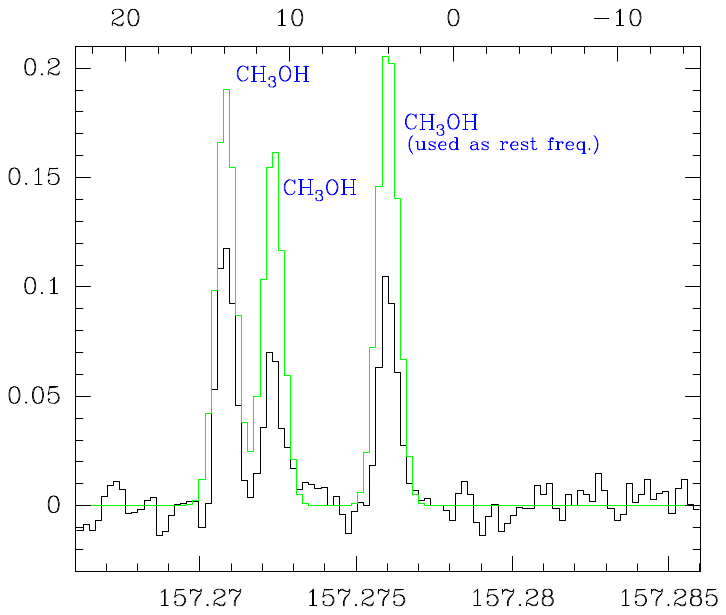}
  \end{minipage} \\
\vspace{-5.5cm}
\hspace{-2cm}
  \begin{minipage}[h]{0.32\textwidth}
    \includegraphics[width=2.5\textwidth]{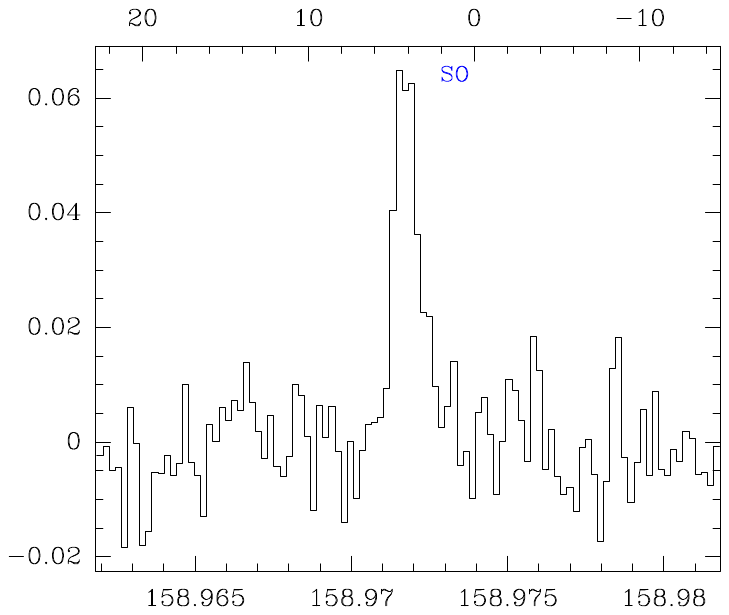}
  \end{minipage}
  \hspace{0.6cm}
  \begin{minipage}[h]{0.32\textwidth}
    \includegraphics[width=2.5\textwidth]{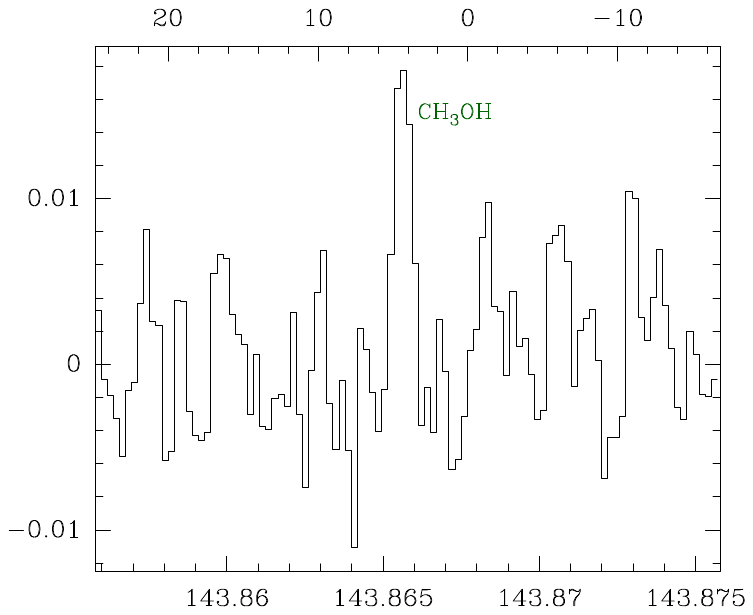}  
  \end{minipage}
 \vspace{-1.5cm} 
\caption{Continued. 
}
\end{figure*}

\newpage
\begin{figure*}[h]
\centering 
\vspace{-4cm}
\hspace{-2cm}
  \begin{minipage}[h]{0.32\textwidth}
    \includegraphics[width=2.5\textwidth]{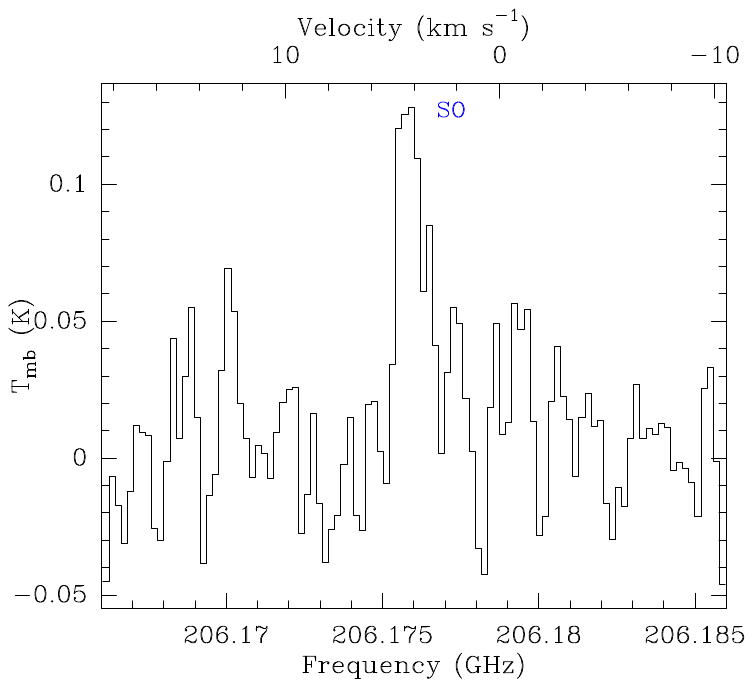}
  \end{minipage}
  \hspace{0.6cm}
  \begin{minipage}[h]{0.32\textwidth}
    \includegraphics[width=2.5\textwidth]{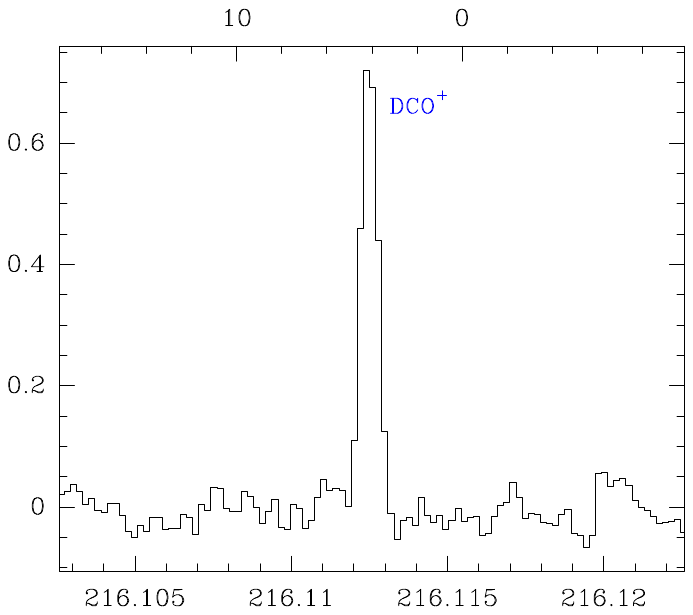}  
  \end{minipage}
  \hspace{0.6cm}
  \begin{minipage}[h]{0.32\textwidth}
    \includegraphics[width=2.5\textwidth]{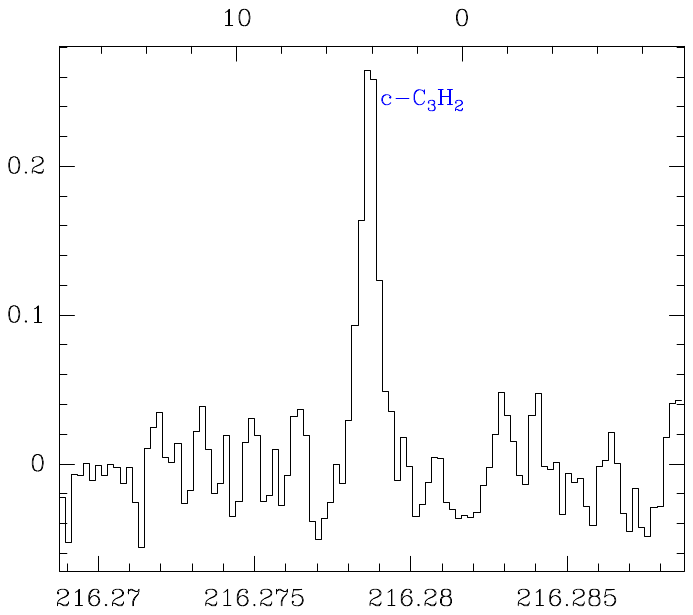}
  \end{minipage} \\
\vspace{-5.5cm}
\hspace{-2cm}
  \begin{minipage}[h]{0.32\textwidth}
    \includegraphics[width=2.5\textwidth]{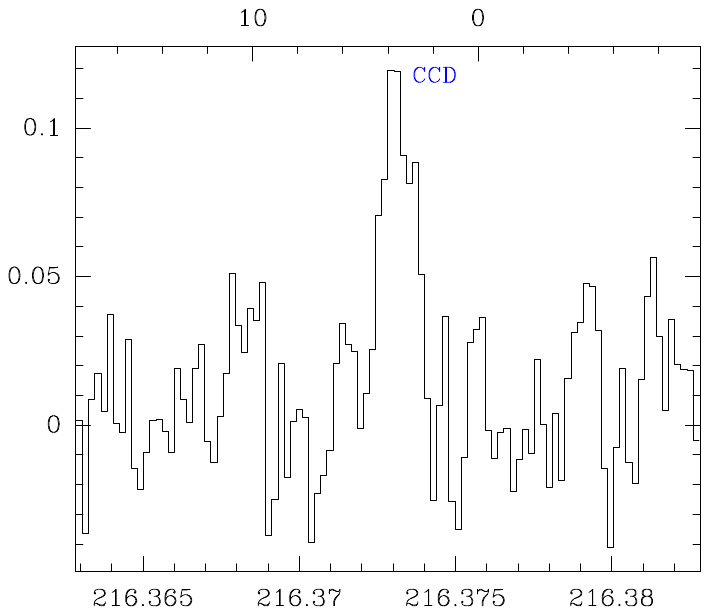}
  \end{minipage}
  \hspace{0.6cm}
  \begin{minipage}[h]{0.32\textwidth}
    \includegraphics[width=2.5\textwidth]{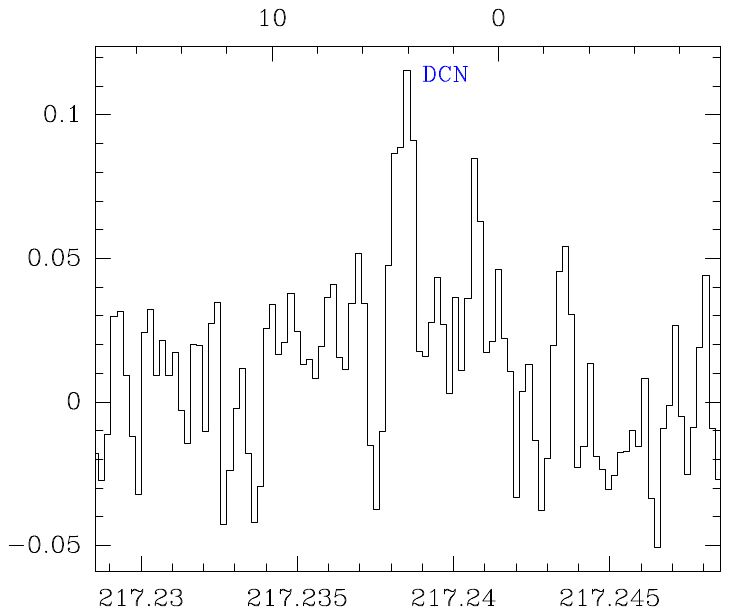}  
  \end{minipage}
  \hspace{0.6cm}
  \begin{minipage}[h]{0.32\textwidth}
    \includegraphics[width=2.5\textwidth]{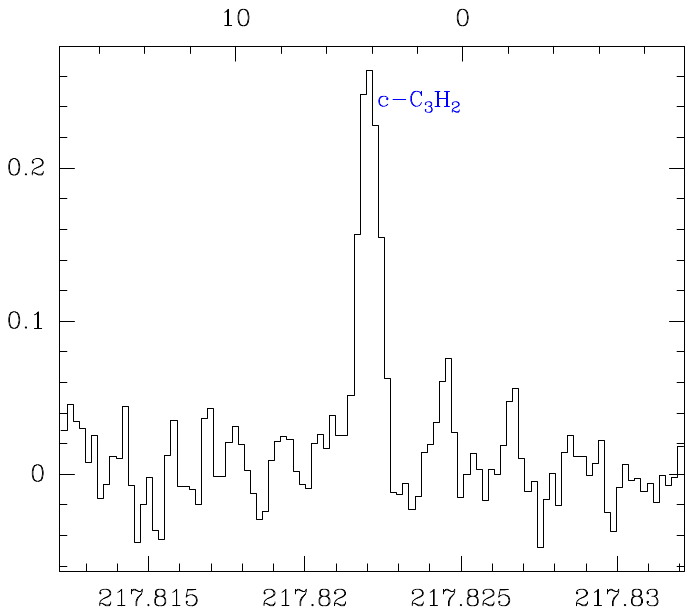}
  \end{minipage} \\
\vspace{-5.5cm}
\hspace{-2cm}
  \begin{minipage}[h]{0.32\textwidth}
    \includegraphics[width=2.5\textwidth]{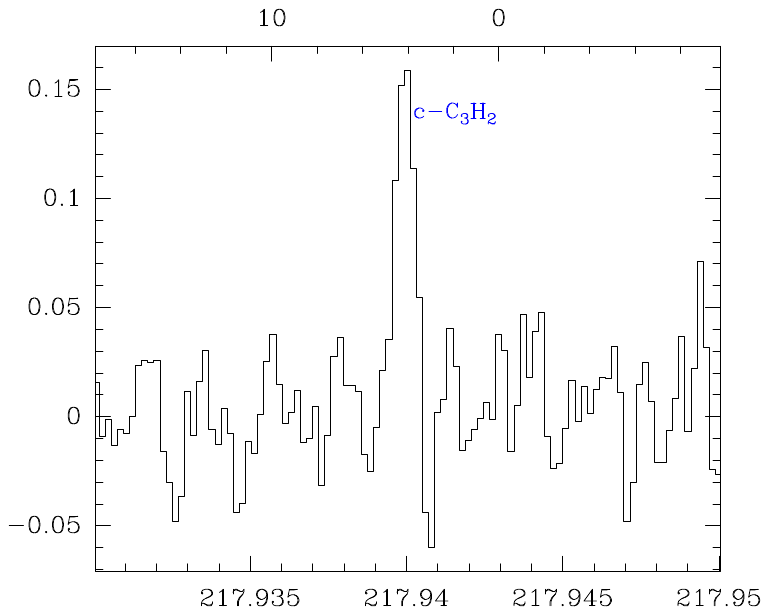}
  \end{minipage}
  \hspace{0.6cm}
  \begin{minipage}[h]{0.32\textwidth}
    \includegraphics[width=2.5\textwidth]{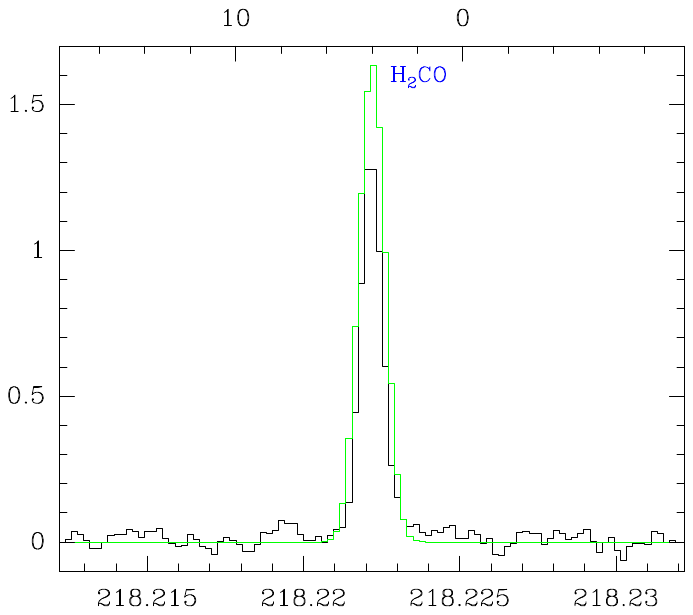}  
  \end{minipage}
  \hspace{0.6cm}
  \begin{minipage}[h]{0.32\textwidth}
    \includegraphics[width=2.5\textwidth]{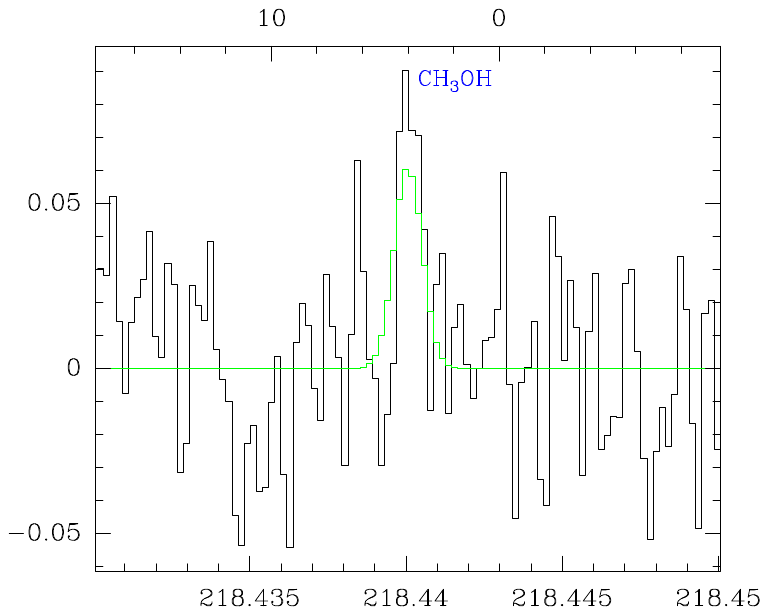}
  \end{minipage} \\
\vspace{-5.5cm}
\hspace{-2cm}
  \begin{minipage}[h]{0.32\textwidth}
    \includegraphics[width=2.5\textwidth]{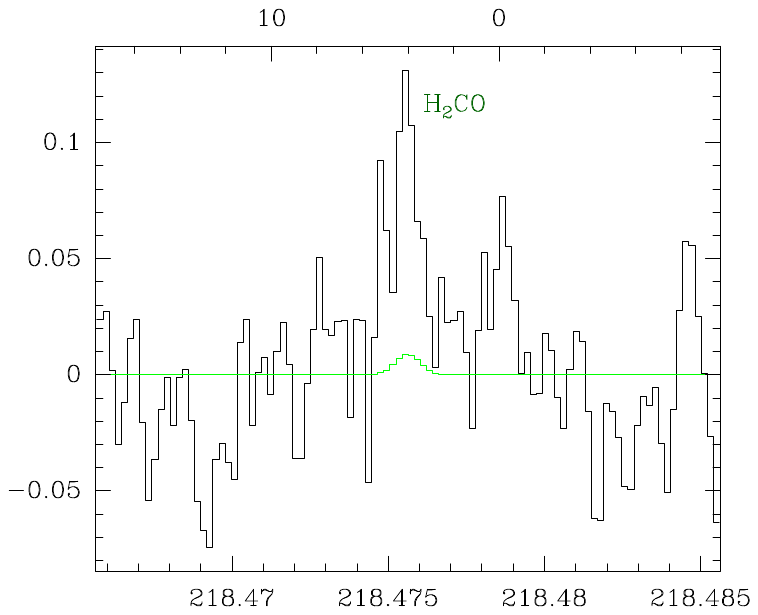}
  \end{minipage}
  \hspace{0.6cm}
  \begin{minipage}[h]{0.32\textwidth}
    \includegraphics[width=2.5\textwidth]{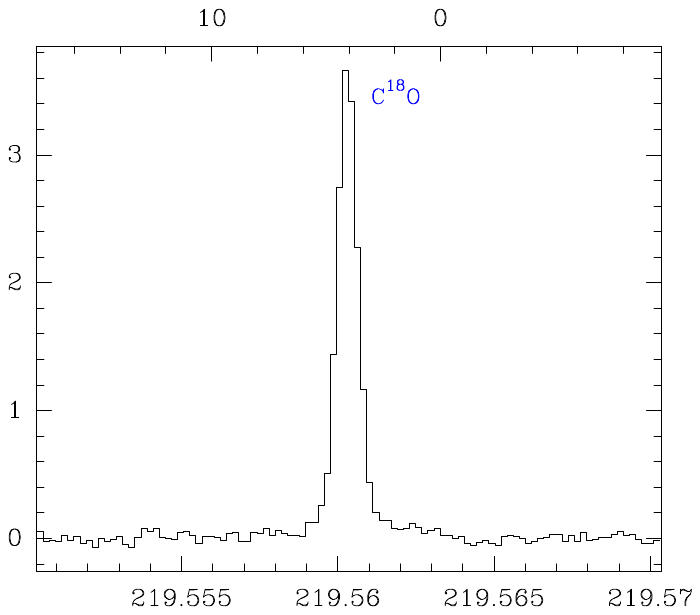}  
  \end{minipage}
  \hspace{0.6cm}
  \begin{minipage}[h]{0.32\textwidth}
    \includegraphics[width=2.5\textwidth]{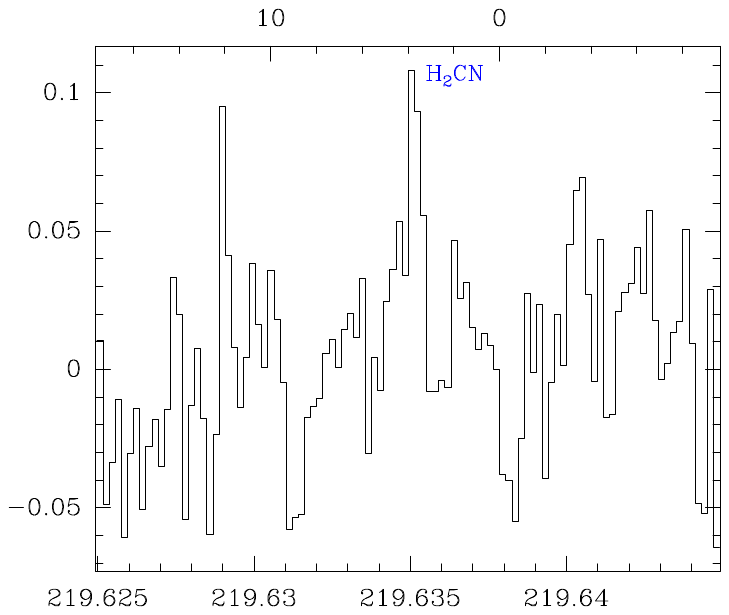}
  \end{minipage} 
 \vspace{-1.85cm} 
\caption{Same as Fig.~\ref{fig:survey-73ghz-small} but for the tuning at 202\,GHz.
}
\label{fig:survey-202ghz-small}
\end{figure*}

\addtocounter{figure}{-1}
\begin{figure*}[h]
\centering 
\vspace{-4cm}
\hspace{-2cm}
  \begin{minipage}[h]{0.32\textwidth}
    \includegraphics[width=2.5\textwidth]{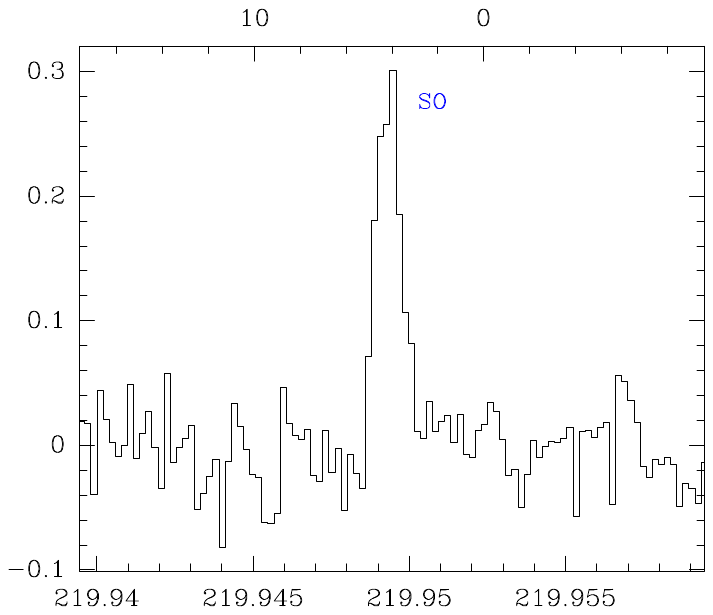}
  \end{minipage}
  \hspace{0.6cm}
  \begin{minipage}[h]{0.32\textwidth}
    \includegraphics[width=2.5\textwidth]{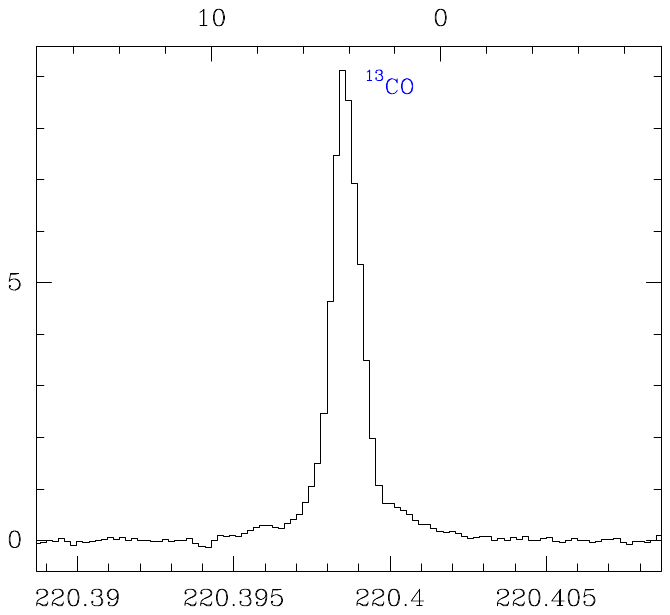}  
  \end{minipage}
  \hspace{0.6cm}
  \begin{minipage}[h]{0.32\textwidth}
    \includegraphics[width=2.5\textwidth]{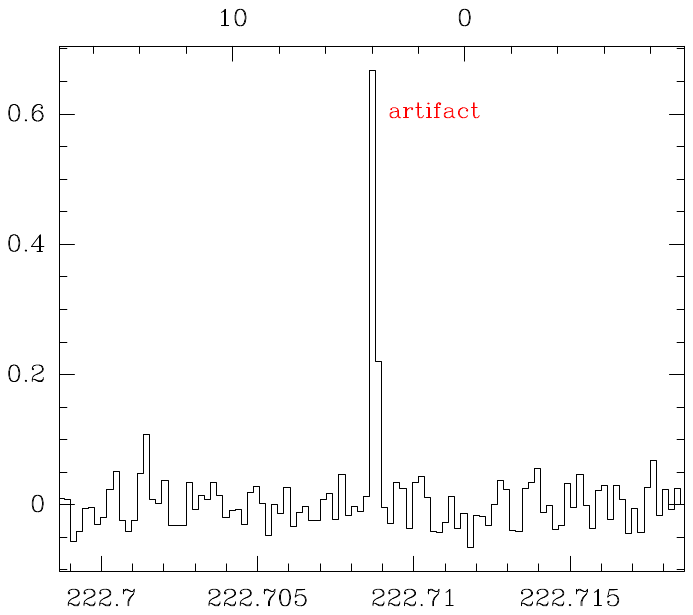}
  \end{minipage} \\
\vspace{-1.5cm} 
\caption{Continued. 
}
\end{figure*}

\newpage
\begin{figure*}[h]
\centering 
\vspace{-4cm}
\hspace{-2cm}
  \begin{minipage}[h]{0.32\textwidth}
    \includegraphics[width=2.5\textwidth]{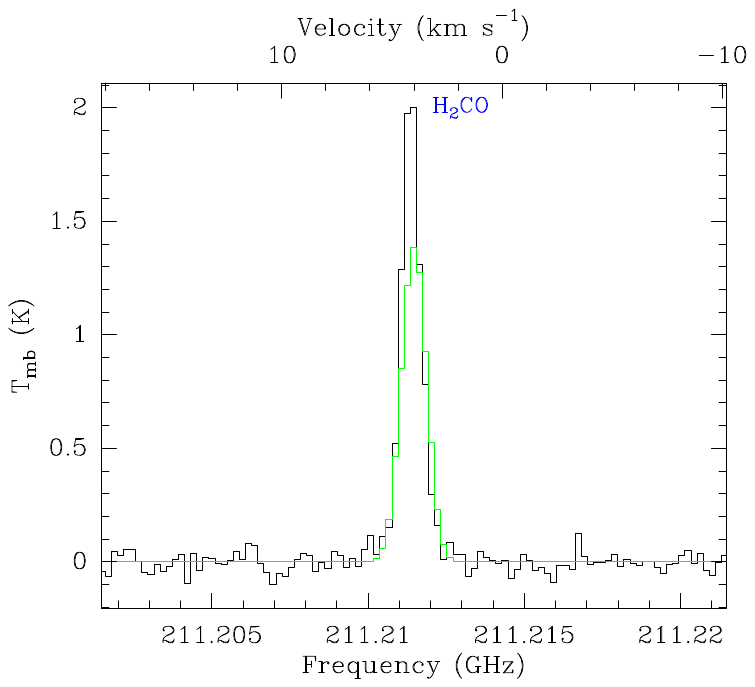}
  \end{minipage}
  \hspace{0.6cm}
  \begin{minipage}[h]{0.32\textwidth}
    \includegraphics[width=2.5\textwidth]{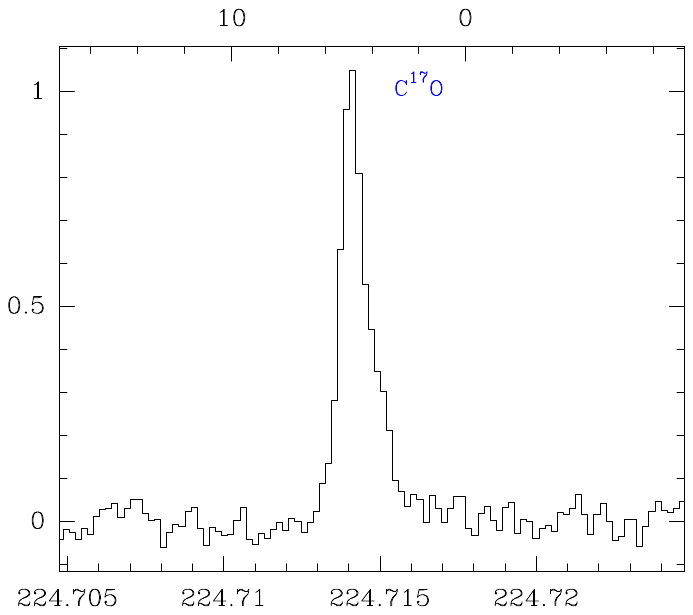}  
  \end{minipage}
  \hspace{0.6cm}
  \begin{minipage}[h]{0.32\textwidth}
    \includegraphics[width=2.5\textwidth]{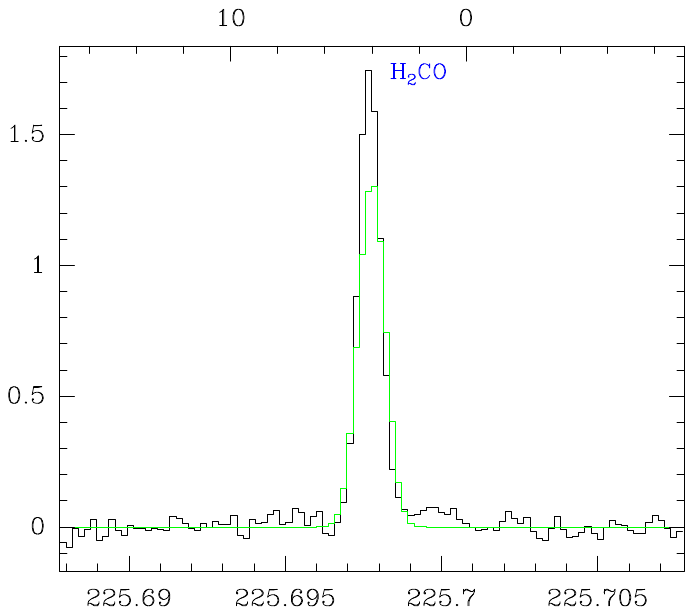}
  \end{minipage} \\
\vspace{-5.5cm}
\hspace{-2cm}
  \begin{minipage}[h]{0.32\textwidth}
    \includegraphics[width=2.5\textwidth]{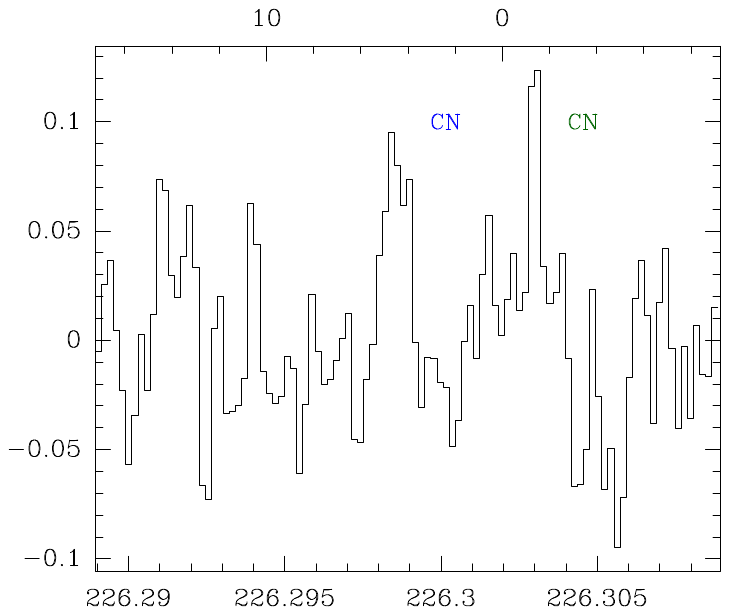}
  \end{minipage}
  \hspace{0.6cm}
  \begin{minipage}[h]{0.32\textwidth}
    \includegraphics[width=2.5\textwidth]{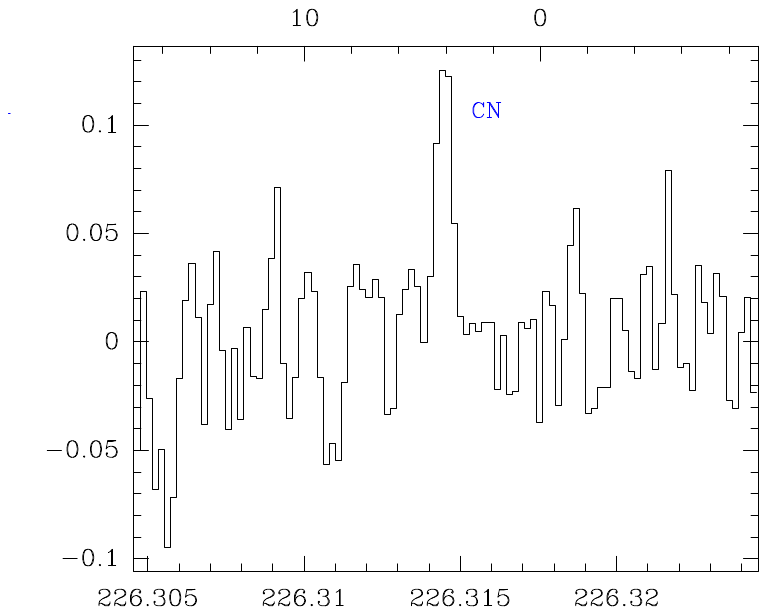}  
  \end{minipage}
  \hspace{0.6cm}
  \begin{minipage}[h]{0.32\textwidth}
    \includegraphics[width=2.5\textwidth]{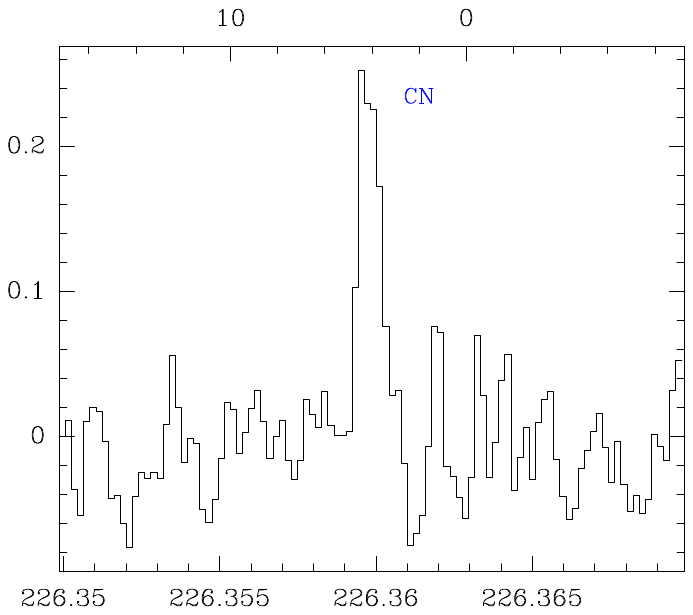}
  \end{minipage} \\
\vspace{-5.5cm}
\hspace{-2cm}
  \begin{minipage}[h]{0.32\textwidth}
    \includegraphics[width=2.5\textwidth]{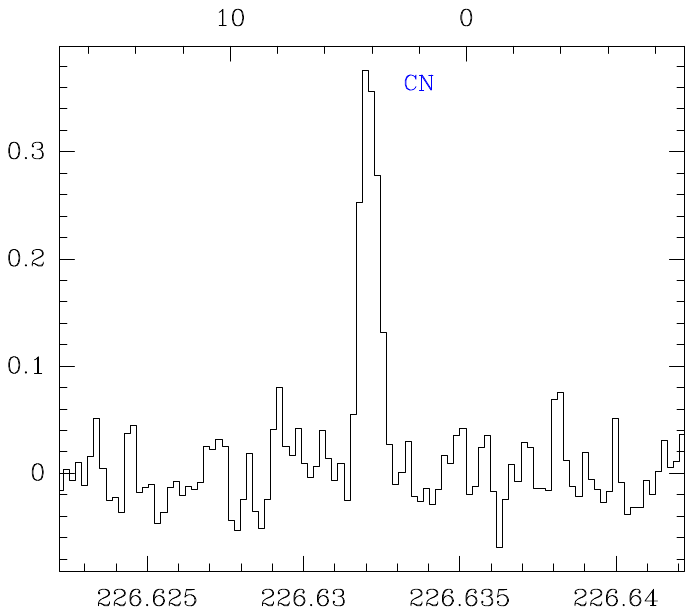}
  \end{minipage}
  \hspace{0.6cm}
  \begin{minipage}[h]{0.32\textwidth}
    \includegraphics[width=2.5\textwidth]{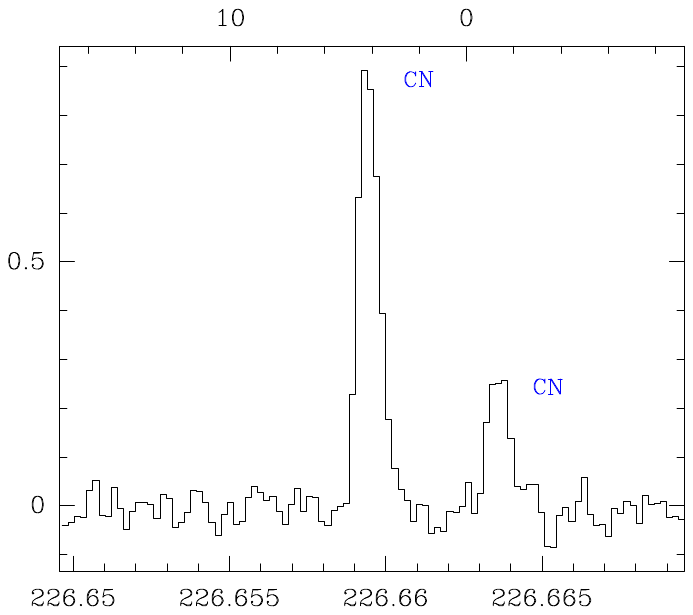}  
  \end{minipage}
  \hspace{0.6cm}
  \begin{minipage}[h]{0.32\textwidth}
    \includegraphics[width=2.5\textwidth]{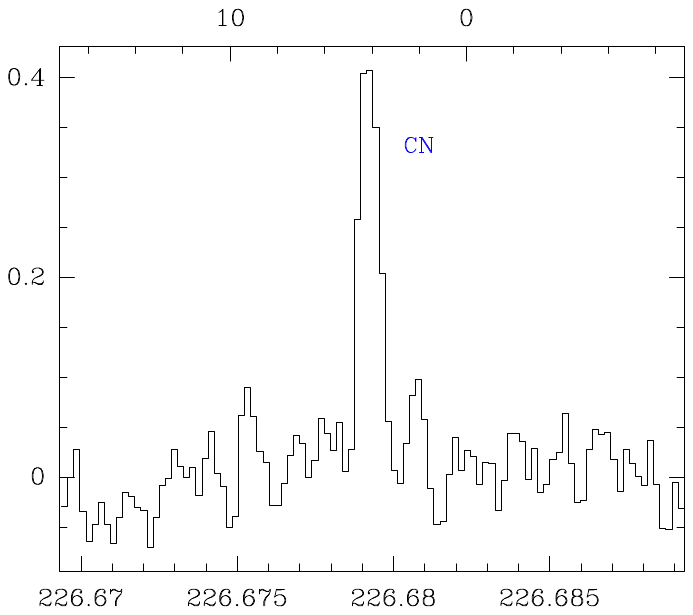}
  \end{minipage} \\
\vspace{-5.5cm}
\hspace{-2cm}
  \begin{minipage}[h]{0.32\textwidth}
    \includegraphics[width=2.5\textwidth]{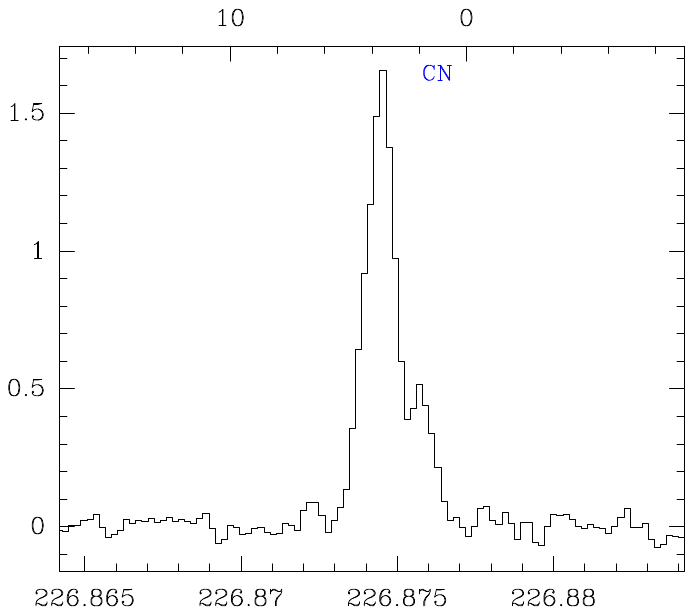}
  \end{minipage}
  \hspace{0.6cm}
  \begin{minipage}[h]{0.32\textwidth}
    \includegraphics[width=2.5\textwidth]{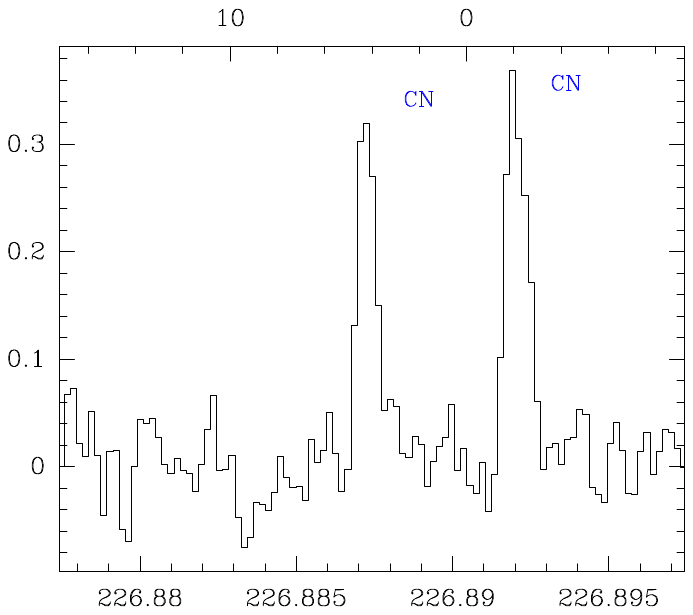}  
  \end{minipage}
  \hspace{0.6cm}
  \begin{minipage}[h]{0.32\textwidth}
    \includegraphics[width=2.5\textwidth]{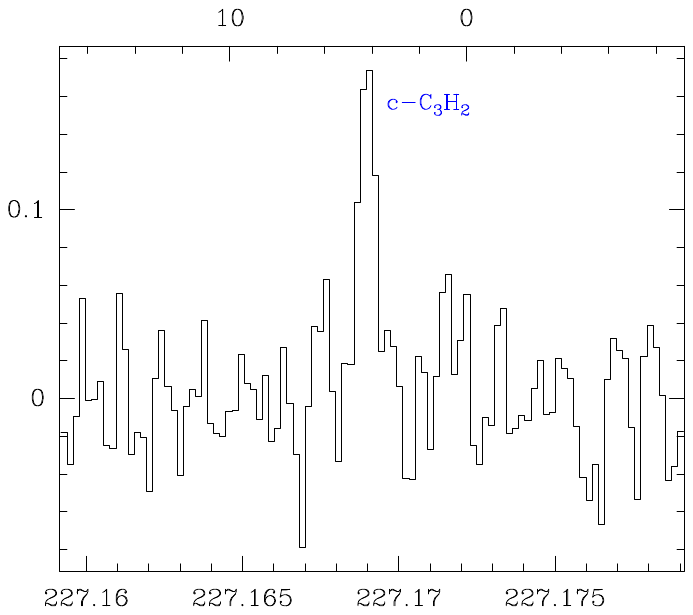}
  \end{minipage} 
 \vspace{-1.85cm} 
\caption{Same as Fig.~\ref{fig:survey-73ghz-small} but for the tuning at 210\,GHz. Possible contamination from the image band (indicating rest frequency and peak temperature in K in the image band) are labelled in red.
}
\label{fig:survey-210ghz-small}
\end{figure*}

\addtocounter{figure}{-1}
\begin{figure*}[h]
\centering 
\vspace{-4cm}
\hspace{-2cm}
  \begin{minipage}[h]{0.32\textwidth}
    \includegraphics[width=2.5\textwidth]{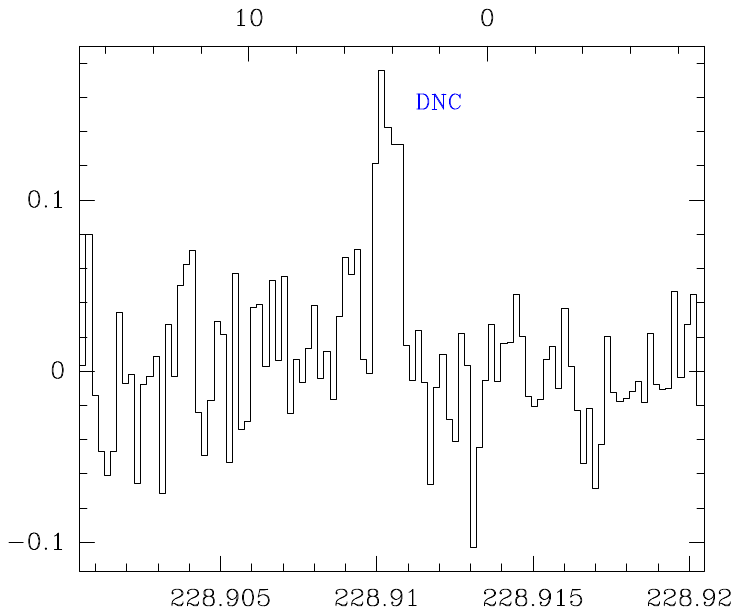}
  \end{minipage}
  \hspace{0.6cm}
  \begin{minipage}[h]{0.32\textwidth}
    \includegraphics[width=2.5\textwidth]{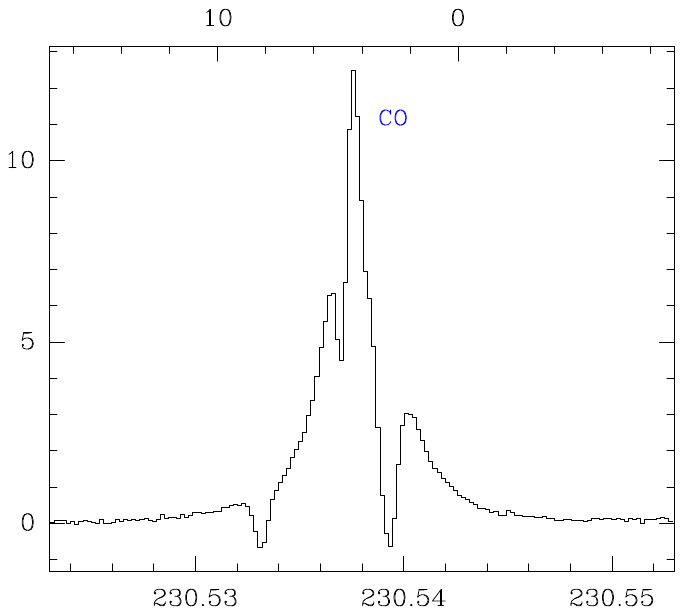}  
  \end{minipage}
  \hspace{0.6cm}
  \begin{minipage}[h]{0.32\textwidth}
    \includegraphics[width=2.5\textwidth]{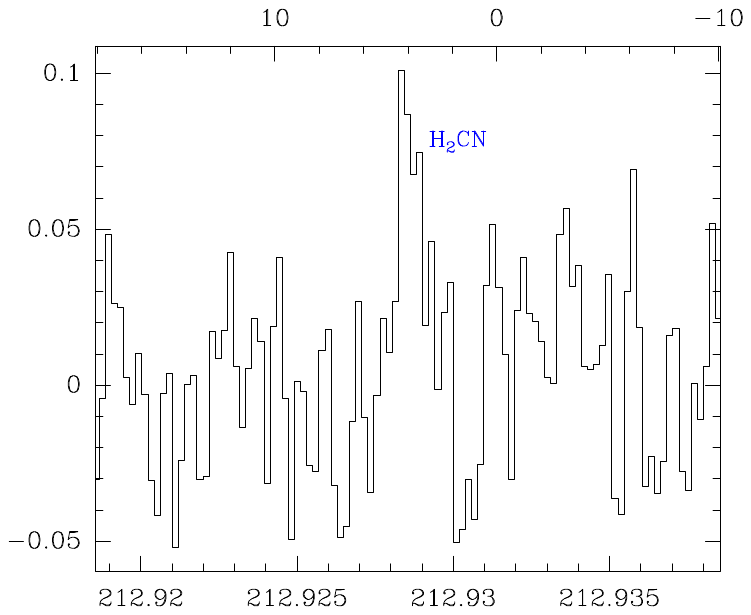}
  \end{minipage} \\
\vspace{-5.5cm}
\hspace{-2cm}
  \begin{minipage}[h]{0.32\textwidth}
    \includegraphics[width=2.5\textwidth]{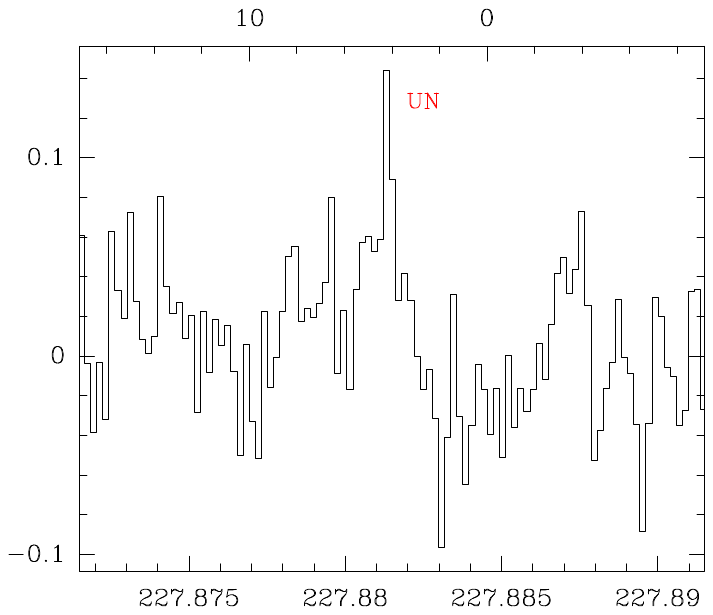}
  \end{minipage}
  \hspace{0.6cm}
  \begin{minipage}[h]{0.32\textwidth}
    \includegraphics[width=2.5\textwidth]{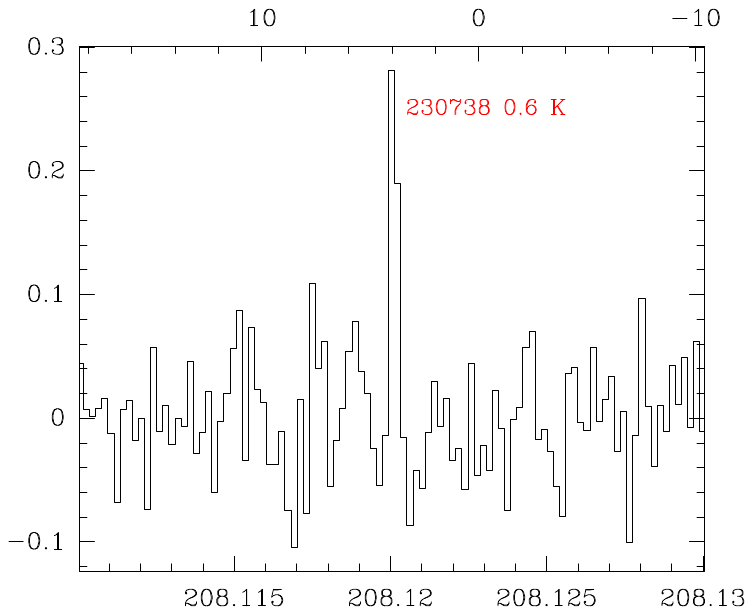}  
  \end{minipage}
  \hspace{0.6cm}
  \begin{minipage}[h]{0.32\textwidth}
    \includegraphics[width=2.5\textwidth]{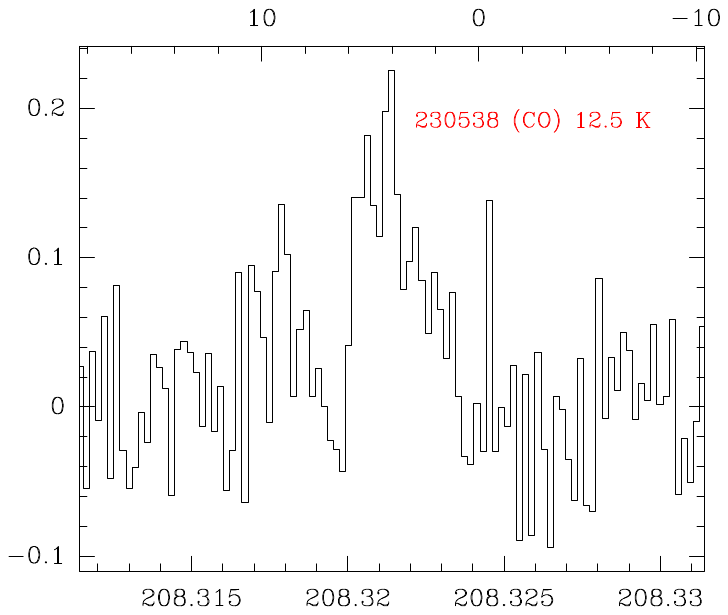}
  \end{minipage} \\
 \vspace{-1.5cm} 
\caption{Continued.}
\end{figure*}

\newpage
\begin{figure*}[h]
\centering 
\vspace{-4cm}
\hspace{-2cm}
  \begin{minipage}[h]{0.32\textwidth}
    \includegraphics[width=2.5\textwidth]{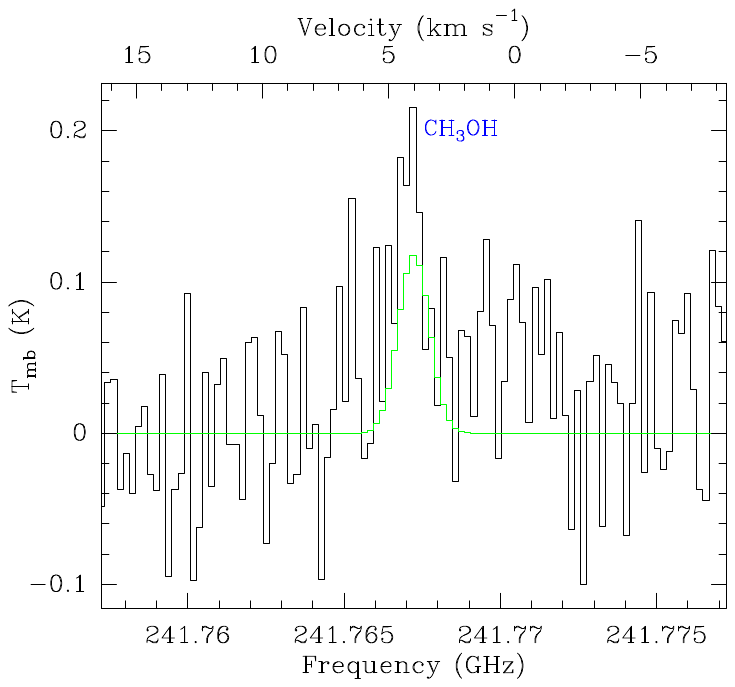}
  \end{minipage}
  \hspace{0.6cm}
  \begin{minipage}[h]{0.32\textwidth}
    \includegraphics[width=2.5\textwidth]{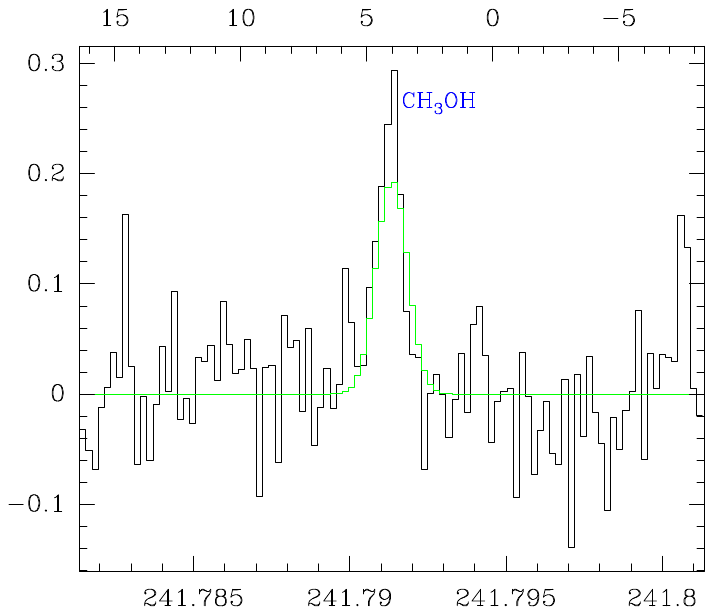}  
  \end{minipage}
  \hspace{0.6cm}
  \begin{minipage}[h]{0.32\textwidth}
    \includegraphics[width=2.5\textwidth]{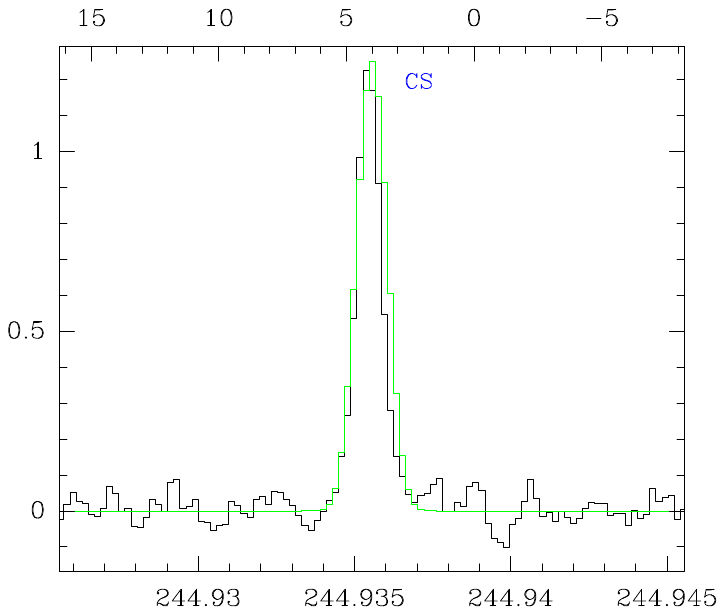}
  \end{minipage} \\
\vspace{-5.5cm}
\hspace{-2cm}
  \begin{minipage}[h]{0.32\textwidth}
    \includegraphics[width=2.5\textwidth]{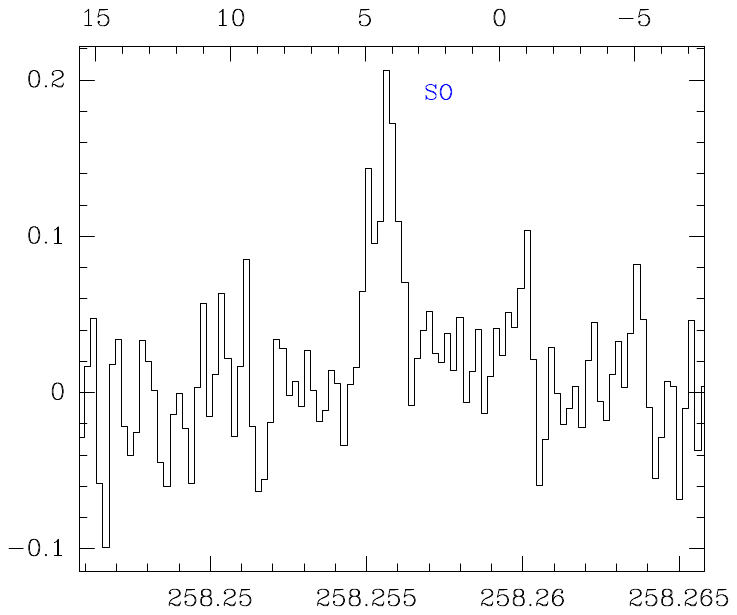}
  \end{minipage}
  \hspace{0.6cm}
  \begin{minipage}[h]{0.32\textwidth}
    \includegraphics[width=2.5\textwidth]{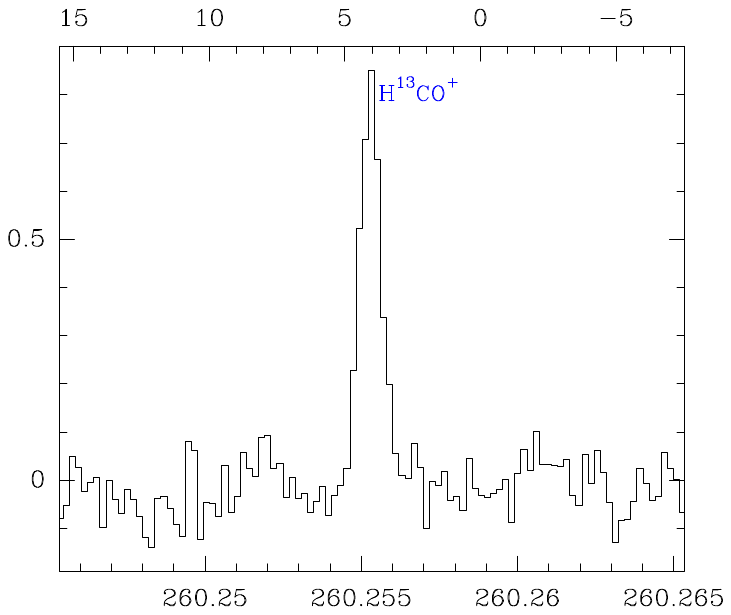}  
  \end{minipage}
  \hspace{0.6cm}
  \begin{minipage}[h]{0.32\textwidth}
    \includegraphics[width=2.5\textwidth]{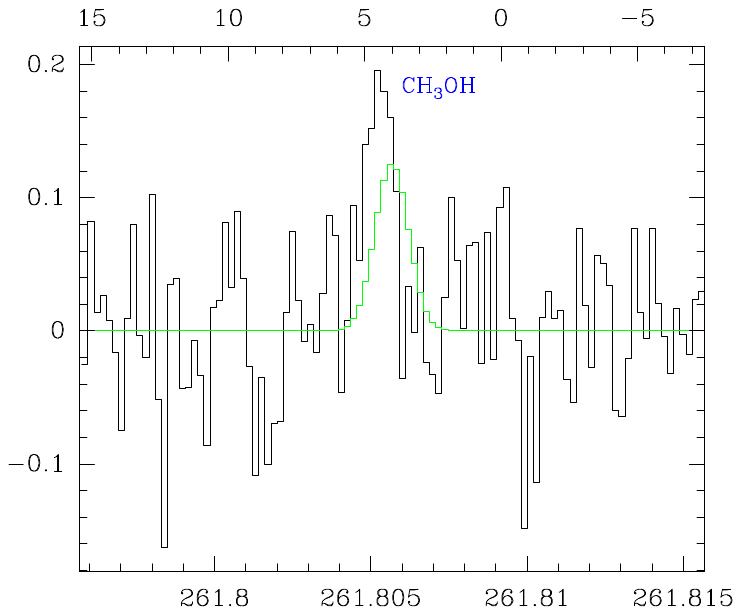}
  \end{minipage} \\
\vspace{-5.5cm}
\hspace{-2cm}
  \begin{minipage}[h]{0.32\textwidth}
    \includegraphics[width=2.5\textwidth]{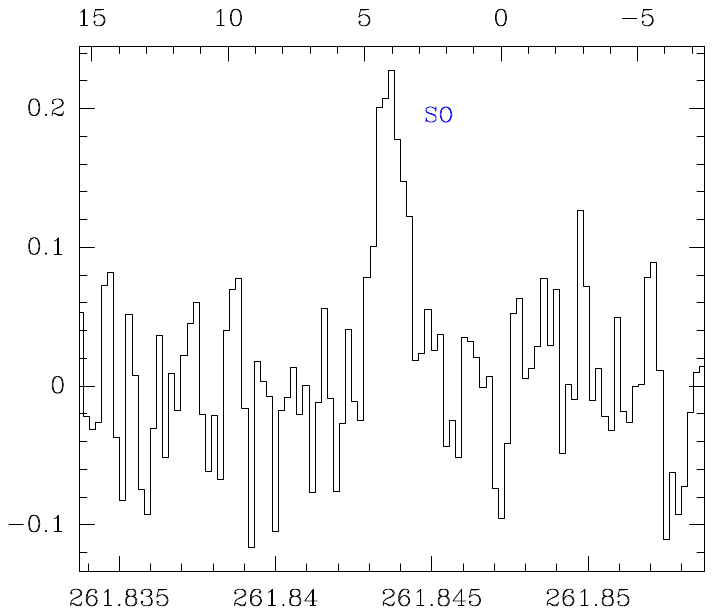}
  \end{minipage}
  \hspace{0.6cm}
  \begin{minipage}[h]{0.32\textwidth}
    \includegraphics[width=2.5\textwidth]{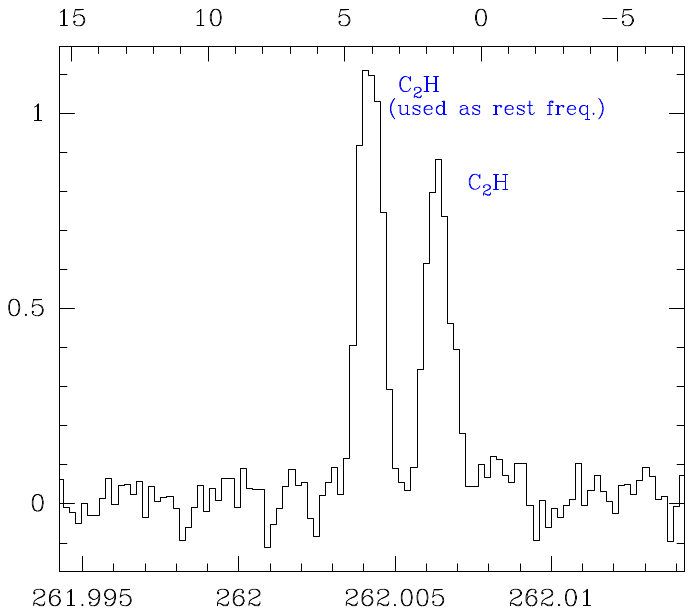}  
  \end{minipage}
  \hspace{0.6cm}
  \begin{minipage}[h]{0.32\textwidth}
    \includegraphics[width=2.5\textwidth]{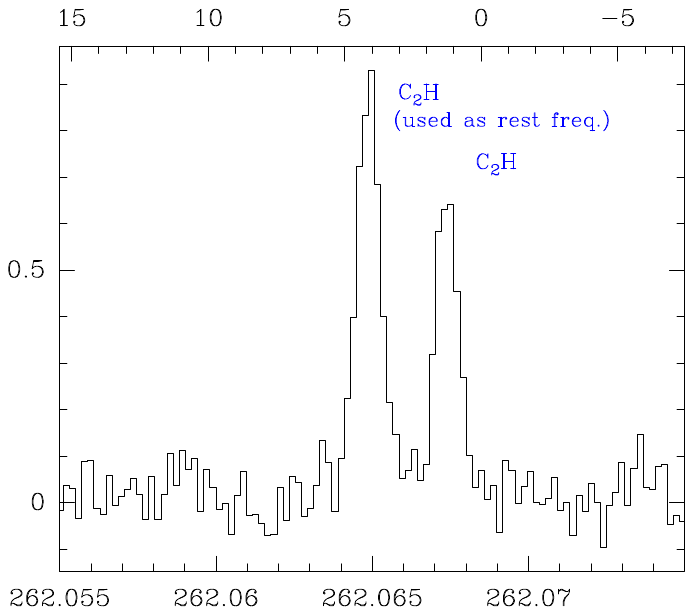}
  \end{minipage} \\
\vspace{-5.5cm}
\hspace{-2cm}
  \begin{minipage}[h]{0.32\textwidth}
    \includegraphics[width=2.5\textwidth]{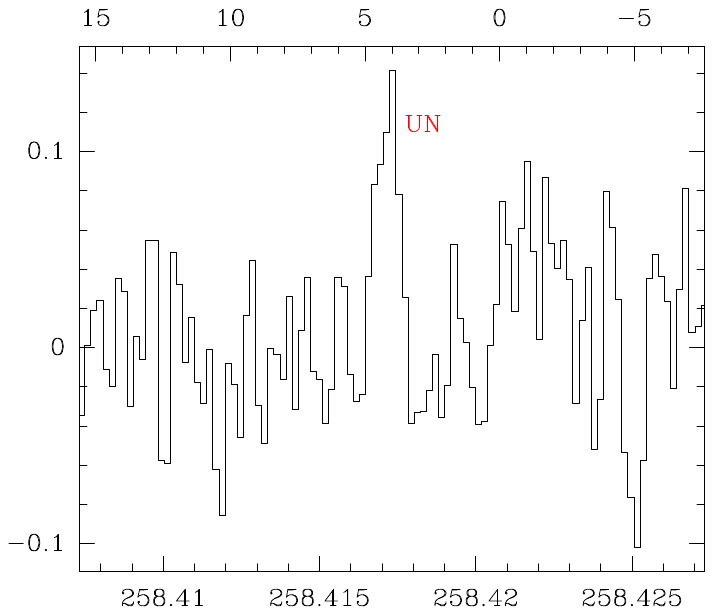}
  \end{minipage}
  \hspace{0.6cm}
  \begin{minipage}[h]{0.32\textwidth}
    \includegraphics[width=2.5\textwidth]{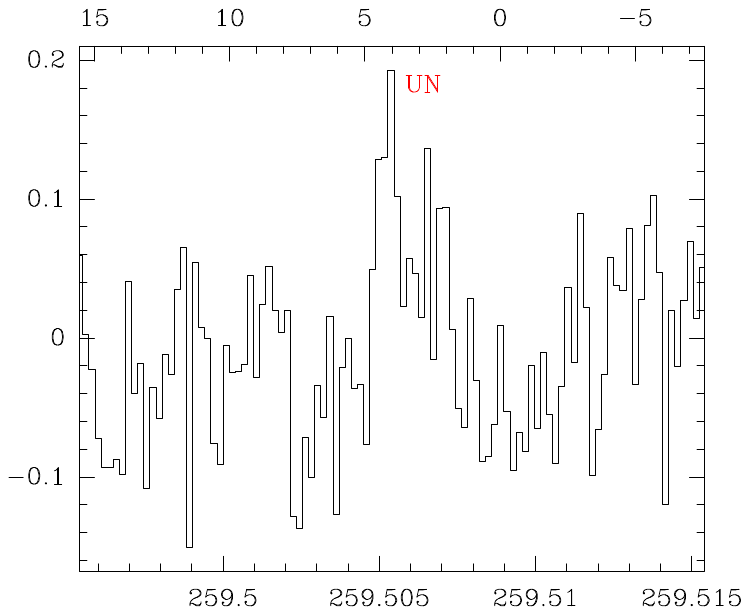}  
  \end{minipage}
 \vspace{-1.5cm} 
\caption{Same as Fig.~\ref{fig:survey-73ghz-small} but for the tuning at 242\,GHz.
}
\label{fig:survey-242ghz-small}
\end{figure*}

\newpage
\begin{figure*}[h]
\centering 
\vspace{0.5cm}
\hspace{-2cm}
  \begin{minipage}[h]{0.32\textwidth}
    \includegraphics[width=2.5\textwidth]{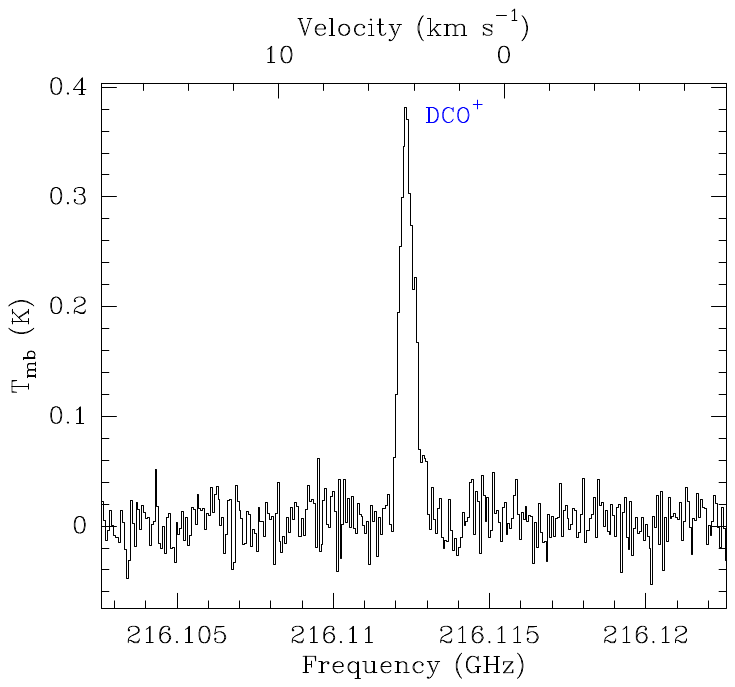}
  \end{minipage}
  \hspace{0.6cm}
  \begin{minipage}[h]{0.32\textwidth}
    \includegraphics[width=2.5\textwidth]{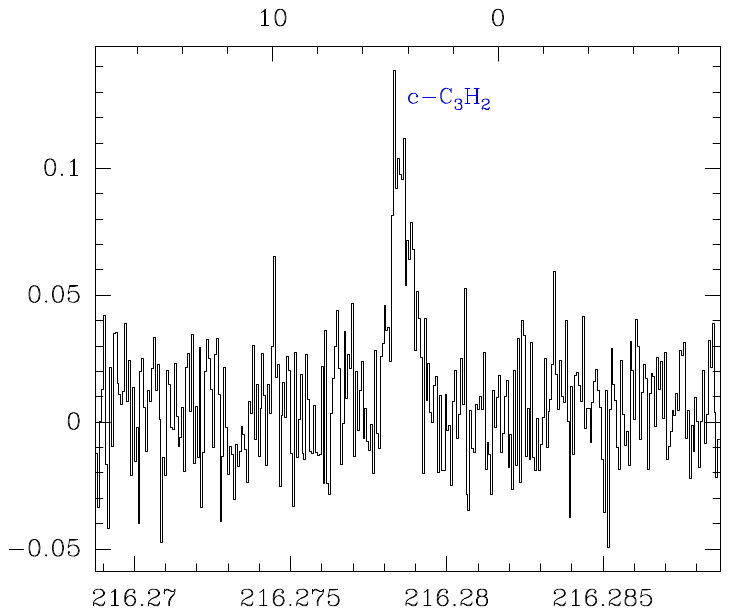}  
  \end{minipage}
  \hspace{0.6cm}
  \begin{minipage}[h]{0.32\textwidth}
    \includegraphics[width=2.5\textwidth]{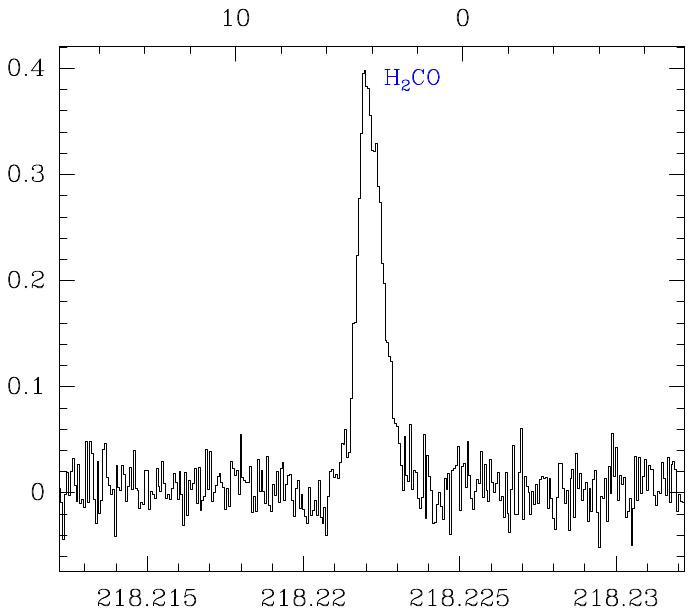}
  \end{minipage} \\
\vspace{-5.5cm}
\hspace{-2cm}
  \begin{minipage}[h]{0.32\textwidth}
    \includegraphics[width=2.5\textwidth]{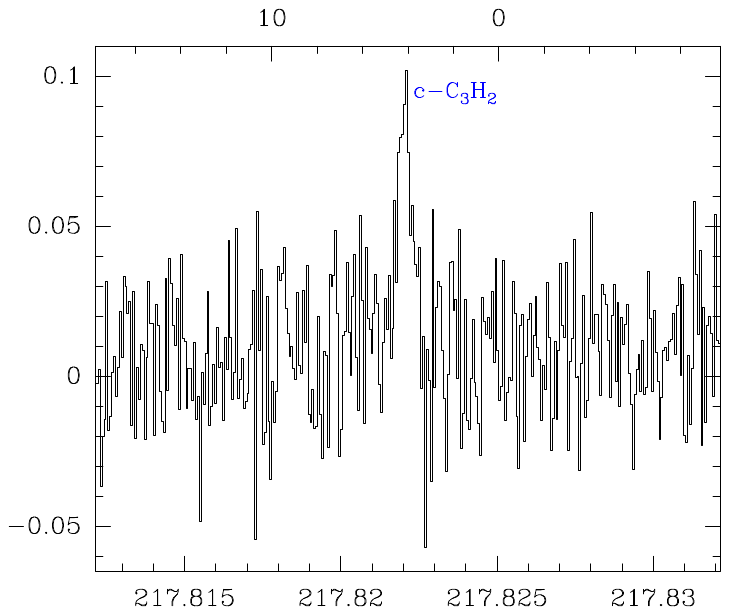}
  \end{minipage}
  \hspace{0.6cm}
  \begin{minipage}[h]{0.32\textwidth}
    \includegraphics[width=2.5\textwidth]{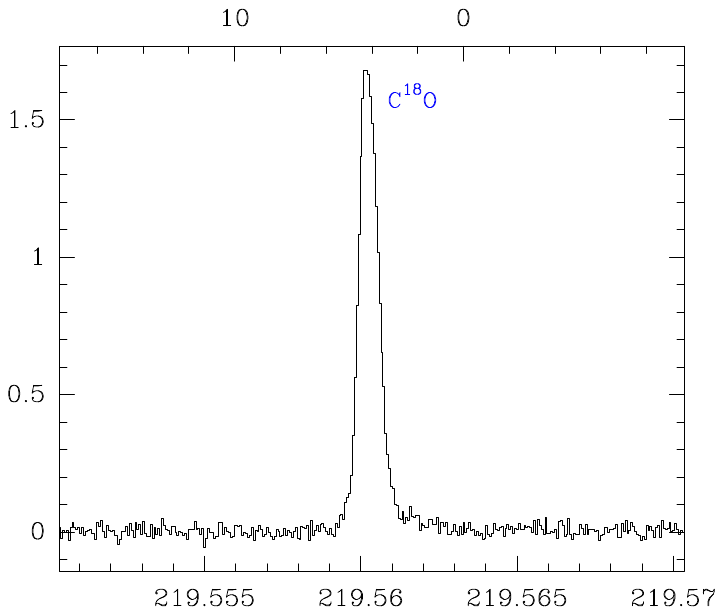}  
  \end{minipage}
  \hspace{0.6cm}
  \begin{minipage}[h]{0.32\textwidth}
    \includegraphics[width=2.5\textwidth]{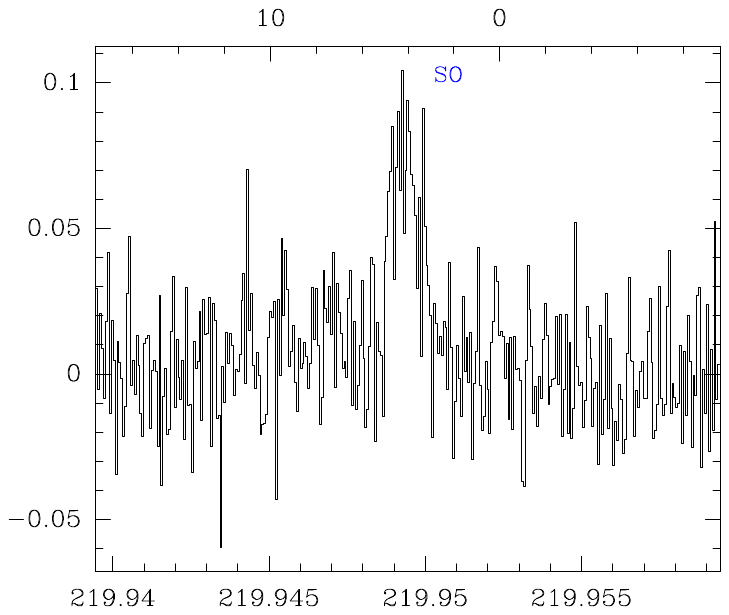}
  \end{minipage} \\
\vspace{-5.5cm}
\hspace{-2cm}
  \begin{minipage}[h]{0.32\textwidth}
    \includegraphics[width=2.5\textwidth]{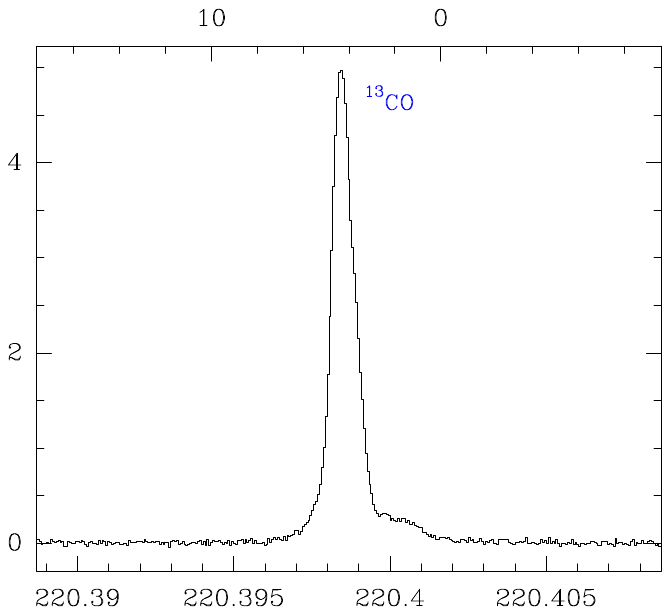}
  \end{minipage}
  \hspace{0.6cm}
  \begin{minipage}[h]{0.32\textwidth}
    \includegraphics[width=2.5\textwidth]{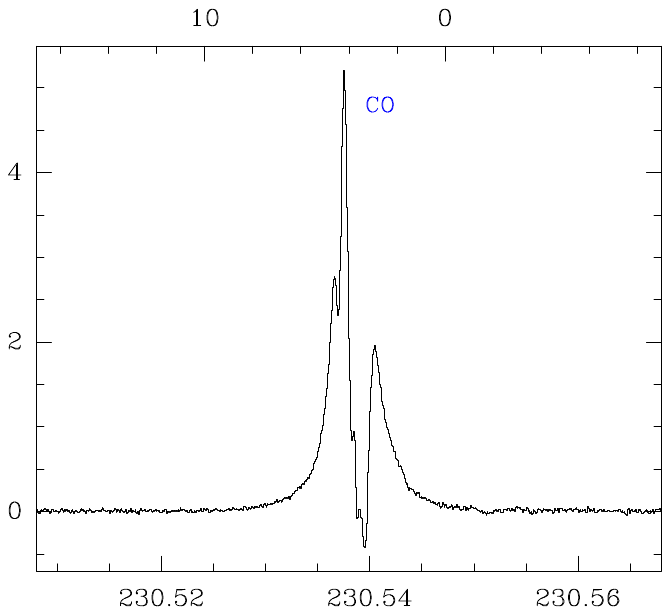}  
  \end{minipage}
  \hspace{0.6cm}
  \begin{minipage}[h]{0.32\textwidth}
    \includegraphics[width=2.5\textwidth]{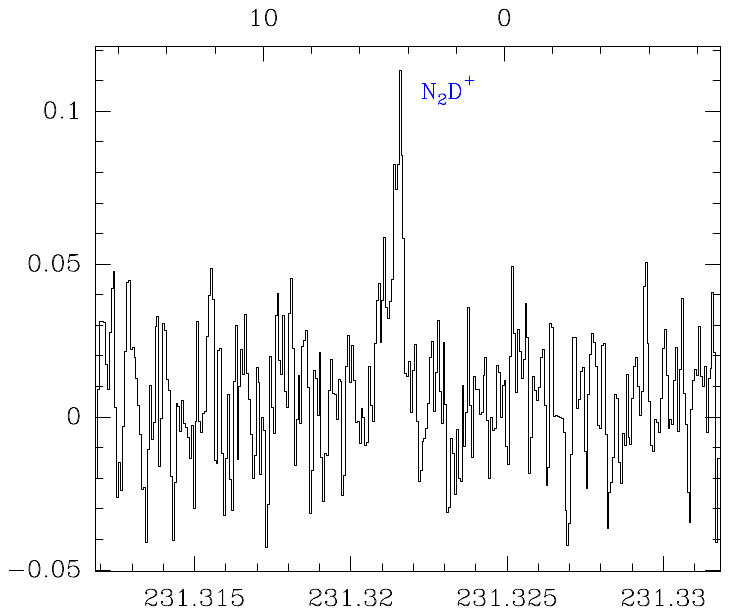}
  \end{minipage} \\
 \vspace{-1.5cm} 
\caption{Spectra of lines detected towards V1057~Cyg with the APEX~12-m telescope with nFLASH, displayed in main-beam temperature scale. Molecule names for identified lines are labelled in blue.}
\label{fig:survey-219.2ghz-small-apex}
\end{figure*}

\newpage
\begin{figure*}[h]
\centering 
\vspace{-4cm}
\hspace{-2cm}
  \begin{minipage}[h]{0.32\textwidth}
    \includegraphics[width=2.5\textwidth]{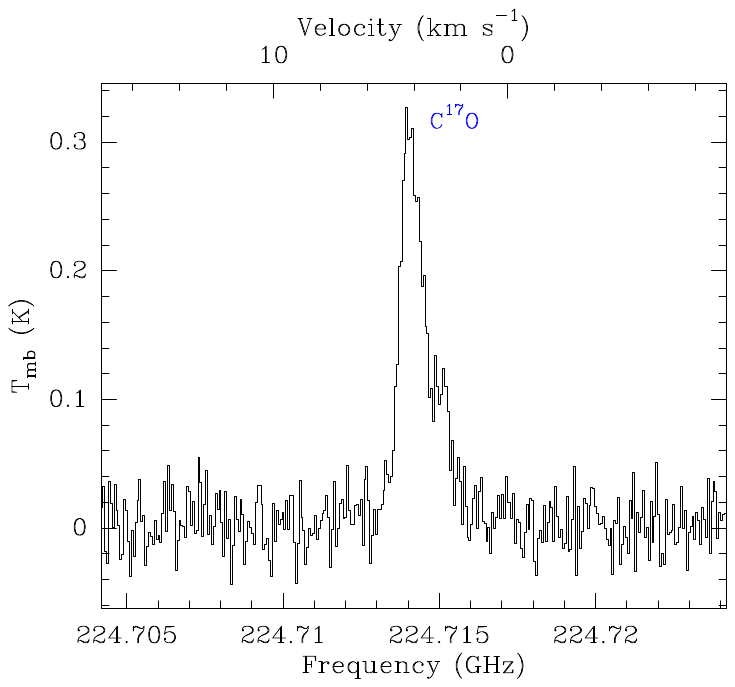}
  \end{minipage}
  \hspace{0.6cm}
  \begin{minipage}[h]{0.32\textwidth}
    \includegraphics[width=2.5\textwidth]{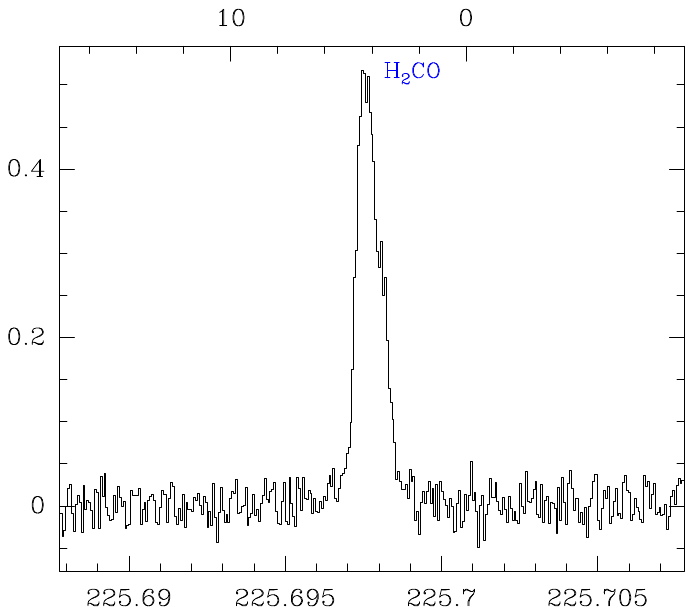}  
  \end{minipage}
  \hspace{0.6cm}
  \begin{minipage}[h]{0.32\textwidth}
    \includegraphics[width=2.5\textwidth]{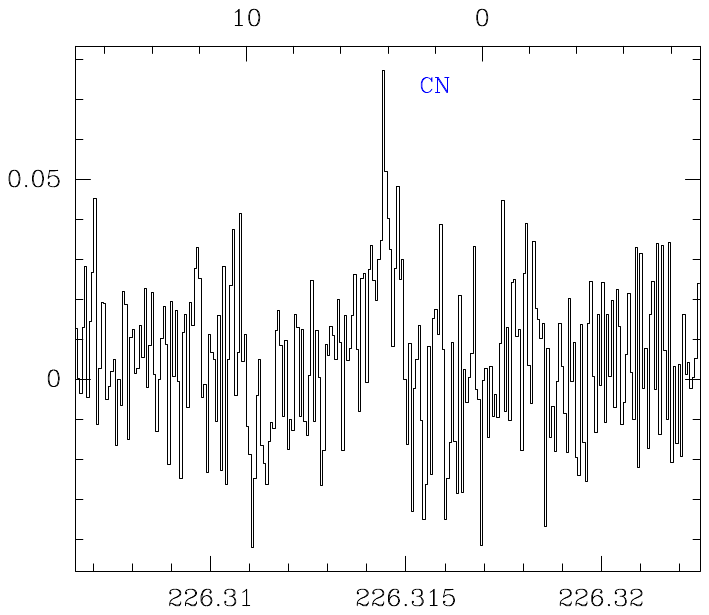}
  \end{minipage} \\
\vspace{-5.5cm}
\hspace{-2cm}
  \begin{minipage}[h]{0.32\textwidth}
    \includegraphics[width=2.5\textwidth]{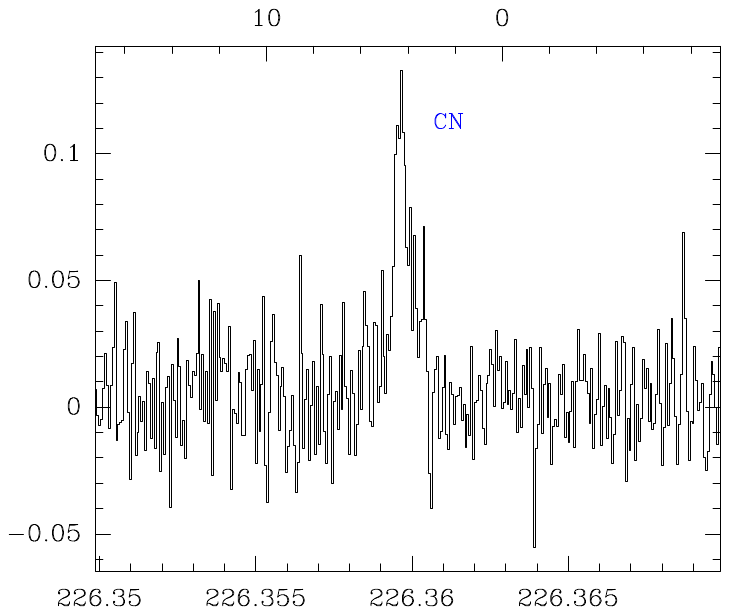}
  \end{minipage}
  \hspace{0.6cm}
  \begin{minipage}[h]{0.32\textwidth}
    \includegraphics[width=2.5\textwidth]{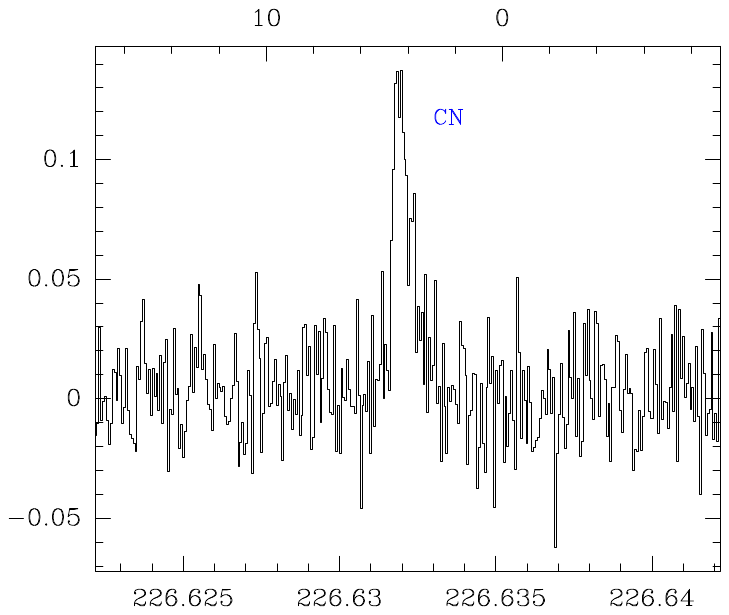}  
  \end{minipage}
  \hspace{0.6cm}
  \begin{minipage}[h]{0.32\textwidth}
    \includegraphics[width=2.5\textwidth]{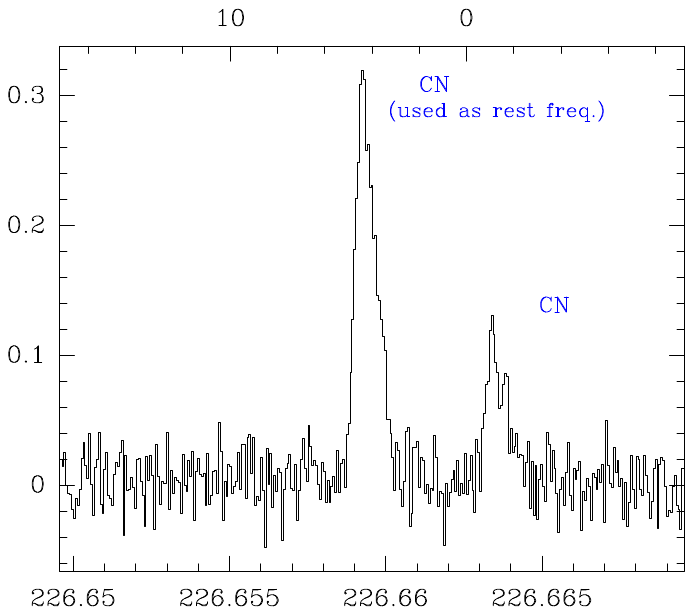}
  \end{minipage} \\
\vspace{-5.5cm}
\hspace{-2cm}
  \begin{minipage}[h]{0.32\textwidth}
    \includegraphics[width=2.5\textwidth]{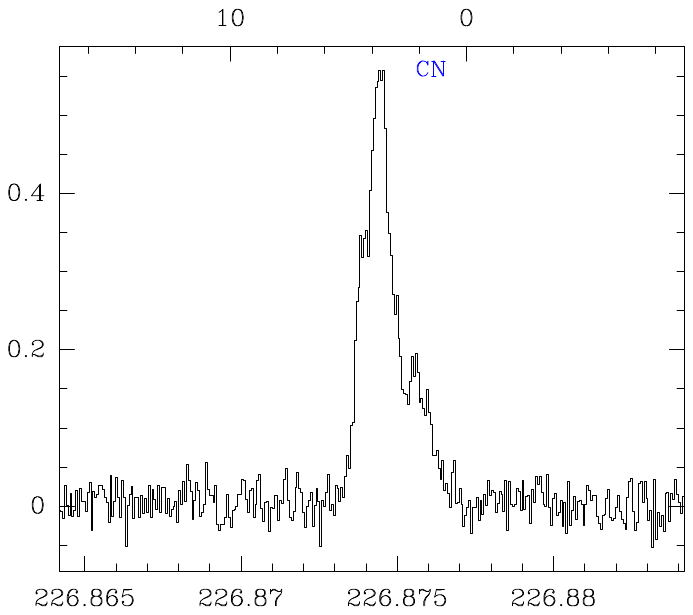}
  \end{minipage}
  \hspace{0.6cm}
  \begin{minipage}[h]{0.32\textwidth}
    \includegraphics[width=2.5\textwidth]{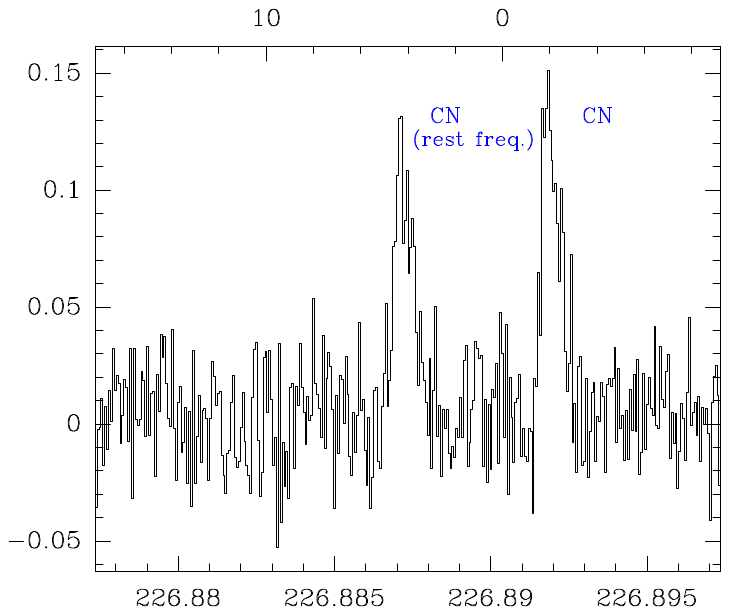}  
  \end{minipage}
  \hspace{0.6cm}
  \begin{minipage}[h]{0.32\textwidth}
    \includegraphics[width=2.5\textwidth]{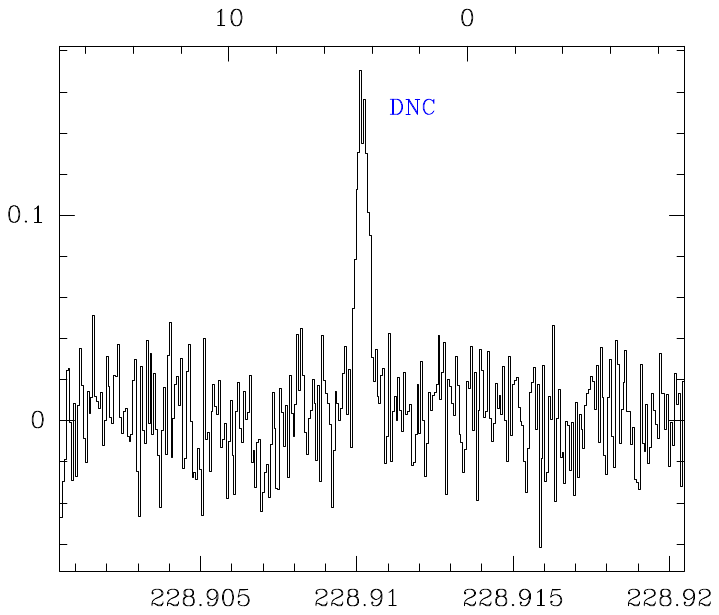}
  \end{minipage} \\
\vspace{-5.5cm}
\hspace{-2cm}
  \begin{minipage}[h]{0.32\textwidth}
    \includegraphics[width=2.5\textwidth]{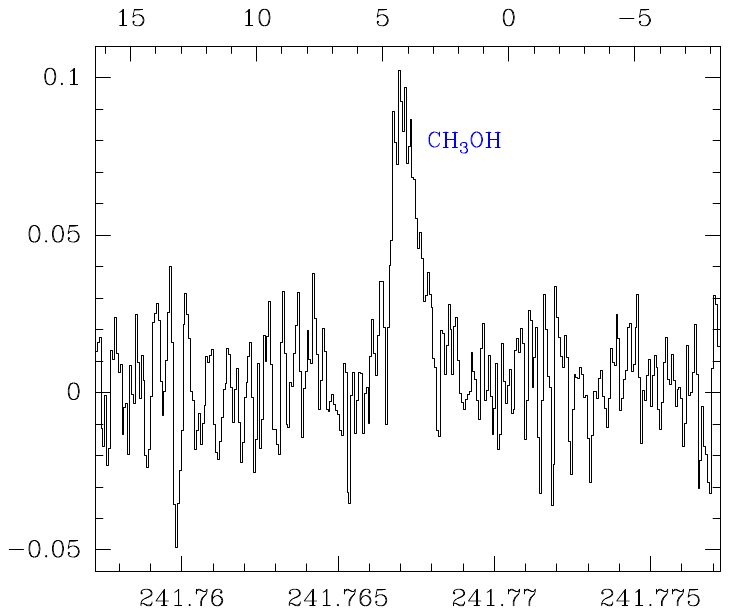}
  \end{minipage}
  \hspace{0.6cm}
  \begin{minipage}[h]{0.32\textwidth}
\includegraphics[width=2.5\textwidth]{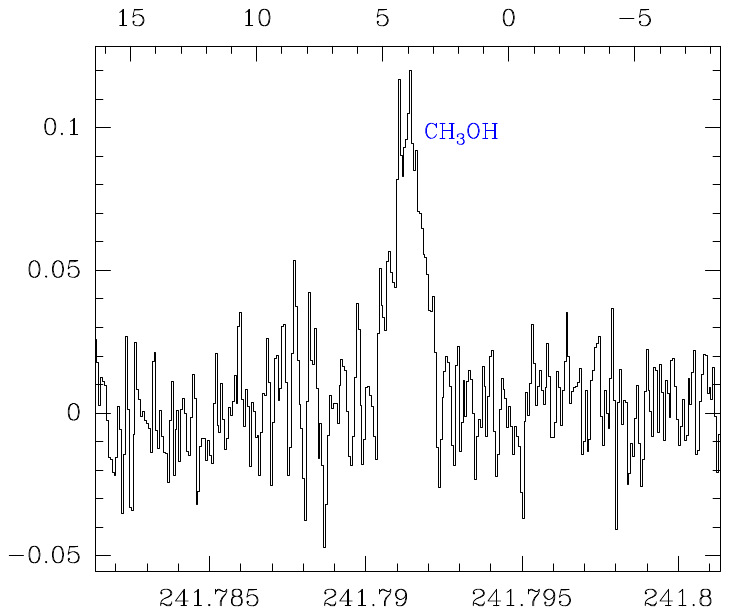}  
  \end{minipage}
  \hspace{0.6cm}
  \begin{minipage}[h]{0.32\textwidth}
    \includegraphics[width=2.5\textwidth]{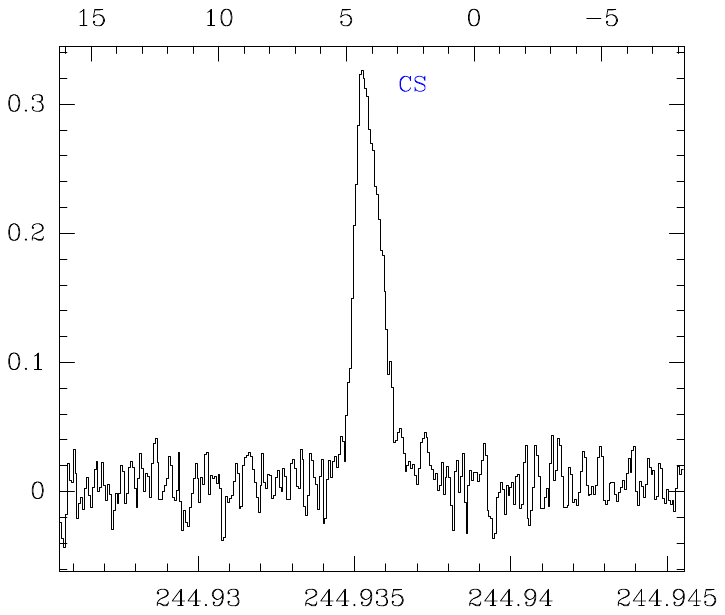}
  \end{minipage}
 \vspace{-1.5cm} 
\caption{Same as Fig.~\ref{fig:survey-219.2ghz-small-apex} but for the 227\,GHz tuning.}
\label{fig:survey-227.2ghz-small-apex}
\end{figure*}

\addtocounter{figure}{-1}
\begin{figure*}[h]
\centering 
\vspace{-4cm}
\hspace{-2cm}
  \begin{minipage}[h]{0.32\textwidth}
    \includegraphics[width=2.5\textwidth]{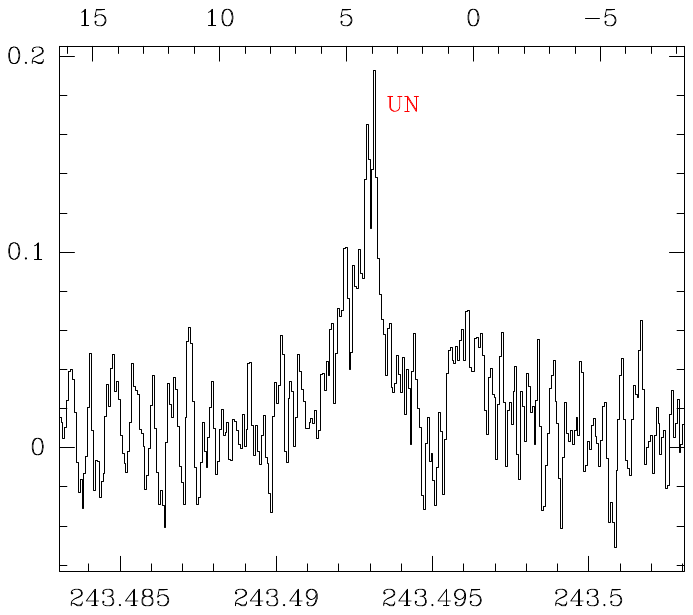}
  \end{minipage}
  \vspace{-1.5cm} 
\caption{Continued.}
\end{figure*}


\newpage
\begin{figure*}[h]
\centering 
\vspace{-4cm}
\hspace{-2cm}
  \begin{minipage}[h]{0.32\textwidth}
    \includegraphics[width=2.5\textwidth]{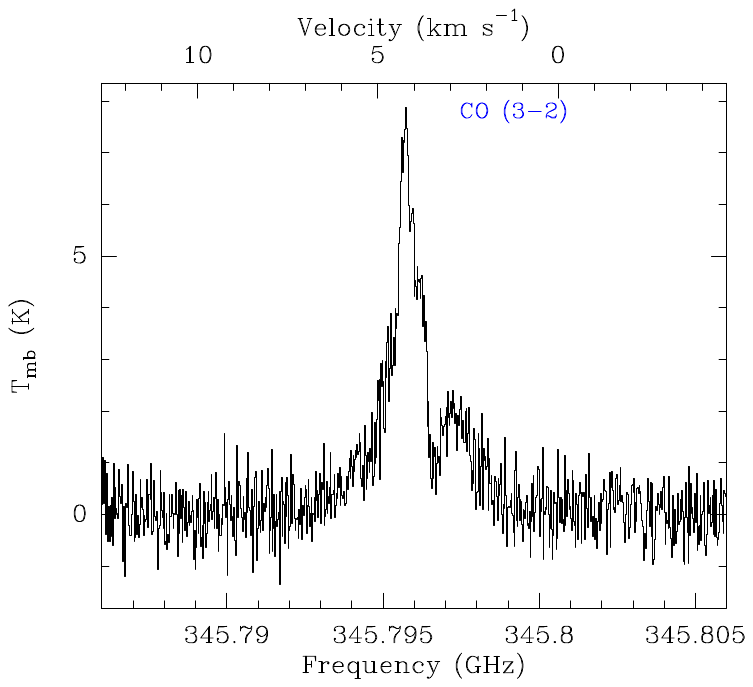}
  \end{minipage}
  \hspace{0.6cm}
  \begin{minipage}[h]{0.32\textwidth}
    \includegraphics[width=2.5\textwidth]{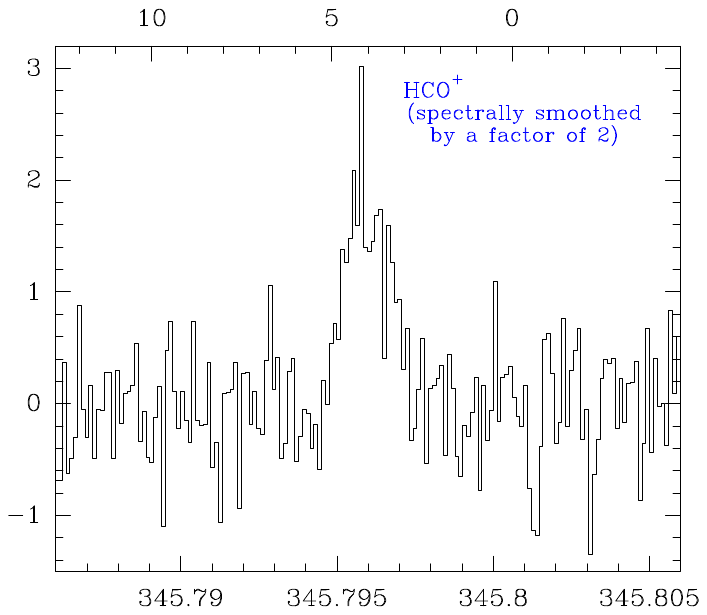}  
  \end{minipage}
 \vspace{-1.5cm} 
\caption{Spectra obtained at the position of V1057~Cyg (i.e.,~0$\arcsec$,0$\arcsec$) using the APEX~12-m telescope with LAsMA in the otf mode, displayed in main-beam temperature scale. The names of identified lines are labelled in blue.}
\label{fig:survey-apex-lasma-2021-11}
\end{figure*}

\begin{figure*}[h]
\centering 
\vspace{-4cm}
\hspace{-2cm}
  \begin{minipage}[h]{0.32\textwidth}
    \includegraphics[width=2.5\textwidth]{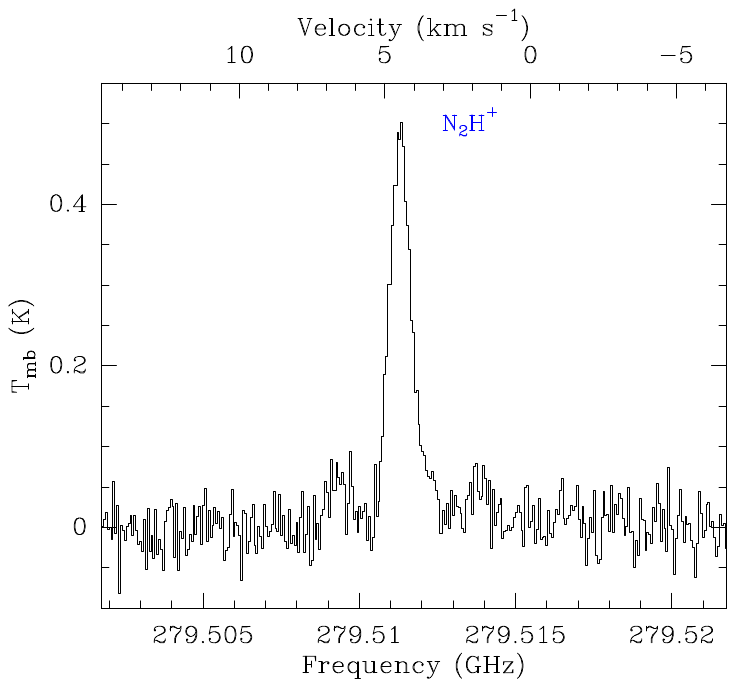}
  \end{minipage}
  \hspace{0.6cm}
  \begin{minipage}[h]{0.32\textwidth}
    \includegraphics[width=2.5\textwidth]{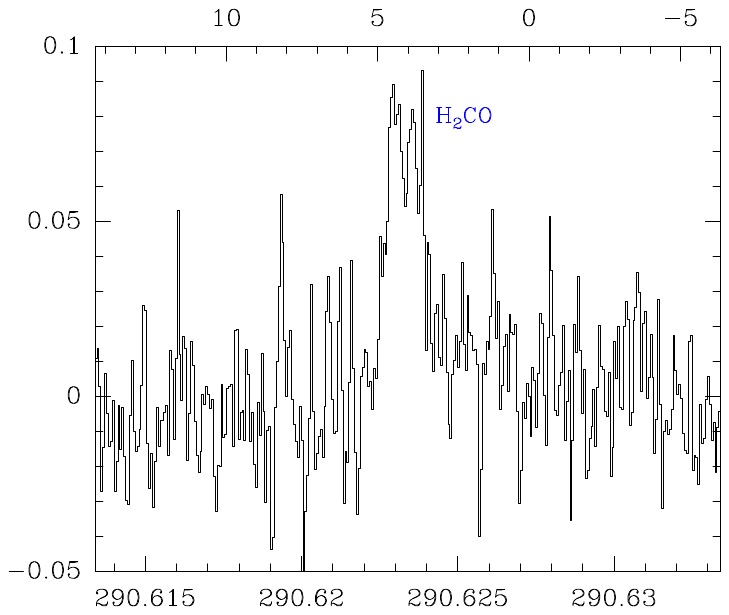}  
  \end{minipage}
  \hspace{0.6cm}
  \begin{minipage}[h]{0.32\textwidth}
    \includegraphics[width=2.5\textwidth]{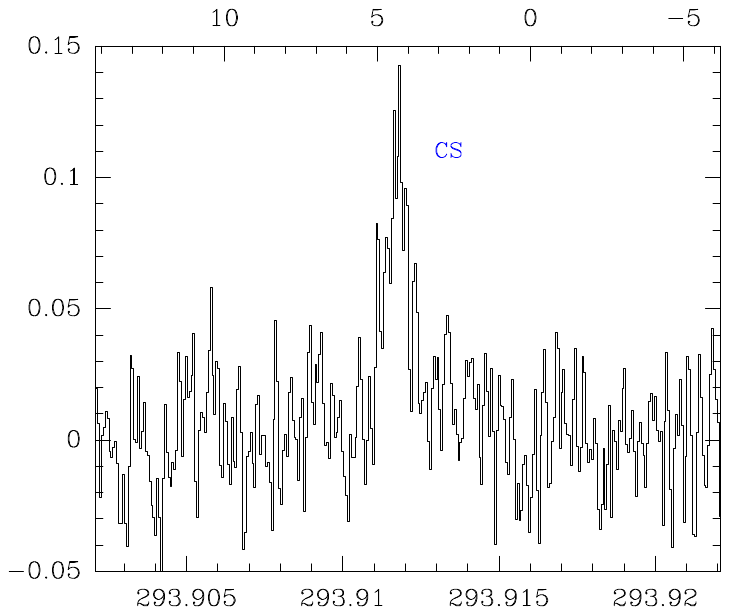}
  \end{minipage} 
 \vspace{-1.5cm} 
 
\caption{Spectra obtained towards V1057~Cyg with the APEX~12-m telescope with SEPIA, displayed in main-beam temperature scale. The names of identified lines are labelled in blue.}
\label{fig:survey-apex-sepia}
\end{figure*}

\end{appendix}

\end{document}